%% file: review_main_rev.tex
\documentclass[physics,article,accept,moreauthors]{Definitions/mdpi}

\newcommand{\bea}{\begin{eqnarray}}
\newcommand{\eea}{\end{eqnarray}}

\def\qq{\langle\bar qq\rangle}

\usepackage{psfrag}
\usepackage{slashed}
\usepackage{float}
\usepackage[caption=false]{subfig}
\usepackage{amsmath,amssymb,amsfonts,amsthm,latexsym}

\usepackage{graphicx}
\usepackage{grffile}
\graphicspath{{figs_sec2/}}
\graphicspath{{figs_sec3/}}
\graphicspath{{figs_sec4/}}
\graphicspath{{figs_sec5/}}

\usepackage{import}

\usepackage{ulem}
\usepackage{psfrag}
\usepackage{slashed}
\usepackage{float}
\usepackage[caption=false]{subfig}
\usepackage{amsmath,amssymb,amsfonts,amsthm,latexsym}
\usepackage{color}

\definecolor{verde}{rgb}{0.0, 0.5, 0.0}
\definecolor{violeta}{rgb}{0.54, 0.17, 0.89}

\firstpage{1}
\pubvolume{xx}
\issuenum{1}
\articlenumber{5}
\pubyear{2021}
\copyrightyear{2021}
\history{ }

\Title{Strong-interaction matter under extreme conditions from chiral quark
models with nonlocal separable interactions}

\Author{D. G\'omez Dumm$^{1,*}$, J.P. Carlomagno$^{1}$ and N.N. Scoccola $^{2,3}$}

\AuthorNames{Firstname Lastname, Firstname Lastname and Firstname Lastname}

\address{
$^{1}$ \quad IFLP, CONICET - Departamento de F\'{\i}sica,
Facultad de Ciencias Exactas, Universidad Nacional de La Plata, C.C. 67, 1900 La Plata, Argentina
\\
$^{2}$ \quad Physics Department, Comisi\'on Nacional de Energ\'{\i}a At\'omica,
Av.Libertador 8250, 1429 Buenos Aires, Argentina
\\
$^{3}$ \quad CONICET, Rivadavia 1917, (1033) Buenos Aires, Argentina
}

\corres{Correspondence: dumm@fisica.unlp.edu.ar }

\abstract{We review the current status of the research on effective
nonlocal NJL-like chiral quark models with separable interactions, focusing
on the application of this approach to the description of the properties of
hadronic and quark matter under extreme conditions. The analysis includes
the predictions for various hadron properties in vacuum, as well as the
study of the features of deconfinement and chiral restoration phase
transitions for systems at finite temperature and/or density. We also
address other related subjects, such as the study of phase transitions for
imaginary chemical potentials, the possible existence of inhomogeneous phase
regions, the presence of color superconductivity, the effects produced by
strong external magnetic fields, and the application to the description of
compact stellar objects.}


\begin{document}

\import{./}{Sec1-rev.tex}

\import{./}{Sec2-rev.tex}

\import{./}{Sec3-rev.tex}

\import{./}{Sec4-rev.tex}

\import{./}{Sec5-rev.tex}

\import{./}{Sec6-rev.tex}

\import{./}{App1-rev.tex}

\reftitle{References}

\import{./}{refs-allsec-rev.tex}
\end{document}

%% file: Sec1-rev.tex

\section{Introduction}
\label{sec1}

The detailed understanding of the behavior of strong-interaction
matter under extreme conditions of temperature and/or density has
attracted great attention in the past decades. This is not only an
issue of fundamental interest but has also important implications
on the description of the early evolution of the
Universe~\cite{Schwarz:2003du} and on the study of the interior of
compact stellar objects~\cite{Page:2006ud,Lattimer:2015nhk}. It is
widely believed that as the temperature $T$ and/or the baryon
chemical potential $\mu_B$ increase, one finds a transition from a
hadronic phase, in which chiral symmetry is broken and quarks are
confined, to a phase in which chiral symmetry is partially
restored and/or quarks are deconfined. In fact, the problems of
how and when these transitions occur have been intensively
investigated, both from theoretical and experimental points of
view. On the experimental side, the properties of
strong-interaction matter are being studied by large research
programs at the Relativistic Heavy Ion Collider (RHIC,
Brookhaven), as well as at the Large Hadron Collider (LHC) and the
Super Proton Synchrotron (SPS) in CERN. Experiments at these
facilities allow for the exploration of the properties of hot and
dense matter created in collisions of ultra-relativistic heavy
ions~\cite{Braun-Munzinger:2015hba,Busza:2018rrf}. The quark-gluon
plasma produced at high energies at RHIC and LHC contains almost
equal amounts of matter and antimatter, and serves to probe the
region of high temperatures and low chemical potentials in the
$\mu_B-T$ phase diagram. In addition, the variation of collision
energies at RHIC through the beam energy scan (BES)
program~\cite{Bzdak:2019pkr} has enabled the systematical
exploration of the phase structure of strong-interaction matter at
nonzero chemical potential. These studies will be complemented in
the next future by experiments at the facilities FAIR in Darmstadt
and NICA in Dubna, reaching in this way experimental access to the
bulk of the phase diagram. It should be stressed that additional
information about the behavior of dense quark matter systems can
be provided by the wealth of data on the properties of compact
stars, obtained through neutron star and gravitational wave
observations from e.g.~the Neutron Star Interior Composition
Explorer (NICER), gravitational-wave observatories such as LIGO or
VIRGO, etc.~\cite{Baym:2017whm}. From the theoretical point of
view, addressing this subject requires to deal with quantum
chromodynamics (QCD) in nonperturbative regimes. One way to cope
with this problem is through lattice QCD (LQCD)
calculations~\cite{Karsch:2001vs,Ding:2015ona}. However, in spite
of the significant improvements produced over the years, this {\it
ab initio} approach is not yet able to provide a full
understanding of the QCD phase diagram, owing to the well known
``sign problem''~\cite{Karsch:2001cy} that affects LQCD
calculations at finite chemical potential. Thus, our present
theoretical understanding of the strong-interaction matter phase
diagram largely relies on the use of effective models of low
energy QCD that show consistency with LQCD results at $\mu_B
\simeq 0$, and can be extrapolated into regions not accessible by
lattice calculation techniques.

One of the most popular approaches to an effective description of QCD
interactions is the quark version of the Nambu$-$Jona-Lasinio (NJL)
model~\cite{Nambu:1961tp,Nambu:1961fr}, in which quark fields interact
through local four-point vertices that satisfy chiral symmetry constraints.
This type of model provides a mechanism for the spontaneous breakdown of
chiral symmetry and the formation of a quark condensate, and has been
widely used to describe the features of chiral restoration at finite
temperature and/or
density~\cite{Vogl:1991qt,Klevansky:1992qe,Hatsuda:1994pi}. Thermodynamic
aspects of confinement, while absent in the original NJL model, can be
implemented by a synthesis with Polyakov-loop (PL)
dynamics~\cite{Polyakov:1978vu,Fukushima:2017csk}. The resulting
Polyakov-Nambu-Jona-Lasinio (PNJL) model
\cite{Meisinger:1995ih,Fukushima:2003fw,Megias:2004hj,Ratti:2005jh,Roessner:2006xn,Mukherjee:2006hq,Sasaki:2006ww}
allows one to study the chiral and deconfinement transitions in a common
framework.

As an improvement over local models, chiral quark models that include
nonlocal separable interactions have also been
considered~\cite{Schmidt:1994di,Burden:1996nh,Bowler:1994ir,Ripka:1997zb}.
Since these approaches can be viewed as nonlocal extensions of the NJL
model, here we denote them generically as ``nonlocal NJL'' (nlNJL) models.
In fact, nonlocal interactions arise naturally in the context of several
successful approaches to low-energy quark
dynamics~\cite{Schafer:1996wv,Roberts:1994dr}, and lead to a momentum
dependence in quark propagators that can be made
consistent~\cite{Noguera:2005ej,Noguera:2008cm} with lattice
results~\cite{Parappilly:2005ei,Furui:2006ks}. Moreover, it can be seen that
nonlocal extensions of the NJL model do not show some of the known
inconveniences that can be found in the local theory. Well-behaved nonlocal
form factors can regularize loop integrals in such a way that anomalies are
preserved~\cite{RuizArriola:1998zi} and charges are properly quantized. In
addition, one can avoid the introduction of various sharp cutoffs to deal
with higher order loop integrals~\cite{Blaschke:1995gr,Plant:2000ty},
improving in this way the predictive power of the models. At the same time,
the separable character of the interactions makes possible to keep much of
the simplicity of the standard NJL model, in comparison with more rigorous
analytical approaches to nonperturbative QCD. Various applications of nlNJL
models to the description of hadron properties at zero temperature and
density can be found e.g.~in
Refs.~\cite{Plant:1997jr,Broniowski:1999dm,Praszalowicz:2001wy,Praszalowicz:2001pi,
Praszalowicz:2003pr,Dorokhov:2003kf,Kotko:2008gy,Kotko:2009ij,Kotko:2009mb,Nam:2012vm,
Nam:2017gzm,GomezDumm:2012qh,Dumm:2013zoa,Golli:1998rf,Broniowski:2001cx,Rezaeian:2004nf}.
In addition, this type of model has been applied to the description of the
chiral restoration transition at finite temperature and/or
density~\cite{Szczerbinska:1999iz,Blaschke:2000gd,General:2000zx,GomezDumm:2001fz,GomezDumm:2004sr}.
The coupling of the quarks to the PL in the framework of nlNJL models gives
rise to the so-called nonlocal Polyakov$-$Nambu$-$Jona-Lasinio (nlPNJL)
models~\cite{Blaschke:2007np,Contrera:2007wu,Hell:2008cc}, in which quarks
move in a background color field and interact through covariant nonlocal
chirally symmetric couplings. It has been found that, under certain
conditions, it is possible to derive the main features of nlPNJL models
starting directly from QCD~\cite{Kondo:2010ts}.

The aim of this article is to present an overview of the existing results
obtained within nonlocal NJL-like models concerning the description of
hadronic and quark matter at both zero and finite temperature and/or
density. The paper is organized as follows. In Sec.~\ref{sec2} we introduce
a two-flavor nlNJL model and discuss its main theoretical features. We also
study the extension of this approach to nonzero temperature and chemical
potential, including the coupling of fermions to the Polyakov loop. Then,
after introducing some possible model parameterizations, results for the
thermodynamics of strong-interaction matter, as well as for the
corresponding phase diagram both for real and imaginary quark chemical
potentials, are presented and discussed. We finish this section by quoting
some results for the thermal dependence of various scalar and pseudoscalar
meson properties. In Sec.~\ref{sec3}, the inclusion of explicit vector and
axial vector interactions in the context of two flavor nlNJL models is
considered, analyzing their impact on the corresponding phenomenological
predictions. In Sec.~\ref{sec4} we analyze the extension of the above
introduced nlPNJL model to $N_f=3$ flavors, incorporating a dynamical
strange quark. This requires the inclusion of explicit flavor symmetry
breaking terms to account for the relatively large $s$-quark mass and also
the addition of a flavor mixing term related to the $U(1)_A$ anomaly. After
describing the main features of this three-flavor nlPNJL model, results for
the thermal dependence of various quantities as well as predictions for the
phase diagram are given. In Sec.~\ref{sec5} we discuss some further
developments and applications of nlPNJL models. These include the
description of superconducting phases, the application to the physics of
compact stars, the study of the existence of inhomogeneous phases and the
analysis of the effects of external strong magnetic fields on hadron
properties and phase transitions. Finally, in Sec.~\ref{sec6} we present our
conclusions. We also add a brief appendix that includes some conventions and
basic formulae.

%% file: Sec2-rev.tex

\section{Two-flavor nonlocal NJL models}
\label{sec2}

In this section we review the features of nlNJL models that include just the
two lightest quark flavors, $u$ and $d$. We begin by addressing the analysis
of model properties in vacuum. Then we consider systems at finite
temperature and chemical potential, discussing the predictions for
thermodynamic properties and the characteristics of phase transitions.

\subsection{Two-flavor nlNJL model at vanishing temperature and chemical potential}
\label{sec2.1}

\subsubsection{Effective action}
\label{sec2.1.1}

We start by studying the features of nlNJL models at zero temperature and
chemical potential. Let us consider the Euclidean action given
by~\cite{Noguera:2008cm}
\begin{equation}
S_{E} \ = \ \int d^{4}x\ \left\{  \bar{\psi}(x)\left(  -i\rlap/\partial
+m_{c}\right)  \psi(x)-
\frac{G_{S}}{2}\Big[j_S(x)j_S(x) + \vec \jmath_P(x)\cdot\vec \jmath_P(x) + j_R(x)j_R(x)\Big]\right\} \ ,
\label{ch2.1.1-eq1}
\end{equation}
where $\psi$ stands for the $u$, $d$ quark field doublet and $m_c$
is the current quark mass. Throughout this article we will work in
the isospin limit, thus we assume the current mass to be the same
for $u$ and $d$ quarks. The nonlocal character of the model arises
from the quark-antiquark currents $j_{S}(x)$, $j_{P}(x)$ and
$j_{R}(x)$, given by
\begin{eqnarray}
j_S(x)   &=& \int d^4z\; \mathcal{G}(z)\,
\bar\psi\left(x+\frac{z}{2}\right) \psi\left(x-\frac{z}{2}\right)
\ , \nonumber\\
\vec \jmath_P(x)  &=& \int d^4z\; \mathcal{G}(z)\,
\bar\psi\left(x+\frac{z}{2}\right) i\, \gamma_5\, \vec \tau \,
\psi\left(x-\frac{z}{2}\right)
\ , \nonumber\\
j_R(x)       &  =& \int d^{4}z\ \mathcal{F}(z)\ \bar{\psi}\left(
x+\frac{z}{2}\right) \
\frac{i{\overleftrightarrow{\rlap/\partial}}}{2\ \varkappa_{p}}\
\psi\left(
x-\frac{z}{2}\right)\ ,
\label{ch2.1.1-eq2}
\end{eqnarray}
where $u(x^{\prime}){\overleftrightarrow{\partial}}v(x) =
u(x^{\prime})\partial_{x}v(x)-\partial_{x^{\prime}}u(x^{\prime})v(x)$. The
functions $\mathcal{G}(z)$ and $\mathcal{F}(z)$ in Eq.~(\ref{ch2.1.1-eq2})
are effective covariant form factors that encode the effects of the
underlying low energy QCD interactions. Chiral invariance requires that the
quark currents $j_{S}(x)$ and $j_{P\,a}(x)$, $a=1,2,3$, carry the same form
factor $\mathcal{G}(z)$, while the $j_{R}(x)j_{R}(x)$ coupling is found to
be self-invariant under chiral transformations. The local version of the
(quark-level) NJL model is obtained  by taking $\mathcal{G}(z) =
\delta^{(4)}(z)$ and $\mathcal{F}(z) = 0$, together with a proper
regularization prescription.

It can be shown that the presence of the nonlocal form factor in the scalar
current $j_{S}(x)$ leads to a momentum-dependent quark effective mass, as
expected from lattice QCD calculations~\cite{Parappilly:2005ei}. On the
other hand, the coupling involving the current $j_{R}(x)$ is related to the
quark wave function renormalization (WFR). Whereas in a local theory
this coupling would simply lead
---at the mean field level--- to a redefinition of fermion fields, in
the nonlocal scheme it leads to a momentum-dependent  WFR of the
quark propagator, in consistency with LQCD
analyses~\cite{Parappilly:2005ei}. To simplify the notation, the
same coupling constant $G_{S}$ has been taken for all interaction
terms. Note, however, that the relative strength between the
$j_{R}(x)j_{R}(x)$ coupling and the other terms is controlled by
the mass parameter $\varkappa_{p}$ in Eq.~(\ref{ch2.1.1-eq2}). The
form factors ${\cal G}(z)$ and ${\cal F}(z)$ are dimensionless and
can be normalized to ${\cal G}(0) = {\cal F}(0) = 1$ without loss
of generality.

It is worth stressing that the currents in Eq.~(\ref{ch2.1.1-eq2}) are not
to be directly identified with the color octet quark currents entering the
QCD Lagrangian. The former are color singlet quantities that contain part of
the effective gluon mediated interaction, and are introduced as a convenient
way of expressing the interaction terms in a chiral quark model. In
addition, it is important to notice that the couplings in
Eq.~(\ref{ch2.1.1-eq1}) are not intended to be the most general nonlocal
current-current interactions compatible with QCD symmetries. In fact, this
is just the simplest model that, within the mean field approximation, leads
to an effective quark propagator including a momentum-dependent wave
function renormalization $Z(p)$ and a momentum-dependent mass $M(p)$.
Clearly, other quark-antiquark and/or quark-quark current interactions can
be added in order to properly account for the description of vector mesons
physics, color superconductivity effects, etc. Some possible extensions of
the model will be discussed in the next sections of this article. In
addition, it is worth taking into account that in Eqs.~(\ref{ch2.1.1-eq2})
we have chosen a particular way of introducing nonlocal form factors in the
quark currents. As discussed in Refs.~\cite{Schmidt:1994di,Burden:1996nh},
this scheme is based on a separable approach to the effective  one gluon
exchange (OGE) interactions. Alternatively, a scheme based on the features
of the instanton liquid model (ILM) has been introduced in
Ref.~\cite{Bowler:1994ir}. In that approach a nonlocal form factor is
associated to each quark field, in such a way that e.g.\ the scalar nonlocal
current $j_{S}(x)$ reads
\begin{equation}
j_{S}^{\mathrm{ILM}} (x)  \  = \
\int d^{4}y \, d^{4}z\ \mathcal{R}(y-x)\, \mathcal{R}(x-z)\ \bar{\psi}\left(y\right)
\psi\left(z\right)\ .
\label{ch2.1.1-eq3}
\end{equation}
Whereas both OGE and ILM-inspired schemes are equivalent at the mean- field
level, the treatment of fluctuations is somewhat different. A comparison of
both approaches can be found in Ref.~\cite{GomezDumm:2006vz}. In this review
we mostly concentrate on the OGE-based scheme, which is the more widely used
to study the behavior of quark and hadronic matter at finite temperature
and/or density. However, when possible, references to the results obtained
within the alternative ILM-based scheme will be included.

In order to deal with meson degrees of freedom, it is convenient to perform
a standard bosonization of the theory given by the action in
Eq.~(\ref{ch2.1.1-eq1}). This can be done by considering the corresponding
partition function $\mathcal{Z}=\int{\rm D}\bar{\psi}\,{\rm D}%
\psi\,\exp(-S_{E})$, and introducing auxiliary fields $\sigma_1(x)$,
$\sigma_2(x)$ and $\pi_a(x)$, $a =1,2,3$, which account for scalar and
pseudoscalar mesons. Integrating out the quark fields one gets
\begin{equation}
\mathcal{Z}\ =\ \int {\rm D}\sigma_{1}\,{\rm D}\sigma_{2}\ {\rm D}%
\vec{\pi}\ \exp\!\big(\!-S_{E}^{\mathrm{bos}}\big)\ ,
\label{ch2.1.1-eq4}
\end{equation}
where the bosonized action is given by
\begin{equation}
S_{E}^{\mathrm{bos}} \ = \ -\ln\,\det {\cal D} +\frac{1}{2G_{S}}\int\frac{d^{4}q}{(2\pi )^{4}}\
\Big[\sigma_{1}(q)\ \sigma_{1}(-q)+\vec{\pi}(q)\cdot\vec{\pi
}(-q)+\sigma_{2}(q)\ \sigma_{2}(-q)\Big]\ .
\label{ch2.1.1-eq5}
\end{equation}
Here, the operator ${\cal D}$ reads, in momentum space,
\begin{eqnarray}
{\cal D}(p,p^{\prime})  &  =  & (\rlap/p+m_{c})\,(2\pi)^{4}\,\delta
^{(4)}(p-p^{\prime})+g\left(  \frac{p+p^{\prime}}{2}\right)  \ \Big[
\sigma_{1}(p-p^\prime)+i\gamma_{5}\vec{\tau}\cdot\vec{\pi}(p-p^\prime)\Big] \nonumber\\
& & -\ f\left(  \frac{p+p^{\prime}}%
{2}\right)  \ \frac{\rlap/p+{\rlap/p}^{\prime}}{2\ \varkappa_{p}}\
\sigma _{2}(p-p^\prime) \ ,
\label{ch2.1.1-eq6}
\end{eqnarray}
where $g(p)$ and $f(p)$ stand for Fourier transforms of the form factors
$\mathcal{G}(z)$ and $\mathcal{F}(z)$, namely
\begin{equation}
g(p) \ = \ \int d^4z \ e^{-i p\cdot z} \ \mathcal{G}(z)\ , \qquad \qquad
f(p) \ = \ \int d^4z \ e^{-i p\cdot z} \ \mathcal{F}(z)\ .
\label{ch2.1.1-eq6b}
\end{equation}
Note that Lorentz invariance implies that they can only be functions of
$p^{2}$.

Now it is assumed that the fields $\sigma_1$ and $\sigma_2$ have nontrivial
translational invariant mean field values $\bar\sigma_1$ and
$\varkappa_p\,\bar \sigma_2$, while the mean field values of the
pseudoscalar fields $\pi_{i}$ are taken to be zero. Thus, one can write
\begin{eqnarray}
\sigma_{1}(x) \ = \ \bar{\sigma}_{1}+\delta\sigma_{1}(x) \ , \qquad
\sigma_{2}(x) \ = \ \varkappa_{p}\ \bar{\sigma}_{2}+\delta\sigma_{2}(x)\ ,
\qquad \vec{\pi}(x) \ = \ \delta\vec{\pi}(x) \ .
\label{ch2.1.1-eq7}
\end{eqnarray}
A more general case, in which the bosonic fields are expanded around
inhomogeneous ground states, is considered in Sec.~\ref{sec5.3}. Replacing
in the bosonized effective action, and expanding in powers of the meson
fluctuations $\delta\sigma_1$, $\delta\sigma_2$ and $\delta\vec\pi$, one
gets
\begin{eqnarray}
S_{E}^{\mathrm{bos}}\ =\
S_{E}^{\mathrm{MFA}}\, + \, S_{E}^{\mathrm{quad}}\, +\ \dots
\label{ch2.1.1-eq8}
\end{eqnarray}
The mean field action per unit volume reads~\cite{Noguera:2008cm}
\begin{equation}
\frac{S_{E}^{\mathrm{MFA}}}{V^{(4)}}\ = \ -\, 2N_{c}\int\frac{d^{4}p}{(2\pi)^{4}%
}\ \mbox{tr}\, \ln \mathcal{D}_{0}(p) \,
+ \, \frac{\bar{\sigma }_{1}^{2}}{2G_{S}} \, + \, \frac{\varkappa_{p}^{2}\
\bar{\sigma}_{2}^{2}}{2G_{S}}\ ,
\label{ch2.1.1-eq9}
\end{equation}
where $N_c=3$ is the number of colors, and the trace is taken on Dirac space.
The mean field quark propagator $\mathcal{S}_{0}(p)$ is given by
\begin{equation}
\mathcal{S}_{0}(p)\ =\  \mathcal{D}_{0}(p)^{-1} \ = \ \frac{Z(p)}{\rlap/p+M(p)}\ ,
\label{ch2.1.1-eq10}
\end{equation}
with
\begin{eqnarray}
Z(p) \ = \ \left(  1-\bar{\sigma}_{2}\ f(p)\right)  ^{-1} \ , \qquad\qquad
M(p) \ = \ Z(p)\left(  m_{c}+\bar{\sigma}_{1}\ g(p)\right)\ .
\label{ch2.1.1-eq11}
\end{eqnarray}
The quadratic piece in  Eq.~(\ref{ch2.1.1-eq8}) can be written as
\begin{equation}
S_{E}^{\mathrm{quad}} \ = \ \frac{1}{2}\int\frac{d^{4}q}{(2\pi)^{4}}
\left[
G_{\sigma}(q^{2})\ \delta\sigma(q)\ \delta\sigma(-q) \, + \,
G_{\sigma^{\prime}}(q^{2})\ \delta\sigma^{\prime}(q)\ \delta\sigma^{\prime}(-q) \,
+ \, G_{\pi}(q^{2})\ \delta\vec{\pi}(q)\cdot\delta\vec{\pi}(-q)\right]  \ ,
\label{ch2.1.1-eq12}
\end{equation}
where $\delta\sigma$ and $\delta\sigma^{\prime}$ are related to
$\delta\sigma_{1}$ and $\delta\sigma_{2}$ by
\begin{eqnarray}
\delta\sigma & = & \cos\theta\ \delta\sigma_{1}-\sin\theta\ \delta\sigma_{2}
\ , \nonumber \\
\delta\sigma^{\prime} & = & \sin\theta\ \delta\sigma_{1}+\cos\theta \ \delta\sigma_{2}\ ,
\label{ch2.1.1-eq13}
\end{eqnarray}
the mixing angle $\theta$ being defined in such a way that there is no
$\sigma-\sigma^{\prime}$ mixing at the level of the quadratic
action. The function $G_{\pi}(q^{2})$ introduced in
Eq.~(\ref{ch2.1.1-eq12}) is given by~\cite{Noguera:2008cm}
\begin{equation}
G_{\pi}(q^{2})\ =\ \frac{1}{G_{S}}-\,8\, N_{c}\int\frac{d^{4}p}{(2\pi)^{4}}%
\ g(p)^2\,\frac{Z(p^{+})Z(p^{-})}{D(p^{+})D(p^{-})}\,
\left[p^{+}\cdot p^{-}+M(p^{+})M(p^{-})\right]\ ,
\label{ch2.1.1-eq15}
\end{equation}
with $p^{\pm}=p\pm q/2\,$ and $D(p)=p^{2}+M(p)^{2}$. For the
$\sigma-\sigma^{\prime}$ system one has
\begin{equation}
G_{\sigma \choose {\sigma'}}(q^{2})\ =\ \frac{G_{\sigma_{1}\sigma_{1}}(q^{2})+G_{\sigma_{2}%
\sigma_{2}}(q^{2})}{2}\mp\sqrt{\left[
G_{\sigma_{1}\sigma_{2}}(q^{2})\right] ^{2}\!+\!\left[
\frac{G_{\sigma_{1}\sigma_{1}}(q^{2})-G_{\sigma_{2}\sigma
_{2}}(q^{2})}{2}\right]^{2}} \ \ ,
\label{ch2.1.1-eq16}
\end{equation}
where
\begin{eqnarray}
G_{\sigma_{1}\sigma_{1}}(q^{2})  &=& \frac{1}{G_{S}}-\,8\,
N_{c}\int \frac{d^{4}p}{(2\pi)^{4}}\ g(p)^2\, \frac{Z(p^{+})Z(p^{-})}{D(p^{+})D(p^{-}%
)}\left[  p^{+}\cdot p^{-}-M(p^{+})M(p^{-})\right]\ ,
\nonumber\\
G_{\sigma_{2}\sigma_{2}}(q^{2})  &=& \frac{1}{G_{S}}+\,\frac{8\, N_{c}%
}{\varkappa_{p}^{2}}\ \int\frac{d^{4}p}{(2\pi)^{4}}\
p^{2}\,f(p)^2\, \frac
{Z(p^{+})Z(p^{-})}{D(p^{+})D(p^{-})}\nonumber \\
& & \times \,\Big[  p^{+}\cdot p^{-} - M(p^{+})M(p^{-}) + \frac{{p^{+}}^{\,2}\,
{p^{-}}^{\,2}-(p^{+}\cdot p^{-})^{2}}{2\,p^{2}}\Big]\ ,
\nonumber\\
G_{\sigma_{1}\sigma_{2}}(q^{2})  &=& -\frac{8\, N_{c}}{\varkappa_{p}}%
\ \int\frac{d^{4}p}{(2\pi)^{4}}\ g(p)\, f(p)\,\frac{Z(p^{+})Z(p^{-})}%
{D(p^{+})D(p^{-})}\ p\cdot\left[
p^{-}M(q^{+})+p^{+}M(q^{-})\right]\ .
\label{ch2.1.1-eq17}
\end{eqnarray}

\subsubsection{Mean field approximation and chiral condensates}
\label{sec2.1.2}

The mean field values $\bar{\sigma}_1$ and $\bar\sigma_2$ can be determined
by minimizing the mean field action $S_{E}^{\mathrm{MFA}}$. This leads to
the coupled ``gap equations''
\begin{eqnarray}
\bar{\sigma}_{1}\, - \, 8N_{c}\ G_{S}\int\frac{d^{4}p}{(2\pi)^{4}}\ g(p)\ \frac {Z(p)M(p)}{D(p)}
& =& 0\ ,
\label{ch2.1.2-eq1a}
\\
\bar{\sigma}_{2}\, + \, 8N_{c}\ G_{S}\int\frac{d^{4}p}{(2\pi)^{4}}\
\frac{p^{2}}{\varkappa_{p}^{2}}\ f(p)\ \frac{Z(p)}{D(p)}  &=& 0 \ .
\label{ch2.1.2-eq1b}
\end{eqnarray}
It is worth noticing that the loop integrals in these equations will be
convergent, as long as the nonlocal form factors are well behaved in the
ultraviolet limit. Thus, there is no need to introduce sharp momentum
cutoffs in this kind of model.

The chiral condensates, given by the vacuum expectation values $\langle\bar
u u\rangle$ and $\langle\bar d d\rangle$, can be easily obtained by
performing the variation of
$\mathcal{Z}^{\mathrm{MFA}}=\exp(-S_{E}^{\mathrm{MFA}})$ with respect to the
$u$ and $d$ current quark masses. Up and down quark condensates will be
equal, owing to isospin symmetry. The corresponding analytical expressions
are finite in the chiral limit, while they turn out to be ultraviolet
divergent for nonzero values of current quark masses. Such a divergence can
be regularized through the subtraction of the perturbative vacuum
contribution. One has in this way
\begin{equation}
\langle\,\bar{q}q\,\rangle \ = \ -\,4N_{c}\int\frac{d^{4}p}{(2\pi)^{4}}\ \left(
\frac{Z(p)M(p)}{D(p)}-\frac{m_{c}}{p^{2}+m_{c}^{2}}\right)  \ , \qquad q =
u,\ d\ .
\label{ch2.1.2-eq2}
\end{equation}

\subsubsection{Meson properties}
\label{sec2.1.3}

Taking into account the quadratic piece of the Euclidean action in
Eq.~(\ref{ch2.1.1-eq12}), it is seen that the scalar and pseudoscalar meson
masses can be obtained by solving the equations
\begin{eqnarray}
G_{M}(-m_{M}^{2})\ =\ 0\ ,
\label{ch2.1.3-eq1}
\end{eqnarray}
where $M = \sigma$, $\sigma'$, $\pi$. The on-shell quark-meson coupling
constants $g_{Mq\bar{q}}$ and the associated meson renormalization constants
$Z_M$ will be given by
\begin{equation}
Z_M^{-1}\ = \ g_{Mq\bar{q}}^{-2} \ = \ \frac{dG_{M}%
(q^2)}{dq^{2}}\bigg|_{q^{2}=-m_{M}^{2}}\ .
\label{ch2.1.3-eq2}
\end{equation}

Let us consider the pion weak decay constant $f_{\pi}$, which is given by
the matrix element of the axial vector quark current between the vacuum and
a one-pion state,
\begin{equation}
\langle0|\bar \psi(x)\,
\gamma_\mu\gamma_5\frac{\tau_a}{2}\,\psi(x)|\tilde{\pi}_b(q)\rangle \ =\
-\, i\,f_{\pi}\,e^{\,iq\cdot x}\,
\delta_{ab}\, q_\mu\ ,
\label{ch2.1.3-eq3}
\end{equation}
at the pion pole $q^2 = - m_\pi^2$. Here the pion field has been
renormalized as $\tilde\pi_a = Z_\pi^{-1/2}\, \delta\pi_a\,$. In order to
obtain an explicit expression for the above matrix element one has to
``gauge'' the effective action $S_{E}$ by introducing a set of axial gauge
fields $\mathcal{W}_{a\,\mu}(x)$. For a local theory this gauging
procedure is usually done by performing the replacement
\begin{equation}
\partial_{\mu}\ \rightarrow\ \partial_{\mu}-\frac{i}{2}\, \gamma_{5}\, \vec{\tau
}\cdot{\vec{\mathcal{W}}}_{\mu}(x)\ .
\label{ch2.1.3-eq4}
\end{equation}
In the present case ---owing to the nonlocality of the involved
fields--- one has to perform additional replacements in the
interaction terms~\cite{Ripka:1997zb,Broniowski:1999bz}, namely
\begin{eqnarray}
\psi(x-z/2)\ &\rightarrow&\ W_{A}\left(  x,x-z/2\right)  \ \psi (x-z/2)\ , \nonumber
\\
\psi^{\dagger}(x+z/2)\ &\rightarrow&\ \psi^{\dagger}(x+z/2)\ W_{A}\left(
x+z/2,x\right)\ .
\label{ch2.1.3-eq5}
\end{eqnarray}
Here $x$ and $z$ are the variables appearing in the definitions of the
nonlocal currents given in Eq.~(\ref{ch2.1.1-eq2}), and the function
$W_{A}(x,y)$ is defined by
\begin{equation}
W_{A}(x,y)\ =\ \mathrm{P}\;\exp\left[ - \frac{i}{2}\int_{x}^{y}ds_{\mu}%
\ \gamma_{5}\ \vec{\tau}\cdot{\vec{\mathcal{W}}}_{\mu}(s)\right]
\ ,
\label{ch2.1.3-eq6}
\end{equation}
where $s$ runs over an arbitrary path connecting $x$ with $y$, and $P$
denotes path ordering.

Once the gauged effective action is built, the matrix elements in
Eq.~(\ref{ch2.1.3-eq3}) can be obtained by taking the derivative of this
action with respect to the renormalized meson fields and the axial gauge
fields, evaluated at ${\vec{\mathcal{W}}}_{\mu}=0$, $\vec{\tilde\pi} = 0$.
After a rather lengthy calculation one gets~\cite{Noguera:2008cm}
\begin{equation}
f_{\pi} \ = \ \frac{g_{\pi q\bar{q}}}{m_{\pi}^{2}}\;
m_{c}\; F_{0}(-m_{\pi}^{2})\ ,
\label{ch2.1.3-eq7}
\end{equation}
where
\begin{equation}
F_{0}(q^{2}) \ = \ 8\, N_{c}\int\frac{d^{4}p}{(2\pi)^{4}}\ g(p)\;\frac
{Z(p^{+})Z(p^{-})}{D(p^{+})D(p^{-})}\ \left[  p^{+}\cdot p^{-}+M(p^{+}%
)M(p^{-})\right]\ .
\label{ch2.1.3-eq8}
\end{equation}
It is important to notice that in this calculation the integration along the
above mentioned arbitrary path turns out to be trivial; hence, the result
becomes path-independent. To leading order in the chiral expansion, it can
be seen~\cite{Noguera:2005ej,GomezDumm:2006vz} that
Eqs.~(\ref{ch2.1.1-eq15}), (\ref{ch2.1.2-eq1a}), (\ref{ch2.1.2-eq2}),
(\ref{ch2.1.3-eq1}), (\ref{ch2.1.3-eq2}) and (\ref{ch2.1.3-eq7}) lead to the
quark-level Goldberger-Treiman and Gell-Mann-Oakes-Renner relations
\begin{equation}
f_{\pi}\; g_{\pi q\bar q} \ = \ \frac{M(0)}{Z(0)}
\end{equation}
and
\begin{equation}
f_{\pi}^2\, m_\pi^2 \ = \ -\, m_c \, \langle\bar u u + \bar d d\rangle\ .
\label{ch2.1.3-eq9}
\end{equation}

It is worth mentioning that, besides the main properties discussed above,
many other features of mesons have been studied in the framework of both
OGE- and ILM-based nonlocal approaches, see
Refs.~\cite{Plant:1997jr,Broniowski:1999dm,Praszalowicz:2001wy,
Praszalowicz:2001pi,Praszalowicz:2003pr,Dorokhov:2003kf,
Kotko:2008gy,Kotko:2009ij,Kotko:2009mb,Nam:2012vm,Nam:2017gzm,GomezDumm:2012qh,Dumm:2013zoa}.
In addition, the possible description of baryons has been considered in
Refs.~\cite{Golli:1998rf,Broniowski:2001cx,Rezaeian:2004nf}.

\subsection{Extension to finite temperature and chemical potential. Inclusion of the Polyakov loop}
\label{sec2.2}


Having described the main features of the nlNJL approach in vacuum, we turn
to discuss how to extend the model in order to study the behavior of
strong-interaction matter at nonzero temperature $T$ and/or  quark chemical
potential $\mu = \mu_B/3$. The temperature can be introduced in a standard
way by using the imaginary time Matsubara formalism. This is done by
performing the integration over Euclidean time appearing in the effective
action  on a restricted finite interval $0 \leq x_4 \leq 1/T$, and imposing
proper boundary conditions on the
fields~\cite{Kapusta:2006pm,Bellac:2011kqa}. Alternatively, the real-time
formalism can be applied~\cite{Loewe:2011qc,Loewe:2013zaa}. On the other
hand, the chemical potential can be introduced through the replacement
$\partial_4 \rightarrow
\partial_4 - \mu$ in the effective action. In the case of the nonlocal
models under consideration, to obtain the appropriate conserved currents
this replacement has to be complemented with a modification of the nonlocal
currents in Eq.~(\ref{ch2.1.1-eq2})~\cite{Carter:1998ji,GomezDumm:2001fz}.
In practice, with the conventions used in this article (see Appendix), this
implies that the momentum dependence of the form factors $g(p)$ and $f(p)$
has to be modified by changing $p_4 \to p_4 + i \mu$.

The nonlocal NJL approach described so far is found to provide a basic
understanding for the mechanisms governing both the spontaneous breakdown of
chiral symmetry and the dynamical generation of massive quasiparticles from
almost massless current quarks, in close contact with QCD. However, it does
not  account for some important features expected from the underlying QCD
interactions. In particular, the model predicts the existence of colored
quasiparticles in regions of $T$ and $\mu$ where they should be suppressed
by confinement. A quite successful way to deal with this problem, originally
proposed in the context of the local NJL model, is to include a coupling of
the quark fields to the Polyakov loop
(PL)~\cite{Meisinger:1995ih,Fukushima:2003fw,Megias:2004hj,Ratti:2005jh,Roessner:2006xn,Mukherjee:2006hq,Sasaki:2006ww},
which can be taken as an order parameter for the confinement/deconfinement
transition. Indeed, in pure gauge QCD the traced Polyakov loop $\Phi$ can be
associated with the spontaneous breaking of the global Z$_3$ center symmetry
of color SU(3). The value $\Phi =0$ corresponds to the symmetric, confined
phase, while one has $\Phi = 1$ in the limit where quark asymptotic freedom
has been achieved~\cite{Pisarski:2000eq}. Although the traced PL strictly
serves as an order parameter for the confinement-deconfinement phase
transition only in pure gauge QCD, it is still useful as an indicator for a
rapid crossover transition even in the presence of quarks. The incorporation
of the couplings between dynamical quarks and the PL promotes the nlNJL
model to the nlPNJL
model~\cite{Blaschke:2007np,Contrera:2007wu,Hell:2008cc}.

In the nlPNJL approach the quarks are assumed to move in a constant
background field $\phi =  \textsl{g} \, G^a_{4} \lambda_a/2$, where
$\lambda_a$ ($a=1,\dots,8$) are the Gell-Mann matrices and $G^a_{\mu}$
denotes the SU(3) color gauge fields. Then the traced Polyakov loop is given
by $\Phi=\frac{1}{3} {\rm Tr}\, \exp( i\phi/T)$. It is usual to work in the
so-called Polyakov gauge, in which the matrix $\phi$ is given a diagonal
representation $\phi = {\rm diag}(\phi_r,\phi_g,\phi_b) = \phi_3 \lambda_3 +
\phi_8 \lambda_8$ (here, $r$, $g$ and $b$ refer to red, green and blue
colors). This leaves only two independent variables, $\phi_3$ and $\phi_8$,
in terms of which one has
\begin{equation}
  \Phi \ = \ \frac{1}{3}\left[
 \, 2\,\exp\left(\frac{i}{\sqrt{3}}\frac{\phi_8}{T}\right)\cos\left(\frac{\phi_3}{T}\right)
+\, \exp\left(-\frac{2i}{\sqrt{3}}\frac{\phi_8}{T}\right)\right]\ .
\label{ch2.2-eq1}
\end{equation}

The effective gauge field self-interactions can be incorporated by
introducing a mean field Polyakov loop potential $U(\Phi,\Phi^\ast,T)$. We
consider for this potential two alternative functional forms that are
commonly used in the literature. The first one, based on a Ginzburg-Landau
ansatz, is a polynomial function given
by~\cite{Pisarski:2000eq,Ratti:2005jh}
\begin{equation}
{\cal{U}}_{\rm poly}(\Phi ,\Phi^\ast,T) \ = \
T^4
\left[
-\,\frac{b_2(T)}{2}\, \Phi^\ast \Phi\,
-\,\frac{b_3}{6}\,
\Big( \Phi^3 + {\Phi^{\ast\,}}^3 \Big)
+\,\frac{b_4}{4}\,
\big( \Phi^\ast \Phi \big)^2
\right]\ ,
\label{ch2.2-eq2}
\end{equation}
where
\begin{equation}
b_2(T) \ = \ a_0 +a_1 \left(\dfrac{T_0}{T}\right) +
a_2\left(\dfrac{T_0}{T}\right)^2 +
a_3\left(\dfrac{T_0}{T}\right)^3\ .
\label{ch2.2-eq3}
\end{equation}
The parameters in these expressions can be fitted to pure gauge lattice QCD
data so as to properly reproduce the corresponding equation of state and
 PL potential. This yields~\cite{Ratti:2005jh}
\begin{eqnarray}
& & a_0 = 6.75\ ,\qquad a_1 = -1.95\ ,\qquad a_2 = 2.625\ , \nonumber \\
& & a_3 = -7.44 \ ,\qquad b_3 = 0.75 \ ,\qquad b_4 = 7.5 \ .
\label{ch2.2-eq4}
\end{eqnarray}
A second usual form is based on the logarithmic expression of the Haar
measure associated with the SU(3) color group  path integral.
The potential reads in this case~\cite{Roessner:2006xn}
\begin{equation}
{\cal{U}}_{\rm log}(\Phi ,\Phi^*,T) \ = \ T^4\,
\left\{-\,\frac{1}{2}\, a(T)\,\Phi^* \Phi \;+
\;b(T)\, \ln\left[1 - 6\, \Phi^\ast \Phi\, + 4\, \Big(\Phi^3 +
{\Phi^{\ast\,}}^3\Big) - 3\, \big(\Phi^\ast \Phi\big)^2\right]\right\} \ ,
\label{ch2.2-eq5}
\end{equation}
where the coefficients $a(T)$ and $b(T)$ are parameterized as
\begin{equation}
a(T) \ = \ a_0 +a_1 \left(\dfrac{T_0}{T}\right) + a_2\left(\dfrac{T_0}{T}\right)^2
\ ,
\qquad
b(T) \ = \ b_3\left(\dfrac{T_0}{T}\right)^3 \ .
\label{ch2.2-eq6}
\end{equation}
Once again the values of the constants can be fitted to pure gauge lattice
QCD results. This leads to~\cite{Roessner:2006xn}
\begin{equation}
a_0 = 3.51\ ,\qquad a_1 = -2.47\ ,\qquad a_2 = 15.2\ ,\qquad b_3 = -1.75\ .
\label{ch2.2-eq7}
\end{equation}
The dimensionful parameter $T_0$ in Eqs.~(\ref{ch2.2-eq3}) and
(\ref{ch2.2-eq6}) corresponds in principle to the deconfinement
transition temperature in the pure Yang-Mills theory, $T_0 =
270$~MeV. However, it has been
argued~\cite{Schaefer:2007pw,Schaefer:2009ui} that in the presence
of light dynamical quarks this temperature scale has to be
modified. Effects of this change are discussed below.

The coupling between the quarks and the background gauge field $\phi$ is
implemented through a standard gauge covariant derivative. In the nlPNJL
model this has to be supplemented by the modification of the nonlocal
currents, as discussed in Sec.~\ref{sec2.1.3} in relation to weak gauge
field interactions. Thus, at the mean field level, the quark contribution to
the grand canonical thermodynamic potential including the coupling
to the PL can be obtained from the  mean field Euclidean action in
Eq.~(\ref{ch2.1.1-eq9}) by making the replacements
\begin{eqnarray}
p^2 & \rightarrow & (\rho_{n,\vec{p}}^c)^2=\vec{p}^{\,2} + \left[(2n+1)\pi T + i
\mu  - \phi_c\right]^2
\label{ch2.2-eq8a}
\end{eqnarray}
and
\begin{eqnarray}
N_c \int\frac{d^4p}{(2\pi)^4}\; F(p)& \rightarrow & T
\sum_{n=-\infty}^{\infty}\ \sum_{c=r,g,b}\
\int\frac{d^3p}{(2\pi)^3} \; F(\rho^c_{n,\vec{p}})\ ,
\label{ch2.2-eq8}
\end{eqnarray}
where $n$ labels the Matsubara frequencies (see Appendix).

Before quoting the explicit form of the  mean field thermodynamic potential,
let us discuss some restrictions concerning $\phi_8$. In the case of
vanishing chemical potential the mean field traced Polyakov loop is expected
to be a real quantity, owing to the charge conjugation properties of the QCD
lagrangian. Since $\phi_3$ and $\phi_8$ are real valued, this condition
implies $\phi_8=0$. In principle, this does not need to be valid for a
finite chemical potential; hence, in that case the Polyakov loop could lead
formally to a complex-valued action.  Now, even for a complex Euclidean
action, one can search for the configuration with the largest weight in the
path integral, which can be referred to as the mean field
configuration~\cite{Roessner:2006xn}. One way to establish such a lowest
order approximation is to use the real part of the thermodynamic potential
to obtain the mean field ``gap equations''. Demanding the thermal
expectation values $\langle \Phi \rangle$ and $\langle \Phi^\ast \rangle$ to
be real quantities~\cite{Dumitru:2005ng,Roessner:2006xn}, this means $ \Phi
= \Phi^\ast$ for the mean field configurations. Thus, assuming $\phi_3$ and
$\phi_8$ to be real valued, one has once again $\phi_8=0$ and the mean field
thermodynamic potential $\Omega^{\rm MFA}$ becomes a real quantity even for
nonzero $\mu$. In general, this assumption will be considered in the
analyses discussed in this work. However, it is worth noticing that the
inclusion of beyond mean field corrections for Im$\,\Omega$ induced by the
temporal gauge fields could cause in general a splitting between $\langle
\Phi \rangle$ and $\langle \Phi^\ast \rangle$~\cite{Ratti:2006wg,
Roessner:2006xn, Ratti:2007jf}. On the other hand, by taking
$\Phi={\Phi^\ast}$ within the mean field approximation one avoids the sign
problem (which is also found, at finite density, in the local PNJL
model~\cite{Fukushima:2006uv,Nishimura:2014kla}). Alternatively, other
approaches, such as e.g.\ the Lefschetz thimble
method~\cite{Tanizaki:2015pua} would need to be implemented so as to
correctly perform the path integrals, even at the mean field level. Finally,
it is important to notice that for the theoretically interesting case of
 an imaginary chemical potential, to be discussed in
Sec.~\ref{sec2.6}, the restriction to  a real-valued $\Phi$ does not
hold, and both $\phi_3$ and $\phi_8$ are allowed to be nonzero.

Taking into account the above discussion, the  mean field
thermodynamic potential for the nlPNJL model is found to be given
by~\cite{Pagura:2011rt,Pagura:2013rza}
\begin{equation}
\Omega^{\rm MFA} \ = \ -4\,T\,\sum_{n,\,c}\,\int \frac{d^3 p}{(2\pi)^3} \,
\ln\left[\frac{(\rho_{n,\vec p}^c)^2+M(\rho_{n,\vec
p}^c)^2}{Z(\rho_{n,\vec p}^c)^2}\right] +
\frac{\bar\sigma_1^2+\varkappa_p^2\,\bar\sigma_2^2}{2\,G_S}\, + \,
\mathcal{U}(\Phi,\Phi^\ast,T)\ .
\label{ch2.2-eq9}
\end{equation}
This expression turns out to be divergent and has to be regularized.
Following the prescription in Ref.~\cite{GomezDumm:2004sr}, a finite
thermodynamic potential $\Omega^{\rm MFA}_{\rm reg}$ can be defined as
\begin{equation}
\Omega^{\rm MFA}_{\rm reg} \ = \
\Omega^{\rm MFA}-\Omega^{\rm free}_q \, + \,\Omega^{\rm free}_{q,{\rm reg}} \,
+ \,\Omega_0 \ ,
\label{ch2.2-eq10}
\end{equation}
where $\Omega^{\rm free}_q$ is obtained from the quark contribution to
$\Omega^{\rm MFA}$ by setting $\bar \sigma_1=\bar \sigma_2=0$, and
$\Omega^{\rm free}_{q,{\rm reg}}$ is the regularized expression for the
quark thermodynamical potential in the absence of fermion interactions,
\begin{equation}
\Omega^{\rm free}_{q,{\rm reg}} \ = \ -4\,T\int\frac{d^3p}{(2\pi)^3}\sum_{c=r,g,b}\
\sum_{s=\pm 1}\textrm{Re} \ln\left[ 1+
\exp\left(-\frac{\epsilon_p+s \  ( \mu + i \phi_c) }{T}\right)\right]\ ,
\label{ch2.2-eq11}
\end{equation}
where $\epsilon_p=\sqrt{\vec{p}^{\phantom{i}2}+m^2}$. Note that the r.h.s.\
of Eq.~(\ref{ch2.2-eq10}) also includes a constant $\Omega_0$, which is
fixed by the condition that $\Omega^{\rm MFA}_{\rm reg}$ vanishes for
$T=\mu=0$.

In general, the mean field values $\bar \sigma_1$ and $\bar \sigma_2$, as
well as the values of $\phi_3$ and $\phi_8$, can be obtained from a set of
four coupled ``gap equations'' that follow from the minimization of the
regularized thermodynamic potential, namely
\begin{equation}
\frac{\partial\Omega^{\rm MFA}_{\rm reg}}{\partial\bar\sigma_{1}} \ = \ 0
\ ,\quad
\frac{\partial\Omega^{\rm MFA}_{\rm reg}}{\partial\bar\sigma_{2}} \ = \ 0
\ ,\quad
\frac{\partial\Omega^{\rm MFA}_{\rm reg}}{\partial\phi_3}\ = \ 0
\ ,\quad
\frac{\partial\Omega^{\rm MFA}_{\rm reg}}{\partial\phi_8}\ = \ 0 \ .
\label{ch2.2-eq12}
\end{equation}
We recall that, in the framework of the nlPNJL model studied here, either
for vanishing or real $\mu$ one has $\phi_8=0$. Thus, the last of
Eqs.~(\ref{ch2.2-eq12}) will be only required in the case of a nonzero
imaginary chemical potential (see Sec.~\ref{sec2.6}).

Once the mean field values have been determined, the behavior of other
relevant quantities as functions of the temperature and chemical potential
can be obtained. In particular, here we concentrate on the chiral quark
condensate $\langle\bar{q}q\rangle$, which is given by
\begin{equation}
\langle \bar q q\rangle \ = \
\frac{\partial\Omega^{\rm MFA}_{\rm reg}}{\partial m_c}\ ,
\label{ch2.2-eq13}
\end{equation}
and the chiral susceptibility $\chi_{\rm ch}$, defined as
\begin{equation}
\chi_{\rm ch} \ = \ \frac{\partial\,\langle\bar qq\rangle}{\partial m_c}\ .
\label{ch2.2-eq14}
\end{equation}

To characterize the deconfinement transition, it is usual to introduce the
associated PL susceptibility $\chi_\Phi$, defined by
\begin{equation}
\chi_\Phi \ = \ \frac{d \Phi}{dT}\ .
\label{ch2.2-eq15}
\end{equation}
However, as seen below, there are regions of the $\mu-T$ phase diagram where
this quantity  may not be adequate to determine the occurrence of the
transition. Alternatively, it can be assumed that the deconfinement region
is characterized by a value of $\Phi$ lying in some intermediate range,
e.g.\ between $0.3$ and $0.5$~\cite{Contrera:2010kz}.

\subsection{Form factors, parametrizations and numerical results for $T=\mu=0$}
\label{sec2.3}


To fully specify the nonlocal NJL model under consideration one has to fix
the model parameters as well as the form factors $g(p)$ and $f(p)$ that
characterize the nonlocal interactions. Several forms for these functions
have been considered in the literature. For definiteness, here we will
concentrate mostly on three particular functional forms, which lead to the
parameterizations that we call PA, PB and PC, defined as follows.

\begin{itemize}

\item In parametrization PA we consider the simple case in which one takes
\begin{equation}
g(p) \ = \ \exp(-p^{2}/\Lambda_{0}^{2})\ ,\qquad\qquad
f(p) \ = \ 0 \ ,
\end{equation}
i.e.~$Z(p)=1$. In general, the exponential form of $g(p)$ ensures a fast
convergence of loop integrals, and has been often used in the literature. On
the other hand, the results obtained with this parametrization can be
related to those of early studies in which the quark propagator WFR was
ignored (see e.g.\
Refs.~\cite{Bowler:1994ir,GomezDumm:2001fz,GomezDumm:2006vz,Golli:1998rf,GomezDumm:2004sr}).

\item In the second parametrization, PB, it is assumed that both $g(p)$ and $f(p)$
are given by Gaussian functions, namely
\begin{equation}
g(p) \ = \ \exp(-p^{2}/\Lambda_{0}^{2})\ ,\qquad\qquad
f(p) \ = \ \exp(-p^{2}/\Lambda _{1}^{2})\ .
\label{ch2.3-eq1}
\end{equation}
Note that the range (in momentum space) of the nonlocality in each channel
is determined by the parameters $\Lambda_{0}$ and $\Lambda_{1}$,
respectively. From Eqs.~(\ref{ch2.1.1-eq11}) one has
\begin{eqnarray}
M(p)   & = & Z(p)  \, \left[  m_{c}+\bar{\sigma}_{1}\
\exp(-p^{2}/\Lambda_{0}^{2})\right]\ ,
\nonumber\\
Z(p)   &  = & \left[  1-\bar{\sigma}_{2}\
\exp(-p^{2}/\Lambda_{1}^{2})\right]^{-1}\ .
\label{ch2.3-eq2}
\end{eqnarray}

\item Finally, the third parametrization, PC, has been taken from
Refs.~\citep{Noguera:2005ej,Noguera:2008cm}. In this case the functions $M(p)$
and $Z(p)$ are written as
\begin{eqnarray}
M(p)  &  = & m_{c}+\alpha_{m}\, f_{m}(p)\ ,\nonumber\\
Z(p)  &  = & 1+\alpha_{z}\, f_{z}(p) \ ,
\label{ch2.3-eq3}
\end{eqnarray}
where $\alpha_m = M(0)-m_c$, $\alpha_z = Z(0)-1$. The form of $f_m(p)$ and
$f_z(p)$ is proposed taking into account the results from lattice QCD
calculations in the Landau gauge. One has
\begin{equation}
f_{m}(p)=\left[  1+\left(  p^{2}/\Lambda_{0}^{2}\right)^{3/2}\right]^{-1}\
,\qquad\qquad f_{z}(p)=\left[  1+\left( p^{2}/\Lambda_{1}^{2}\right) \right]^{-5/2}\ .
\label{ch2.3-eq4}
\end{equation}
The analytical expression for $f_{m}(p)$ has been originally proposed in
Ref.~\citep{Bowman2003}, while that of $f_{z}(p)$ has been chosen so as to
reproduce lattice results, ensuring at the same time the convergence of
quark loop integrals. Some alternative parametrization of this type,
suggested from vector meson dominance in the pion form factor, can be found
in Ref.~\citep{RuizArriola:2003bs}. In terms of the functions $f_{m}(p)$,
$f_{z}(p)$ and the constants $m_{c}$, $\alpha_{m}$ and $\alpha_{z}$, the
form factors $g(q)$ and $f(q)$ are given by [see Eqs.~(\ref{ch2.1.1-eq11})]
\begin{equation}
g(p) \ = \ \frac{1+\alpha_{z}}{1+\alpha_{z}f_{z}(p)}\ \frac{\alpha_{m}
f_{m}(p)-m_{c}\,\alpha_{z}f_{z}(p)}{\alpha_{m}-m_{c}\,\alpha_{z}}
\ ,\qquad\qquad
f(p) \ = \ \frac{1+\alpha_{z}}{1+\alpha_{z}f_{z}(p)}\ f_{z}(p)\ ,
\label{ch2.3-eq5}
\end{equation}
while for the mean field values $\bar\sigma_1$ and $\bar\sigma_2$ one has
\begin{equation}
\bar\sigma_1  \  = \ \frac{\alpha_{m}-m_{c}\alpha_{z}}{1+\alpha_{z}}\ ,
\qquad\qquad \bar{\sigma}_{2} \ = \ \frac{\alpha_{z}}{1+\alpha_{z}}\ .
\label{ch2.3-eq6}
\end{equation}
\end{itemize}

The numerical values of the model parameters corresponding to the above
parametrizations are quoted in Table~\ref{tab1}. In all three cases they
lead to the empirical values of the pion mass and decay constant,
$f_{\pi}=92.4$~MeV and $m_{\pi}=139$~MeV. In addition, for parametrizations
PA and PB the input value $\langle \bar qq\rangle = -(240\ {\rm MeV})^3$ is
taken. In the case of PB, which includes a wave function renormalization
$Z(p)$, it is also required that the parameters lead to $Z(0)=0.7$, as
suggested by lattice calculations~\citep{Parappilly:2005ei,Furui:2006ks}.
Finally, for parametrization PC the condition $Z(0)=0.7$ is also imposed,
and the effective cutoff scales $\Lambda_{0}$ and $\Lambda_{1}$ are taken in
such a way that the functions $f_{m}(p)$ and $Z(p)$ reproduce reasonably
well the lattice QCD data given in Ref.~\citep{Parappilly:2005ei}.

\begin{table}[hbt]
\par\begin{centering}
\begin{tabular}{ccccc}
\hline \hline
                          &        & \hspace{.5cm} PA \hspace{.5cm}  & \hspace{.5cm} PB \hspace{.5cm}  & \hspace{.5cm} PC \hspace{.5cm} \\ \hline
$m_{c}$                   &  MeV   &                 5.78               &           5.70                     &               2.37             \\
$G_{s}\Lambda_{0}^{2}$    &        &                20.65               &           32.03                    &               20.82            \\
$\varkappa_{P}$           & GeV    &                 -                  &         4.18                   &                6.03             \\
$\Lambda_{0}$             & MeV    &                752               &            814                  &                 850            \\
$\Lambda_{1}$             & MeV    &                -                   &            1034                  &                 1400              \\
\hline \hline
\end{tabular}
\par\end{centering}
\vspace*{0.5cm}
\caption{Model parameters for the alternative parameterizations.}%
\label{tab1}%
\end{table}

\begin{figure}[h!bt]
\centerline{\includegraphics[width=0.87\textwidth]{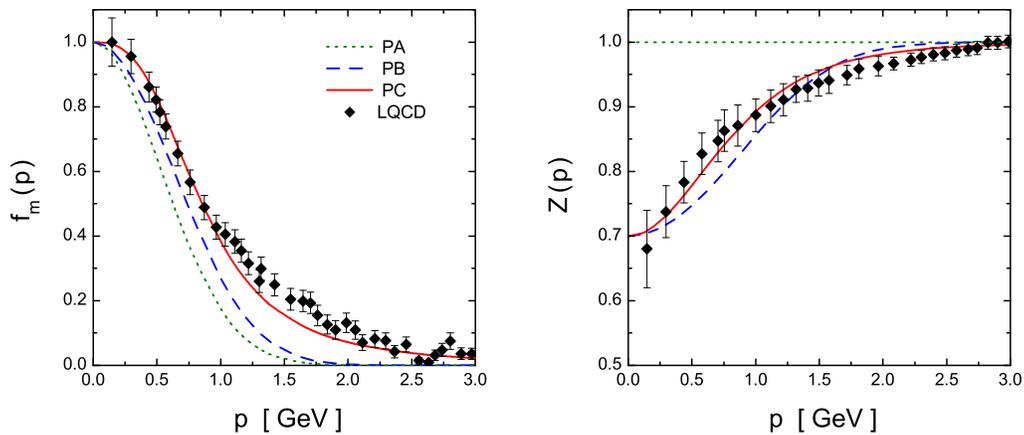}}
\caption{(Color online) Values of $f_{m}(p)$ (left panel) and
$Z(p)$ (right panel) for parametrizations PA, PB and PC, and
lattice QCD results from Ref.~\citep{Parappilly:2005ei}.}%
\label{Fig1}%
\end{figure}

In Fig.~\ref{Fig1} we show the quark mass functions $f_{m}(p)$ and the quark
WFR functions $Z(p)$ for the three above scenarios, including for comparison
lattice QCD results quoted in Ref.~\citep{Parappilly:2005ei}. The main
reason for taking $f_{m}(p)$, defined by the first of
Eqs.~(\ref{ch2.3-eq3}), instead of $M(p)$ is that lattice calculations
obtained by the groups in Ref.~\citep{Parappilly:2005ei} and
Ref.~\cite{Furui:2006ks}  ---which use different inputs--- lead to quite
similar results for $f_{m}(p)$, in spite of showing substantial differences
in the functions $M(p)$. On the other hand, the  quark WFR is much less
sensitive to the choice of lattice input parameters; in fact, the two
mentioned groups provide similar results for the behavior of $Z(p)$. From
the figure it is seen that in the case of PA and PB the functions
$f_{m}(p)$, which are based on exponential forms, decrease faster with the
momentum than lattice data. However, as seen below, many observables are not
significantly affected by the parametrization, and the choice of Gaussian
form factors can be taken as a reasonable approximation.

In Table~\ref{tab2} we quote the numerical results obtained for several
quantities in the framework of the considered nlNJL model. By construction,
parameterizations PA and PB lead to a quark condensate lying within the
phenomenologically accepted range $(200\ {\rm MeV})^3 \lesssim
-\langle\bar{q}q\rangle\lesssim (260\ {\rm
MeV})^3$~\citep{Dosch:1997wb,Giusti:1998wy}, while for PC the absolute value
of the condensate is found to be somewhat larger. On the other hand, the
current quark mass $m_c$ in parametrization PC is quite smaller than in the
case of PA and PB. In this regard, it is worth noticing that both the chiral
condensate and the current quark masses are scale dependent objects. In
particular, the above mentioned phenomenological range for the condensate
corresponds to a renormalization scale  $\mu_R\simeq 1$~GeV. In
parametrization PC some parameters have been determined so as to obtain a
good approximation to LQCD results for the quark propagator, which also
depend on the renormalization point. In fact, lattice values in
Ref.~\citep{Parappilly:2005ei} have been obtained taking a renormalization
scale $\mu_R=3$ GeV. One might wonder whether the fact that this
renormalization point differs from the one usually used to quote the values
of the condensates can account for the fact that the PC prediction falls
outside the empirical range. This issue is discussed in
Ref.~\cite{Noguera:2008cm}, taking into account that the current quark mass
and the condensate are linked by the Gell-Mann-Oakes-Renner relation given
by Eq.~(\ref{ch2.1.3-eq9}). It is found that a rescaling of the quark
condensate in parametrization PC down to $\mu_R=1$~GeV would lead to
$-\langle\bar {q}q\rangle^{1/3}\,\sim\,280$~MeV, close to the upper limit of
the phenomenologically allowed range.

\begin{table}[h]
\par
\begin{centering}
\begin{tabular}{cccccc}
\hline \hline
                           &        & \hspace{.5cm} PA \hspace{.5cm}  &   \hspace{.5cm} PB \hspace{.5cm}  & \hspace{.5cm} PC \hspace{.5cm} \\
\hline
$\bar{\sigma}_{1}$         &   MeV  &                 424                &                    529               &                     442        \\
$\bar{\sigma}_{2}$         &        &                 -                  &                   $-0.43$ \            &                   $-0.43$ \     \\
$-\langle \bar q q\rangle^{1/3}$        &   MeV  &                 \ $240^\ast$            &                   \ $240^\ast$            &                     326        \\
\hline
$g_{\pi q\bar{q}}$         &        &                 4.62               &                    5.74              &                     4.74       \\
$g_{\sigma q\bar{q}}^{(0)}$&        &                  5.08              &                    5.97              &                     4.60       \\
$g_{\sigma q\bar{q}}^{(1)}$&        &                  -                 &                  $-0.77$ \            &                    $-0.26$ \    \\
\hline
$m_{\pi}$                  &   MeV  &                \ $139^*$             &               \ $139^*$            &                 \ $139^*$     \\
$m_{\sigma}$               &   MeV  &                 683                &                    622               &                     552        \\
$f_{\pi}$                  &   MeV  &                \ $92.4^*$           &                \ $92.4^*$           &                 \ $92.4^*$    \\
$\Gamma_{\sigma\pi\pi}$    &   MeV  &                  347               &                   263                &                     182        \\
\hline \hline
\end{tabular}
\par\end{centering}
\vspace*{0.5cm}
\caption{Numerical results for various phenomenological
quantities. Input values are marked with an asterisk.}%
\label{tab2}%
\end{table}

It is seen from Table~\ref{tab2} that the mass and decay width of the sigma
meson show some dependence on the parametrization, although this dependence
is less significant than in the case of the local NJL
model~\citep{Nakayama:1991ue}. In general, it can be said that the
predictions for $m_\sigma$ and $\Gamma_{\sigma\pi\pi}$ are compatible with
empirical data, taking into account the large experimental errors. Present
estimations from the Particle Data Group lead to values of the $\sigma$
meson mass in the range from 400 to 550~MeV, and to a total width
$\Gamma_\sigma$ between 400 and 700~MeV~\cite{Zyla:2020zbs}. In the case of
the $\sigma^{\prime}$ state, the mass is not given in Table~\ref{tab2} since
the corresponding two-point function shows no real poles at a low energy
scale in which the effective model could be trustable.

We conclude this subsection by mentioning that an alternative functional
form for $g(p)$ has been used in
Refs.~\cite{Hell:2008cc,Hell:2011ic,Hell:2009by}. The basic difference with
the functions in PA and PB is that in those articles the exponential form is
used  only up to certain matching scale $\Gamma \simeq 0.8$~GeV. Beyond this
scale the exponential is replaced by a function of the form $\sim
\alpha_s(p^2)/p^2$ [$\alpha_s(p^2)$ being the QCD running coupling
constant], which is based on the large momentum behavior predicted by QCD.
This type of form factor does not exclude that some quantities like e.g.\
the quark condensates can be weakly divergent, and the introduction of a
cutoff at very high momenta ($\sim 20$~GeV) is still required. In any case,
the results obtained using this alternative scheme are not significantly
different from those arising from the parameterizations given above. One
important issue addressed in Ref.~\cite{Hell:2009by} is the validity of the
assumption that nonlocal form factors are separable. To get an idea of the
uncertainty introduced by this approximation, the solution of the
Schwinger-Dyson (gap) equation resulting from a nonseparable interaction
$g(q-p)$ is compared with the one associated with the separable form
$g(q)g(p)$. For simplicity, Gaussian functions are used for the momentum
distributions $g(p)$ both in the full Schwinger-Dyson expression and in the
case of the separable ansatz. The numerical analysis shows that the
corresponding results turn out to be coincident up to a 95$\%$ level (see
Fig.~1 and Sec.~II-E1 of Ref.~\cite{Hell:2009by} for details). Another
relevant point discussed in Ref.~\cite{Hell:2009by} is that since the
integrals in momentum space needed for the calculation of the different
relevant quantities include $n$ powers of the momentum $p$ ($n=2$ and $n=3$
for integrals in three and four dimensions, respectively), the details of
the low-momentum behavior of form factors in the integrands do not affect
dramatically the numerical results.

\subsection{Results for finite temperature and vanishing chemical potential}
\label{sec2.4}

In this subsection we analyze, in the context of the above presented nlPNJL
approach, the characteristics of the deconfinement and chiral restoration
transitions at finite temperature and vanishing chemical potential. We
consider the parametrizations PA, PB and PC defined in the previous section,
as well as the polynomial and logarithmic Polyakov loop potentials
introduced in Sec.~\ref{sec2.2}, taking the reference temperature $T_0$ as a
parameter. The corresponding numerical results are obtained by solving
numerically Eqs.~(\ref{ch2.2-eq12}), with $\phi_8=0$.

\begin{figure}[h!bt]
\centerline{
\includegraphics[width=0.8\textwidth,angle=0]{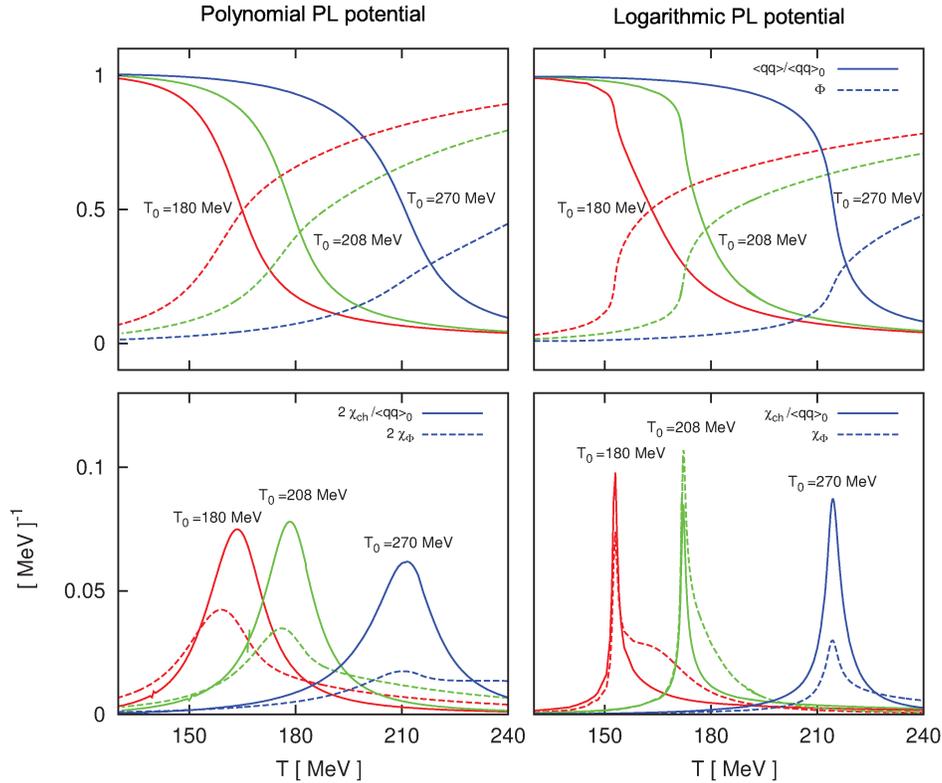}
}\caption{(Color online) Order parameters (upper panels) and
susceptibilities (lower panels) at $\mu=0$ for values of $T_0$ corresponding
to $0$, $2$ and $3$ dynamical quarks. Left (right) panels show the results
obtained using the polynomial (logarithmic) PL potential. All results
correspond to parametrization PC. Note that in the case of the polynomial
potential the susceptibilities are multiplied by a factor 2.}
\label{ch2.4-fig1}
\end{figure}

In Fig.~\ref{ch2.4-fig1} we show the results for the normalized quark
condensates and the traced Polyakov loop (upper panels), as well as the
corresponding susceptibilities (lower panels), for the lattice inspired
parametrization PC~\cite{Pagura:2013rza}. Left and right panels correspond
to the  polynomial PL potential in Eq.~(\ref{ch2.2-eq2}) and the logarithmic
PL potential in Eq.~(\ref{ch2.2-eq5}), respectively. Regarding the parameter
$T_0$, three characteristic values have been considered. As stated in
Sec.~\ref{sec2.2}, the proposed PL potentials are such that $T_0$ turns out
to be the deconfinement transition temperature obtained from LQCD in the
pure gauge theory, viz.~$T_0 = 270$~MeV. However, as discussed in
Refs.~\cite{Schaefer:2007pw,Schaefer:2009ui}, this value should be modified
when the color gauge fields are coupled to dynamical fermions. In the case
of two and three active flavors, it is found that $T_0$ gets shifted to
208~MeV and 180~MeV, respectively~\cite{Schaefer:2007pw,Schaefer:2009ui}. In
Fig.~\ref{ch2.4-fig1} we show the results corresponding to these two values,
and for comparison the case $T_0 = 270$~MeV is also included. From the
figure it is clearly seen that both the chiral restoration and deconfinement
transitions proceed at smooth crossovers. In addition, it is found that both
transitions occur at approximately the same critical temperature, as
indicated by the peaks of the corresponding susceptibilities. The curves for
the susceptibilities also show that the transitions turn out to be smoother
in the case of the polynomial PL potential.

In the case of the logarithmic PL potential, it can be noticed that as long
as $T_0$ decreases the chiral susceptibility tends to become asymmetric
around the peak, being somewhat broader on the high temperature side. While
this could be considered as an indication of some shift in the chiral
restoration critical temperature, even for $T_0 = 180$ MeV the splitting
between the main peak and the potential second broad peak is less than
10~MeV. For the polynomial PL potential it is seen that the peaks of the
susceptibilities are somewhat separated, though this splitting is not larger
than a few MeV for the considered $T_0$ range. It is important to point out
that this strong entanglement between chiral restoration and deconfinement
transitions, which also holds for parametrizations PA and PB, occurs in a
natural way within nonlocal models, and is in agreement with lattice QCD
results. On the contrary, this feature is usually not observed in local PNJL
models, where both transitions appear to be typically separated by about 20
MeV, or even more (see e.g.~Refs.~\cite{Fu:2007xc,Costa:2008dp}). In those
models the splitting can be reduced e.g.\ by including an eight-quark
coupling~\cite{Sakai:2009dv} or by considering an ``entangled scalar
interaction'', in which the effective four-quark coupling constant is a
function of $\Phi$~\cite{Sakai:2010rp,Sasaki:2011wu}. It is also worth
noticing that while nlPNJL models predict in general steeper transitions
than the local PNJL  model, this feature should not be understood as a
consequence of the nonlocality. In fact, the enhancement of the steepness
arises from the feedback between chiral restoration and deconfinement
transitions. This is supported by the results found in the above mentioned
``entangled'' PNJL scheme: if one includes a $\Phi$-dependent coupling that
leads e.g.~to a common critical temperature of about $175$~MeV, it is found
that the transitions become steeper, resembling those obtained within
nonlocal PNJL models.

\begin{figure}[h!tb]
\centerline{
\includegraphics[width=0.7\textwidth]{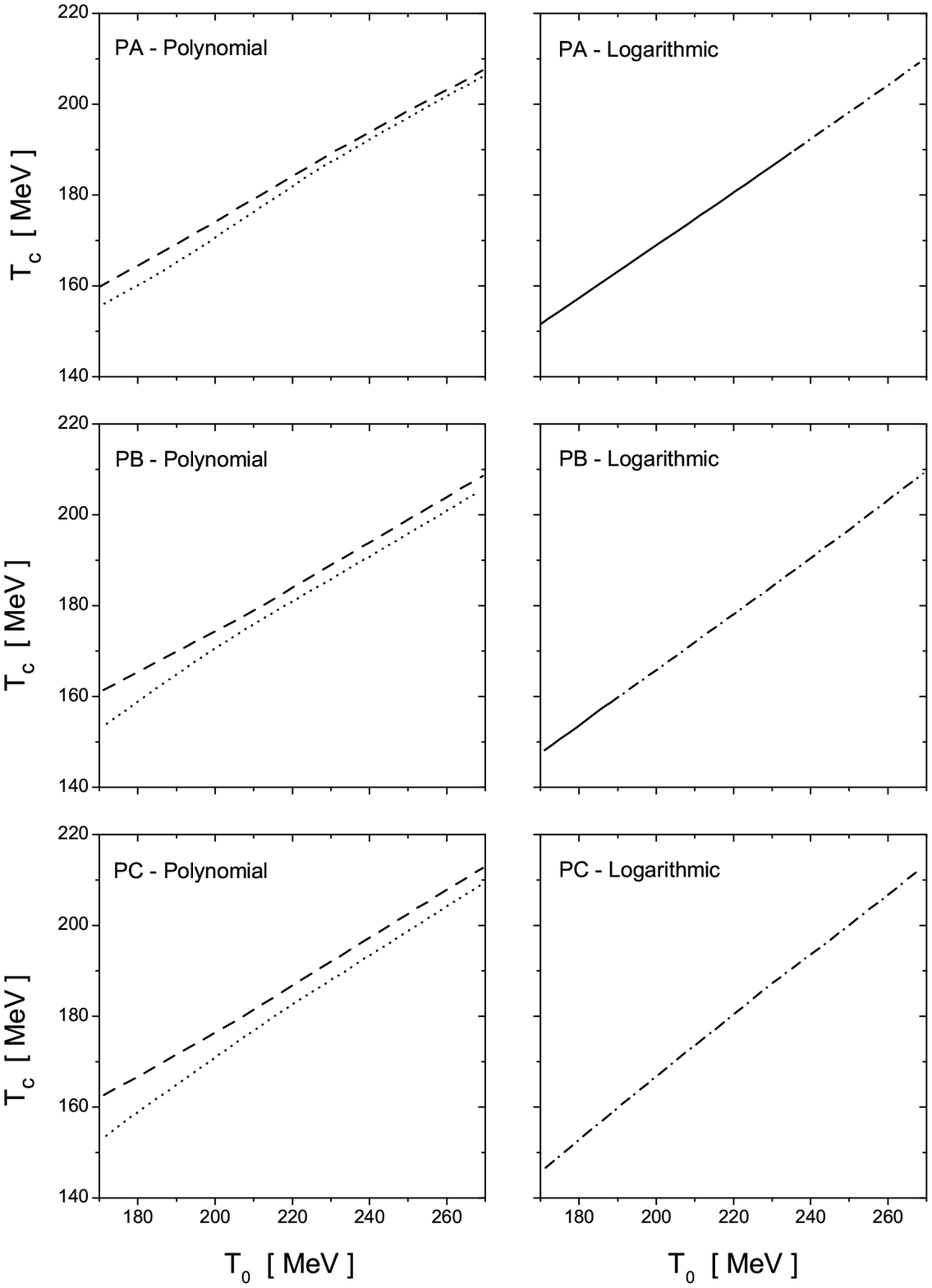}
} \caption{Critical temperatures as functions of $T_0$ for parametrizations
PA (upper panels), PB (central panels) and PC (lower panels), considering
polynomial (left) and logarithmic (right) PL potentials. Solid lines
correspond to first order chiral transitions, while dashed (dotted) lines
correspond to chiral (deconfinement) crossover-like transitions. Dash-dotted
lines stand for the cases in which both crossover-like transitions
coincide.} \label{rev-ch2.4-fig2}
\end{figure}

On the other hand, although for the cases considered above the critical
temperatures of deconfinement and chiral restoration transitions are
basically coincident, the character of the transitions may be different from
one another. This is shown in Fig.~\ref{rev-ch2.4-fig2}, where we quote the
behavior of the critical temperatures as functions of
$T_0$~\cite{Pagura:2013rza}. In the case of the polynomial PL potential
(left panels) the transitions are always crossover-like. The critical
temperatures for chiral restoration, $T_{\rm ch}$ (dashed lines), are
slightly different from those of deconfinement, $T_\Phi$ (dotted lines). The
splitting is in all cases lower than 10~MeV and decreases as $T_0$ grows.
For the logarithmic PL potential it is found that both transitions occur at
the same temperature, and can be either crossover-like (dash-dotted lines)
or first order (full lines). For parametrizations PA and PB they become of
first order for values of $T_0$ below $\sim 235$~MeV and 190~MeV,
respectively, while ---as already mentioned--- for the lattice-inspired
parameterization PC they proceed as smooth crossovers for all considered
values of $T_0$. It should be stressed that for $T_0 = 208$~MeV
(corresponding to the $N_f = 2$ models addressed throughout this section)
the resulting critical temperatures are in good agreement with the
values obtained from $N_f= 2$ lattice QCD calculations, namely $T_{\rm
ch}^{\rm LQCD} = 173\pm 8$~MeV~\cite{Karsch:2003jg}. Indeed, for the
polynomial potential the chiral restoration transition is found to
occur at $T_{\rm ch} = 178$, 178 and 180~MeV for PA, PB and PC,
respectively, while for the logarithmic potential the
corresponding critical temperatures are $T_{\rm ch} = 173$, 171 and $173$
MeV. Regarding lattice QCD results, there has been some debate concerning
the nature of both transitions for the case of two light flavors. While most
studies~\cite{Karsch:1994hm,Bernard:1996iz,Iwasaki:1996ya,Aoki:1998wg,AliKhan:2000wou}
favor a second order transition in the chiral limit, there are also some
results which indicate that it could be of first
order~\cite{DElia:2005nmv,Bonati:2009yg}. The analysis of nlPNJL models
discussed here appears to favor the second order scenario.

\hfill

Another relevant feature of the chiral restoration and deconfinement
transitions is their dependence on the strength of the explicit symmetry
breaking induced by the finite quark masses. The study of this property
provides a further test of the reliability of effective models, since the
results can be compared with available lattice QCD calculations. It fact,
for vanishing chemical potential, it has been shown that several chiral
effective models~\cite{Berges:1997eu,Dumitru:2003cf,Braun:2005fj} are not
able to reproduce the behavior of critical temperatures found in lattice QCD
when one varies the parameters that explicitly break chiral symmetry, viz.\
the current quark masses, or the pion mass in the case of meson models.

\begin{figure}[h!bt]
 \centerline{
 \includegraphics[width=0.7\textwidth]{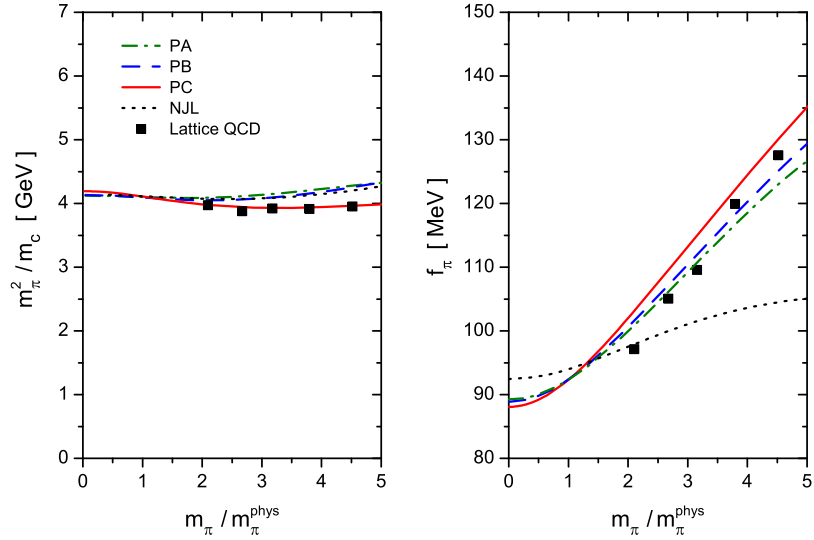}
}
\caption{(Color online) Pion properties at $T=0$ as functions of the pion
mass in local and nonlocal NJL-like models. Left and right panels correspond
to the ratio $m_\pi^2/m_c$ and the pion decay constant $f_\pi$,
respectively. Lattice results are taken from Ref.~\cite{Noaki:2008iy}, while
the local NJL model parametrization is that given in
Refs.~\cite{Roessner:2006xn,Kahara:2009sq}.}
\label{rev-ch2.4-fig3}
\end{figure}

This issue has been investigated in the framework of two-flavor nlPNJL
models in Ref.~\cite{Pagura:2012ku}. As a first step, the behavior of the
pion mass and the pion decay constant at vanishing temperature were compared
with those obtained in the local NJL model and in lattice QCD. Results are
shown in Fig.~\ref{rev-ch2.4-fig3}. As usual in LQCD literature, the
relevant quantities are plotted as functions of $m_\pi$ instead of $m_c$.
The main reason for this is that $m_\pi$ is an observable, i.e., a
renormalization scale-independent quantity, whereas $m_c$ is scale
dependent. Hence, the value of $m_c$ is subject to possible ambiguities
related to the choice of the renormalization point, as mentioned in
Sec.~\ref{sec2.3}. The left panel of Fig.~\ref{rev-ch2.4-fig3} shows the
behavior of the ratio $m_\pi^2/m_c$ as a function of $m_\pi/m_\pi^{\rm
phys}$, with $m_\pi^{\rm phys} = 139$~MeV. In order to account for the above
mentioned renormalization point ambiguities, the corresponding quark masses
have been normalized so as to yield the lattice value $m_{u,d}^{\rm
\overline{MS}}\simeq 4.45$~MeV at the physical point~\cite{Noaki:2008iy}.
From the figure one observes that both NJL and  nlNJL models reproduce
qualitatively the results from lattice QCD, showing a particularly good
agreement in the case of the  nlNJL model for parametrization PC. However,
the situation is different in the case of $f_\pi$ (right panel of
Fig.~\ref{rev-ch2.4-fig3}): while the predictions from nonlocal models
follow a steady increase with $m_\pi$, in agreement with lattice results,
the local NJL model in general fails to reproduce this behavior. Moreover,
it can be seen that the discrepancy cannot be cured even if one allows the
coupling $G_S$ to depend on the current quark mass~\cite{Kahara:2009sq} (the
curves in Fig.~\ref{rev-ch2.4-fig3} correspond to constant values of $G_S$,
both for local and nonlocal models). In this way, these results can be
considered as a further indication in favor of the inclusion of nonlocal
interactions as a step towards a more realistic description of low momenta
QCD dynamics.

\begin{figure}[h!bt]
 \centerline{
\includegraphics[width=0.8\textwidth]{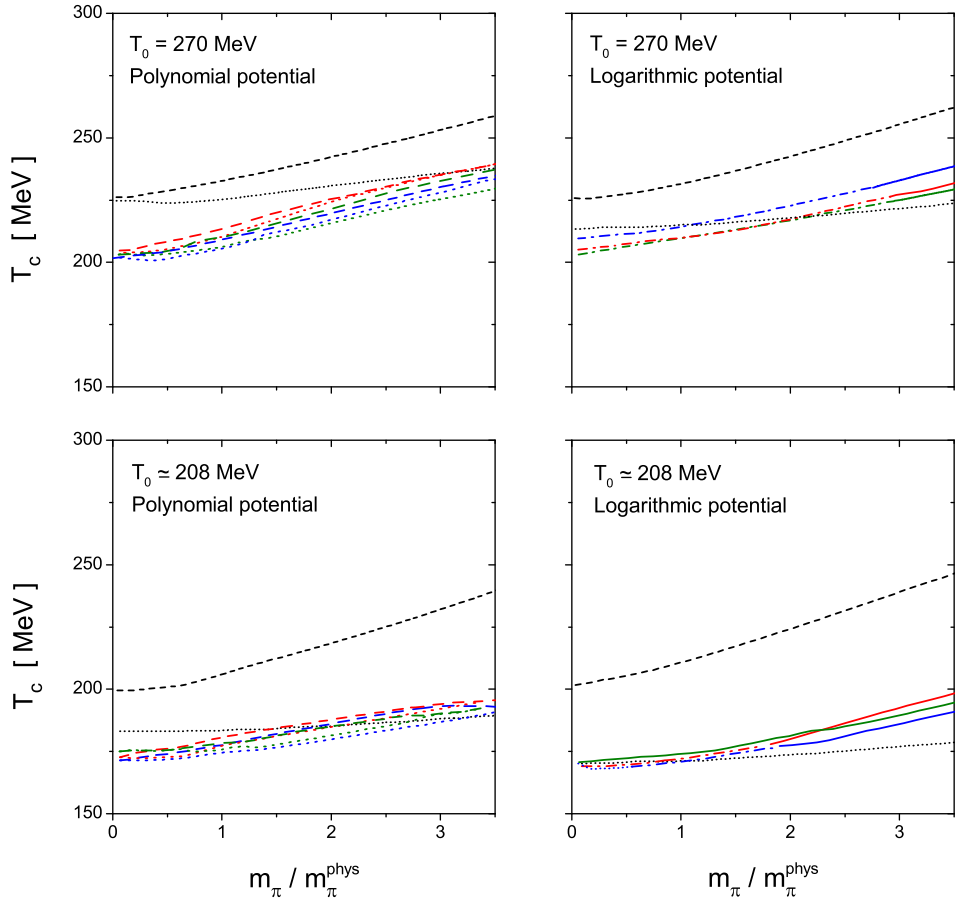}
}\caption{(Color online)
Critical temperatures as functions of the pion mass for PNJL and nlPNJL
models. Left: results for a polynomial PL potential. Dashed and dotted lines
correspond to chiral restoration and deconfinement transition temperatures
$T_{\rm ch}$ and $T_\Phi$, respectively. Right: results for a logarithmic PL
potential. Solid and dash-dotted lines correspond to first order and
crossover-like transitions, respectively, in which $T_{\rm ch}=T_\Phi$. In all
four panels, green, blue and red lines correspond to nlPNJL model
parametrizations PA, PB and PC, respectively, while black short-dashed
(short-dotted) lines correspond to PNJL results for $T_{\rm ch}$
($T_\Phi$).}
\label{rev-ch2.4-fig4}
\end{figure}

The above results also indicate that within nonlocal NJL models the change
in the amount of the explicit symmetry breaking can be accounted for by
varying only the current quark mass, i.e., other model parameters do not
appear to change significantly with $m_c$. Having this in mind, the features
of deconfinement and chiral restoration transitions have been studied
in nlPNJL models for different values of $m_c$, keeping the coupling
constants and cutoff scales unchanged. The results for the critical
temperatures $T_{\rm ch}$ and $T_\Phi$ as functions of $m_\pi$ are displayed
in Fig.~\ref{rev-ch2.4-fig4}, where, for comparison, we also quote typical
curves obtained in the framework of the local PNJL model (here we have
considered the parameterization in Ref.~\cite{Roessner:2006xn}). Left and
right panels correspond to polynomial and logarithmic PL potentials,
respectively. Concerning the value of $T_0$, the results in the upper
panels correspond to the pure gauge value $T_0 = 270$~MeV, while the curves
in the lower panels are obtained considering the coupling of color gauge
fields to $N_f = 2$ active quark flavors, plus some dependence on the
current quark mass, as proposed in Ref.~\cite{Schaefer:2007pw}. In fact,
this dependence is found to be rather mild, and one gets in practice $T_0
\simeq 208$~MeV for the whole range covered by the graphs in the figure.

Before discussing the results obtained for nlPNJL models, let us comment
those corresponding to the local PNJL model. From Fig.~\ref{rev-ch2.4-fig4}
it is seen that already at the physical value $m_\pi = m_\pi^{\rm phys}$ the
model predicts a noticeable splitting between $T_{\rm ch}$ (short-dashed
lines) and $T_{\Phi}$ (short-dotted lines). In addition, the growth of
$T_{\rm ch}$ with $m_\pi$ is found to be more pronounced than that of
$T_{\Phi}$, which implies that the splitting between both critical
temperatures becomes larger if $m_\pi$ is increased. This is not supported
by existing lattice results~\cite{Karsch:2000kv,Bornyakov:2009qh}, which
indicate that both transitions occur at approximately the same temperature
up to values of $m_\pi$ even larger than those considered here. Comparing
left and right panels, it is seen that the splitting is larger for the PNJL
model that includes  a logarithmic Polyakov-loop potential.

We turn now to the curves obtained within nonlocal PNJL
models~\cite{Pagura:2012ku}. By looking at the graphs in
Fig.~\ref{rev-ch2.4-fig4}, it is observed that all parameterizations lead to
qualitatively similar results. Moreover, contrary to the situation in the
local PNJL model, in nlPNJL models both the chiral restoration and
deconfinement transitions occur at basically the same temperature for all
considered values of $m_\pi$. Comparing the results for the two alternative
PL potentials it is seen that the main qualitative difference is given by
the fact that in the case of the logarithmic potential (right panels of
Fig.~\ref{rev-ch2.4-fig4}) the character of the transition changes from
crossover-like to first order when the pion mass exceeds a critical value.
Crossover and first order transition regions are indicated by dashed-dotted
and solid lines, respectively.

Let us analyze in more detail the pion mass dependence of the
critical temperatures. For $T_0 = 270$~MeV, it is seen that  the
results from nlPNJL models can be quite accurately adjusted through
linear functions
\begin{equation}
T_c (m_\pi) \ = \ A \ m_\pi + B \ ,
\end{equation}
where $T_c$ denotes either $T_{\rm ch}$ or $T_\Phi$. This is in agreement
with the findings of LQCD calculations in
Refs.~\cite{Karsch:2000kv,Bornyakov:2009qh}. The slope parameter $A$ can be
fitted for all considered nlPNJL model parameterizations and Polyakov loop
potentials in Fig.~\ref{rev-ch2.4-fig4}, leading to values in the range
$0.06-0.07$~\cite{Pagura:2012ku}. For comparison, most lattice calculations
find $A \lesssim
0.05$~\cite{Karsch:2007dt,Karsch:2000kv,Bornyakov:2005dt,Cheng:2006qk},
while according to the analyses in
Refs.~\cite{Ejiri:2009ac,Bornyakov:2009qh} the value could be somewhat above
this bound. Thus, the slope parameter predicted by nlPNJL models appears to
be compatible with lattice estimates. This can be contrasted with the
results obtained within other effective chiral models, where one finds a
strong increase of the chiral restoration temperature with
$m_\pi$~\cite{Berges:1997eu,Dumitru:2003cf,Braun:2005fj}. For example,
within the chiral quark model studied in Ref.~\cite{Braun:2005fj} one has
$A=0.243$. Finally, let us consider the results in the lower panels of
Fig.~\ref{rev-ch2.4-fig4}, which correspond to $T_0 \simeq 208$~MeV. As
shown in the figure, the lowering of $T_0$ leads to an overall decrease of
the transition temperatures, which keep the rising linear dependence on
$m_\pi$. However, the slope parameter is found to be reduced by about
$15-20$\%. In addition, it is found that in all cases the transition becomes
steeper, which leads to lower values of the pion mass threshold at which it
starts to be of first order. For example, for parametrization PC it is found
that the transition becomes of first order already at $m_\pi \simeq 500$~MeV
(i.e., somewhat above the range shown in Fig.~\ref{rev-ch2.4-fig4}) in the
case of the polynomial PL potential, and about one half of this value for
the logarithmic one. Lattice QCD results also predict the onset of a first
order phase transition for $m_\pi$ larger than some critical value, which is
found to be of the order of a few GeV~\cite{Saito:2011fs}. In any case, the
estimation of this critical mass is rather uncertain in nlPNJL models,
depending crucially on the form of the PL potential.

\begin{figure}[h!bt]
\centerline{
\includegraphics[width=0.7\textwidth]{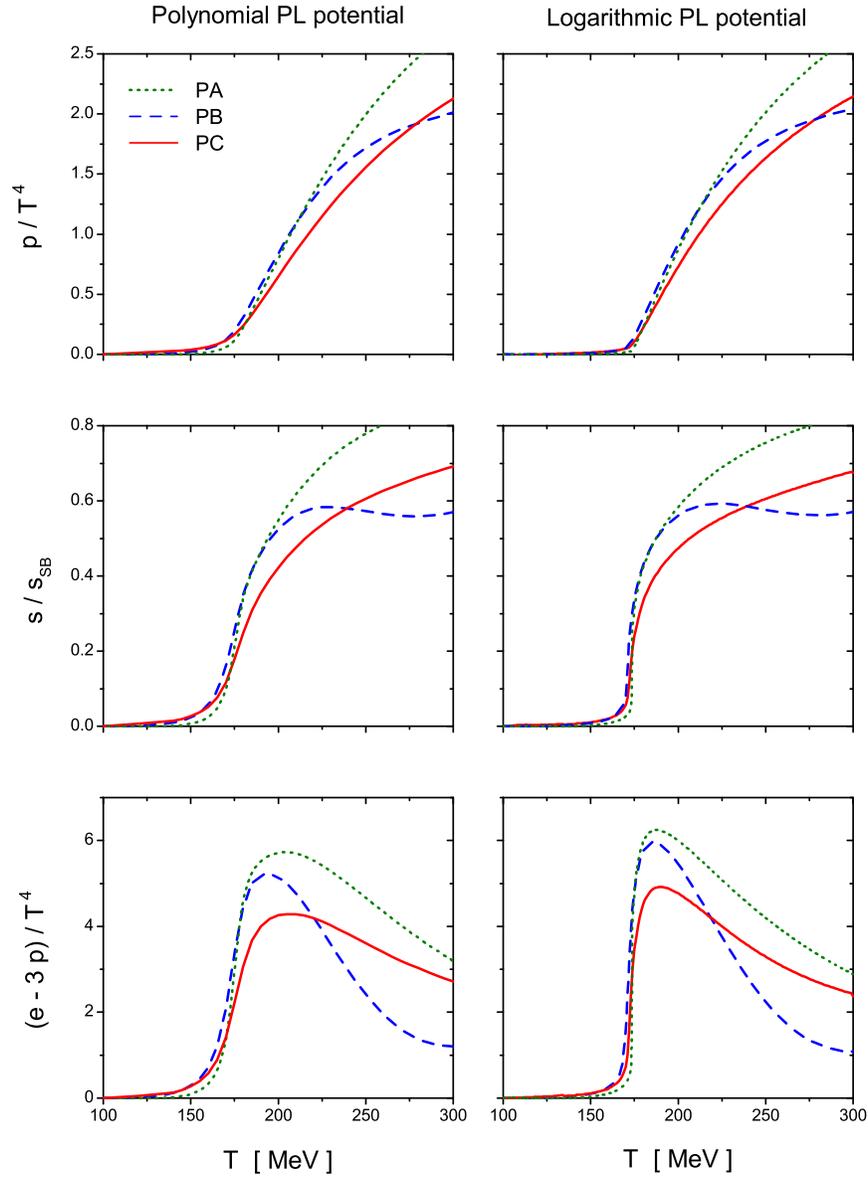}
}\caption{(Color online)
Normalized pressure $p/T^4$ ( top), normalized entropy $s/s_{SB}$
(center) and trace anomaly $(e- 3p)/T^4$ ( bottom) as functions of the
temperature, for nlPNJL model parametrizations PA, PB and PC. Left (right)
panels correspond to the polynomial (logarithmic) PL potential. In all cases
$T_0 = 208$~MeV has been used.}
\label{rev-ch2.4-fig6}
\end{figure}

\hfill

We conclude this  subsection by  quoting nlPNJL model results
for some selected thermodynamical quantities, viz.\ the normalized pressure
$p/T^4$, the normalized entropy $s/s_{SB}$ and the trace anomaly $(e-
3p)/T^4$. These can be obtained from the regularized thermodynamic potential
through the relations
\begin{equation}
p \ = \ -\, \Omega^{\rm MFA}_{\rm reg} \ , \qquad s \ = \ \frac{\partial p}{\partial T} \ ,
\qquad e \ = \ -\, p\, +\, T\, s\ .
\end{equation}
The massless Stefan-Boltzmann limit for the entropy for $N_f = 2$ and $N_c =
3$ is given by $s_{SB} = 74 \pi^2 T^3/45$. The numerical results obtained
for the parametrizations introduced in Sec.~\ref{sec2.3} are shown in
Fig.~\ref{rev-ch2.4-fig6}, where left and right panels correspond to the
polynomial and logarithmic PL potentials, respectively. Most of these
quantities have been also calculated within the Polyakov-quark-meson
model~\cite{Schaefer:2007pw}, showing a thermal behavior similar to the one
observed in Fig.~\ref{rev-ch2.4-fig6} for parametrization PC. It is worth
noticing that the curves for PB show some oscillation at about $T\sim
250$~MeV, which is not observed for the other parametrizations. This is
particularly clear for the case of the normalized entropy. In fact, as
mentioned in Refs.~\cite{Benic:2012ec,Marquez:2014kla}, in the absence of
the couplings between the quarks and the PL, thermodynamic instabilities
might appear in the context of nonlocal models for some particular form
factors. Although the couplings to the PL largely reduce the impact of these
instabilities on thermodynamic quantities, it is observed that in the case
of PB they still lead to sizeable effects. A detailed
analysis~\cite{Benic:2013eqa} shows that the oscillatory behavior is due to
the pole structure of the WFR function $Z(p)$ that arises from the
exponential shape of the form factor $f(p)$. From this point of view it can
be concluded that the power-like behavior for $f(p)$ used in parametrization
PC appears to be a more convenient choice (the instabilities are also not
found in the case of parametrization PA, for which $Z(p) = 1$). Another
point analyzed in Ref.~\cite{Benic:2013eqa} in connection with the couplings
leading to quark WFR concerns the effect of medium-induced Lorentz symmetry
breaking. As a general conclusion, it is found that this effect does not
modify significantly the phase transition features and the behavior of
thermodynamic functions.

A final aspect to be commented regards the steepness of the curves in the
transition region. As stated above, it is seen that for the case of the
polynomial potential the transition is somewhat smoother than for the
logarithmic one. However, it is worth mentioning that the observed behavior
may be softened after the inclusion of mesonic corrections to the Euclidean
action, since when the temperature is increased light meson degrees of
freedom should get excited before quarks excitations arise. For nlPNJL
models this has been analyzed in
Refs.~\cite{Blaschke:2007np,Hell:2008cc,Radzhabov:2010dd}. In any case, the
critical temperatures should not be modified by the incorporation of meson
fluctuations.

\subsection{Results for finite temperature and (real) chemical potential}
\label{sec2.5}

In this subsection we analyze the main features of the phase diagram in the
$\mu-T$ plane in the context of the above discussed nlPNJL models. For now
we consider the chemical potential to be a real quantity, as it should be
for a physical system. However, let us recall that, as mentioned in
Sec.~\ref{sec1}, the region of nonzero real $\mu$ is not fully accessible
from first principle lattice QCD calculations. On the contrary, for a purely
imaginary chemical potential these calculations become feasible, providing a
further test of the predictive capacity of effective models for low energy
QCD. We come back to this issue in Sec.~\ref{sec2.6}. In addition, we stress
that here scalar field VEVs and quark condensates are assumed to be
translational invariant quantities. The possible existence of inhomogeneous
phase regions is studied in Sec.~\ref{sec5.3}.

\begin{figure}[h!bt]
\centerline{
\includegraphics[angle=0,width=0.9\textwidth]{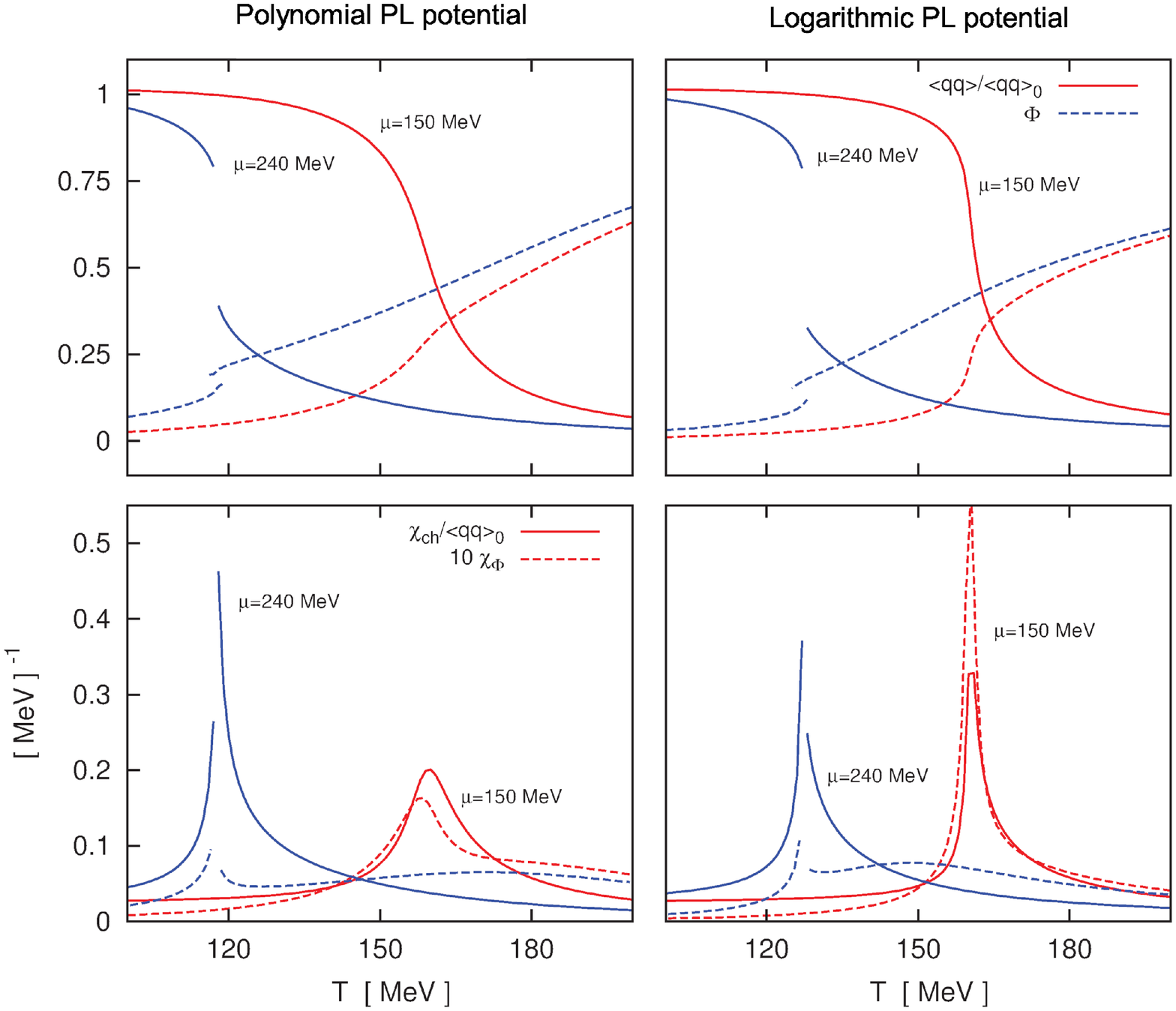}
} \caption{(Color online)
Order parameters and susceptibilities as functions of the temperature for
two representative values of the quark chemical potential. Left (right)
panels show the results for polynomial (logarithmic) PL potentials, with
$T_0=208$. All curves correspond to parametrization PC.}
\label{rev-ch2.5-fig1}
\end{figure}

Let us start by analyzing the behavior of deconfinement and chiral
restoration order parameters as functions of the temperature for various
chemical potentials. The results obtained in nlPNJL models are illustrated
in Fig.~\ref{rev-ch2.5-fig1}, where we display the normalized chiral
condensate $\langle \bar qq\rangle / \langle \bar qq\rangle_{T = 0}$ and the
traced Polyakov loop $\Phi$ (upper panels), as well as their associated
susceptibilities (lower panels), for parametrization PC. Left (right) panels
correspond to polynomial (logarithmic) PL potentials, with $T_0 = 208$~MeV.
The representative values $\mu = 150$ and 240~MeV, which lead to different
types of transitions, have been chosen.

It is seen from Fig.~\ref{rev-ch2.5-fig1} that for $\mu=150$ MeV there is a
critical temperature at which the chiral condensate decreases quite fast,
signalling the restoration of chiral symmetry. At about the same temperature
$\Phi$ starts to grow, indicating the beginning of the deconfinement
transition. Both transitions are found to be crossover-like for this
chemical potential. By looking at the susceptibilities, it is seen that for
the logarithmic PL potential both peaks coincide, while in the case of the
polynomial potential there is a small difference between the critical
temperatures associated to both transitions. Such a behavior is very similar
to the one found at $\mu=0$. By increasing the chemical potential, one
arrives at a certain value $\mu_{\rm CEP}$ for which the chiral condensate
starts to be discontinuous, i.e., the chiral restoration transition becomes
of first order. The point in the $\mu - T$ phase diagram at which this
happens is known as ``critical end point'' (CEP). As shown in
Ref.~\cite{Contrera:2010kz}, for nlPNJL models the transition is of second
order at this point. The behavior of the order parameters for $\mu >
\mu_{\rm CEP}$ is illustrated in Fig.~\ref{rev-ch2.5-fig1} considering the
case $\mu=240$~MeV. It is seen that the first order character of the chiral
restoration transition also induces a discontinuity in the order parameter
for deconfinement. One observes, however, that whereas the PL susceptibility
presents a divergent behavior at this point, the order parameter $\Phi$
still remains  being quite close to zero. Therefore, as mentioned in
Sec.~\ref{sec2.2}, in this region of the phase diagram it is convenient to
introduce an alternative definition for the deconfinement critical
temperature $T_\Phi$. One reasonable way of determining this temperature is
by requiring that $\Phi$ reaches a value in the range between $0.3$ and
$0.5$, which could be assumed to be large enough so as to denote
deconfinement~\cite{Contrera:2010kz}. Note that the transition is smooth,
and occurs at higher temperatures than the chiral restoration one. This
implies the existence of a phase in which quarks remain confined ($\Phi
\lesssim 0.3$) even though chiral symmetry is already restored. The latter
is usually referred to as a ``quarkyonic''
phase~\cite{McLerran:2007qj,McLerran:2008ua,Abuki:2008nm}.

\begin{figure}[h!bt]
 \centerline{
 \includegraphics[width=0.82\textwidth]{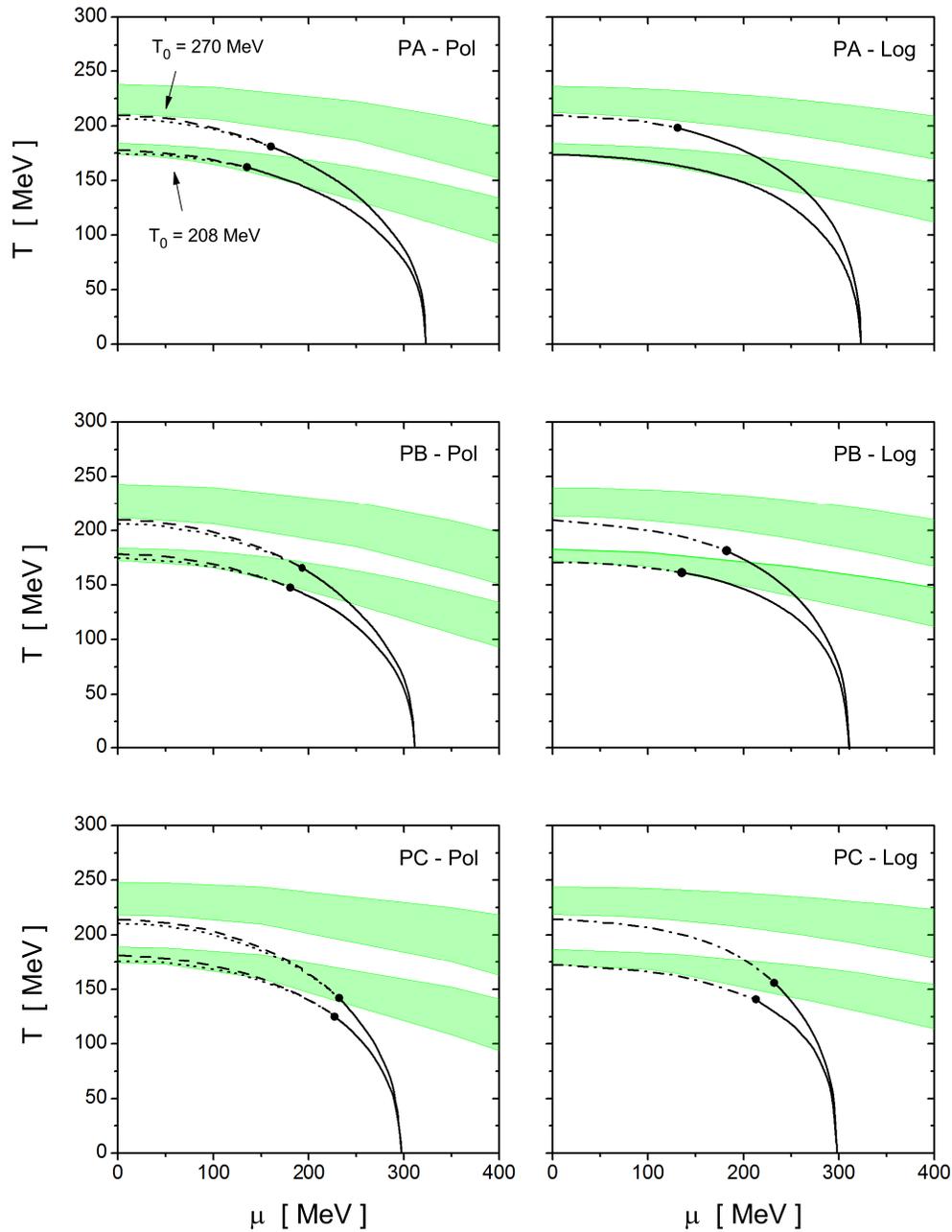}
}
\caption{(Color online)
Phase diagrams for parametrizations PA ( top), PB (center) and PC
( bottom), considering polynomial (left) and logarithmic (right) PL
potentials, with both $T_0 = 270$ and 208~MeV. Solid lines correspond to
first order chiral transitions, while dashed (dotted) lines correspond to
chiral (deconfinement) crossover-like transitions. Dash-dotted lines stand
for the cases in which both crossover-like transitions coincide. The green
bands indicate the regions in which $0.3 \leq \Phi \leq 0.5$.}
\label{rev-ch2.5-fig2}
\end{figure}

The phase diagrams corresponding to the three nonlocal NJL model
parameterizations introduced in Sec.~\ref{sec2.3} are displayed in
Fig.~\ref{rev-ch2.5-fig2}. Upper, central and lower panels show the results
for PA, PB and PC, respectively, while left (right) panels correspond to the
polynomial (logarithmic) potentials. In each panel, both the diagrams
associated to $T_0=270$ MeV and $T_0=208$ MeV are shown. Solid and dashed
lines indicate the critical temperatures for first order and crossover-like
chiral restoration transitions, dotted lines correspond to the peaks of the
PL susceptibility, and the green bands indicate the regions in which $0.3
\leq \Phi \leq 0.5$. It is observed that, as in the $\mu=0$ case, the
deconfinement and chiral restoration transitions occur at basically the same
critical temperature in the whole range of values of $\mu$ for which the
chiral transition is crossover-like. In fact, for the logarithmic potential
both transitions overlap, which is indicated by the dash-dotted lines in the
right panels of Fig.~\ref{rev-ch2.5-fig2}. For the polynomial potential, the
splitting between the critical temperatures $T_{\rm ch}$ and $T_\Phi$ does
not exceed the values obtained for $\mu=0$, namely  $|T_{\rm ch} - T_\Phi|
\lesssim 6$~MeV. In principle, the dotted lines (peaks of the PL
susceptibility) could also be extended to $\mu > \mu_{\rm CEP}$; at
$\mu=\mu_{\rm CEP}$ they are found to suffer a discontinuity, after which
they fall within the green bands. These lines are not shown in
Fig.~\ref{rev-ch2.5-fig2} since, as stated, in that region we find it
preferable to define the deconfinement transition temperatures through the
bands where $\Phi$ lies in the range from 0.3 to 0.5.

Concerning the character of the transitions, it is seen that the case of PA
and a logarithmic PL potential, with $T_0=208$ MeV, is the only one for
which the transition is always of first order. For all other
parametrizations and PL potentials, there is a CEP at which the chiral
transition changes its character from crossover to first order as $\mu$
grows. Once the transition becomes of first order (i.e.\ for $\mu > \mu_{\rm
CEP}$) the critical temperatures for chiral restoration start to decrease
quite fast as $\mu$ increases, reaching the $T=0$ axis at a certain value
value $\mu_{c}(0)$. Beyond that critical chemical potential, the system lies
in a phase where chiral symmetry is approximately restored for all values of
$T$. However, as mentioned above, for temperatures below the green bands
quarks are still confined and in these regions the system is in the
quarkyonic phase. Similar phase transition features have been found within
other effective approaches for low energy QCD, as e.g.~the quark-meson and
Polyakov-quark-meson models, both at the mean field
level~\cite{Schaefer:2007pw} and after the inclusion of meson
fluctuations~\cite{Herbst:2010rf,Herbst:2013ail,Pawlowski:2014zaa}. For
those models it is seen that the inclusion of beyond mean field corrections
can affect significantly the location of the CEP in the $\mu-T$ plane.

\begin{figure}[h!bt]
 \centerline{
 \includegraphics[width=0.7\textwidth]{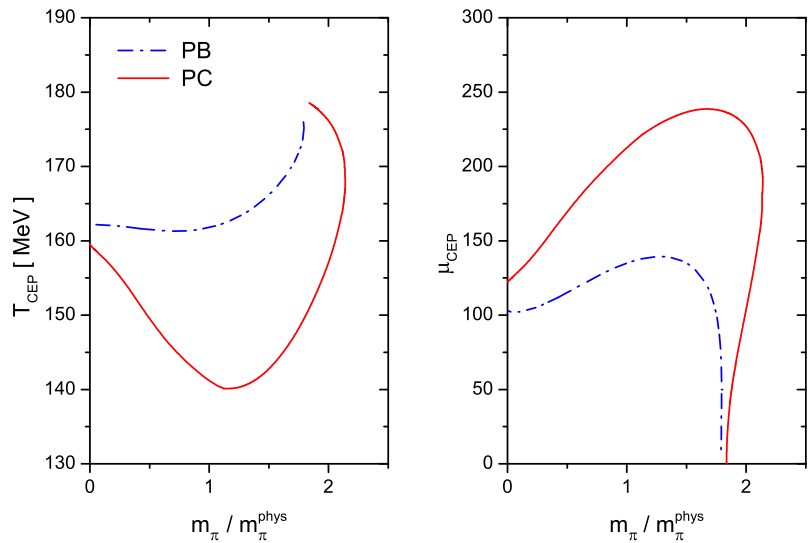}
}
\caption{(Color online) CEP temperature and chemical potential as functions of the pion
mass. The results correspond to the logarithmic PL potential, with
$T_0=208$~MeV.}
\label{rev-ch2.5-fig3}
\end{figure}

The position of the most relevant points in the phase diagrams are quoted in
Table~\ref{tab4}. Given a value of $T_0$, the main difference between
parametrizations PA, PB and PC resides in the location of the CEP. In
general, parametrization PA ---which does not account for WFR effects---
leads to lower values of $\mu_{\rm CEP}$. Comparing the results of PB and PC
we observe that the latter leads to somewhat lower values of $T_{\rm CEP}$
and higher values of $\mu_{\rm CEP}$. The position of the critical end point
is also quite sensitive to the amount of explicit breakdown of chiral
symmetry (i.e., the  size of current quark masses). This is illustrated in
Fig.~\ref{rev-ch2.5-fig3}, where we plot the values of $T_{\rm CEP}$ (left
panel) and $\mu_{\rm CEP}$ (right panel) as functions of the ratio
$m_\pi/m_\pi^{\rm phys}$. The results correspond to the logarithmic PL
potential, with $T_0=208$ MeV. As discussed in Sec.~\ref{sec2.4}, the values
of $m_\pi$ are obtained by varying the value of the current quark mass $m_c$
while keeping fixed  other model parameters. Note that for a pion mass
$m_\pi \sim 1.75\,m_\pi^{\rm phys}$ a second CEP appears at low chemical
potentials. This is due to the fact that the phase transition at $\mu=0$
becomes of first order, and consequently a crossover line connects the two
CEPs. As $m_\pi$ increases this crossover line gets shortened, until at
$m_\pi/m_\pi^{\rm phys} \sim 2$ the two CEPs meet. Beyond that value the
whole transition line becomes of first order. Although it is likely that the
predicted values for these critical pion masses are too low, it would be
interesting to verify if this behavior of the CEP position as a function of
the amount of explicit symmetry breaking is supported by other model
calculations.

\begin{table}
\begin{center}%
\begin{tabular}{c cc c cc c cc}
\hline \hline \\
  & \multicolumn{8}{c}{$T_0=208$} \vspace*{.1cm} \\ \hline \vspace*{-.1cm} \\
 & \multicolumn{2}{c}{PA} &&  \multicolumn{2}{c}{PB} &&  \multicolumn{2}{c}{PC}  \vspace*{.1cm} \\
 \cline{2-3}\cline{5-6}\cline{8-9}  \vspace*{-.2cm}  \\
              & \hspace*{.1cm} Pol \hspace*{.1cm}   & \hspace*{.1cm}   Log \hspace*{.1cm}
              && \hspace*{.1cm} Pol \hspace*{.1cm}   & \hspace*{.1cm}   Log \hspace*{.1cm}
               && \hspace*{.1cm} Pol \hspace*{.1cm}   & \hspace*{.1cm}    Log \hspace*{.1cm}
                \\
\hline
$T_\chi(0)$       &  178 & 174 && 178 & 171 && 181 & 173 \\
$T_\Phi(0)$        &  175 & 174 && 174 & 171 && 175 & 173\\
$\mu_{\rm CEP}$    &  135 & $-$ && 180 & 135 && 227 & 213 \\
$T_{\rm CEP}$      &  162 & $-$ && 147 & 162 && 125 & 141 \\
$\mu_c(0)$     &   322 & 322 &&  312 & 312 &&  298 & 298 \\
 \hline \hline \\
  & \multicolumn{8}{c}{$T_0=270$}\vspace*{.1cm} \\ \hline \vspace*{-.1cm} \\
 & \multicolumn{2}{c}{PA} &&  \multicolumn{2}{c}{PB} &&  \multicolumn{2}{c}{PC}  \vspace*{.1cm} \\
      \cline{2-3}\cline{5-6}\cline{8-9} \vspace*{-.2cm}  \\
              & \hspace*{.1cm} Pol \hspace*{.1cm}   & \hspace*{.1cm}   Log \hspace*{.1cm}
              && \hspace*{.1cm} Pol \hspace*{.1cm}   & \hspace*{.1cm}   Log \hspace*{.1cm}
               && \hspace*{.1cm} Pol \hspace*{.1cm}   & \hspace*{.1cm}    Log \hspace*{.1cm}
                \\
\hline
$T_\chi(0)$       & 210 & 210  && 210 & 210 && 214 & 215\\
$T_\Phi(0)$       & 206 & 210 && 206 & 210 && 210 & 215\\
$\mu_{\rm CEP}$     & 160 & 132  && 193 & 182 && 232 & 235 \\
$T_{\rm CEP}$      & 181 & 198  && 165 & 182 && 142 & 154 \\
$\mu_c(0)$     &  322 & 322 && 312 & 312 && 298 & 298 \\
\hline \hline
\hfill
\end{tabular}
\caption{Positions of some characteristic points in the $\mu - T$ phase
diagram for various nlPNJL model parametrizations. All values are given in
MeV.}
\label{tab4}
\end{center}
\end{table}

\subsection{Extension to imaginary chemical potential}
\label{sec2.6}

In this section we address the extension of previous analyses to the case of
imaginary chemical potential. As mentioned above, this deserves significant
theoretical interest, since for imaginary $\mu$ lattice QCD calculations do
not suffer from the sign
problem~\cite{Alford:1998sd,D'Elia:2002gd,deForcrand:2002hgr,Wu:2006su}, and the
corresponding results can be compared with the predictions arising from
effective models. It is seen that lattice data, as well as analyses based on
the exact renormalization group equations~\cite{Braun:2009gm}, suggest a
close relation between the deconfinement and chiral restoration transitions
for imaginary chemical potentials. In addition, the behavior of physical
quantities in the region of imaginary chemical potential is expected to have
implications on the QCD phase diagram at finite real values of
$\mu$~\cite{Karbstein:2006er,Bonati:2015bha,Bellwied:2015rza,Borsanyi:2020fev}.

As shown by Roberge and Weiss (RW)~\cite{Roberge:1986mm}, the QCD
thermodynamic potential in the presence of an imaginary chemical potential
$\mu = i \theta T$ is invariant under the so-called extended $Z_3$ symmetry
transformations, which are  given by a combination of a $Z_3$ transformation
of the quark and gauge fields and a shift $\theta \rightarrow \theta +
2k\pi/3$. The RW symmetry is a remnant of the $Z_3$ symmetry that exists in
the pure gauge theory. In QCD with dynamical quarks, if the temperature is
larger than a certain value $T_{RW}$ it can be seen that three $Z_3$ vacua
appear. They can be classified according to the corresponding Polyakov loop
phases, viz. $\varphi$, $\varphi+2 \pi/3$ and $\varphi+4 \pi/3$. It should
be stressed that, as mentioned in Sec.~\ref{sec2.2}, for imaginary chemical
potential the restriction of having $\Phi \, \in \, \mathbb{R}$ in order to
get a real thermodynamic potential is lost. Thus, both $\phi_3$ and $\phi_8$
can be nonvanishing. For $T > T_{RW}$ there is a first order phase
transition at $\theta = \pi/3$ mod $2\pi/3$ ---known as the  ``Roberge-Weiss
transition''--- in which the vacuum jumps to one of its $Z_3$ images. The
point at the end of the RW transition line in the $(T, \theta)$ plane, i.e.
$(T, \theta)  = (T_{RW}, \pi/3)$, is known as the ``RW end point''. The
order of the RW transition at the RW end point has been subject of
considerable interest in the last years in the framework of lattice
QCD~\cite{D'Elia:2007ke,D'Elia:2009qz,deForcrand:2010he,Bonati:2010gi,Bonati:2018fvg,Goswami:2018qhc},
due to the implications it might have on the QCD phase diagram at finite
real chemical potential. According to $N_f=2$ LQCD calculations, it appears
that the RW end point is first order for realistically small values of the
current quark mass.

It is not difficult to show that the RW symmetry is also present in nlPNJL
models~\cite{Pagura:2011rt,Kashiwa:2011td}. This can be done as follows. The
last two terms of the thermodynamic potential in the r.h.s.~of
Eq.~(\ref{ch2.2-eq9}) are clearly invariant under the transformations
\begin{equation}
\Phi(\theta) \ \rightarrow \ \Phi(\theta) \exp(-i\, 2\,k\, \pi/3) \ , \qquad
\qquad
\theta \ \rightarrow \ \theta + 2\, k\, \pi/3 \ .
\label{ch2.6-eq1}
\end{equation}
To check the invariance of the first term, notice that the first of
the transformations in Eq.~(\ref{ch2.6-eq1}) can be obtained through
\begin{equation}
\phi_3(\theta) \ \rightarrow \ \phi_3(\theta) \ , \qquad \qquad
\phi_8(\theta) \ \rightarrow \ \phi_8(\theta) - 2 \, k\,\pi\,T/\sqrt{3}\ .
\label{ch2.6-eq2}
\end{equation}
Taking into account Eqs.~(\ref{ch2.6-eq1}) and (\ref{ch2.6-eq2}), from the
definition of $(\rho_{n, \vec{p}}^c)^2$ in Eq.~(\ref{ch2.2-eq9}) it is easy
to see that any sum of the form $\sum_{c=r,g,b}F[(\rho_{n, \vec{p}}^c)^2]$,
where $F$ is an arbitrary function, is invariant under the extended $Z_3$
symmetry transformations. The invariance of the terms introduced
through the regularization of the thermodynamic potential can be shown in a
similar way. As a further evidence of the $Z_3$ invariance of the model, it
is interesting to study the behavior of the order parameters as functions of
$T$ and $\theta$. For this analysis it is useful to introduce an ``extended
traced Polyakov loop'' $\Psi$, defined by $\Psi = \exp(i\theta)\Phi$. The
latter is invariant by construction under the transformations in
Eq.~(\ref{ch2.6-eq1}), and its phase $\psi$ can be taken as order parameter
of the RW transition~\cite{Sakai:2008um}.

\begin{figure}[h!bt]
\centerline{
\includegraphics[width=0.8\textwidth]{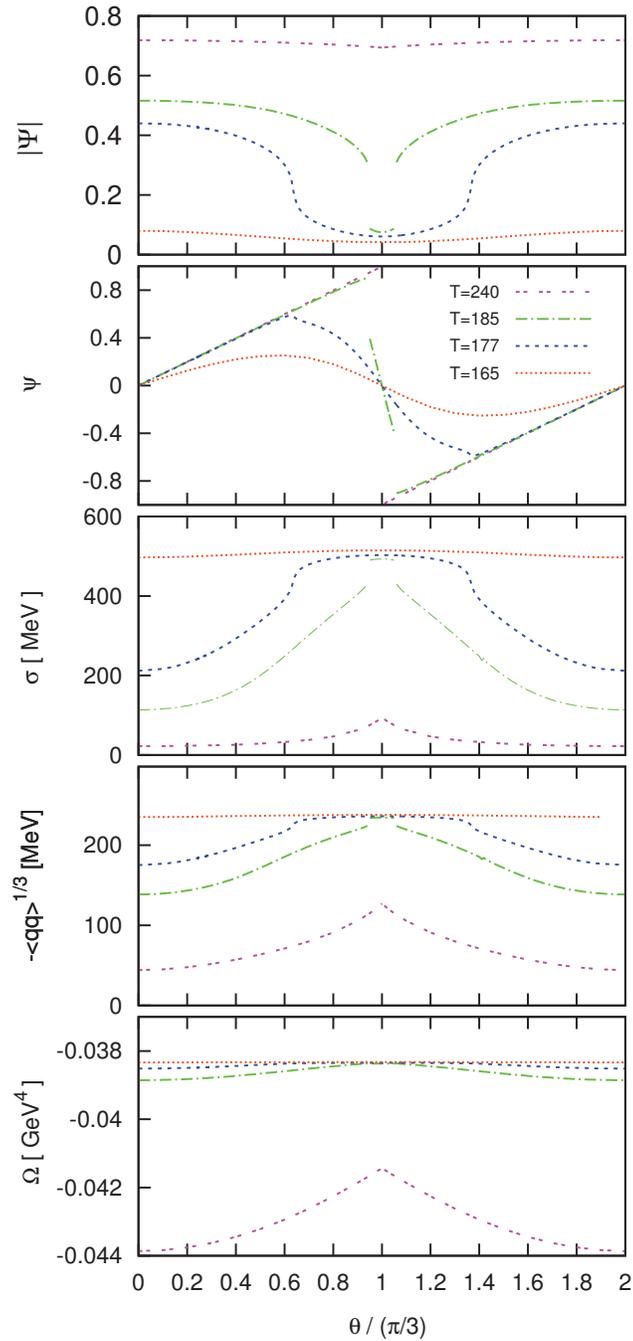}
} \caption{(Color online) Results for some relevant quantities as functions
of $\theta/(\pi/3)$, for fixed values of the temperature. The observed
periodicity is due to the fact that the nlPNJL model is invariant under the
extended $Z_3$ symmetry transformations. The results correspond to PB  and a
logarithmic PL potential, with $T_0=208$ MeV.}
\label{periodic}
\end{figure}

In what follows we discuss some of the results obtained using the nlPNJL
model parametrizations introduced in
Sec.~\ref{sec2.3}~\cite{Pagura:2011rt,Pagura:2013rza}. Similar results
arising from some alternative parametrization have been quoted in
Ref.~\cite{Kashiwa:2011td}, while results from the Polyakov-quark-meson
model, including meson fluctuations, can be found in
Ref.~\cite{Morita:2011jva}. First, let us keep $T$ constant and verify the
periodicity of thermodynamic quantities as functions of $\theta$. For
definiteness we concentrate for now in the case of the logarithmic
potential. In Fig.~\ref{periodic} we display, from above to below, the
modulus of the extended PL, the phase $\psi$, the mean field value of the
$\sigma_1$ field, the chiral condensate and the thermodynamic potential, as
functions of $\theta/(\pi/3)$ for various values of $T$. The curves
correspond to parametrization PB, for $T_0=208$~MeV. Qualitatively similar
results are obtained for PC with $T_0=208$~MeV, as well as for all three
parametrizations PA, PB and PC with $T_0=270$~MeV. In the case of PA with
$T_0 = 208$~MeV, although the same periodicity is observed, all curves turn
out to be discontinuous for temperatures in the transition region.
For $T>T_{RW}$ one finds the above mentioned RW first order phase transition
at $\theta=\pi/3$, which is signalled by a discontinuity in the phase $\psi$
of the extended Polyakov loop.

\begin{figure}[h!bt]
\centering
\includegraphics[width=0.53\textwidth]{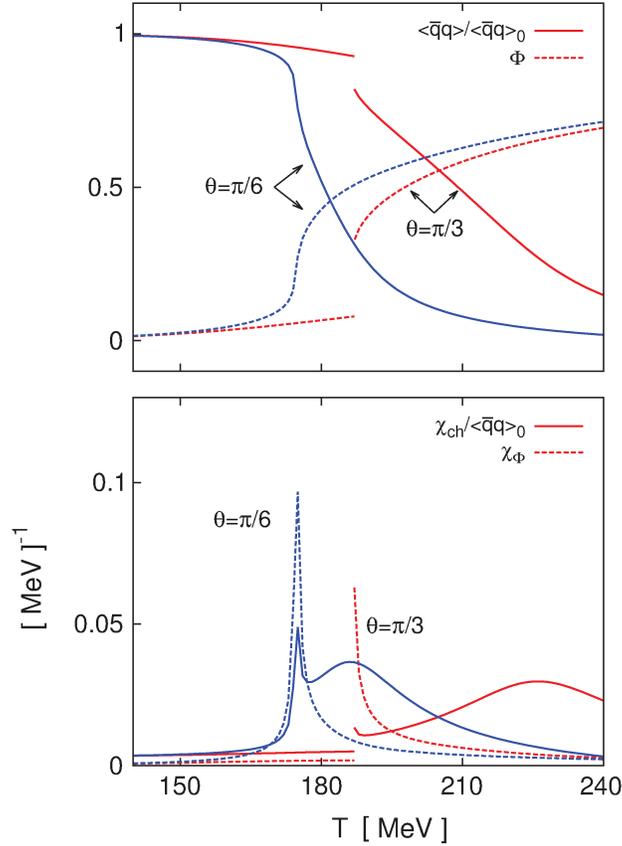}
\caption{(Color online) Order parameters for the deconfinement and chiral restoration
transitions (upper panel) and associated susceptibilities (lower
panel) as functions of $T$, for $\theta=\pi/6$ and $\theta= \pi/3$. The
results correspond to parametrization PB, with $T_0=208$ MeV.}
\label{suc}
\end{figure}

On the other hand, by looking at the behavior of the order parameters and
susceptibilities as functions of the temperature one can find signals of
both deconfinement and chiral symmetry restoration transitions. This is
clearly seen in Fig.~\ref{suc}, where we show the curves for the normalized
quark condensate, the traced PL and the corresponding susceptibilities
$\chi_{\rm ch}$ and $\chi_\Phi$ as functions of $T$, taking $\theta$ fixed
at the representative values $\theta=\pi/6$ and $\pi/3$. The results
correspond once again to parametrization PB with $T_0=208$ MeV. For
$\theta=\pi/6$, both the deconfinement and chiral restoration transitions
are found to be crossover-like. Moreover, it can be seen that the
susceptibilities associated to both order parameters show peaks at a common
temperature. This can be interpreted as a signal indicating an entanglement
between the transitions. However, notice that $\chi_{\rm ch}$ shows at a
larger temperature an additional, broader peak. For $\theta=\pi/3$, one
observes at $T\simeq 190$~MeV a jump in $\Phi$ that can be understood as a
first order deconfinement phase transition. However, in the case of the
chiral condensate the corresponding gap is relatively small and, moreover,
beyond this discontinuity one still finds the broad peak in $\chi_{\rm ch}$.
Consequently, we find it reasonable to identify the chiral restoration
temperature through the maximum of this broad peak. We conclude that for
$\theta=\pi/3$ only the deconfinement transition is of first order, while
the chiral restoration still proceeds as a crossover. For PC with $T =
208$~MeV and for all three parametrizations with $T_0=270$~MeV the
situation is very similar, whereas for PA with $T_0=208$~MeV the transition
is of first order for any value of $\theta$ in the considered range of
temperatures.

\begin{figure}[h!bt]
\centerline{
\includegraphics[width=1.\textwidth]{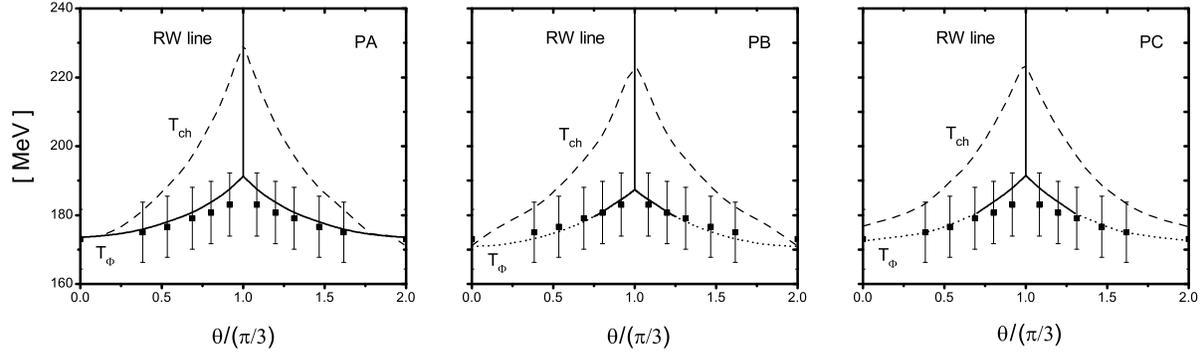}
}
\caption{Critical temperatures as functions of $\theta$ for parametrizations
PA, PB and PC. First order transitions are indicated with solid lines, while
dotted and dashed lines correspond to deconfinement and chiral restoration
transitions, respectively, in the regions where they are crossover-like.
Vertical solid lines indicate the first order RW transition. The fat dots
correspond to lattice QCD results}
\label{phdiag_muim208}
\end{figure}

In Fig.~\ref{phdiag_muim208} we quote the critical temperatures as functions
of $\theta$ (normalized to $\pi/3$) for parametrizations PA, PB and
PC~\cite{Pagura:2011rt}. The results correspond to the logarithmic PL
potential, with $T_0=208$~MeV. For comparison, values obtained from LQCD
calculations taken from Ref.~\cite{Wu:2006su} are also shown. They have an
error of about $10\%$ due the uncertainty in the determination of $T_{\rm
ch}(\theta=0)$. From the figure it is seen that the predictions of the
nlPNJL model for the deconfinement critical temperatures are compatible with
LQCD data. Moreover, the values of $T_{RW}$ obtained within the model are
found to be 191, 188 and 191~MeV for PA, PB and PC, respectively, in good
agreement with the value $T_{RW} = 185(9)$~MeV arising from $N_f=2$ LQCD
calculations~\cite{Wu:2006su} (for $N_f=2+1$, LQCD results seem to favor a
slightly larger value~\cite{Bonati:2016pwz}).

\begin{figure}
\centerline{
\includegraphics[width=0.45\textwidth]{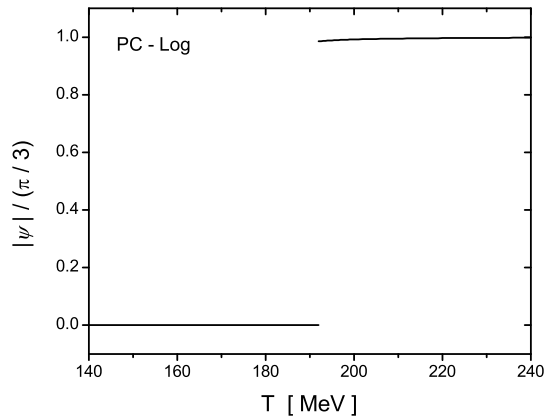}
}
\caption{Phase of the extended PL loop $\Psi=\exp(i\theta)\Phi$ as a
function of $T$ for $\theta=\pi/3$. The results correspond to
parametrization PC and the logarithmic PL potential, with $T_0 = 208$~MeV.}
\label{rev-ch2.6-fig4}
\end{figure}

It is also seen that there is a splitting between chiral restoration and
deconfinement critical temperatures, which gets larger when $\theta$
increases. As already mentioned, there is a critical value $\theta_{\rm CEP}
\sim 0.7 \times \pi/3$ above which the deconfinement transition is of first
order for all three parametrizations. Therefore, it is found that in all
cases the transition lines are of first order when they reach the RW end
point. This implies that the RW end point is a triple point, and the RW
transition is also of first order there. This can be clearly seen in
Fig.~\ref{rev-ch2.6-fig4}, which shows the behavior of the phase $\psi$ as a
function of $T$, for $\theta=\pi/3$.

Up to now we have discussed the results obtained considering a logarithmic
PL potential. For the polynomial potential, the main novel qualitative
feature is that there is no first order deconfinement transition. As a
consequence, for this PL potential the RW transition at the end point is of
second order for all considered parametrizations. This appears to be in
contradiction with LQCD results reported in
Refs.~\cite{D'Elia:2007ke,D'Elia:2009qz,deForcrand:2010he,Bonati:2010gi},
which indicate that such a transition is of first order for a physical pion
mass. Concerning the predictions for $T_{RW}/T_{\rm ch}(\theta =
0)$, they are somewhat larger than those found for the logarithmic PL
potential. On the other hand, the ratios $T_{\rm ch}(\theta =
\pi/3)/T_{\rm ch}(\theta = 0)$ for $T_0=208$~MeV ($T_0=270$~MeV) are similar
(larger) to those obtained using the logarithmic potential.

\subsection{Meson properties at finite temperature}
\label{sec2.7}

To conclude this section, in this last subsection we concentrate on the
thermal behavior of some $\sigma$ and $\pi$ meson properties. In the context
of nlPNJL models, the relevant theoretical expressions can be obtained from
those given in Sec.~\ref{sec2.1} using the Matsubara formalism, as described
in Sec.~\ref{sec2.2}. In this way, the finite temperature extension of the
quadratic term in the Euclidean action [see Eq.~(\ref{ch2.1.1-eq12})] reads
\begin{equation}
 S_E^{\mathrm{quad}} \ = \ \frac{1}{2}\sum_{M=\sigma,\sigma',\pi}
 \ \int_{q,m} G_M(\vec q^{\, 2},
\nu_m^2)\;  \delta M(q_m)\, \delta M(-q_m) \ ,
\label{ch2.7-eq1}
\end{equation}
where $q_m=(\vec q, \nu_m)$, with $\nu_m=2\pi mT$. Here the notation
\begin{equation}
\int_{q,m} \ \equiv \ T \sum_{m=-\infty}^\infty\int\frac{d^3q}{(2\pi)^3}
\end{equation}
has been used. For charged and neutral pions one has
\begin{eqnarray}
G_{\pi}(\vec q^{\, 2}, \nu_m^2) & = & \frac{1}{G_S}-8\sum_c\int_{p,n} g\Big(p_{nc} +
  \frac{q_m}{2}\Big)^2 \
  \frac{Z(p_{nc}+q_m)Z(p_{nc})}{D(p_{nc}+q_m)D(p_{nc})}
\nonumber\\
      && \times \left[p_{nc}^2 + p_{nc}\cdot q_m + M(p_{nc}+q_m) M(p_{nc})\right]\ ,
\label{ch2.7-eq2}
\end{eqnarray}
where $p_{nc} = (\vec p, (2n+1)\pi T  - \phi_c)$ and
$D(p)=p^2+M(p)^2$. In the case of the $\sigma$ and $\sigma'$ mesons, the
$G_M$ functions are given by
\begin{eqnarray}
 G_{{\sigma \choose {\sigma'}}}(\vec q^{\, 2}, \nu_m^2) & = &
 \frac{G_{\sigma_1\sigma_1}(\vec q^{\, 2}, \nu_m^2)+G_{\sigma_2\sigma_2}(\vec q^{\, 2}, \nu_m^2)}{2}\nonumber\\
 &&\mp\, \sqrt{\left[G_{\sigma_1\sigma_2}(\vec q^{\, 2}, \nu_m^2)\right]^2 +
 \left[\frac{G_{\sigma_1\sigma_1}(\vec q^{\, 2}, \nu_m^2)-G_{\sigma_2\sigma_2}(\vec q^{\, 2},
 \nu_m^2)}{2}\right]^2}\ ,
\label{ch2.7-eq3}
\end{eqnarray}
where
\begin{eqnarray}
G_{\sigma_1\sigma_1}(\vec q^{\, 2}, \nu_m^2)&=&\frac{1}{G_S}-8\sum_c\int_{p,n}\,g\left(p_{nc} +
                                            \frac{q_m}{2}\right)^2
                                            \frac{Z(p_{nc}+q_m)Z(p_{nc})}{D(p_{nc}+q_m)D(p_{nc})}
                                    \nonumber\\
      && \times \left[p_{nc}^2 + p_{nc}\cdot q_m - M(p_{nc}+q_m) M(p_{nc})\right],
\nonumber\\
 G_{\sigma_2\sigma_2}(\vec q^{\, 2},
 \nu_m^2)&=&\frac{1}{G_S}+\frac{8}{\varkappa_p^2}\sum_c\int_{p,n}\,\left(p_{nc}+\frac{q_m}{2}\right)^2
 \, f\left(p_{nc} + \frac{q_m}{2}\right)^2
 \frac{Z(p_{nc}+q_m)Z(p_{nc})}{D(p_{nc}+q_m)D(p_{nc})}
\nonumber\\
&&\times\left[p_{nc}^2 + p_{nc}\cdot q_m - M(p_{nc}+q_m) M(p_{nc})+
\frac{p^2_{nc}(p_{nc}+q_m)^2 - (p^2_{nc} + p_{nc}\cdot q_m)^2}{2
(p_{nc}+q_m/2)^2}\right]\ ,
\nonumber\\
 G_{\sigma_1\sigma_2}(\vec q^{\, 2}, \nu_m^2)&=&-\frac{8}{\varkappa_p^2}\sum_c\int_{p,n}\,g\left(p_{nc} + \frac{q_m}{2}\right)f\left(p_{nc} + \frac{q_m}{2}\right)
\frac{Z(p_{nc}+q_m)Z(p_{nc})}{D(p_{nc}+q_m)D(p_{nc})}\nonumber\\
&&\times \left( p_{nc} + \frac{q_m}{2}\right) \cdot \left[ p_{nc} \, M(p_{nc}+ q_m) + (p_{nc} + q_m)
M(p_{nc})\right]\ .
\label{ch2.7-eq4}
\end{eqnarray}

As in the $T=0$ case discussed in Sec.~\ref{sec2.1}, the meson masses can be
found by looking for the poles of the corresponding propagators. In the
present case they are given by the solutions of the equations
\begin{equation}
 G_M(-m_M^2,0) \ = \ 0 \ .
\label{ch2.7-eq5}
\end{equation}
The masses obtained in this way correspond to the spatial screening
masses associated with Matsubara zero modes. They determine a behavior $\sim
\exp(-m_M\, r)$ in configuration space, i.e., the reciprocals $m_M^{-1}$
describe the persistence lengths of zero modes in equilibrium with the
thermal bath. In fact, these masses are the quantities that are usually
studied in LQCD calculations~\cite{Karsch:2003jg}. It should be noted that
one has a screening mass for each Matsubara mode.

Another relevant physical quantity to be studied is the pion decay constant
$f_\pi$. Its finite temperature behavior provides another way to
characterize the chiral restoration transition. To obtain the corresponding
expression at finite temperature we replace $F_0$ in Eq.~(\ref{ch2.1.3-eq7})
by
\begin{eqnarray}
      F_0(\vec q^{\, 2}, \nu_m^2) &=& 8\sum_c\int_{p,n} { g\left(p_{nc} +
      \frac{q_m}{2}\right)}
                                            \frac{Z(p_{nc}+q_m)Z(p_{nc})}{D(p_{nc}+q_m)D(p_{nc})}
                                    \nonumber\\
      && \times\ \left[p_{nc}^2 + p_{nc}\cdot q_m + M(p_{nc}+q_m)
      M(p_{nc})\right]\ ,
\label{ch2.7-eq6}
\end{eqnarray}
to be evaluated at $(\vec q^{\, 2}, \nu_m^2) = (-m_\pi^2,0)$.

\begin{figure}[h!bt]
\centerline{
\includegraphics[width=0.77\textwidth]{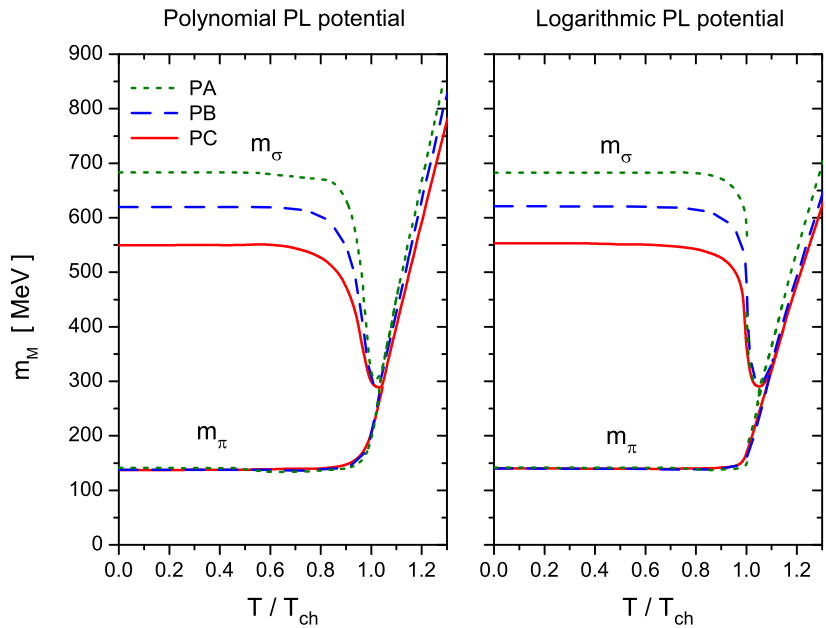}
}
\caption{(Color online) Masses of $\sigma$ and $\pi$ mesons as functions
of $T/T_{\rm ch}$ for polynomial (left) and logarithmic (right) PL
potentials, with $T_0 = 208$~MeV. In each panel the results for
parametrizations PA, PB and PC are shown.}
\label{rev-ch2.7-fig1}
\end{figure}

Let us analyze the numerical results obtained for the above meson properties
in the context of nlPNJL models. We start by discussing the behavior of
$\sigma$ and $\pi$ meson masses. As mentioned in Sec.~\ref{sec2.3}, at $T=0$
one gets $m_\sigma = 683$, 622 and 554~MeV for parametrizations PA, PB and
PC, respectively, while the value $m_\pi(T=0)=139$~MeV has been taken in all
cases as one of the inputs for the determination of the model parameters. In
Fig.~\ref{rev-ch2.7-fig1} we show the behavior of $\sigma$ and $\pi$ masses
as functions of $T/T_{\rm ch}$, for polynomial (left) and logarithmic
(right) PL potentials, taking $T_0 = 208$~MeV. In each panel the curves
corresponding to PA, PB and PC are displayed. Starting from $T=0$, it is
seen that if the temperature is increased the masses remain almost constant
up approximately the chiral restoration critical temperature. Close to that
temperature the $\sigma$ mass shows a sudden drop and the $\pi$ mass starts
to increase, and both curves meet at $T\gtrsim T_{\rm ch}$. Then the
masses of both chiral partners grow together towards their asymptotic value
$2\pi T$, associated with an uncorrelated $q\bar q$
pair~\cite{Florkowski:1993br,Eletsky:1988an}. By analyzing the curves in
more detail it can be seen that for the largest temperature considered in
the graph such a limit has not been reached yet. This is due to the fact
that for $T/T_{\rm ch} \sim 1.4$ the contribution of the PL parameter
$\phi_3$ to the quark screening masses still is nonnegligible. In fact,
after the restoration of chiral symmetry the curves in the figure grow
according to $m_M = 2(\pi T - \phi_3)$, behaving as straight lines with
approximately the same slope. Finally, it should be noted that in the right
panel (corresponding to the logarithmic potential) there is a discontinuity
in the masses for parametrization PA. This is associated with the fact that
the chiral restoration transition is of first order in that case. Except for
this particularity, the thermal behavior of the $\sigma$ and $\pi$ masses is
qualitatively similar in all {considered cases}. Results for $T_0=270$~MeV,
which also include medium induced Lorentz braking effects, can be found in
Ref.~\cite{Benic:2013eqa}. They turn out to be very similar to those shown
in Fig.~\ref{rev-ch2.7-fig1}.

\begin{figure}[h!bt]
\centerline{
 \includegraphics[width=0.85\textwidth]{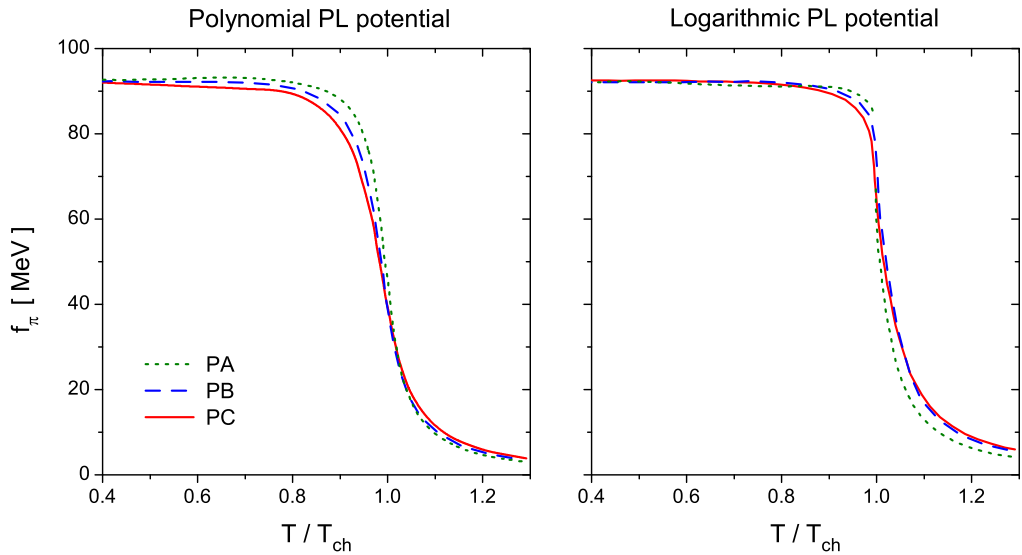}
}
\caption{(Color online) Pion decay constant $f_\pi$ as a function
of $T/T_{\rm ch}$ for polynomial (left) and logarithmic (right) PL
potentials, with $T_0 = 208$ MeV. In each panel the results for
parametrizations PA, PB and PC are shown.}
\label{rev-ch2.7-fig2}
\end{figure}

Finally, in Fig.~\ref{rev-ch2.7-fig2} we show the temperature dependence of
the pion decay constant $f_\pi$, for the three parametrizations under
consideration. Left and right panels correspond to polynomial and
logarithmic PL potentials, respectively. It is found that, again, the only
case that presents a distinct behavior is that of PA with the logarithmic PL
potential, where the curves show a discontinuity at the transition
temperature.

%% file: Sec3-rev.tex
\section{Two-flavor nonlocal NJL model including vector and axial vector {quark currents}}
\label{sec3}

In this section we study an extension of the previously considered nlNJL
models in which vector and axial vector quark current-current interactions
are included. This allows for a description of vector and axial vector meson
phenomenology. The cases of both zero and finite temperature and/or quark
chemical potential are discussed.

\subsection{Formalism at vanishing temperature and {chemical potential}}
\label{sec3.1}

\subsubsection{Effective action and { mean field} equations}
\label{sec3.1.1}

As stated, {in what follows} we analyze a two-flavor nlNJL chiral
quark model that includes nonlocal vector and axial vector
quark-antiquark currents. This {type of model has been studied}
e.g.\ in
Refs.~\cite{Villafane:2016ukb,Contrera:2012wj,Contrera:2016rqj,Carlomagno:2019yvi}.
We consider an effective Euclidean action given
by~\cite{Villafane:2016ukb}
\begin{eqnarray}
\hspace{-0.5cm}
S_E &=& \int d^4x \ \left\{ \bar\psi(x)(-\,i \slashed
\partial + m_c)\psi(x)-\frac{G_S}{2}\Big[ j_S(x)j_S(x) + \vec \jmath_P(x)\cdot\vec \jmath_P(x) + j_R(x)j_R(x)\Big] \right.
\nonumber\\
    &&\left.
    -\,\frac{G_V}{2}\Big[
\vec \jmath_{V\mu}(x)\cdot\vec \jmath_{V\mu}(x) + \vec \jmath_{A\,\mu}(x)\cdot\vec \jmath_{A\,\mu}(x)\Big]
 -\frac{G_0}{2}\; j^{\,0}_{V\mu}(x)\, j^{\,0}_{V\mu}(x)
 -\frac{G_5}{2}\; j^{\,0}_{A\, \mu}(x)\,j^{\,0}_{A\, \mu}(x)\right\}\ .
\label{ch3.1.1-eq1}
\end{eqnarray}
As in Sec.~\ref{sec2} we work in the isospin limit; thus, we
assume the current mass to be the same for $u$ and $d$ quarks. The
model includes the {four-fermion} couplings studied in
Sec.~\ref{sec2.1.1} [see Eqs.~(\ref{ch2.1.1-eq1}) and
(\ref{ch2.1.1-eq2})], plus additional vector and axial vector
pieces. The new quark-antiquark currents read
\begin{eqnarray}
\vec \jmath_{V \mu}(x) &=& \int d^4z\; {\cal H}(z)\,
\bar\psi\left(x+\frac{z}{2}\right)\vec\tau \gamma_\mu
\psi\left(x-\frac{z}{2}\right)
\ , \nonumber\\
\vec \jmath_{A\, \mu}(x) &=& \int d^4z\; {\cal H}(z)\,
\bar\psi\left(x+\frac{z}{2}\right)\vec \tau \gamma_\mu \gamma_5
\psi\left(x-\frac{z}{2}\right)
\ , \nonumber\\
j^{\,0}_{V\mu}(x)    &=& \int d^4z\; {\cal H}_0(z)\,
\bar\psi\left(x+\frac{z}{2}\right) \gamma_\mu
\psi\left(x-\frac{z}{2}\right)
\ , \nonumber\\
j^{\,0}_{A\,\mu}(x)    &=& \int d^4z\; {\cal H}_5(z)\,
\bar\psi\left(x+\frac{z}{2}\right) \gamma_\mu\gamma_5
\psi\left(x-\frac{z}{2}\right)\ ,
\label{ch3.1.1-eq1}
\end{eqnarray}
where the functions ${\cal H}(z)$, ${\cal H}_0(z)$ and ${\cal H}_5(z)$ are
covariant nonlocal form factors. Notice that, in order to guarantee chiral
invariance, the couplings involving the currents $\vec \jmath_{V\mu}$ and
$\vec \jmath_{A\,\mu}$ have to carry the same form factor ${\cal H}(z)$.

As discussed in Sec.~\ref{sec2.1.1}, for the study of meson phenomenology it
is convenient to perform a bosonization of the fermionic theory. In the
present case, this is done by introducing the auxiliary bosonic fields
$\sigma_1(x)$, $\sigma_2(x)$ and $\pi_a(x)$ previously considered in
Sec.~\ref{sec2}, as well as new fields $v^0_{\mu}(x)$, $v^a_\mu(x)$ (vector)
and $a^0_{\mu}(x)$, $a^a_\mu (x)$ (axial vector), where indices $a$ run from
1 to 3. After integrating out the fermion fields the partition function can
be written as
\begin{eqnarray}
\mathcal{Z} &=& \int {\rm D}\sigma_1\, {\rm D}\sigma_2\,
{\rm D}\,\vec{\pi}\, {\rm D}\,v^0_{\mu}\,
{\rm D}\,a^0_{\mu}\, {\rm D}\, \vec{v}_{\mu}\,
{\rm D}\,\vec{a}_{\mu} \; \exp\big(\!-S_E^{\rm bos}\big) \ ,
\label{ch3.1.1-eq2}
\end{eqnarray}
where $S_E^{\rm bos}$ stands for the Euclidean bosonized action.
In momentum space, the latter is given by
\begin{eqnarray}
S^{\rm bos}_{E} &=& -\ln\,\det {\cal D}  +\int
\dfrac{d^{4}q}{(2\pi)^{4}}
\left\lbrace\dfrac{1}{2G_S}\left[\sigma_1(q)\sigma_1(-q)+
\vec{\pi}(q)\cdot\vec{\pi}(-q)+\sigma_2(q)\sigma_2(-q)\right]\right. \nonumber\\
 & &  \hspace{-0.5cm}
+\left. \dfrac{1}{2G_V} \left[
\vec{v}_{\mu}(q)\cdot\vec{v}_{\mu}(-q)+
\vec{a}_{\mu}(q)\cdot\vec{a}_\mu(-q)\right] +
\dfrac{1}{2G_0}\;v^0_{\mu}(q) v^0_\mu(-q) +
\dfrac{1}{2G_5}\;a^0_{\mu}(q) a^0_\mu(-q) \right\rbrace\ ,
\label{ch3.1.1-eq3}
\end{eqnarray}
where the operator ${\cal D}(p,p^\prime)$ reads
\begin{eqnarray}
\label{A} {\cal D}(p,p^\prime)&=& (2\pi)^4
\delta^{(4)}(p-p^\prime)(\rlap/p + m_c) +
                 g(\bar p)\, \bigg[\sigma_1(p-p^\prime)+i \gamma_5
                 \vec{\tau} \cdot \vec{\pi}(p-p^\prime)\bigg]
                 \nonumber\\
             & &
             -\ f(\bar p)\,\dfrac{\rlap/\bar p }{\varkappa_p}\,\sigma_2(p-p^\prime)\
             +\ h(\bar p)\, \gamma_{\mu}\bigg[
             \vec{\tau} \cdot \vec{v}_{\mu}(p-p^\prime) +
             \gamma_5\, \vec{\tau} \cdot \vec{a}_{\mu}(p-p^\prime)\bigg]
                 \nonumber\\
             & &
             +\ h_0(\bar p)\,\gamma_{\mu}\; v^0_{\mu}(p-p^\prime)
             +\ h_5(\bar p)\,\gamma_{\mu}\gamma_5\; a^0_{\mu}(p-p^\prime)
             \ ,
\label{ch3.1.1-eq4}
\end{eqnarray}
with $\bar p \equiv (p+p')/2$. The functions $f(p)$, $g(p)$, $h(p)$,
$h_0(p)$, and $h_5(p)$ stand for the Fourier transforms of the form factors
entering the nonlocal currents. Without loss of generality, the coupling
constants can be chosen in such a way that the form factors are normalized
to $f(0) = g(0) = h(0) = h_0(0) = h_5(0) = 1$. Next, the bosonic fields can
be expanded around their vacuum expectation values, $\phi(x) = \bar \phi +
\delta\phi(x)$. {As done in} Sec.~\ref{sec2.1.1}, on the basis of charge,
parity and Lorentz symmetries, it is assumed that in vacuum only
$\sigma_1(x)$ and $\sigma_2(x)$ have nontrivial translational invariant mean
field values. {These are denoted by} $\bar{\sigma}_1$ and
$\varkappa_p\,\bar\sigma_2$, respectively, {while} vacuum expectation values
of the remaining bosonic fields are taken to be zero. {Notice that,} as
discussed below, at nonvanishing chemical potential Lorentz symmetry is
broken, and the mean field expectation value of $v_4^0(x)$ can also be
nonzero. Following similar steps as those described in Sec.~\ref{sec2.1.1},
the bosonized effective action in Eq.~(\ref{ch3.1.1-eq3}) can be expanded in
powers of meson fluctuations as
\begin{equation}
S_{E}^{\rm bos}\ =\ S_{E}^{\rm MFA}\; + \;S_{E}^{\rm
quad}\;+\;\dots\
\label{ch3.1.1-eq5}
\end{equation}
It turns out that the mean field piece $S_{E}^{\rm MFA}$ is the same as the
one given in Eq.~(\ref{ch2.1.1-eq9}). Therefore, the corresponding gap
equations coincide with Eqs.~(\ref{ch2.1.2-eq1a}) and (\ref{ch2.1.2-eq1b}).
As found in
Sec.~\ref{sec2.1.1}, the mean field quark propagator is given by
\begin{equation}
\mathcal{S}_{0}(p)\ = \mathcal{D}_{0}(p)^{-1}\ =\ \frac{Z(p)}{\rlap/p+M(p)}\ ,
\label{ch3.1.1-eq5a}
\end{equation}
with
\begin{eqnarray}
Z(p) = \left(  1-\bar{\sigma}_{2}\ f(p)\right)  ^{-1} \ ,
\qquad\qquad M(p) = Z(p)\left(  m_{c}+\bar{\sigma}_{1}\
g(p)\right)\ .
\label{ch3.1.1-eq5b}
\end{eqnarray}

\subsubsection{Meson masses and decay constants}
\label{sec3.1.2}

In this subsection we discuss the analytical expressions to be
used for the calculation of basic meson phenomenological
quantities, such as masses and decay constants. It is important to
notice that pion observables, already calculated within the
nonlocal NJL approach in Sec.~\ref{sec2.1.3}, need to be revisited
owing to the mixing between pion and axial vector fields.

As discussed in Sec.~\ref{sec2.1.3}, the meson masses can be obtained from
the piece of the Euclidean action that is quadratic in the bosonic fields,
$S_E^{\rm quad}$. In the present case one has
\begin{eqnarray}
S_E^{\rm quad} &=& \dfrac{1}{2} \int \frac{d^4 q}{(2\pi)^4}\
\Big\{ G_\sigma(q^2)\, \delta\sigma(q)\, \delta\sigma(-q)  +
G_{\sigma^\prime}(q^2)\, \delta\sigma^\prime(q)\,\delta\sigma^\prime(-q) \nonumber\\
                && +\;G_{\pi}(q^{2}) \, \delta\vec\pi(q)\cdot\delta\vec\pi(-q) -
                 i\,G_{\pi a}(q^{2})\Big[q_\mu\,\delta\vec{a}_{\mu}(-q) \cdot
                \delta\vec{\pi}(q)-q_\mu\,\delta\vec{a}_{\mu}(q)
                \cdot \delta\vec{\pi}(-q)\Big] \nonumber \\
                & & +\; G_{0\,\mu\nu}(q^2)\,\delta v^0_{\mu}(q)\,\delta
                v^0_{\nu}(-q) + G_{5\,\mu\nu}(q^2)\,\delta a^0_{\mu}(q)\,\delta
                a^0_{\nu}(-q)\nonumber \\
                & & + \;G_{v\,\mu\nu}(q^2)\,\delta\vec{v}_\mu(q)\cdot\delta\vec{v}_\nu(-q) +
G_{a\,\mu\nu}(q^2)\,\delta\vec{a}_\mu(q)\cdot\delta\vec{a}_\nu(-q)\Big\}\ ,
\label{ch3.1.3-eq1}
\end{eqnarray}
where the functions $G_M(q^2)$, $M = \sigma,\sigma',\pi,\dots$ include
one-loop integrals arising from the fermionic determinant in the bosonized
action. The analysis of the scalar meson sector is similar to the one
discussed in Sec.~\ref{sec2.1.1}, i.e., the mass eigenstates $\delta\sigma$
and $\delta\sigma'$ are defined as linear combinations of $\delta\sigma_1$
and $\delta\sigma_2$ as in Eq.~(\ref{ch2.1.1-eq13}). Moreover, the
expressions for $G_\sigma(q^2)$ and $G_{\sigma'}(q^2)$ are the same as those
given by Eqs.~(\ref{ch2.1.1-eq16}-\ref{ch2.1.1-eq17}). To analyze the vector
meson sector one has to take into account the tensors $G_{v\,\mu\nu}$,
$G_{a\, \mu\nu}$, $G_{0\,\mu\nu}$ and $G_{5\,\mu\nu}$. From the expansion of
the fermionic determinant one gets~\cite{Villafane:2016ukb}
\begin{eqnarray}
G_{v\,\mu\nu}(q^2) &=&
G_{\rho}(q^2)\left(\delta_{\mu\nu}-\dfrac{q_{\mu}q_{\nu}}{q^2}\right)+
L_{+}(q^2)\dfrac{q_{\mu}q_{\nu}}{q^{2}}\ ,  \nonumber\\
G_{a\,\mu\nu}(q^2) &=& G_{{\rm
a}_1}(q^2)\left(\delta_{\mu\nu}-\dfrac{q_{\mu}q_{\nu}}{q^{2}}\right)+
L_{-}(q^{2})\dfrac{q_{\mu}q_{\nu}}{q^{2}}\ ,
\label{ch3.1.3-eq3}
\end{eqnarray}
where
\begin{eqnarray}
\hspace{-4mm} G_{\rho \choose {\rm a}_1}(q^{2})&=&
\dfrac{1}{G_V}-8N_c \int
                               \dfrac{d^{4}p}{(2\pi)^{4}} \, h(p)^2\,
                               \dfrac{Z(p^+)Z(p^-)}{D(p^{+})D(p^{-})} \,
                               \left[\dfrac{p^{2}}{3}+\dfrac{2(p\cdot
                               q)^{2}}{3q^{2}}-
                               \dfrac{q^{2}}{4}\pm M(p^{-})M(p^{+})\right],
\label{ch3.1.3-eq4}
\\
\hspace{-4mm} L_{\pm}(q^{2})&=&\dfrac{1}{G_V}-8N_c \int
\dfrac{d^{4}p}{(2\pi)^{4}} \, h(p)^2\,
                 \dfrac{Z(p^+)Z(p^-)}{D(p^{+})D(p^{-})} \,
                 \left[p^{2}-\dfrac{2(p\cdot q)^{2}}{q^{2}}+\dfrac{q^{2}}{4}\pm
                 M(p^{-})M(p^{+})\right] ,
\label{ch3.1.3-eq5}
\end{eqnarray}
with $p^\pm = p \pm q/2$. The functions $G_{\rho,{\rm
a}_1}(q^{2})$ and $L_{\pm}(q^{2})$ correspond to the transverse
and longitudinal projections of the vector and axial vector
fields, describing meson states with spin 1 and 0, respectively.
Thus, the masses of the physical $\rho^0$ and $\rho^\pm$ vector
mesons (which are degenerate in the isospin limit) can be obtained
by solving the equation
\begin{equation}
G_\rho(-m_\rho^2)\ =\ 0 \ .
\label{ch3.1.3-eq6}
\end{equation}
In addition, in order to obtain the physical states, the vector
meson fields have to be normalized through
\begin{equation}
\delta v^a_\mu(q) \ = \ Z_\rho^{1/2}\;{\tilde v}^a_\mu(q)\ ,
\label{ch3.1.3-eq7}
\end{equation}
where
\begin{equation}
Z_\rho^{-1} \ = \ g_{\rho q\bar q}^{-2} \ = \
\frac{dG_\rho(q^2)}{dq^2}\bigg\vert_{q^2=-m_\rho^2} \ .
\label{ch3.1.3-eq8}
\end{equation}
Here, $g_{\rho q\bar q}$ can be viewed as an effective $\rho$ meson-quark
effective coupling constant. Regarding the isospin zero channels, it is easy
to see that the expressions for $G_{0\,\mu\nu}(q^2)$ can be obtained from
those for $G_{v\,\mu\nu}(q^2)$, just replacing $G_V\to G_0$ and $h(p)\to
h_0(p)$. The $I=0$, $J^P=1^-$ state can be naturally associated with the
$\omega$ vector meson, denoting by $G_{\omega}(q^2)$ the form factor
corresponding to the transverse part of $G_{0\,\mu\nu}(q^2)$. Thus, the
$\omega$ mass and wave function renormalization can be obtained as in
Eqs.~(\ref{ch3.1.3-eq7}) and (\ref{ch3.1.3-eq8}). Similar relations apply to
the $I=0$ axial vector sector, where $G_{5\,\mu\nu}(q^2)$ can be obtained
from $G_{a\,\mu\nu}(q^2)$ by replacing $G_V\to G_5$ and $h(p)\to h_5(p)$.
The lightest physical state associated to this sector ($I=0$, $J^P=1^+$) is
the $f_1$ axial vector meson. Hence, we denote by $G_{f_1}(q^2)$ the form
factor corresponding to the transverse part of $G_{5\,\mu\nu}(q^2)$.

In the case of the pseudoscalar sector, it is seen from
Eq.~(\ref{ch3.1.3-eq1}) that there is a mixing between the pion fields and
the longitudinal part of the axial vector
fields~\cite{Ebert:1985kz,Bernard:1993rz}. The mixing term includes a loop
function $G_{\pi a}(p^2)$, while the term quadratic in $\delta\pi$ is
proportional to $G_\pi(p^2)$. These functions are given by
\begin{eqnarray}
G_{\pi}(q^2) & = & \dfrac{1}{G_S} \, - \, 8N_c \int
\dfrac{d^{4}p}{(2\pi)^{4}}\; g(p)^2\,
                 \dfrac{Z(p^+)Z(p^-)}{D(p^{+})D(p^{-})}\,
                 \left[p^{+}\cdot p^-\,+\,M(p^{+})\,M(p^{-})\right] \ , \nonumber \\
G_{\pi a}(q^2) & = &  \dfrac{8N_{C}}{q^{2}}\, \int
\dfrac{d^{4}p}{(2\pi)^{4}}\, g(p)\,h(p)\,
                 \dfrac{Z(p^+)Z(p^-)}{D(p^{+})D(p^{-})}\,
                 \left[p^{+} \,M(p^{-})-p^{-} \,M(p^{+})\right]\cdot p \ ,
\label{ch3.1.3-eq9}
\end{eqnarray}
where once again we have used the definitions {$p^\pm = p \pm q/2$}.
The physical states $\vec{\tilde a}_\mu$ and $\vec{\tilde \pi}$ can be now
obtained through the relations~\cite{Ebert:1985kz,Bernard:1993rz}
\begin{eqnarray}
\delta\pi^b(q)     &=& Z^{1/2}_\pi\;{\tilde \pi}^b(q) \ ,\nonumber\\
\delta a^b_{\mu}(q) &=& Z^{1/2}_a\;{\tilde a}^b_\mu(q) \, + \, i \,
\lambda(q^2) \,  q_\mu \, Z^{1/2}_\pi\;{\tilde \pi}^b(q) \ ,
\label{ch3.1.3-eq10}
\end{eqnarray}
where the mixing function $\lambda(q^2)$, defined in such a way
that the cross terms in the quadratic expansion vanish, is given
by
\begin{equation}
\lambda(q^2) \ = \ \dfrac{G_{\pi a}(q^2)}{L_-(q^2)} \ .
\label{ch3.1.3-eq11}
\end{equation}
Thus, the pion mass can be calculated from $G_{\tilde \pi}(-m_\pi^2) = 0$, where
\begin{equation}
G_{\tilde{\pi}}(q^2) \ = \ G_{\pi}(q^2)-\,q^2\,\lambda(q^2)\, G_{\pi a}(q^2)\ ,
\label{ch3.1.3-eq12}
\end{equation}
while the pion WFR is obtained from
\begin{equation}
Z_\pi^{-1} \ = \ g_{\pi q\bar q}^{-2} \ = \
\frac{dG_{\tilde \pi}(q^2)}{dq^2}\bigg\vert_{q^2=-m_\pi^2} \ .
\label{ch3.1.3-eq13}
\end{equation}
In the case of the ${\rm a}_1$ axial vector mesons ($I=1$
triplet), since the transverse parts of the $a_\mu^b$ fields do
not mix with the pions, the corresponding mass and WFR can be
calculated using relations analogous to those quoted for the
vector meson sector, namely Eqs.~(\ref{ch3.1.3-eq7}) and
(\ref{ch3.1.3-eq8}), with $G_{{\rm a}_1}(p^2)$ given by
Eq.~(\ref{ch3.1.3-eq4}).

On the other hand, the pion weak decay constant $f_\pi$ can be determined
following similar steps as those described in Sec.~\ref{sec2.1.3}, now
taking into account the existence of {$\pi- a$} mixing. One gets in
this way~\cite{Villafane:2016ukb}
\begin{equation}
f_\pi = \dfrac{g_{\pi q \bar{q}}}{m_\pi^2} \, m_c \left[ F_0(q^2) + \lambda(q^2) \,
F_1 (q^2)\right]\bigg\vert_{q^2=-m_\pi^2} \ ,
\label{ch3.1.3-eq19}
\end{equation}
where
\begin{eqnarray}
F_0 (q^2) &=& 8N_c\,\int \dfrac{d^4 p}{(2\pi)^4}\,g(p)\,
\dfrac{Z(p^+)Z(p^-)}{D(p^+)D(p^-)}
              \left[ p^+\cdot p^- + M(p^+)\,M(p^-)\right] \ ,\nonumber\\
F_1 (q^2) &=& 8N_c\, \int \dfrac{d^4 p}{(2\pi)^4}\,h(p)\,
\dfrac{Z(p^+)Z(p^-)}{D(p^+)D(p^-)} \;
               q \cdot \left[ p^+\,M(p^-) - p^- \,M(q^+)\right]\ .
\label{ch3.1.3-eq20}
\end{eqnarray}
It is important to notice that the result for $f_\pi$ does not depend on the
path chosen for the transport function in Eq.~(\ref{ch2.1.3-eq6}). In the
absence of axial-vector meson fields, the mixing term in
Eq.~(\ref{ch3.1.3-eq19}) vanishes and this expression reduces to the one
given in Eq.~(\ref{ch2.1.3-eq7}).

Concerning the vector and axial vector meson sector, after some lengthy
calculations one can obtain analytical expressions for other physical
quantities, such as the vector and axial vector decay constants $f_v$ and
$f_{\rm a}$ and the decay widths for the processes $\rho\to \pi\pi$ and
${\rm a}_1\to\rho\pi$. The corresponding results can be found in
Refs.~\cite{Villafane:2016ukb,Carlomagno:2019yvi}.

\subsection{Extension to finite T and $\mu$ in the mean field approximation}
\label{sec3.2}


We extend now the analysis of this model to a system at finite temperature
and chemical potential. Following the prescriptions described in
Sec.~\ref{sec2.2}, the grand canonical thermodynamic potential in the mean
field approximation is found to be given
by~\cite{Kashiwa:2011td,Contrera:2012wj,Contrera:2016rqj,Carlomagno:2019yvi}
\begin{equation}
\Omega^{\rm MFA}_{\rm reg} \ = \ \Omega^{\rm MFA} \, - \,
\Omega^{\rm free}_q \, + \, \Omega^{\rm free}_{q,\rm reg} \,  + \, \Omega_0 \ ,
\label{ch3.2-eq1}
\end{equation}
where
\begin{eqnarray}
\Omega^{\rm MFA} \, - \, \Omega^{\rm free}_q \ &=& \ - \,4 T
\sum_{c=r,g,b} \ \sum_{n=-\infty}^{\infty} \int \frac{d^3\vec
p}{(2\pi)^3} \ \ln \left[ \frac{ (\tilde \rho_{n, \vec{p}}^c)^2 +
M(\rho_{n,\vec{p}}^c)^2}{Z(\rho_{n, \vec{p}}^c)^2\ [(\rho_{n,
\vec{p}}^c)^2 + m_c^2]}\right] \nonumber
\\ && \ + \, \frac{\bar\sigma_1^2 +
\varkappa_p^2\; \bar\sigma_2^2}{2\,G_S} \, - \,
\frac{\bar\omega^2}{2\,G_0} \, + \,
\mathcal{U}(\Phi,\Phi^\ast,T)  \ .
\label{ch3.2-eq2}
\end{eqnarray}
{Here, $\bar\omega$ stands for the mean field value of the isoscalar
field $v_4^0(x)$ (which vanishes for $\mu =0$), while the function
$\mathcal{U}(\Phi,\Phi^\ast,T)$ is an effective Polyakov loop potential that
accounts for color gauge field self-interactions (see Sec.~\ref{sec2.2}).}
The expression for $\Omega^{\rm free}_{q,\rm reg}$ in Eq.~(\ref{ch3.2-eq1})
has been given in Eq.~(\ref{ch2.2-eq11}), and {$\Omega_0$ is a constant
that fixes $\Omega^{\rm MFA}_{\rm reg}=0$ for $T = \mu=0$.

Notice that in Eq.~(\ref{ch3.2-eq2}) we have also introduced the generalized
momenta $\rho_{n,\vec{p}}^c$ and $\tilde \rho_{n,\vec{p}}^c$. The former has
been defined in the same way as in Sec.~\ref{sec2.2},} namely
\begin{equation}
\Big({\rho_{n,\vec{p}}^c} \Big)^2 \ = \ {\vec{p}}\ \! ^2 + \Big[ (2 n +1 )\pi  T + i\mu - \phi_c
\Big]^2  \ .
\label{ch3.2-eq4}
\end{equation}
{A similar definition applies to $\tilde \rho_{n, \vec{p}}^c\,$, in which
$\mu$ is replaced} by a shifted chemical potential $\tilde \mu$ given
by~\cite{Contrera:2012wj}
\begin{equation}
\tilde{\mu} \ = \ \mu - g({\rho}_{n,\vec{p}}^c)\
Z({\rho}_{n,\vec{p}}^c)\ \bar{\omega} \ .
\label{ch3.2-eq5}
\end{equation}
This shift arises from the presence of the {nonvanishing} mean field
value $\bar\omega$, associated {to the $\omega$ vector meson.}

{As discussed in Sec.~\ref{sec2.2},} the mean field values of the meson
fields and the traced Polyakov loop $\Phi$ can be calculated by minimizing
the regularized thermodynamic potential. In the present case,
Eqs.~(\ref{ch2.2-eq12}) have to be supplemented by the condition
\begin{eqnarray}
\frac{\partial\Omega^{\rm MFA}_{\rm reg}}{\partial\bar \omega} \ = \ 0 \ .
\label{ch3.2-eq7}
\end{eqnarray}
On the other hand, the quark condensates and the susceptibilities
$\chi_{\rm ch}$ and $\chi_\Phi$, associated with chiral
restoration and deconfinement transitions, can be obtained as
indicated in Sec.~\ref{sec2.2}, see
Eqs.~(\ref{ch2.2-eq13}-\ref{ch2.2-eq15}).

\subsection{Model parametrization and numerical results for zero $T$ and $\mu$}
\label{sec3.3}


As {stated in Sec.~\ref{sec2.3}, to obtain numerical predictions for
physical quantities it is necessary to specify the model parameters and
nonlocal form factors}. For definiteness and simplicity, we take $h(p) =
h_0(p) = g(p)$. In fact, this can be justified from the assumption of a
similar effective form for the quark currents carrying angular momenta $J=0$
and $J=1$, together with an approximate degeneracy between the
vector-isovector and vector-isoscalar couplings. The axial vector-isoscalar
sector can be studied separately, since it decouples from the rest of the
Lagrangian. Following Ref.~\cite{Villafane:2016ukb} we consider $h_5(p) =
g(p)$ just to get an estimation for the constant $G_5$ from phenomenology.

For the form factors $g(p)$ and $f(p)$ we consider here an
exponential momentum dependence as the one given by
Eq.~(\ref{ch2.3-eq1}), which corresponds to parametrization PB of
the model studied in Sec.~\ref{sec2}. Namely, we take
\begin{equation}
g(p) = {\rm exp}(-p^2 / \Lambda_0^2)\ , \qquad\qquad f(p) = {\rm
exp}(-p^2 / \Lambda_1^2)\ .
\label{ch3.3-eq1}
\end{equation}
As stated, the exponential functions ensure a fast ultraviolet
convergence of quark loop integrals. In any case, from the
analysis in the previous section it is seen that {most numerical
results} are qualitatively similar for different form factor
functions, such as those in parametrizations PB and PC. As
mentioned in Sec.~\ref{sec2.3}, the form factors introduce two
additional parameters $\Lambda_0$ and $\Lambda_1$, which act as
effective momentum cutoff scales. The other six free parameters
entering the action in Eq.~(\ref{ch3.1.1-eq1}) are the current
quark mass $m_c$, the coupling constants $G_S$, $G_V$, $G_0$ and
$G_5$, and the parameter $\varkappa_p$ in the term that includes
the derivative current $j_R(x)$.

\begin{figure}[h]
\centering
\subfloat{\includegraphics[width=0.43\textwidth]{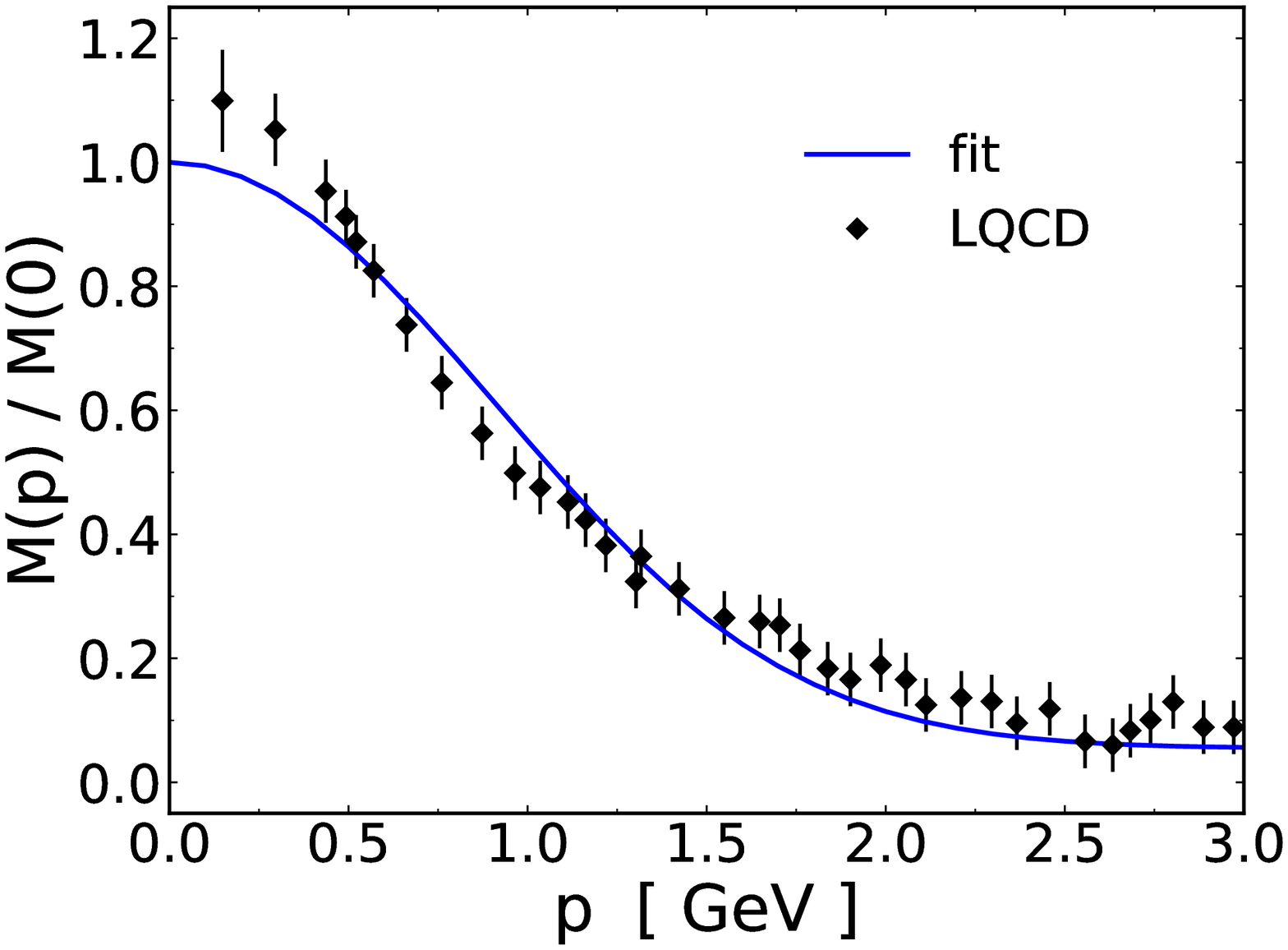}}
\subfloat{\includegraphics[width=0.43\textwidth]{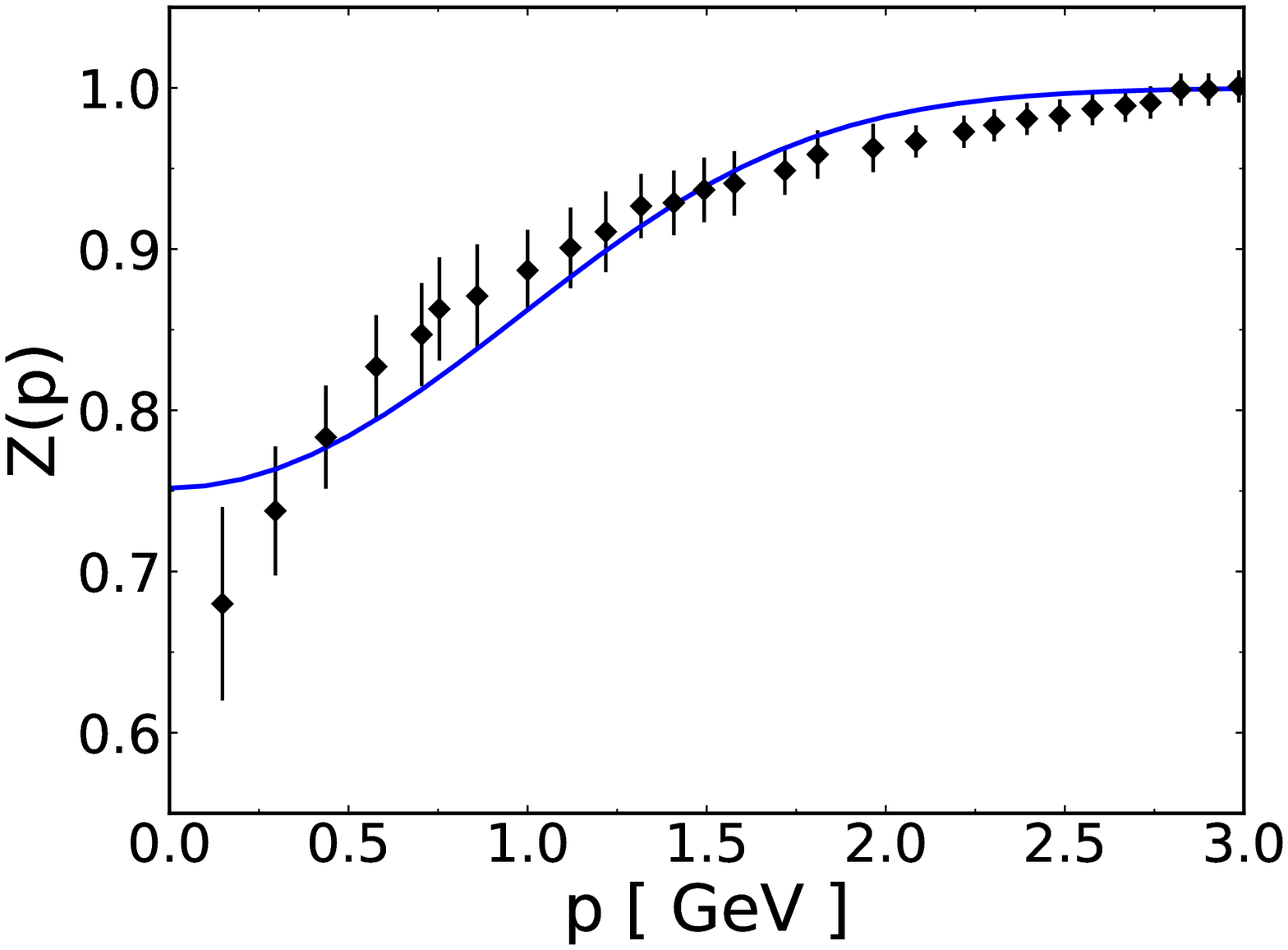}}
\caption{(Color online) Fit to lattice data from
Ref.~\cite{Parappilly:2005ei} for the functions $M(p)$ and $Z(p)$.
Values up to $p=3$~GeV have been considered.} \label{fig:ff}
\end{figure}

As discussed in the previous section, to get a close resemblance to QCD
features one can require the model to reproduce the results obtained from
lattice QCD calculations for the functions $M(p)$ and $Z(p)$ that
characterize the effective quark propagators. Since these functions are
determined by the shape of the form factors in Eq.~(\ref{ch3.3-eq1}), from a
fit to LQCD data it is possible to obtain a set of values for $Z(0)$ and the
parameters $\Lambda_0$ and $\Lambda_1$. The analysis in
Ref.~\cite{Carlomagno:2019yvi}, taking LQCD results from
Ref.~\cite{Parappilly:2005ei}, leads to $Z(0) = 0.75$, $\Lambda_0 =
1092$~MeV and $\Lambda_1 = 1173$~MeV. The corresponding curves for the
functions $M(p)$ and $Z(p)$, together with LQCD data, are shown in
Fig.~\ref{fig:ff}. It can be seen that the fit is somewhat
inaccurate for low momenta (where errors from LQCD calculations are larger).
However, as mentioned at the end of Sec.~\ref{sec2.2}, since volume
integrals in momentum space include in general $n$ powers of momentum $p$
($n=2$ and $n=3$ for integrals in three and four dimensions, respectively),
these differences in the low-momentum behavior of the form factors are not
expected to have a significant effect on the numerical results.

Given the form factor functions, it is possible to determine the remaining
model parameters from some set of input phenomenological quantities.
Following Ref.~\cite{Carlomagno:2019yvi}, we consider the case in which one
takes as inputs the $\pi$ and $\rho$ meson masses and the pion weak decay
constant $f_{\pi}$, together with the already mentioned value $Z(0) = 0.75$.
This lead to the model parameter values quoted in Table~\ref{tab:param}.
Regarding the coupling constant $G_0$, corresponding to the isoscalar vector
sector, as in Refs.~\cite{Contrera:2012wj,Carlomagno:2019yvi} we leave this
parameter free, using the ratio $\eta = G_0/G_V$ to tune the model. In fact,
the qualitative effect of this ratio gets increased in the case of a finite
chemical potential, where the mean field value $\bar\omega$ is in general
nonzero and the isoscalar vector term can contribute significantly to the
mean field thermodynamic potential. Finally, the coupling constant $G_5$
can be in principle determined from the $f_1$ meson mass. However, the
observed value of this mass is well above 1~GeV, which makes the
determination quite uncertain. This is discussed in
Ref.~\cite{Villafane:2016ukb}, where $G_5\sim G_V$ is obtained.

\begin{table}[h!bt]
\begin{center}
\begin{tabular}{c c c}
\hline \hline
\multicolumn{2}{c}{\ } &  Value \\
\hline
$m_c$ & MeV & 2.26
  \\
$G_S \Lambda_0^2$ & & 27.78
 \\
$G_V \Lambda_0^2$ & & 23.91
  \\
$G_5$ & & $\sim G_V\ \;$
  \\
$\varkappa_p$ & GeV & 4.265
  \\
$\Lambda_0$ & MeV &  1092
  \\
$\Lambda_1$ & MeV &  1173
\\
\hline \hline
\end{tabular}
\caption{Model parameter values.}
\label{tab:param}
\end{center}
\end{table}

Once the model parametrization has been established, numerical predictions
for different quantities can be obtained. The results for the mean field
values of scalar fields, quark condensates and quark-meson effective
couplings are summarized in Table~\ref{tab:propa}. In addition, in
Table~\ref{tab:propb} we quote some numerical predictions for meson
properties, together with the corresponding experimental estimates. In
general, it is seen that the values obtained for meson masses, mixing angles
and decay constants are in reasonable agreement with phenomenological
expectations. As in the case of the parametrization PC considered in
Sec.~\ref{sec2}, one finds relatively low values for $m_c$, and a somewhat
large value for the light quark condensate. On the other hand, one gets
$-\langle \bar qq\rangle m_c \simeq 8.1\times 10^{-5}$~GeV$^4$, which is
consistent with the scale-independent result obtained from the
Gell-Mann-Oakes-Renner relation in Eq.~(\ref{ch2.1.3-eq9}), viz.\ $-\langle
\bar qq\rangle m_c = f_\pi^2 m_\pi^2/2 \simeq 8.3\times 10^{-5}$~GeV$^4$.

\begin{table}[h]
\begin{center}
\begin{tabular}{ccc}
\hline \hline
 &&  Model  \\
\hline $\bar\sigma_1$ & MeV & 648
 \\
$\bar\sigma_2$ & &  $-0.331\ $
\\
$-\langle \bar q q \rangle ^{1/3}$ & MeV &  330
\\
$g_{\pi q \bar q}$ & & 7.07
\\
$g_{\rho q \bar q}$ & & 4.16
\\
$g_{{\rm a_1} q \bar q}$ & & 3.74
\\
\hline \hline
\end{tabular}
\caption{{Numerical results for various phenomenological
quantities.}} \label{tab:propa}
\end{center}
\end{table}
\begin{table}[h]
\begin{center}
\begin{tabular}{cccc}
\hline \hline
 & & Model & Phenomenology \\ \hline
$m_{\pi}$     &  MeV & 139$^*$   & 139         \\
$m_{\sigma}$  &  MeV & 794   & 400$-$550         \\
$m_{\rho}$    &  MeV & 775$^*$   & 775         \\
$m_{\rm a_1}$ &  MeV & 1204  & $1230\pm 40$    \\
$f_\pi$       &  MeV & 92.4$^*$ & 92.4         \\
$f_v$         &      & 0.173 & 0.200           \\
$\Gamma_{\rho\to \pi\pi}$ & MeV & 121  & 149   \\
$\Gamma_{{\rm a}_1\to \rho\pi}$ & MeV & 185 & 150$-$360 \\
\hline \hline
\end{tabular}
\caption{Numerical results for various meson properties. Quantities marked
with an asterisk have been taken as inputs.}
\label{tab:propb}
\end{center}
\end{table}

It is worth pointing out that effective theories based on quark
current-current interactions usually present a threshold above which the
constituent quarks can be simultaneously on shell. This threshold, which
depends on the model parametrization and regularization prescriptions, is
typically of the order of $1$~GeV. One noticeable feature of the
parametrization in Table~\ref{tab:param} is that it leads to a threshold of
about $1.25$~GeV, which turns out to be above the value obtained for the
$\rm a_1$ meson mass. This prevents the unphysical situation of a
possible decay of the $\rm a_1$ meson into two on-shell quarks.

\subsection{Phase transitions in the $\mu - T$ plane}
\label{sec3.4}

Through the study of the behavior of the order parameters $\langle
\bar qq\rangle$ and $\Phi$ it is possible to determine the regions
of the phase diagram in which the chiral symmetry is either broken
or approximately restored, as well as those in which the system
lies either in confined or deconfined states. In this subsection
we identify these regions in the $\mu -T$ plane, together with the
features of the existing phase transitions. The numerical results
correspond to the parametrization in Sec.~\ref{sec3.3},
considering the polynomial Polyakov loop potential in
Eq.~(\ref{ch2.2-eq2}) and $T_0 = 208$~MeV.

\begin{figure}[h!bt]
\centering
\includegraphics[width=0.5\textwidth]{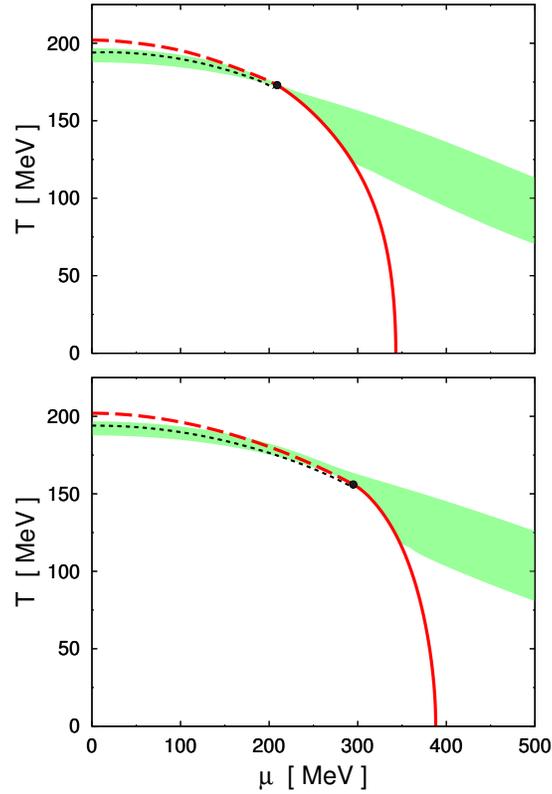}
\caption{(Color online) $\mu-T$ phase diagrams for the parametrization given
in Sec.~\ref{sec3.3}, taking $\eta = 0$ (upper panel) and $\eta=0.5$ (lower
panel). The results correspond to a polynomial PL potential, with
$T_0=208$~MeV. First order chiral restoration transitions are indicated by
solid lines, while dashed and dotted lines correspond to chiral restoration
and deconfinement crossover-like transitions, respectively. The green bands
indicate the regions in which $0.3 \leq \Phi \leq 0.5$.}
\label{fig:QCDpd}
\end{figure}

In general, the situation is found to be similar to the one observed in the
case of the model with no vector and axial vector quark current interactions
analyzed in Sec.~\ref{sec2.5}. The results for the phase diagram are
sketched in Fig.~\ref{fig:QCDpd}, where the values $\eta = 0$ and $\eta =
0.5$ have been considered~\cite{Carlomagno:2019yvi}. At vanishing chemical
potential, it is seen that the system undergoes crossover-like chiral
restoration and deconfinement transitions. The corresponding critical
temperatures are close to each other, viz.\ $T_{\rm ch}(0) = 202$~MeV and
$T_\Phi(0) = 194$~MeV. {They are found to be somewhat larger than those
obtained for the model in Sec.~\ref{sec2}, due to the different
parametrizations considered}. Now, taking a fixed temperature and increasing
the chemical potential, it is seen that for $T\lesssim T_{\rm ch}(0)$ both
crossover-like transitions occur at some approximately common critical value
$\mu = \mu_c(T)$. This is shown in Fig.~\ref{fig:QCDpd}, where chiral
restoration and deconfinement curves are indicated with dashed and dotted
lines, respectively. Both crossover lines end at a critical end point (CEP)
of coordinates $(\mu_{\rm CEP},T_{\rm CEP})$. For $T < T_{\rm CEP}$, it is
seen that at a critical chemical potential the quark condensate shows a
discontinuity, signaling a first order phase transition (solid lines in
Fig.~\ref{fig:QCDpd}). This gap in $\langle \bar qq\rangle$ induces also a
jump in the traced Polyakov loop $\Phi$. However, this should not
necessarily be interpreted as a first order deconfinement transition. In
fact, for low temperatures it is seen that the values of $\Phi$ are
relatively low at both sides of the discontinuity, indicating that the
system is still in a confined phase even if chiral symmetry has been
approximately restored (we recall that $\Phi =0$ and $\Phi = 1$ indicate
maximum confinement and deconfinement, respectively). As discussed in
Sec.~\ref{sec2.5}, one can alternatively define the deconfinement transition
region as the one in which $\Phi$ takes a value in the range between $0.3$
and $0.5$. For the model under consideration, this is shown by the green
shaded bands in Fig.~\ref{fig:QCDpd}. As mentioned in Sec.~\ref{sec2.5}, the
region in which $\Phi$ is below this range and chiral symmetry is already
approximately restored is usually referred to as a quarkyonic
phase~\cite{McLerran:2007qj,McLerran:2008ua,Abuki:2008nm}.

The effect of the value of the parameter $\eta$ on the phase diagram can be
observed by comparing the upper and lower panels of Fig.~\ref{fig:QCDpd}. As
stated, this parameter measures the relative strength of the $J=1$ isoscalar
couplings in comparison with their isovector counterparts. As expected, it
is seen that the effect of the isoscalar piece increases with the
chemical potential. In particular, when $\eta$ is enhanced the CEP moves
towards lower temperatures and higher chemical potentials, and the critical
value $\mu_c(T = 0)$ gets larger. The results for various critical
temperatures and chemical potentials, considering $\eta = 0$, 0.3 and 0.5
are summarized in Table~\ref{tab:cep}. Qualitative similar results for the
phase diagram have been found in Ref.~\cite{Contrera:2012wj} for a
logarithmic PL potential. In addition, a more complete analysis of the
existence and position of the CEP within nonlocal models that {include the
couplings between isoscalar $J=1$ quark currents} can be found in
Ref.~\cite{Contrera:2016rqj}. Regarding the case of imaginary chemical
potential, it has been argued~\cite{Kashiwa:2011td} that the isoscalar
vector interactions can have some effect on the location of the RW end point
and the pattern of transition lines in its neighborhood.

\begin{table}[h]
\begin{center}
\begin{tabular*}{0.4\textwidth}{@{\extracolsep{\fill}} cccc }
\hline \hline
                 & $\eta = 0$ & $\eta = 0.3$ & $\eta = 0.5$ \\
\hline
$T_{\rm ch}(0)$  & 202 & 202 & 202 \\
$T_{\Phi}(0)$    & 194 & 194 & 194 \\
$T_{\rm CEP}$    & 173 & 166 & 156 \\
$\mu_{\rm CEP}$  & 209 & 252 & 295 \\
$\mu_c(0)$       & 343 & 366 & 388 \\
\hline \hline
\end{tabular*}
\caption{Critical temperatures and chemical potentials for various values
of $\eta$. Values are given in~MeV}
\label{tab:cep}
\end{center}
\end{table}

\subsection{Thermal behavior of meson properties}
\label{sec3.5}

\begin{figure}[H]
\centering
\includegraphics[width=0.5\textwidth]{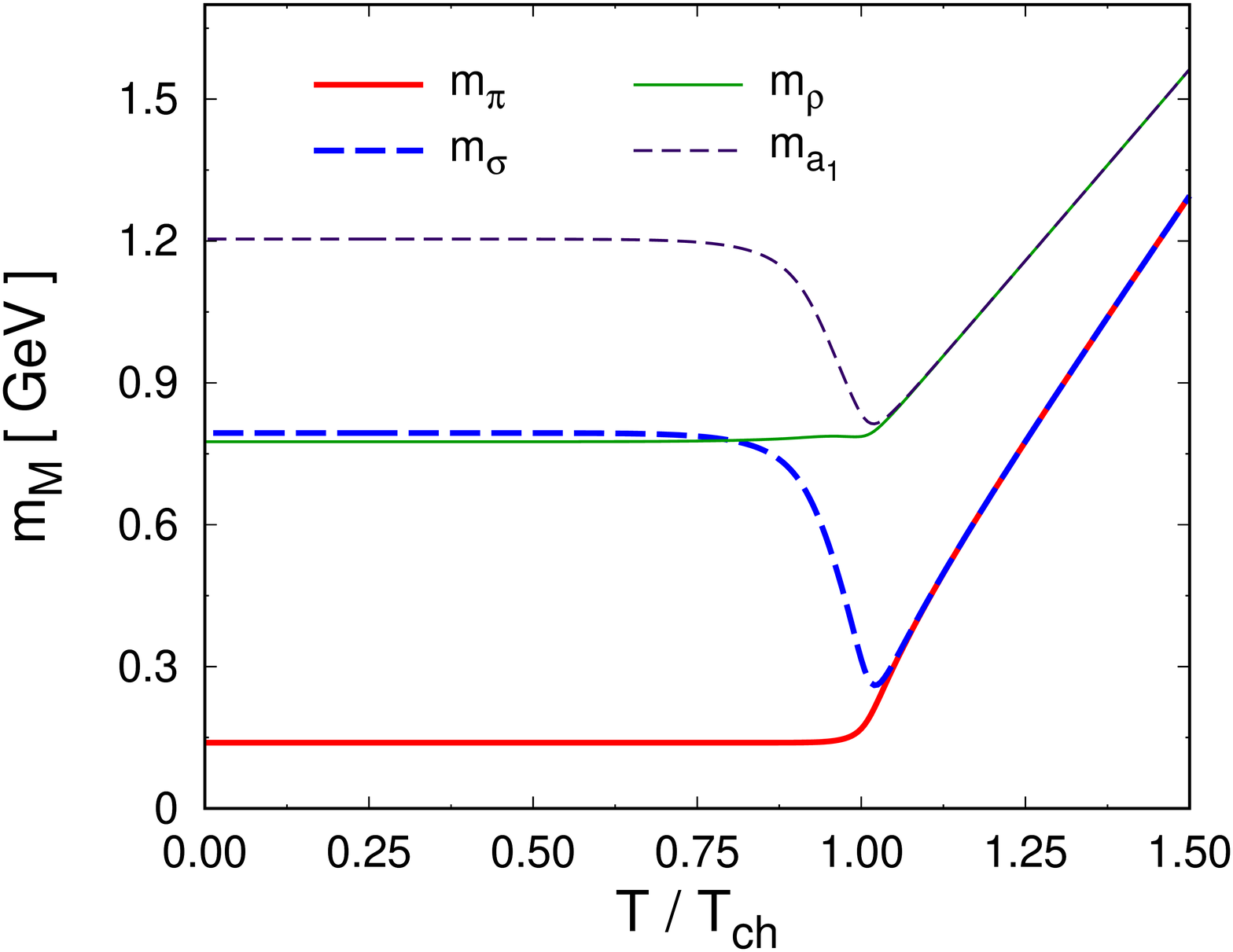}
\caption{Meson masses as functions of the temperature.}
\label{fig:masses}
\end{figure}

The thermal evolution of meson masses and decay constants in the context of
the nlPNJL models studied in this section can be obtained by following
similar steps as those outlined in Sec.~\ref{sec2.7}. In
Fig.~\ref{fig:masses} we show the numerical results obtained in
Ref.~\cite{Carlomagno:2019yvi} for the masses of $J=0$ chiral partners
$\sigma$ and $\pi$, as well as $J=1$ mesons $\rho$ and a$_1$. It is seen
that $\pi$ and $\rho$ meson masses (solid lines) remain approximately
constant up to the critical temperature $T_{\rm ch}$, while $\sigma$ and
a$_1$ meson masses (dashed lines) start to drop somewhat below $T_{\rm ch}$.
As expected from chiral restoration, right above the critical temperature
chiral partner masses become degenerate. When the temperature is further
increased, the masses rise continuously, showing that they are basically
dominated by thermal energy.

\begin{figure}[H]
\centering
\includegraphics[width=0.5\textwidth]{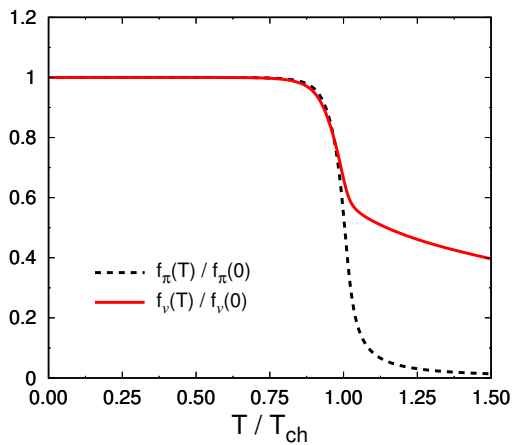}
\caption{Normalized $f_\pi$ and $f_v$ decay form factors as functions of the
temperature.}
\label{fig:decay}
\end{figure}

Finally, the thermal behavior of $f_\pi$ and $f_v$ decay form factors is
shown in Fig.~\ref{fig:decay}. It is seen that both quantities remain
constant up to approximately the chiral restoration critical temperature,
and then they show a sudden drop. For large temperatures both $f_\pi$ and
$f_v$ tend to zero, at different rates.

%% file: Sec4-rev.tex
\section{Three-flavor nonlocal NJL models}
\label{sec4}

In this Section we review the extension of nonlocal NJL models to three
dynamical quark flavors. This has been firstly addressed in
Ref.~\cite{Scarpettini:2003fj} for the case of vanishing temperature and
chemical potential, and then extended to finite temperature ---including the
coupling to the Polyakov loop--- in Ref.~\cite{Contrera:2007wu}. In those
works, the scheme based on the instanton liquid model has been considered
(see the discussion in Sec.~\ref{sec2.1.1}). In the case of the nlNJL
approach inspired on one gluon exchange interactions, three-flavor models
have been firstly analyzed in Refs.~\cite{Hell:2009by,Contrera:2009hk}.

\subsection{Three-flavor nlNJL model at vanishing temperature and chemical potential}
\label{sec4.1}

\subsubsection{Effective action}
\label{sec4.1.1}

Let us consider a three-flavor extension of the model studied in
Sec.~\ref{sec2}, which includes quark WFR interactions. The corresponding
Euclidean effective action, proposed in Ref.~\cite{Carlomagno:2013ona},
reads
\begin{eqnarray}
S_E &=& \int d^4x \ \left\{
\bar \psi (x)(- i \slashed \partial + \hat
m)\psi(x)-\frac{G_S}{2}\left[
j^a_S(x)j^a_S(x)+j^a_P(x)j^a_P(x)+j_R(x)j_R(x)\right] \right.
\nonumber\\
    &&\left. -\frac{H}{4} A_{abc}\left[
j^a_S(x)j^b_S(x)j^c_S(x)-3j^a_S(x)j^b_P(x)j^c_P(x)\right] \right\}
\ ,
\label{ch_su3.1-eq1}
\end{eqnarray}
where $\psi$ is the $u$, $d$, $s$ quark triplet and $\hat m={\rm
diag}(m_u,m_d,m_s)$ is the current quark mass matrix. As in the
two-flavor case, we consider the isospin symmetry limit, assuming
$m_u=m_d$. The nonlocal quark-antiquark currents are now given by
\begin{eqnarray}
j^a_S(x) &=& \int d^4z\; \mathcal{G}(z)\, \bar \psi
\left(x+\frac{z}{2}\right)\lambda_a\,
\psi\left(x-\frac{z}{2}\right)
\ , \nonumber\\
j^a_P(x) &=& \int d^4z\;  \mathcal{G}(z)\, \bar
\psi\left(x+\frac{z}{2}\right)i\, \lambda_a \gamma_5\,
\psi\left(x-\frac{z}{2}\right)
\ , \nonumber\\
j_R(x)   &=& \int d^4z\;  \mathcal{F}(z)\, \bar
\psi\left(x+\frac{z}{2}\right) \frac{i
\overleftrightarrow{\slashed \partial}}{2\varkappa_p} \,
\psi\left(x-\frac{z}{2}\right)\ , \label{ch_su3.1-eq2}
\end{eqnarray}
where $\mathcal{G}(z)$ and $ \mathcal{F}(z)$ are covariant form factors, as
described in Sec.~\ref{sec2.3}, and $\lambda_a$, $a=0,\dots,8$, are the
standard eight Gell-Mann matrices, plus
$\lambda_0=\sqrt{2/3}\;\mathbf{1}_{3\times 3}$. As in the two-flavor case,
the couplings involving scalar and pseudoscalar currents have to carry the
same form factor to guarantee chiral invariance, while the coupling
$j_R(x)j_R(x)$, responsible for quark WFR, is self-invariant under chiral
SU(3) transformations. The relative weight of this self-invariant term is
controlled by the parameter $\varkappa_p$. In addition, the model accounts
for flavor mixing through a 't Hooft-like term, in which the constants
$A_{abc}$ are defined by
\begin{equation}
A_{abc} \ = \ \frac{1}{3!}\epsilon_{ijk}\epsilon_{mnl}(\lambda_a)_{im}
(\lambda_b)_{jn}(\lambda_c)_{kl} \ .
\end{equation}
The role of {this six-fermion coupling}, which {is obtained from}
a determinant in flavor space, is to break the U(1)$_A$ symmetry,
not observed in nature.

As stated in the previous sections, in order to deal with meson degrees of
freedom it is convenient to perform a bosonization of the fermionic theory.
In the present case this can be done by introducing scalar fields
$\sigma_a(x)$ and $\zeta(x)$ and pseudoscalar fields $\pi_a(x)$, together
with auxiliary fields $S_a(x)$, $P_a(x)$ and $R(x)$, with $a=0,\dots ,8$.
After integrating out the fermion fields one obtains a partition function of
the form
\begin{eqnarray}
\mathcal{Z} & = & \int
{\rm D}\sigma_a\, {\rm D}\pi_a\, {\rm D}\zeta\ \
{\rm det}\, {\cal D} \ \int {\rm D}S_a\, {\rm D}P_a\, {\rm D}R \
\exp\Bigg[ \int d^4x\ \bigg( \sigma_a S_a \, + \, \pi_a P_a
\, + \, \zeta R
\nonumber\\
& & \quad + \; \frac{G_S}{2}(S_aS_a+P_aP_a+R^2) \; + \;
\frac{H}{4}A_{abc} (S_a S_b S_c - 3 S_a P_b P_c) \bigg) \Bigg] \ .
\label{ch_su3.1-eq4}
\end{eqnarray}
Here, the operator ${\cal D}$ ---in momentum space--- is
given by
\begin{eqnarray}
\!\!\!\!
{\cal D}(p,p^\prime)&=&(2\pi)^4 \delta^{(4)}(p-p^\prime)(\slashed p+\hat m)
\nonumber\\
& &
+\, g\left(\frac{p+p^\prime}{2}\right)\Big[\sigma_a(p-p^\prime)+ i\, \gamma_5
\pi_a(p-p^\prime)\Big]\lambda_a\,
-\,f\left(\frac{p+p^\prime}{2}\right)
\frac{\slashed p+\slashed p^\prime}{2\varkappa_p} \zeta(p-p^\prime)\ ,
\label{ch_su3.1-eq5}
\end{eqnarray}
where, as in the two-flavor case, $g(p)$ and $f(p)$ denote the
Fourier transforms of the form factors $\mathcal{G}(z)$ and $
\mathcal{F}(z)$.

\subsubsection{Mean field approximation and chiral {quark-antiquark} condensates}
\label{sec4.1.2}

We consider the mean field approximation (MFA), in which {scalar
and pseudoscalar} fields are expanded around homogeneous vacuum
expectation values, viz.
\begin{eqnarray}
\sigma_a(x) &=& \bar \sigma_a+\delta\sigma_a(x) \ , \nonumber\\
\pi_a(x) &=& \delta\pi_a(x) \ , \nonumber\\
\zeta(x) &=& \bar\zeta+\delta\zeta(x) \ . \label{ch_su3.1-eq6}
\end{eqnarray}
As usual, it {is} assumed that pseudoscalar mean field values
vanish, owing to parity conservation. Moreover, {in the scalar
field sector} only $\bar\sigma_0$, $\bar\sigma_8$ and $\bar\zeta$
are taken to be different from zero, due to charge and isospin
symmetries.

Following the stationary phase approximation, the path integral over the
auxiliary fields can be replaced by the corresponding integrand, evaluated
at the values $\tilde S_a$, $\tilde P_a$, and $\tilde{R}$ that minimize
the argument of the exponential in
Eq.~(\ref{ch_su3.1-eq4})~\cite{Scarpettini:2003fj}. In this way, the
Euclidean action per unit volume reduces to
\begin{equation}
\frac{S_E^{\rm MFA}}{V^{(4)}} \ = \ - \, \int
\frac{d^4p}{(2\pi)^4}\; {\rm Tr}\,\ln {\cal D}_0(p) -
\bar\sigma_a\bar S_a-\bar\zeta\bar R- \frac{G_S}{2}(\bar S_a\bar
S_a+\bar R^2)- \frac{H}{4}A_{abc}\bar S_a\bar S_b\bar S_c \ ,
\label{ch_su3.1-eq7}
\end{equation}
where $\bar S_a$, $\bar P_a$ and $\bar R$ stand for the values of
$\tilde S_a$, $\tilde P_a$ and $\tilde R$ within the MFA. {In the
first term on the right hand side one has
\begin{equation}
{\cal D}_0(p) \ = \ \left[ 1 -  \frac{\bar\zeta}{\varkappa_p} f(p) \right] \slashed p \, +
\, \hat m \, + \, g(p) \Big( \bar \sigma_0 \lambda_0 + \bar \sigma_8 \lambda_8\Big)\ ,
\end{equation}}
and the trace is taken over color, flavor and Dirac indices.

For the neutral fields ($a=0,3,8$) it is convenient to change to a flavor
basis, $\phi_a \to \phi_i$, where $i=u,d,s$, or equivalently $i=1,2,3$. In
this basis, by minimizing the mean field action in Eq.~(\ref{ch_su3.1-eq7})
one obtains the gap equations
\begin{eqnarray}
\bar \sigma_u + G_S\bar S_u+\frac{H}{2}\bar S_d \bar S_s &=& 0 \ , \nonumber\\
\bar \sigma_d + G_S\bar S_d+\frac{H}{2}\bar S_s \bar S_u &=& 0 \ , \nonumber\\
\bar \sigma_s + G_S\bar S_s+\frac{H}{2}\bar S_u \bar S_d &=& 0 \ ,
\label{ch_su3.1-eq8}
\end{eqnarray}
plus an extra equation arising from the $j_R(x)$ current-current
interaction, namely
\begin{equation}
\bar\zeta \, + \, G_S\,\bar R \ = \ 0 \ .
\end{equation}
The mean field values $\bar S_i$ and $\bar R$ are given by
\begin{eqnarray}
\bar S_i &=& -8N_c \int \frac{d^4p}{(2\pi)^4}\ g(p)\;
\frac{Z(p)\, M_i(p)}{D_i(p)} \ , \nonumber\\
\bar R &=& \frac{4 N_c}{\varkappa_p}  \sum_{i=1}^3 \int \frac{d^4p}{(2\pi)^4}\
p^2\, f(p) \; \frac{Z(p)}{D_i(p)} \ ,
\label{ch_su3.1-eq10}
\end{eqnarray}
where we have defined $D_i(p) = p^2 + M_i(p)^2$. As in the two-flavor case
[see Eq.~(\ref{ch2.1.1-eq11})], the functions $M_i(p)$ and $Z(p)$ correspond
to the momentum-dependent effective masses and WFR of quark propagators.
They are related to the nonlocal form factors through
\begin{eqnarray}
M_i(p) &=& Z(p)\, \Big[ m_i\, +\, \bar \sigma_i\; g(p)\Big] \ ,
\nonumber\\
Z(p) &=& \left[ 1\,-\,\frac{\bar
\zeta}{\varkappa_p}\,f(p)\right]^{-1}\ .
\label{ch_su3.1-eq11}
\end{eqnarray}
Thus, for a given set of model parameters and form factors, from
Eqs.~(\ref{ch_su3.1-eq8}-\ref{ch_su3.1-eq11}) one can numerically
obtain the mean field values $\bar \sigma_i$ and $\bar\zeta$. {As
expected from isospin symmetry, one has $\bar S_u = \bar S_d$ and
$\bar\sigma_u = \bar\sigma_d$.}

\subsubsection{Meson masses and decay constants}
\label{sec4.1.3}

As in the two-flavor case (see Sec.~\ref{sec2.1}), to analyze the
properties of meson fields it is necessary to go beyond the MFA,
considering quadratic fluctuations in the Euclidean action,
\begin{eqnarray}
 S_E^{\rm quad} &=& \dfrac{1}{2} \int \frac{d^4 q}{(2\pi)^4} \sum_{M}\  r_M\ G_M(q^2)\  \phi_M(q)\, \bar\phi_M(-q)
\ . \label{ch_su3.1-eq13}
\end{eqnarray}
Here, the meson fluctuations $\delta\sigma_a$ and $\delta\pi_a$ in
Eq.~(\ref{ch_su3.1-eq6}) have been translated to a charge basis $\phi_M$,
$M$ being the scalar and pseudoscalar mesons in the lowest mass nonets, plus
the $\zeta$ field. The coefficient $r_M$ is 1 for charge eigenstates $M={\rm
a}_0^0$, $\sigma$, $f_0$, $\zeta$, $\pi^0$, $\eta$ and $\eta^\prime$, and 2
for $M={\rm a}_0^+$, $K_0^{\ast +}$, $K_0^{\ast 0}$, $\pi^+$, $K^+$ and
$K^0$. Analogously to Eq.~(\ref{ch2.1.3-eq2}), meson masses are given by the
equations
\begin{equation}
G_M(-m_M^2)\ =\ 0 \ . \label{ch_su3.1-eq14}
\end{equation}
In addition, physical states have to be normalized through
\begin{equation}
\tilde{\phi}_M(q)=Z_M^{-1/2}\ \phi_M(q)\ ,
\end{equation}
where the meson renormalization constants $Z_M$ and the associated on-shell
quark-meson coupling constants $g_{Mq\bar{q}}$ are given by
\begin{equation}
Z_M^{-1}=g_{Mq\bar{q}}^{-2} \ = \ \frac{dG_M(q^2)}{dq^2}\bigg\vert_{q^2=-m_M^2}\ .
\end{equation}

The functions $G_M(q^2)$ can be written in terms of the coupling constants
$G_S$ and $H$, the mean field values $\bar S_{u,s}$ and quark loop functions
that prove to be ultraviolet convergent owing to the asymptotic behavior of
the nonlocal form factors. For the pseudoscalar meson sector, the $\pi$ and
$K$ mesons decouple, while the $I=0$ states get mixed. In the case of the
scalar fields, the ${\rm a}_0$ and $K_0^\ast$ mesons decouple, while the
$\zeta$, $\sigma_0$ and $\sigma_8$ states get mixed by a $3\times 3$ matrix
[see Eq.~(\ref{ch_su3.1-eq23}) below].

{The} explicit expressions for the functions $G_M(q^2)$ in the
context of nlNJL models with quark WFR {have been quoted} in
Ref.~\cite{Carlomagno:2013ona}. For the $I\neq 0$ states $\pi$,
${\rm a}_0$, $K$ and $K_0^\ast$ one has
\begin{eqnarray}
G_{\binom{\pi}{{\rm a}_0}}(q^2)&=& (G_S\pm\frac{H}{2}\bar S_s)^{-1}+4\, C_{uu}^{\mp}(q^2)\ ,
\nonumber\\
G_{\binom{K}{K_0^\ast}}(q^2)&=& (G_S\pm\frac{H}{2}\bar S_u)^{-1}+4\, C_{us}^{\mp}(q^2)\ ,
\end{eqnarray}
where the functions $C_{ij}^\mp(q^2)$, with $i,j=u$ or $s$ are
defined as
\begin{eqnarray}
C_{ij}^{\mp}(q^2)&=& - \, 2\, N_c \int \frac{d^4 p}{(2\pi)^4}\ g(p)^2\; \frac{Z(p^+)}{ D_i(p^+)} \ \frac{Z(p^-)}{D_j(p^-)}
\ [ p^+\cdot p^- \pm M_i(p^+)M_j(p^-)]\ ,
\end{eqnarray}
with $p^\pm = p \pm q/2$. At the $I=0$ pseudoscalar sector one has
a mixing between the $\eta_0$ and $\eta_8$ fields. The masses of
the physical states $\eta$ and $\eta'$ can be obtained from the
functions
\begin{equation}
G_{\binom{\eta}{\eta^\prime}}(q^2) \ = \
\frac{G_{88}^-(q^2)+G_{00}^-(q^2)}{2}\mp \sqrt{[G_{80}^-(q^2)]^2+\left(\frac{G_{88}^-(q^2)-G_{00}^-(q^2)}{2}\right)^2}\
,
\end{equation}
where we have used the definitions
\begin{eqnarray}
G_{00}^{\mp}(q^2)&=& \frac{4}{3}\left[2C_{uu}^{\mp}(q^2)+C_{ss}^{\mp}(q^2)+\frac{6 G_S \mp
H \bar S_s\pm 4H\bar S_u}{8G_S^2- 4H^2\bar S_u^2\mp 4 H G_S
\bar S_s} \right]\ ,
\nonumber\\
G_{88}^{\mp}(q^2)&=& \frac{4}{3}\left[
2C_{ss}^{\mp}(q^2)+C_{uu}^{\mp}(q^2) + \frac{6 G_s \mp 2H \bar
S_s\mp4H\bar S_u}{8G_S^2-4H^2\bar S_u^2\mp 4 H G_S \bar S_s}
\right]\ ,
\nonumber\\
G_{80}^{\mp}(q^2)&=& \frac{4}{3}\sqrt{2}\left[
C_{uu}^{\mp}(q^2)-C_{ss}^{\mp}(q^2)\pm\frac{H(\bar S_s-\bar
S_u)}{8G_S^2-4H^2 \bar S_u^2 \mp 4 H G_S \bar S_s} \right] \ .
\end{eqnarray}
The states $\eta$ and $\eta'$ are defined as
\begin{eqnarray}
\eta &=& \eta_8 \cos\theta_\eta - \eta_0 \sin\theta_\eta \ ,
\nonumber\\
\eta^\prime &=& \eta_8 \sin\theta_\eta^\prime + \eta_0
\cos\theta_\eta^\prime \ ,
\end{eqnarray}
where the mixing angles $\theta_\eta$, $\theta_{\eta'}$ are given
by
\begin{equation}
 \tan 2\,\theta_{\eta,\eta^\prime}=\frac{-2G_{80}^-(q^2)}{G_{88}^-(q^2) -G_{00}^-(q^2)}
 \bigg\vert_{q^2=-m^2_{\eta,\eta^\prime}} \ .
 \label{ch_su3.1-eq22}
\end{equation}

Finally, for the $I=0$ scalar sector, the quadratic terms
involving the fields $\zeta$, $\sigma_0$ and $\sigma_8$ are mixed
by the $3\times 3$ matrix
\begin{equation}
\left( \begin{array}{ccc} 4\,C^\zeta(q^2)+G_S^{-1} & \sqrt{\frac{8}{3}}[ 2\,C_{u}^{+\zeta}(q^2) + C_{s}^{+\zeta}(q^2) ] & \frac{4}{\sqrt{3}}[ C_{u}^{+\zeta}(q^2) - C_{s}^{+\zeta}(q^2) ] \\
\sqrt{\frac{8}{3}}[ 2\,C_{u}^{+\zeta}(q^2) + C_{s}^{+\zeta}(q^2) ] & G_{00}^{+}(q^2) & G_{80}^{+}(q^2) \\
\frac{4}{\sqrt{3}}[ C_{u}^{+\zeta}(q^2) - C_{s}^{+\zeta}(q^2) ] & G_{80}^{+}(q^2) & G_{88}^{+}(q^2) \end{array} \right)\ ,
\label{ch_su3.1-eq23}
\end{equation}
where
\begin{eqnarray}
C^\zeta(q^2) & = & \dfrac{ N_c}{\varkappa_p^2}\,  \sum_{i=1}^3 \int\!\!
\frac{d^4p}{(2\pi)^4}\ p^2 f(p)^2\;
\frac{Z(p^+)}{D_i(p^+)}\ \frac{Z(p^-)}{D_i(p^-)}
\nonumber \\
& & \times
\left[ p^+ \cdot p^- + \dfrac{p^{+2}p^{-2}-(p^+ \cdot p^-)^2}{2p^2} -
M_i(p^+)M_i(p^-)\right]\ ,
\nonumber\\
C_{i}^{+\zeta}(q^2) & = & -\dfrac{2\, N_c}{\varkappa_p}\,\int
\frac{d^4p}{(2\pi)^4}\ g(p)\, f(p)\, \frac{Z(p^+)}{D_i(p^+)}\ \frac{Z(p^-)}{D_i(p^-)}
\ p\cdot \left[ p^- \, M_i(p^+)+ p^+ M_i(p^-)\right] \ ,
\end{eqnarray}
with $i=u,s$. The functions $G_\sigma(q^2)$, $G_{f_0}(q^2)$ and
$G_\zeta(q^2)$ are given by the eigenvalues of this matrix. From
the first two one can determine the masses of the $\sigma$ and
$f_0$ physical states, while the function $G_\zeta(q^2)$ turns out
to be positive definite for {real, positive} values of $-q^2$. The
corresponding mixing angles can be obtained in a similar way as
{for} the $\eta$ meson sector, now defining SO(3) rotation
matrices.

The weak decay constants of pseudoscalar mesons can be calculated
following similar steps as those sketched in Sec.~\ref{sec2.1.3}.
Details of this procedure for the three-flavor case can be found
in Refs.~\cite{Scarpettini:2003fj,Hell:2009by,Carlomagno:2013ona}.
The weak decay constants for $\pi$ and $K$ mesons in the isospin
limit are given by
\begin{eqnarray}
f_\pi &=& \frac{g_{\pi q\bar{q}}}{m_\pi^2}\;
F_{uu}(-m_\pi^2) \ ,
\nonumber\\
f_K &=& \frac{g_{K q\bar{q}}}{m_K^2}\; F_{us}(-m_K^2) \ ,
\label{ch_su3.1-eq30}
\end{eqnarray}
where
\begin{eqnarray}
F_{ij}(q^2) & = & 2 N_c  \int \frac{d^4 p}{(2\pi)^4}\;
\left[g(p^+)+g(p^-)-2g(p)\right] \, Z(p) \left[\frac{M_i(p)}{D_i(p)}\, + \, \frac{M_j(p)}{D_j(p)}\right]
\nonumber\\
& & -\,2 N_c \int \frac{d^4p}{(2\pi)^4}\;
(\bar\sigma_i+\bar\sigma_j)\, \left[g(p^+)+g(p^-)-2g(p)\right]\,
g(p)
\nonumber\\
& & \qquad \qquad \times\,
\frac{Z(p^+)}{D_i(p^+)}\;\frac{Z(p^-)}{D_j(p^-)}
\;\left[ p^+\cdot p^- + M_i(p^+)M_j(p^-) \right]
\nonumber\\
& & +\,4 N_c \int \frac{d^4 p}{(2\pi)^4}\; g(p)\;
\frac{\left[M_i(p^+)p^- - M_j(p^-)p^+\right]\cdot
\left[Z(p^-)p^+-Z(p^+)p^-\right]}{D_i(p^+)\,D_j(p^-)}
\ \ ,
\end{eqnarray}
with $p^\pm = p \pm q/2$. It is worth pointing out that the functions
$F_{ij}$ do not depend on the arbitrary path chosen in the transport
functions.

For the $\eta-\eta^{\prime}$ sector, one has
\begin{eqnarray}
f_{\eta}^a &=& \frac{Z_\eta^{1/2}}{m_\eta^2} \left[f_{a8}(q^2)\cos \theta_\eta - f_{a0}(q^2) \sin\theta_\eta\right]\bigg|_{q^2=-m_\eta^2} \ ,
\nonumber \\
f_{\eta^\prime}^a &=& \frac{Z_{\eta^\prime}^{1/2}}{m_{\eta^\prime}^2}\left[f_{a8}(q^2) \sin\theta_{\eta^\prime} + f_{a0}(q^2)\cos\theta_{\eta^\prime}\right]\bigg|_{q^2=-m_{\eta^\prime}^2} \ ,
\end{eqnarray}
where $a=0,8$. The functions $f_{ab}(q^2)$ are related to $F_{ij}(q^2)$
through
\begin{eqnarray}
f_{00}(q^2) &=& \frac{1}{3}\left[ 2F_{uu}(q^2)+F_{ss}(q^2)\right]\ ,\nonumber\\
f_{88}(q^2) &=& \frac{1}{3}\left[ F_{uu}(q^2)+2F_{ss}(q^2)\right]\ ,\nonumber\\
f_{08}(q^2) &=& \frac{\sqrt{2}}{3}\left[ F_{uu}(q^2)-F_{ss}(q^2)\right] \ .
\end{eqnarray}

In order to compare with phenomenological determinations, it is
convenient to consider an alternative parametrization in terms of
two decay constants $f_0$, $f_8$ and two mixing angles $\theta_0$,
$\theta_8$~\cite{Leutwyler:1997yr,Feldmann:1999uf}. Both
parametrizations are related by
\begin{equation}
\left( \begin{array}{cc} f_{\eta}^8 & f_{\eta}^0 \\
f_{\eta^\prime}^8  & f_{\eta^\prime}^0 \end{array} \right) \ = \
\left( \begin{array}{cc} f_8\cos\theta_8 & -f_0\sin\theta_0 \\
f_8\sin\theta_8  & f_0\cos\theta_0 \end{array} \right) \ .
\label{ch_su3.1-eq33}
\end{equation}

\subsection{Extension to finite temperature and chemical potential. Coupling with the Polyakov loop}
\label{sec4.2}

In this subsection we analyze the predictions of three-flavor nonlocal
models for a system at finite temperature $T$ and/or quark chemical
potential $\mu$. The formalism presented in Sec.~\ref{sec4.1} can be
extended to nonzero $T$ and $\mu$ following the prescriptions described in
Sec.~\ref{sec2.2} for the two-flavor case. This includes the couplings
between the quarks and the Polyakov loop and the effective potential
$\mathcal{U}(\Phi,\Phi^\ast,T)$ that accounts for color gauge
self-interactions. The grand canonical thermodynamic potential per unit
volume within the MFA is given
by~\cite{Contrera:2007wu,Hell:2009by,Carlomagno:2013ona}
\begin{equation}
\Omega^{\rm MFA}_{\rm reg} \ = \ \Omega^{\rm MFA} \, -
\,\Omega^{\rm free}_q \,+ \,\Omega^{\rm free}_{q,{\rm reg}}\,
 + \,\Omega_0 \ ,
\label{sec3.2-eq1}
\end{equation}
where
\begin{eqnarray}
\Omega^{\rm MFA} - \Omega^{\rm free}_q & = & -\,2\, \sum_{c,f}\ T
\sum_{n=-\infty}^{\infty} \int \dfrac{d^3p}{(2\pi)^3} \ln
\left[\dfrac{(\rho_{n,\vec p}^c)^2 + M_f(\rho_{n,\vec
p}^c)^2}{Z(\rho_{n,\vec p}^c)^2\, [(\rho_{n,\vec p}^c)^2 + m^2_f]}
\right]
\nonumber \\
 & &
 - \left(\bar\zeta\, \bar R + \dfrac{G_S}{2}\,\bar R^2 +
\dfrac{H}{4}\,\bar S_u\, \bar S_d \, \bar S_s \right) -
\dfrac{1}{2}\, \sum_f \left( \bar \sigma_f \bar S_f
+ \dfrac{G_S}{2}\, \bar S_f^2 \right)\nonumber \\
& & +\,\ \mathcal{U}(\Phi,\Phi^\ast,T) \ ,
\nonumber \\
\Omega^{\rm free}_{q,{\rm reg}} &=& -2\, T \sum_{c,f} \sum_{s=\pm
1} \int \dfrac{d^3p}{(2\pi)^3}\; {\rm Re}\,\ln
\left[1+\exp\left(-\frac{\epsilon_{fp} + s\,(\mu\,+i\, \phi_c)}{T}
\right)\right]\ .
\end{eqnarray}
Here we have used the definition in Eq.~(\ref{ch2.2-eq8a}), viz.\
$(\rho_{n,\vec p}^c)^2 = \vec p^{\,2} + [(2n+1)\pi T + i\mu - \phi_c]^2$,
while the free quark energies are given by $\epsilon_{fp}=\sqrt{\vec p^{\;2}
+ m_f^2}$. The sums over color and flavor indices run over $c=r,g,b$ and
$f=u,d,s$, respectively. As discussed in Sec.~\ref{sec2.3}, the color
background fields $\phi_c$ can be written in terms of a single field
$\phi_3$ (the other independent field, $\phi_8$, is taken to be zero since
in this section we deal with real values of the chemical potential). We
recall that $\Omega_0$ is a constant that fixes to zero the thermodynamic
potential at $T=\mu=0$. From the minimization {of $\Omega^{\rm
MFA}_{\rm reg}$} it is possible to obtain a set of coupled gap equations
that determine the mean field values of the scalar fields,
as well as the traced Polyakov loop, at a given
temperature $T$ and chemical potential $\mu$.

To characterize the chiral and deconfinement phase transitions it is
necessary to define the corresponding order parameters. As stated, the
chiral quark condensates $\langle \bar q q\rangle$ are appropriate order
parameters for the restoration of the chiral symmetry. Their expressions can
be obtained by varying the {regularized thermodynamic potential} with
respect to the current quark masses. In general, {the corresponding quark loop
integrals are divergent}, and can be regularized by subtracting the free
quark contributions as in Eq.~(\ref{ch2.1.2-eq2}). For the three-flavor case
it is usual to define a subtracted chiral condensate normalized to its value
at $T=0$ according to
\begin{equation}
\langle \bar q q\rangle_{\rm sub} \ = \ \dfrac{\langle \bar u u
\rangle \, -\, \frac{m_u}{m_s} \, \langle \bar s s\rangle}{\langle
\bar u u \rangle_{0} \, - \, \frac{m_u}{m_s} \, \langle \bar s s
\rangle_{0}} \ .
\label{qqsub_sec4}
\end{equation}
In addition, from the thermodynamic potential we can also
calculate various thermodynamic quantities such as the {pressure,
the entropy and the energy density, which are given by
\begin{equation}
p\ = \ -\Omega^{\rm MFA}_{\rm reg} \ , \qquad \qquad
s\ = \ -\,\dfrac{\partial \Omega^{\rm MFA}_{\rm reg}}{\partial T}
\ , \qquad \qquad \varepsilon \ = \ \Omega^{\rm MFA}_{\rm reg} +
T\; s \ .
\end{equation}
}

\subsection{Form factors, parametrizations and numerical results for $T=\mu = 0$}
\label{sec4.3}

The three-flavor action in Eq.~(\ref{ch_su3.1-eq1}) includes five
free parameters, namely the current quark masses $m_u$ and $m_s$,
and the coupling constants $G_S$, $H$ and $\varkappa_p$. In
addition, one has to specify the nonlocal form factors $g(p)$ and
$f(p)$. As in the case of two-flavor models, we take into account
two types of functions for these form factors. The first
parametrization, PI, {considers} the often used exponential forms,
which guarantee a fast ultraviolet convergence of the loop
integrals. For the second one, PII, we consider a momentum
dependence based on lattice QCD results for quark
propagators~\cite{Carlomagno:2013ona}. The corresponding explicit
expressions are given by Eqs.~(\ref{ch2.3-eq1}) and
(\ref{ch2.3-eq3}-\ref{ch2.3-eq5}) in Sec.~\ref{sec2.3}, where the
parametrizations are denoted as PB and PC, respectively.

\begin{figure}[h!bt]
\centering
\subfloat{\includegraphics[width=0.45\textwidth]{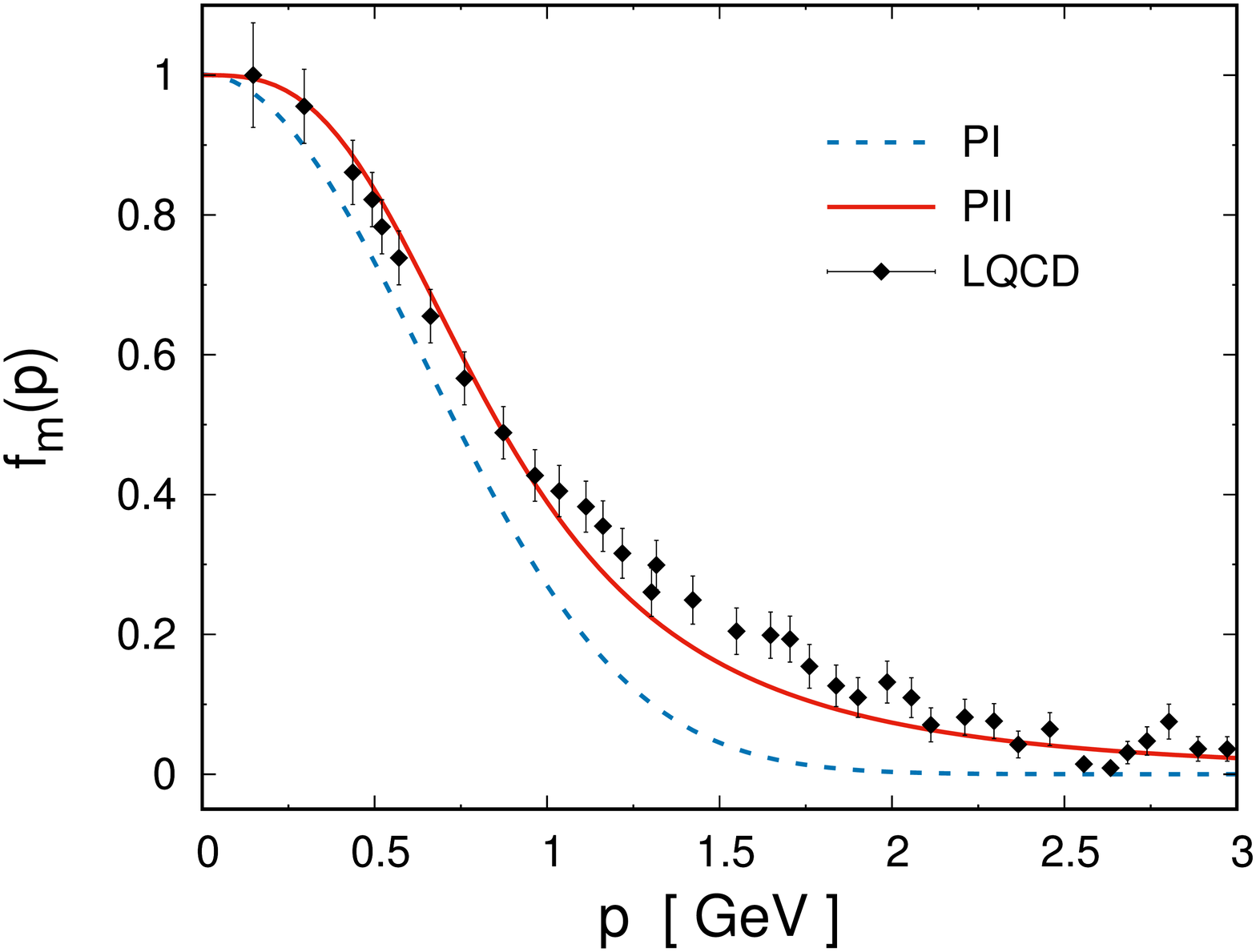}}
\subfloat{\includegraphics[width=0.45\textwidth]{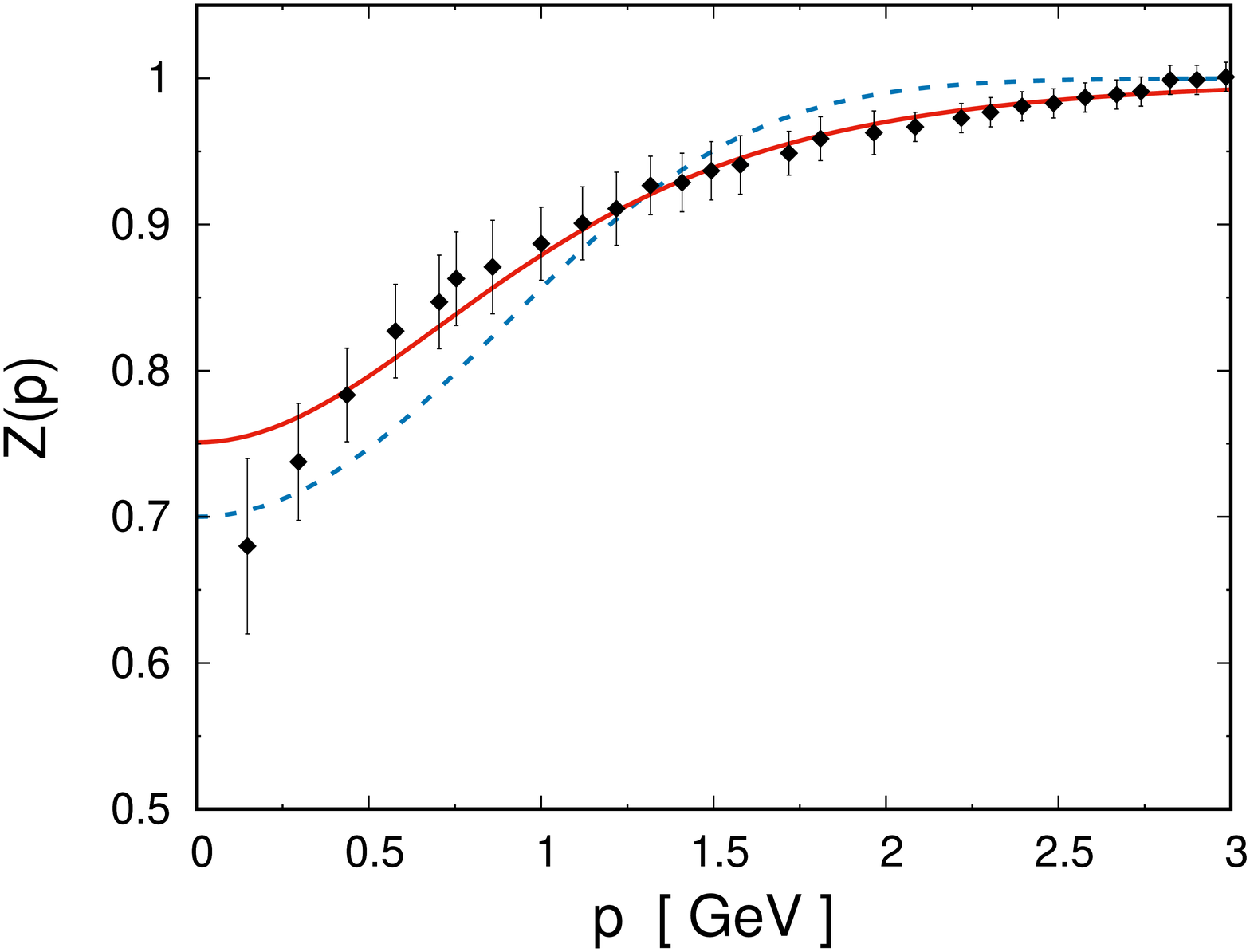}}
\caption{(Color online) Quark propagator functions $f_m(p)$ (left)
and $Z(p)$ (right). Dashed and solid lines correspond to
parametrizations PI, Eq.~(\ref{ch2.3-eq2}), and PII,
Eq.~(\ref{ch2.3-eq3}), respectively. Lattice results from
Ref.~\cite{Parappilly:2005ei} are indicated by the black dots.}
\label{fig:ch4.3-fig1}
\end{figure}

Taking into account the nonlocal form factors, one has to add to the above
mentioned free parameters the effective momentum cutoff scales $\Lambda_0$
and $\Lambda_1$. Then, to determine the full parameter sets, one can require
that the model be able to reproduce the empirical values of some physical
quantities. Following Refs.~\cite{Carlomagno:2013ona,Carlomagno:2018tyk}
these are taken to be the masses of the pseudoscalar mesons $\pi$, $K$ and
$\eta^{\prime}$, and the pion weak decay constant $f_{\pi}$. In the case of
PI~\cite{Carlomagno:2013ona}, the additional three input values are the
current quark mass $m_u$, the value of {the quark wave function
renormalization $Z(p)$ at $p=0$} and the chiral quark condensate $\langle
\bar qq\rangle$, $q =u,d$, which is fixed to $\langle \bar qq\rangle = (-
240\ {\rm MeV})^3$. For the lattice-inspired parametrization PII, input
values for $\Lambda_0$, $\Lambda_1$ and $\alpha_z$ [or, equivalently,
$Z(0)$] are obtained through a fit to LQCD results quoted in
Ref.~\cite{Parappilly:2005ei} for the functions $Z(p)$ and $f_m(p)$ given in
Eqs.~(\ref{ch2.3-eq3}) and (\ref{ch2.3-eq4}). This fit leads to $\Lambda_0 =
861$~MeV, $\Lambda_1 = 1728$~MeV and $\alpha_z =
-0.249$~\cite{Carlomagno:2018tyk}. The numerical values for the model
parameters, for both PI and PII, are quoted in Table~\ref{tab:1-su3}. The
curves corresponding to the functions $f_m(p)$ and $Z(p)$ for both
parametrizations, together with $N_f=2+1$ LQCD results from
Ref.~\cite{Parappilly:2005ei}, are shown in Fig.~\ref{fig:ch4.3-fig1}.

\begin{table}[h!bt]
\begin{center}
\begin{tabular*}{0.3\textwidth}{@{\extracolsep{\fill}} c c c c}
\hline \hline
 & & PI & PII \\
\hline
$m_u$ &MeV & $5.7$ & $2.4$  \\
$m_s$ &MeV & $136$ & $61.5$  \\
$G_S \Lambda_0^2$ && $23.64$ & $14.03$ \\
$H \Lambda_0^5$ && $-526$ & $-159$  \\
$\varkappa_p$ & GeV & $4.36$ & $10.76$  \\
$\Lambda_0$ &MeV & $814$ & $861$  \\
$\Lambda_1$ &MeV & $1032$ & $1728$ \\
\hline \hline
\end{tabular*}
\caption{Values of model parameters for
PI~\cite{Carlomagno:2013ona} and PII~\cite{Carlomagno:2018tyk}.}
\label{tab:1-su3}
\end{center}
\end{table}

After fixing the model parametrization, it is possible to calculate the mean
field values of the scalar fields, as well as to obtain predictions for
various physical quantities. In Table~\ref{tab:2-su3a} we quote the
numerical results for the mean field values $\bar\sigma_u$, $\bar\sigma_s$
and $\bar \zeta$, together with the $u$- and $s$-quark condensates. {In
the same way as} in the two-flavor case (see Tables~\ref{tab1}
and~\ref{tab2}), for PII one obtains relatively low values for the {current
quark masses} and large values for the quark
condensates~\cite{Noguera:2008cm,Hell:2011ic,Carlomagno:2013ona}. As
discussed in Sec.~\ref{sec2.3}, this can be understood by noticing that PII
is based on a fit to LQCD results that correspond to a renormalization scale
of 3~GeV. In any case, for both PI and PII one gets a phenomenologically
adequate quark mass ratio $m_u/m_s \sim 1/25$. Moreover, the product
$-\langle \bar uu\rangle m_u$ is found to be about $8\times
10^{-5}$~GeV$^4$, in agreement with the scale-independent result obtained
from the Gell-Mann-Oakes-Renner relation, Eq.~(\ref{ch2.1.3-eq9}), namely
$-\langle \bar uu\rangle m_u = f_\pi^2 m_\pi^2/2 \simeq 8.3\times
10^{-5}$~GeV$^4$.

\begin{table}[h!bt]
\begin{center}
\begin{tabular*}{0.6\textwidth}{@{\extracolsep{\fill}} c c c c }
\hline \hline
& & PI & PII \\
\hline
$\bar\sigma_u$ & MeV & $529$ & $400$ \\
$\bar\sigma_s$ & MeV & $702$ & $630$ \\
$\bar\zeta / \varkappa_p$ && $-0.429$ & $-0.332$ \\
$-\langle \bar u u \rangle ^{1/3}$ & MeV & $240$ & $325$ \\
$-\langle \bar s s \rangle ^{1/3}$ & MeV & $198$ & $358$ \\
\hline \hline
\end{tabular*}
\caption{Numerical results for mean field values of scalar fields
and chiral quark condensates.}
\label{tab:2-su3a}
\end{center}
\end{table}

{Finally,} in Table~\ref{tab:2-su3} we summarize the numerical results
obtained for meson masses, decay constants and mixing
angles~\cite{Carlomagno:2013ona,Carlomagno:2018tyk}. For comparison, we also
include the corresponding phenomenological estimates. Notice that quantities
chosen as inputs are marked with an asterisk. In the case of the $I=0$
scalar particles, the masses have been obtained by determining the zeroes of
the functions $G_M(p^2)$ arising from the diagonalization of the $3\times 3$
matrix in Eq.~(\ref{ch_su3.1-eq23}). As stated, from the corresponding
numerical calculations it is seen that one of these functions is positive
definite for the whole momentum range described by the model. Therefore, the
corresponding eigenstate cannot be associated to a light physical meson. The
remaining two states can be identified with the $\sigma$ (or $f_0(500)$) and
$f_0(980)$ scalar mesons quoted by the Particle Data
Group~\cite{Zyla:2020zbs}. Analogously, the masses of $K_0^\ast$ charged and
neutral {states} can be identified with the $K_0^\ast(1430)$
mesons~\cite{Zyla:2020zbs}. {By comparing with the phenomenological values,
it is seen that in general the predictions of the model for both PI and PII
are in a reasonable agreement with phenomenological expectations. Some
numerical results for three-flavor nlNJL schemes that consider other form
factor functions and/or parameter sets can be found in
Refs.~\cite{Scarpettini:2003fj,Hell:2009by,Hell:2011ic}.}

\begin{table}[h]
\begin{center}
\begin{tabular*}{0.6\textwidth}{@{\extracolsep{\fill}} c c c c c }
\hline \hline
& & PI & PII & Empirical \\
\hline
$m_{\pi}$ &MeV &  \ \ $139$ $^\ast$ & \ \ $139$ $^\ast$  & $139$ \\
$m_{\sigma}$ &MeV & $599$ & $518$ & 400 - 550 \\
$m_{K}$ &MeV & \ \ $495$ $^\ast$ & \ \ $495$ $^\ast$ & $495$ \\
$m_{K_0^\ast}$ &MeV & $1300$ & $1159$ & $1425$ \\
$m_{\eta}$ &MeV & $527$ & $511$ & $547$ \\
$m_{{\rm a}_0}$ &MeV & $936$ & $968$ & $980$ \\
$m_{\eta^{\prime}}$ &MeV & \ \ $958$ $^\ast$ & \ \ $958$ $^\ast$ & $958$ \\
$m_{{\rm f}_0}$ &MeV & $1300$ & $1280$ & $990$ \\
\hline
$f_{\pi}$ &MeV & \ \ $92.4$ $^\ast$ & \ \ $92.4$ $^\ast$ & $92.4$ \\
$f_K/f_{\pi}$ && $1.17$ & $1.18$ & $1.22$ \\
$f_{\eta}^0/f_{\pi}$ && $0.17$ & $0.27$ & $(0.11$ - $0.51)$ \\
$f_{\eta}^8/f_{\pi}$ && $1.12$ & $1.05$ & $(1.17$ - $1.22)$ \\
$f_{\eta^{\prime}}^0/f_{\pi}$ && $1.09$  & $2.12$ & $(0.98$ - $1.16)$ \\
$f_{\eta^{\prime}}^8/f_{\pi}$ && $-0.48$ & $-0.63$ & $-(0.42$ - $0.46)$\ \\
\hline
$\theta_{0}$ & {deg} & $-8.6$ &  $-7$ & $-(10$ - 12) \\
$\theta_{8}$ & deg & {$-23$} &  $-31$ & $-(25$ - 29) \\
\hline \hline
\end{tabular*}
\caption{Numerical results for various phenomenological
quantities. Input values are marked with an asterisk.}
\label{tab:2-su3}
\end{center}
\end{table}

\subsection{Results for finite temperature and vanishing chemical potential}
\label{sec4.4}

\begin{figure}[h!tb]
\begin{center}
\includegraphics[width=0.59\textwidth]{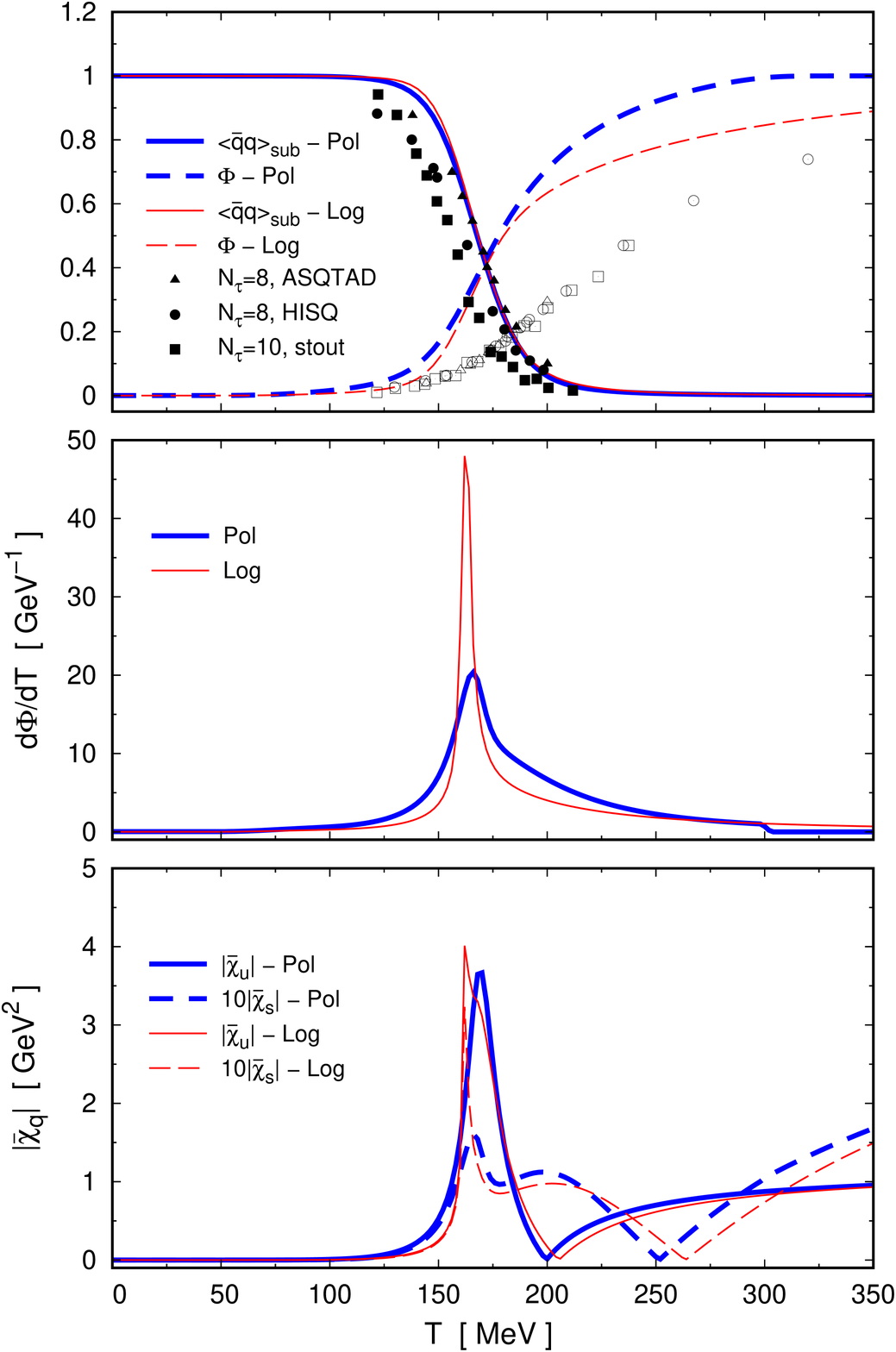}
\end{center}
\caption{(Color online) Order parameters and susceptibilities as functions
of the temperature, for parametrization PII. Triangles, circles and squares
stand for lattice QCD results from
Refs.~\cite{Borsanyi:2010bp,Bazavov:2010sb}.} \label{fig:ch4.4-fig1}
\end{figure}

In the framework of the three-flavor nlPNJL models introduced above, we
present here some results for physical quantities at finite temperature and
$\mu = 0$. We consider the parametrizations PI and PII defined in the
previous subsection, and the Polyakov loop potentials introduced in
Sec.~\ref{sec2.2}. Fig.~\ref{fig:ch4.4-fig1} illustrates the behavior of the
order parameters for deconfinement and SU(2) chiral symmetry restoration
transitions, namely the traced Polyakov loop $\Phi$ and the subtracted
chiral condensate $\langle \bar qq\rangle_{\rm sub}$, as well as the
associated susceptibilities, as functions of the temperature. The curves
correspond to parametrization PII~\cite{Carlomagno:2018tyk}. In the upper
panel, we show the results for $\langle\bar qq\rangle_{\rm sub}$ (solid
lines) and the traced Polyakov loop $\Phi$ (dashed lines). Thin and thick
lines correspond to logarithmic and polynomial PL potentials, respectively,
with $T_0=200$~MeV. For comparison we also include LQCD data quoted in
Refs.~\cite{Borsanyi:2010bp,Bazavov:2010sb}. As expected, it is found that
when the temperature is increased the system undergoes both chiral
restoration and deconfinement transitions, which proceed as smooth
crossovers. As discussed in Sec.~\ref{sec2.4} for the case of two-flavor
nlPNJL models, the curves for the order parameters get steeper for lower
values of $T_0$. Indeed, for the three-flavor model, the transitions are
found to be of first order for $T_0 < 185$~MeV. In the central and lower
panels of Fig.~\ref{fig:ch4.4-fig1} we display the curves for the PL and
chiral susceptibilities, defined by $\chi_\Phi = d\Phi/dT$ and {$\chi_q
= d\langle \bar qq\rangle/dT$} ($q = u,s$), as functions of the temperature
[for clarity, the graphs show the subtracted susceptibilities $\bar \chi_{q}
\equiv \chi_{q}(T) - \chi_q(0)$]. As usual, the deconfinement and chiral
restoration critical temperatures are defined by the peaks of $\chi_\Phi$
and $\chi_u$, respectively. In addition, in the curves for $\chi_s$ it is
possible to identify a second, broad peak that allows one to define an
approximate critical temperature for the restoration of the full SU(3)
chiral symmetry.

\begin{figure}[h]
\begin{center}
\includegraphics[width=0.9\textwidth]{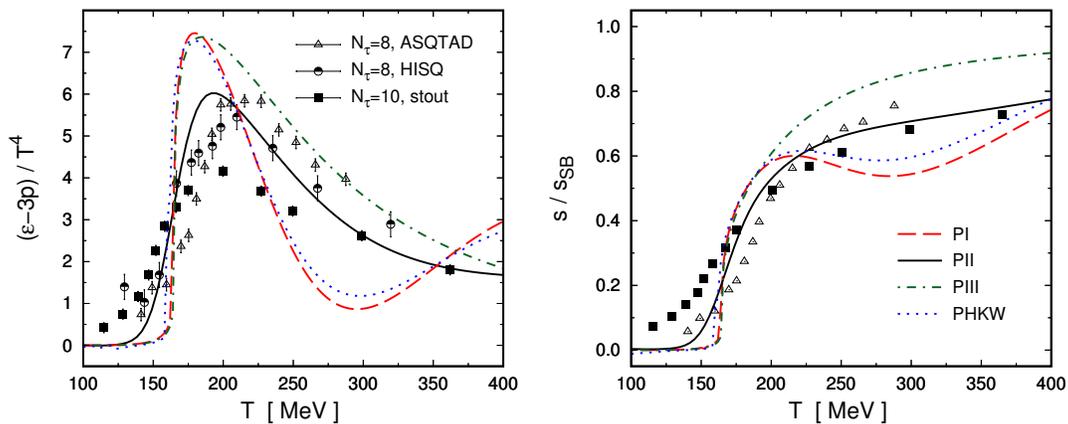}
\end{center}
\caption{(Color online) Normalized interaction energy (left) and entropy
density (right) as functions of the temperature, for different model
parameterizations. Curves correspond to nlPNJL models with logarithmic PL
potentials and $T_0 = 200$~MeV. Squares, circles and triangles stand for
lattice data from Ref.~\cite{Bazavov:2010sb}, Ref.~\cite{Bazavov:2009zn} and
Ref.~\cite{Borsanyi:2010cj}, respectively. PHKW denotes the parametrization
used in Ref.~\cite{Hell:2011ic}.}
\label{fig:ch4.4-fig2}
\end{figure}

It is seen that both the SU(2) chiral restoration and deconfinement
transitions occur essentially at the same critical temperatures, in
agreement with lattice QCD findings. From the curves in
Fig.~\ref{fig:ch4.4-fig1} one gets $T_{\rm ch} \simeq 165$~MeV (see
Table~\ref{tab:4-su3}), while LQCD analyses lead to a transition temperature
of about 160~MeV {for $N_f =
2+1$~\cite{Bazavov:2018mes,Borsanyi:2020fev}}. As in the two-flavor case
(see Table~\ref{tab4}), the above results, which correspond to the lattice
QCD-inspired parameterization PII, are qualitatively similar to those
obtained for parametrization PI, based on Gaussian form factors. To compare
the predictions from both parameterizations it is useful to consider other
thermodynamical quantities, as e.g.~the interaction energy and the entropy.
In Fig.~\ref{fig:ch4.4-fig2} we show the numerical results for the
normalized interaction energy $(\varepsilon - 3p)/T^4$ (left) and the
normalized entropy density $s/s_{SB}$ (right), where $s_{SB} = (32+21
N_f)\pi^2 T^3/45$ is the entropy density in the free-particle
Stefan-Boltzmann limit. Dashed and solid lines correspond to
parameterizations PI and PII, respectively, for the logarithmic PL potential
given by Eq.~(\ref{ch2.2-eq5}). We have included for comparison three sets
of lattice data, taken from
Refs.~\cite{Bazavov:2010sb,Bazavov:2009zn,Borsanyi:2010cj}. As in the case
of two-flavor models (see Fig.~\ref{rev-ch2.4-fig6}), it can be seen that
for {both} the interaction energy and the entropy the curves for PI show a
pronounced dip at about $T\sim 300$~MeV, which is not observed in the case
of PII. In order to trace the source of this effect we have also considered
{a third} parameterization, PIII~\cite{Contrera:2009hk,Hell:2009by}, in
which the form factor $g(p)$ has a Gaussian shape as in PI, but {the
coupling driven by the currents $j_R(x)$ is not included} [i.e.~there is no
{quark WFR, $f(p)=0$, $Z(p) = 1$}]. This parametrization is equivalent to PA
in Sec.~\ref{sec2.3}. {The results from PIII are shown by the dashed-dotted
curves in Fig.~\ref{fig:ch4.4-fig2}, which do not} show the mentioned dip.
This indicates that, as in the two-flavor case, the effect can be attributed
to the exponential behavior of the form factor $f(p)$ {that leads to the
quark WFR in} PI. Moreover, our results can be compared with those obtained
from the parameterization considered in Ref.~\cite{Hell:2011ic}, where the
form factors are introduced so as to fit lattice results for the quark
propagator (as in PII), but $f(p)$ is assumed to have a Gaussian shape. The
curves for the interaction energy and the entropy for this model (indicated
in Fig.~\ref{fig:ch4.4-fig2} as PHKW, dotted lines) are similar to those
obtained for the parameterization PI. Thus, {taking into account lattice
data, it could be stated} that the choice of a power-like behavior for
$f(p)$ as that proposed in Eq.~(\ref{ch2.3-eq5}) turns out to be more
adequate than the exponential one.

\begin{figure}[h!tb]
\begin{center}
\includegraphics[width=0.9\textwidth]{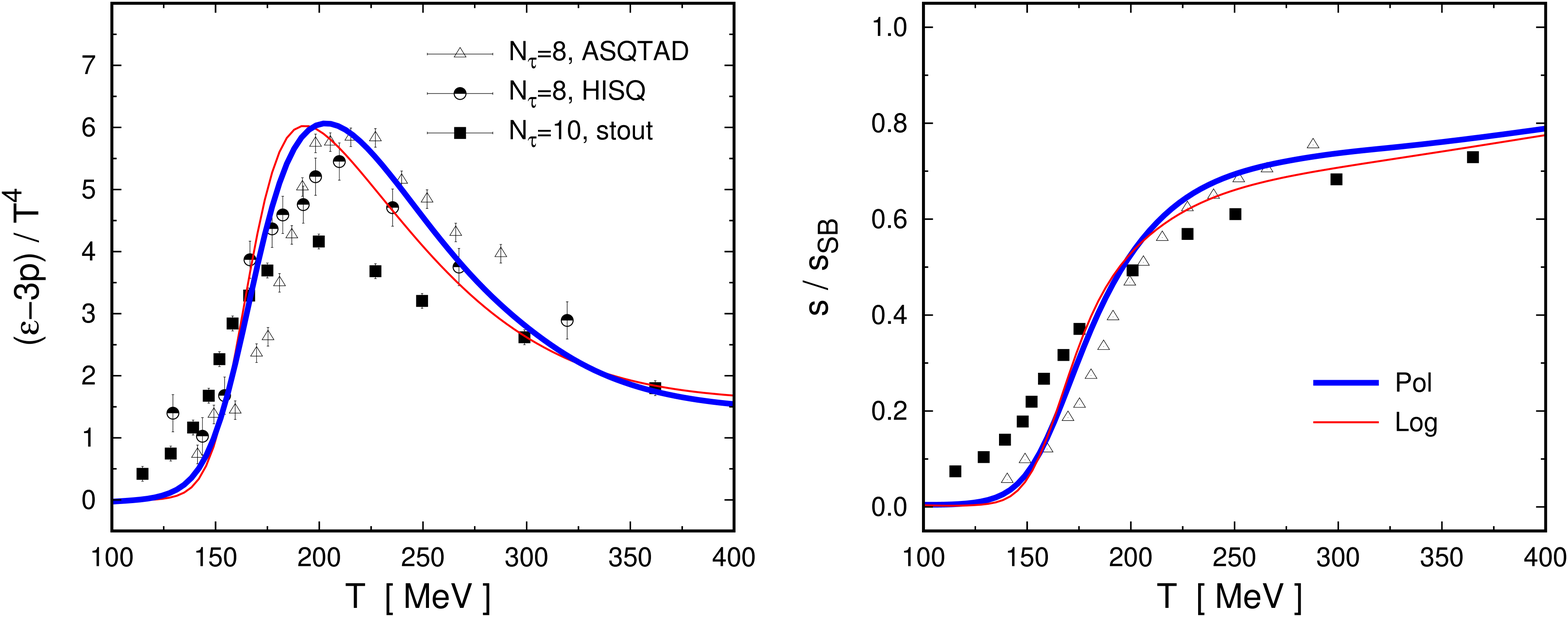}
\end{center}
\caption{(Color online) Normalized interaction energy (left) and entropy
density (right) as functions of the temperature for both the logarithmic and
the polynomial PL potentials. Squares, circles and triangles stand for
lattice data from
Refs.~\cite{Bazavov:2010sb,Bazavov:2009zn,Borsanyi:2010cj}.}
\label{fig:ch4.4-fig3}
\end{figure}

As for the general comparison with lattice QCD results, from the plots in
Fig.~\ref{fig:ch4.4-fig1} it is seen that the transition predicted by nlPNJL
models is somewhat too sharp. {In particular, there is an appreciable
difference with lattice data in} the case of the curves for the traced
Polyakov loop $\Phi$ (dashed lines in the upper panel of
Fig.~\ref{fig:ch4.4-fig1}). This is a general feature of Polyakov NJL-like
models, both local and nonlocal, and also extends to quark-meson models. In
fact, as discussed in
Refs.~\cite{Braun:2007bx,Marhauser:2008fz,Herbst:2013ufa}, the strict
comparison between the curves for the traced PL and lattice data has to be
taken with some care, owing to the difference between the definitions of
$\Phi$ in the continuum and on the lattice. One should expect a coincidence
in the crossover temperatures, which in general appears to be satisfied in
nlPNJL models for the potentials considered here. On the other hand, for PII
the behavior of the normalized interaction energy and the entropy density
(left and right panels in Fig.~\ref{fig:ch4.4-fig2}, respectively) are found
to be in reasonable agreement with LQCD results. The dependence of these
quantities on the PL potential is shown in Fig.~\ref{fig:ch4.4-fig3}, where
the curves for PII considering both logarithmic and polynomial PL potentials
are displayed, together with lattice data from
Refs.~\cite{Bazavov:2010sb,Bazavov:2009zn,Borsanyi:2010cj}. It is worth
mentioning that similar results for the mentioned thermodynamic quantities
and order parameters have been obtained in
Refs.~\cite{Schaefer:2009ui,Haas:2013qwp,Herbst:2013ufa} within the
Polyakov-quark-meson model.

{}From the left panel of Fig.~\ref{fig:ch4.4-fig3} it is seen that both LQCD
data and nlPNJL model numerical results for the interaction energy do not
tend to zero even for large temperatures. This can be interpreted as a
remnant of the strong interaction in the deconfined region, signaling that
the quark-gluon plasma {should not} be understood as a free gas of quarks
and gluons. Concerning the steepness of the transitions, it is worth to
mention that the behavior may be softened after the inclusion of mesonic
corrections to the Euclidean action, since ---as mentioned in
Sec.~\ref{sec2.4}--- when the temperature is increased light meson degrees
of freedom should get excited before quarks excitations
emerge~\cite{Blaschke:2007np,Hell:2008cc,Hell:2009by,Radzhabov:2010dd}. The
incorporation of meson fluctuations should not modify the critical
transition temperatures, which for the parametrizations considered here
turn out to be in good agreement with lattice estimations.

\hfill

To conclude this subsection we briefly mention some results given in
Ref.~\cite{Pagura:2016rit} concerning the magnetic susceptibility of the QCD
vacuum within nonlocal NJL models. A relevant quantity to be analyzed is the
vacuum expectation value of the tensor polarization operator $\langle\bar
\psi\, \sigma_{\mu\nu}\, \psi\rangle = \ {Q_q} F_{\mu\nu}  \, \tau_q $,
where $\sigma_{\mu\nu} = i[\gamma_\mu,\gamma_\nu]/2$ is the relativistic
spin operator, $F_{\mu\nu}$ is the {electromagnetic} field strength
tensor, {$Q_q$} is the quark electric charge and $\tau_q$ is the
so-called tensor coefficient. After a somewhat long but straightforward
calculation, at $T=\mu=0$ the tensor coefficient within the three-flavor
nlPNJL model is found to be given by~\cite{Pagura:2016rit}
\begin{equation}
\tau_q \ = \ 4\, N_c \int \frac{ d^4 p}{(2\pi)^4} \ Z(p) \
\frac{M_q(p) - p^2 M_q'(p) }{\left[ p^2 + M_q(p)^2 \right]^2} \ ,
\label{taumag}
\end{equation}
where $q =u,d,s$ and $M_q'\equiv dM_q(p)/dp^2$. In addition, one can study
the magnetic susceptibilities $\chi_q^{\scriptsize\mbox{(cond)}}$ associated
to the quark condensates. These are given by the relation $\tau_q =
\chi^{\scriptsize\mbox{(cond)}}_q \langle\bar q q\rangle$, for $q =u,d,s$.
In Table~\ref{tab:3-su3} we quote the values obtained for the above
quantities, for parametrization PI. The results show a diamagnetic behavior
for the QCD vacuum, {i.e.~$\chi^{\scriptsize\mbox{(cond)}}_q < 0$}. The
prediction for the $u$-tensor coefficient is in good agreement with the
value obtained from lattice QCD, whereas the quark magnetic susceptibilities
turn out to be of the order of the corresponding LQCD estimates. On the
other hand, the prediction for the $s$-tensor coefficient is significantly
lower than the LQCD value. However, it can be seen~\cite{Pagura:2016rit}
that the nlPNJL result {for $\tau_s$} is strongly dependent on the
chosen parametrization, hence the comparison of numerical values should be
taken with some care in this case.
\begin{table} [h]
\begin{center}
\begin{tabular}{cccc}
\hline \hline
  &           &   PI  &   LQCD   \\
\hline
$\tau_u$      &   MeV &     38.2   & 40             \\
$\tau_s$      &   MeV &      9.7   & 53             \\
$\chi_u^{\scriptsize\mbox{(cond)}}\ \ $ & GeV$^{-2}$ & { $-2.77$\ }   & { $-(2.05\pm 0.09)$} \\
$\chi_s^{\scriptsize\mbox{(cond)}}\ \ $ & GeV$^{-2}$ & { $-1.25$\ }   & { $-(3.40\pm 1.40)$}   \\
\hline \hline
\end{tabular}
\caption{Tensor coefficients and magnetic susceptibilities for
parametrization PI, compared to LQCD results from Ref.~\cite{Bali:2012jv}.}
\label{tab:3-su3}
\end{center}
\end{table}

\begin{figure}[H]
\begin{center}
\includegraphics[width=0.55\textwidth]{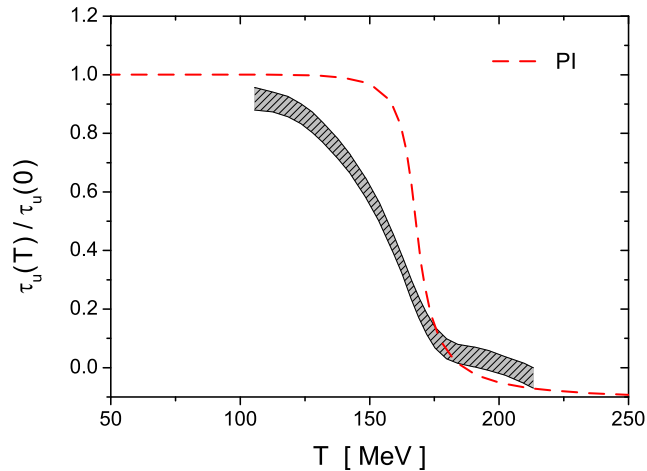}
\end{center}
\caption{(Color online) Normalized $u$-quark tensor coefficient as a function of
the temperature for parametrization PI, considering the polynomial
PL potential. Results from LQCD~\cite{Bali:2012jv} are indicated
by dashed gray band.}
\label{fig:ch4.4-fig4}
\end{figure}

Following the procedure described in Sec.~\ref{sec2.2}, the expression in
Eq.~(\ref{taumag}) can be extended to finite temperature. The thermal
dependence of the $u$-tensor coefficient for parametrization PI and the
polynomial PL potential is shown in Fig.~\ref{fig:ch4.4-fig4}. Results for
other parametrizations and PL potentials are found to be qualitative
similar. A detailed comparison, which also includes the results obtained in
Ref.~\cite{Nam:2013wja} using a ILM-inspired nonlocal model, can be found in
Ref.~\cite{Pagura:2016rit}.
If the temperature is increased from $T=0$, it is seen that the tensor
coefficient remains approximately constant up to the chiral restoration
temperature $T_{\rm ch}\simeq 165$~MeV, where one finds a sudden drop. For
comparison we also show the LQCD estimates from Ref.~\cite{Bali:2012jv}
(gray dashed band). {Once again, the transition predicted by the nlPNJL
model seems to be somewhat too sharp}, although
---as in the case of thermodynamical quantities--- this behavior
may be softened after the inclusion of mesonic corrections to the
Euclidean action.

\subsection{Results at nonzero temperature and chemical potential}
\label{sec4.5}

In this subsection we discuss the features of the phase transitions in the
$\mu - T$ plane in the context of the three-flavor nlPNJL model introduced
in Secs.~\ref{sec4.1}-\ref{sec4.3}. The phase diagram can be sketched by
analyzing the numerical results obtained for the relevant order parameters.
As stated, for the deconfinement and chiral symmetry restoration transitions
we take as order parameters the traced Polyakov loop $\Phi$ and the
subtracted chiral condensate $\langle \bar q q\rangle_{\rm sub}$ [defined in
Eq.~(\ref{qqsub_sec4})], respectively. The associated critical temperatures
$T_{\rm ch}$ and $T_\Phi$ are defined by the position of the peaks in the
chiral susceptibilities in the region where the transition occurs as a
smooth crossover. On the other hand, for relatively low temperatures the
chiral restoration occurs as a first order phase transition at a given
critical chemical potential $\mu_c(T)$. As in the case of the two-flavor
model, even if this leads to a discontinuity in the order parameter $\Phi$,
the latter remains close to zero {after the transition, and it can be
assumed} that the system is still in a confined phase. Thus, we consider the
prescription proposed in Sec.~\ref{sec2.5}, defining a {deconfinement}
transition region in which the PL lies within the range $0.3 \le \Phi \le
0.5$.

\begin{figure}[h!]
\begin{center}
\includegraphics[width=0.45\textwidth]{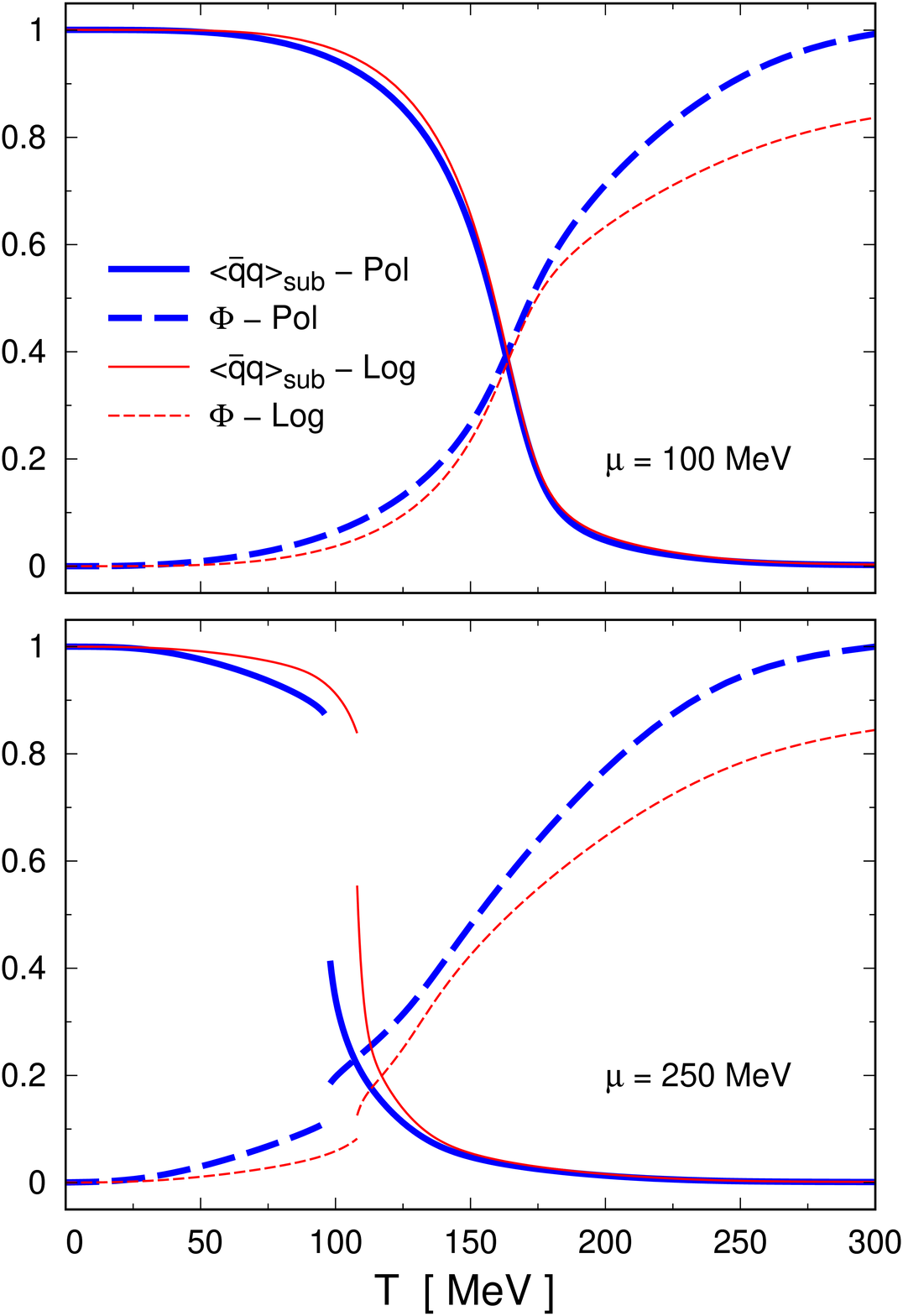}
\end{center}
\caption{(Color online) Subtracted chiral condensate (solid lines) and
traced Polyakov loop (dashed lines) as functions of the temperature. Thin
and thick lines correspond to the logarithmic and polynomial PL potentials,
respectively.}
\label{fig:ch4.5-fig1}
\end{figure}

For all parametrizations and PL potentials under consideration, one finds
that for $T=0$ the chiral restoration occurs through a first order phase
transition at a critical chemical potential $\mu_{c}(0) \sim 300$~MeV. {If
the temperature is increased, the critical chemical potential gets reduced,
and the chiral transition continues to be of first order up to a certain
critical end point (CEP) of coordinates $(\mu_{\rm CEP},T_{\rm CEP})$ (see
Table~\ref{tab:4-su3}). For $T>T_{\rm CEP}$, or $\mu<\mu_{\rm CEP}$, the
chiral restoration phase transition proceeds as a smooth crossover.} In
particular, as already discussed in the previous subsection, at $\mu=0$ the
system undergoes chiral restoration and deconfinement crossover-like
transitions at an approximately common critical temperature $T_{\rm ch} \sim
165$~MeV (see Fig.~\ref{fig:ch4.4-fig1}), in {reasonable} agreement
with LQCD. However, for larger values of the chemical potential
(particularly for $\mu > \mu_{\rm CEP}$) these transitions might occur at
different temperatures. This generic behavior is illustrated in
Fig.~\ref{fig:ch4.5-fig1}, where we quote the order parameters for the
deconfinement transition and the chiral symmetry restoration as functions of
the temperature, {considering} two representative values of the chemical
potential. The results in the figure correspond to parametrization PII. For
$\mu = 100$~MeV (upper panel) the curves are qualitatively similar to those
obtained for $\mu = 0$, see Fig.~\ref{fig:ch4.4-fig1}, while for
$\mu=250$~MeV (larger than $\mu_{\rm CEP}$), there is a jump on $\langle
\bar q q\rangle_{\rm sub}$ at $T\simeq 100$~MeV that signals a first order
chiral restoration transition (lower panel). As discussed above, the values
of $\Phi$ at the discontinuity indicate that right after the transition the
system still remains in a confined phase. In fact, the deconfinement occurs
at larger temperatures where $\Phi$ gets closer to one, say $T\gtrsim
150$~MeV for the chosen value of $\mu$. As already commented in
Sec.~\ref{sec2.5}, the phase in which quarks remain confined (signaled by
$\Phi\lesssim 0.3$) even though chiral symmetry has been restored is usually
referred to as a quarkyonic
phase~\cite{McLerran:2007qj,McLerran:2008ua,Abuki:2008nm}.

\begin{figure}[h!bt]
\begin{center}
\includegraphics[width=0.75\textwidth]{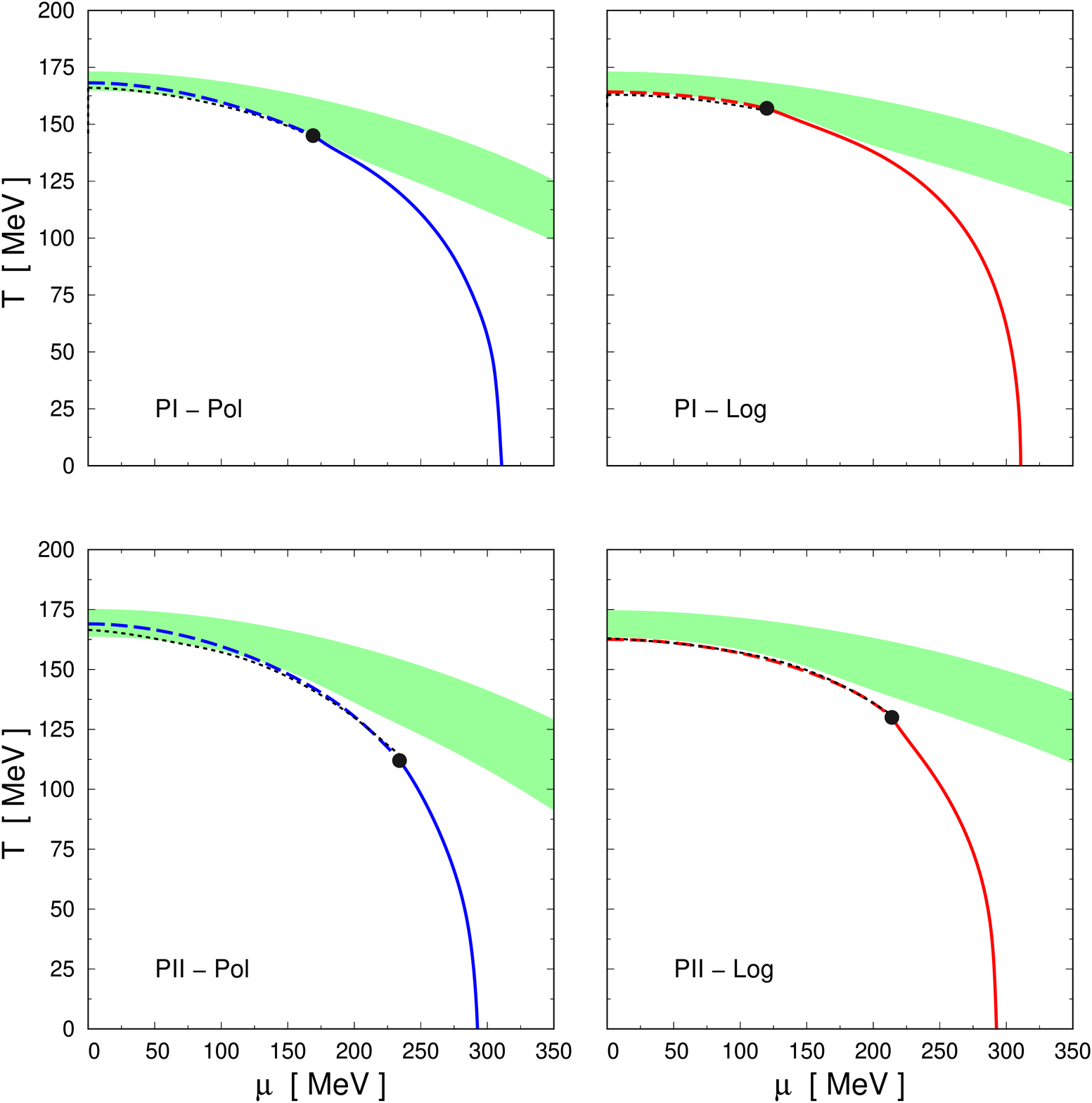}
\end{center}
\caption{(Color online) Phase diagrams corresponding to parametrizations PI
(upper panels) and PII (lower panels), for polynomial (left) and logarithmic
(right) PL potentials. Dashed and solid lines correspond to crossover and
first order chiral restoration transitions, respectively, while critical end
points are indicated by the fat dots. The color shaded areas denote the
deconfinement transition regions, defined by the condition $0.3 \leq \Phi
\leq 0.5$.}
\label{fig:ch4.5-fig2}
\end{figure}

The phase diagrams for three-flavor nlPNJL model parametrizations PI and PII
and both logarithmic and polynomial PL potentials, taking $T_0 = 200$~MeV,
are shown in Fig.~\ref{fig:ch4.5-fig2}. Solid (dashed) lines indicate first
order (crossover) phase transitions for the chiral symmetry restoration,
while the deconfinement transition region (defined by $0.3\leq\Phi\leq 0.5$)
is denoted by the color shaded areas. The fat dots denote the position of
the critical endpoints. Numerical results for CEP coordinates, critical
temperatures and chemical potentials are summarized in Table~\ref{tab:4-su3}
(values are given in MeV). As in the two-flavor case, the location of the
CEP  is found to be quite sensitive to the model parametrization and to the
form of the PL potential. The dependence of CEP coordinates on the axial
anomaly has also been studied~\cite{Hell:2009by}. Although nonnegligible, it
is found to be less strong than that in the local PNJL model. Qualitatively
similar predictions for the structure of phase diagram can be found in
Ref.~\cite{Rennecke:2016tkm} and Ref.~\cite{Schaefer:2011ex} in the context
of quark-meson and Polyakov-quark-meson models, respectively.

\begin{table}[h!bt]
\begin{center}
\begin{tabular*}{0.5\textwidth}{@{\extracolsep{\fill}} cccccc }
\hline \hline
 & \multicolumn{2}{c}{PI} & & \multicolumn{2}{c}{PII} \vspace*{.1cm} \\
 \cline{2-3}\cline{5-6}  \vspace*{-.2cm}  \\
 & \hspace*{.1cm} Pol \hspace*{.1cm}& \hspace*{.1cm} Log \hspace*{.1cm} & & \hspace*{.1cm} Pol \hspace*{.1cm} & \hspace*{.1cm} Log \hspace*{.1cm} \\
\hline
$T_{\rm ch}(0)$ \hspace*{.1cm} & 168 & 164 && 169 & 163 \\
$T_\Phi(0)$  & 166 & 164 && 167 & 163 \\
$\mu_{\rm CEP}$ & 169 & 120 && 234 & 214  \\
$T_{\rm CEP}$   & 145 & 157 && 112 & 130  \\
$\mu_{c}(0)$  &311&311 && 293&293 \\
\hline \hline
\end{tabular*}
\caption{Numerical results for some critical temperatures and chemical
potentials. All values are given in MeV.} \label{tab:4-su3}
\end{center}
\end{table}

\subsection{Meson properties at finite temperature}
\label{sec4.6}

The thermal evolution of meson masses, decay constants and mixing
angles in the context of three-flavor nlPNJL models has been
studied in Ref.~\cite{Carlomagno:2018tyk}. The corresponding
analytical expressions can be obtained from those given in
Sec.~\ref{sec4.1.3} for $T=0$, following the steps sketched in
Secs.~\ref{sec2.2} and \ref{sec2.6}.

\begin{figure}[H]
\centering
\includegraphics[width=0.5\textwidth]{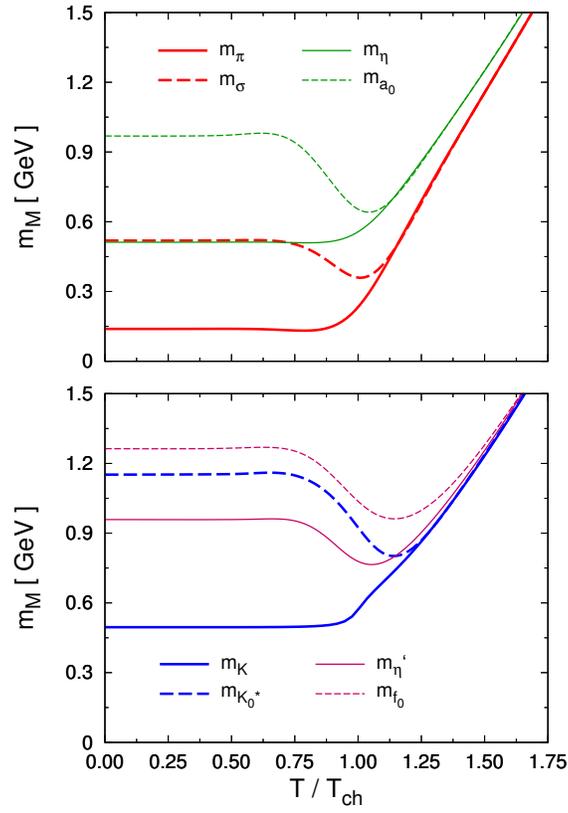}
\caption{(Color online) Scalar (dashed lines) and pseudoscalar (solid lines)
meson masses as functions of the temperature. The results correspond to PII
and a polynomial PL potential.} \label{fig:ch4.6-fig2}
\end{figure}

The behavior is found to be qualitatively similar for all parametrizations
and PL potentials under consideration~\cite{Carlomagno:2018tyk}, and also
analogous to the one obtained within other effective theories, as the
quark-meson model~\cite{Rennecke:2016tkm}. We quote here the results
corresponding to parametrization PII, which ---as shown in
Sec.~\ref{sec4.4}--- provides the best agreement with lattice QCD results
for thermodynamical quantities. For definiteness we consider the polynomial
PL potential, Eq.~(\ref{ch2.2-eq2}), with $T_0 = 200$~MeV. The plots for
meson masses as functions of the temperatures are displayed in
Fig.~\ref{fig:ch4.6-fig2}. As shown in the upper panel, the masses of the
$\pi$ and $\eta$ pseudoscalar mesons (solid lines) remain approximately
constant up to the critical temperature $T_{\rm ch}$, while those of their
scalar meson counterparts {$\sigma$ and ${\rm a}_0$} (dashed lines) start
dropping at somewhat lower temperatures. As expected, right above $T_{\rm
ch}$ the masses of chiral partners become degenerate, while at higher
temperatures they are dominated by thermal energy. In the case of the
$\eta'$ meson and its chiral partner f$_0$, and similarly for $K$ and
$K_0^\ast$ (lower panel of Fig.~\ref{fig:ch4.6-fig2}), the degeneracy is
achieved at {temperatures larger than $T_{\rm ch}$}. This is {due to} the
large current mass of the strange quark, which is expected to shift the
restoration of the full SU(3) chiral symmetry to higher temperatures.

\begin{figure}[H]
\centering
\includegraphics[width=0.48\textwidth]{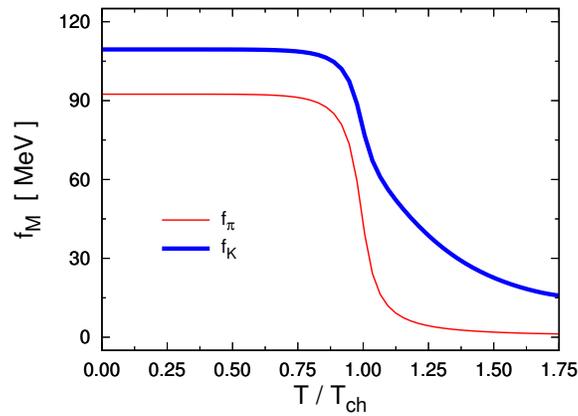}
\caption{(Color online) Pion (thin line) and kaon (thick line) decay
constants as functions of the temperature. The results correspond to PII and
a polynomial PL potential.} \label{fig:ch4.6-fig4}
\end{figure}

Finally, in Fig.~\ref{fig:ch4.6-fig4} we show the numerical results
obtained for the thermal behavior of the pseudoscalar meson decay constants
$f_\pi$ and $f_K$. The curve for $f_\pi$ shows a sudden drop at $T\simeq
T_{\rm ch}$, similar to the one observed in two-flavor nlPNJL models (see
Fig.~\ref{rev-ch2.7-fig2}). In the case of $f_K$ the fall is more moderate,
owing to the large explicit chiral symmetry breakdown caused by the strange
quark current mass.

%% file: Sec5-rev.tex
\section{Further developments and applications}
\label{sec5}

\subsection{Two-flavor superconducting quark matter and $\mu - T$
phase diagram under compact star conditions}
\label{sec5.1}

As mentioned in the Introduction, the analysis of the phase diagram for
{strong-interaction} matter has important applications in astrophysics.
In particular, as we discuss in the next subsection, the region of low
temperatures and moderate baryon chemical potentials is very interesting for
the description of compact star cores~\cite{Page:2006ud,Lattimer:2015nhk}.
Thus, in the framework of nonlocal chiral quark models, it is worth to
analyze the features of the phase diagram under compact star conditions,
i.e., under the requirements of color and electric charge neutrality
together with $\beta$ equilibrium~\cite{Iida:2000ha,AR02}.

An important issue to be addressed in this context is the presence of color
superconducting quark matter~\cite{ARW98}. In fact, a great variety of
possible diquark pairing patterns has been explored in the literature, and
estimations of the order of magnitude of the corresponding pairing gaps have
been obtained~\cite{Alford:2007xm,Casalbuoni:2003wh}. A detailed discussion
on this subject within the local NJL model can be found e.g.\ in
Ref.~\cite{Buballa:2003qv}. For phenomenological applications in compact
stars, one relevant problem is the determination of the number of active
flavors. It turns out that at low temperatures one obtains a sequential
melting pattern of the light and strange quark chiral condensates, which is
rather insensitive to the details of the four-momentum dependence of the
interaction, but crucially dependent on a selfconsistent determination of
the strange quark mass. Results obtained at the mean field level within the
three-flavor {local NJL} model~\cite{Ru05,Bla05,Abuki:2005ms} indicate that
under compact star conditions a two-flavor color superconducting (2SC) phase
is favored over the three-flavor color-flavor-locking (CFL) one. Even if the
third quark flavor occurs at not too high densities to be realized in
compact star interiors, star configurations with CFL quark cores are found
to be hydrodynamically unstable~\cite{Baldo:2002ju}. Studies of neutral 2SC
quark matter consider also the presence of a so-called gapless
phase~\cite{SH03,Aguilera:2004ag}, which is found to occur at intermediate
temperatures and chemical potentials.

\hfill

Color neutrality arises basically from the interactions between matter and
color gauge fields. In NJL-like effective quark models ---which do not
include explicit gauge field dynamics--- the effect of these interactions is
taken into account just through current-current quark couplings. Anyway, it
is possible to account for gauge interactions leading to color neutrality
through the introduction of effective chemical potentials $\mu_{fc}$ for
each quark of flavor $f$ and color $c$. In addition, the conditions of
charge neutrality and $\beta$ equilibrium require the inclusion of
electrons, as well as chemical potentials associated with quark electric
charges.

If the system is in chemical equilibrium, it can be seen that quark chemical
potentials are in general not independent~\cite{BuSh05}. Indeed, taking into
account the gauge symmetry of the theory, it is shown that only one color
chemical potential is needed in order to ensure color charge neutrality. For
a two-flavor quark model, all $\mu_{fc}$ can be written in terms of only
three independent quantities: the baryonic chemical potential $\mu_B$, a
quark electric chemical potential $\mu_{Q_q}$ and a color chemical potential
$\mu_8$. Defining as usual $\mu = \mu_B/3$, the corresponding relations
read~\cite{GomezDumm:2005hy}
\begin{eqnarray}
\mu_{ur} \ = \ \mu_{ug} &=& \mu + \frac23 \mu_{Q_q} + \frac13 \mu_8 \ , \nonumber \\
\mu_{dr} \ = \ \mu_{dg} &=& \mu - \frac13 \mu_{Q_q} + \frac13 \mu_8 \ , \nonumber \\
\mu_{ub} &=& \mu + \frac23 \mu_{Q_q} - \frac23 \mu_8 \ ,\nonumber \\
\mu_{db} &=& \mu - \frac13 \mu_{Q_q} - \frac23 \mu_8 \ .
\label{chemical}
\end{eqnarray}
Notice that red and green quarks remain indistinguishable, owing to
{the existence of a} residual SU(2) color symmetry.

One also has to take into account the presence of electrons and the
condition of $\beta$ equilibrium. The electrons can be easily included as a
gas of free Dirac particles, contributing to the full grand canonical
thermodynamic potential with a new term
\begin{equation}
\Omega_e \ = \ - \frac{1}{12 \pi^2} \left( \mu_e^4 + 2 \pi^2 T^2
\mu_e^2 + \frac{7\pi^4}{15} T^4 \right)\ ,
\end{equation}
where $\mu_e$ is the electron chemical potential. Here the electron mass has
been neglected for simplicity. Finally, the condition of $\beta$ equilibrium
arises from the $\beta$ decay reaction $d \rightarrow u + e + \bar \nu_e$.
Thus, assuming that antineutrinos escape from the stellar core, quark and
electron chemical potentials appear to be related by
\begin{equation}
\mu_{dc} - \mu_{uc} \ = \ - \mu_{Q_q} \ = \ \mu_e \ .
\label{beta}
\end{equation}
Under this condition, all effective fermion chemical potentials can
be written in terms of $\mu$, $\mu_8$ and $\mu_e$.

Now the conditions of electric and color charge neutrality {can be}
imposed by requiring that the electric charge density $\rho_{Q_{\rm tot}}$
and the diagonal color charge densities $\rho_3$ and $\rho_8$ vanish
simultaneously~\cite{AR02}. The condition $\rho_3=0$ is trivially satisfied,
while for the charge densities $\rho_{Q_{\rm tot}}$ and $\rho_8$ one has
\begin{eqnarray}
\rho_{Q_{\rm tot}} & = & \rho_{Q_q}- \rho_e \ = \ \sum_{c=r,g,b}
\left(\frac23 \ \rho_{uc} - \frac13 \ \rho_{dc} \right) - \rho_e \ = \ 0 \ ,
\nonumber \\
\rho_8 & = & \frac{1}{\sqrt3} \sum_{f=u,d} \left( \rho_{fr} +
\rho_{fg} - 2 \rho_{fb} \right) \ = \ 0 \ ,
\label{cod}
\end{eqnarray}
where the quark and electron densities $\rho_{fc}$ and $\rho_e$ can be
obtained from the derivation of the full grand canonical potential with
respect to $\mu_{fc}$ and $\mu_e$, respectively.

\hfill

In the framework of nonlocal NJL-like models, color
superconductivity can be accounted for through the inclusion of
{an interaction that involves} nonlocal quark-quark currents. In
the context of the two-flavor model analyzed in Sec.~\ref{sec2},
this interaction reads~\cite{GomezDumm:2005hy}
\begin{equation}
{\cal L}_{qq} \ = \ -\; \frac{G_D}{2}\sum_{A = 2,5,7}\,\left[j^A_D(x)\right]^\dagger
j^A_D(x) \ , \label{qqint}
\end{equation}
where
\begin{equation}
j^A_D (x) \ = \ \int d^4 z \ {\cal I}(z)\, \bar \psi_C(x+\frac{z}{2}) \
i\, \gamma_5\, \tau_2\, \lambda_A \ \psi(x-\frac{z}{2}) \ .
\label{jdiq}
\end{equation}
Here one has $\psi_C(x) = \gamma_2\gamma_4 \,\bar \psi^T(x)$, while
$\lambda_A$, with $A=2,5,7$, are Gell-Mann matrices acting on color space.

The effective {coupling} in Eq.~(\ref{qqint}) might arise via
Fierz rearrangement from some underlying more fundamental
interactions, and is understood to be used ---at the mean field
level--- in the Hartree approximation. In general, taking into
account the {interactions between quark-antiquark currents} in
Eq.~(\ref{ch2.1.1-eq1}), the ratio of coupling constants $G_D/G_S$
would be determined by these microscopic couplings. For example,
one-gluon exchange interactions, as well as instanton model
interactions, lead to $G_D/G_S = 0.75$. Since the precise
derivation of the effective couplings from QCD is not known, there
is a significant theoretical uncertainty in this {ratio. We
consider here values of $G_D/G_S$} within a phenomenologically
reasonable range between 0.5 and 1, taking for the quark-quark
current in Eq.~(\ref{jdiq}) the same nonlocal form factor as for
the quark-antiquark current, i.e.~${\cal I}(z) = {\cal G}(z)$.

To carry out the bosonization of the fermionic theory, in addition to the
bosonic fields $\sigma$ and $\pi_a$ ($a =1,2,3$) considered in
Sec.~\ref{sec2.1.1} one has to introduce diquark fields $\Delta_A$, with
mean field values $\bar \Delta_A$. Owing to color SU(3) symmetry, without
loss of generality it is possible to take $\bar \Delta_5 = \bar \Delta_7=0$,
$\bar \Delta_2 = \bar \Delta$, {which} leaves a residual symmetry under a
color SU(2) subgroup. The mean field thermodynamic potential per unit volume
can be obtained by making use of the Nambu-Gorkov formalism, as detailed
e.g.~in Refs.~\cite{Duhau:2004pq,GomezDumm:2005hy}, and a new ``gap
equation'' arises by requiring the minimization of this effective
thermodynamic potential with respect to $\bar \Delta$. In this way, under
color neutrality conditions, for each value of $T$ and $\mu$ one should find
the values of $\bar \sigma$, $\mu_e$, $\mu_8$ and $\bar \Delta$ that solve
the corresponding gap equations, supplemented by Eqs.~(\ref{beta})
and~(\ref{cod}).

In what follows we discuss the numerical results obtained in
Ref.~\cite{GomezDumm:2005hy} for a two-flavor nlNJL model. The
parametrization used in that work is similar to PA (see
Sec.~\ref{sec2.3}), i.e.~it involves a Gaussian form factor and
does not include the derivative currents in
Eq.~(\ref{ch2.1.1-eq1}). Moreover, for simplicity the analysis in
Ref.~\cite{GomezDumm:2005hy} does not take into account the
coupling between quarks and the Polyakov loop. In principle, these
simplifications should not imply qualitative changes in the phase
diagram. Even if the inclusion of the couplings involving the
Polyakov loop would increase the transition temperatures at low
chemical potentials, the phase diagram structure should not be
significantly modified in the low temperature region, which is the
most interesting one for compact star applications.

\begin{figure}[hbt]
\begin{center}
\includegraphics[width=0.9\textwidth]{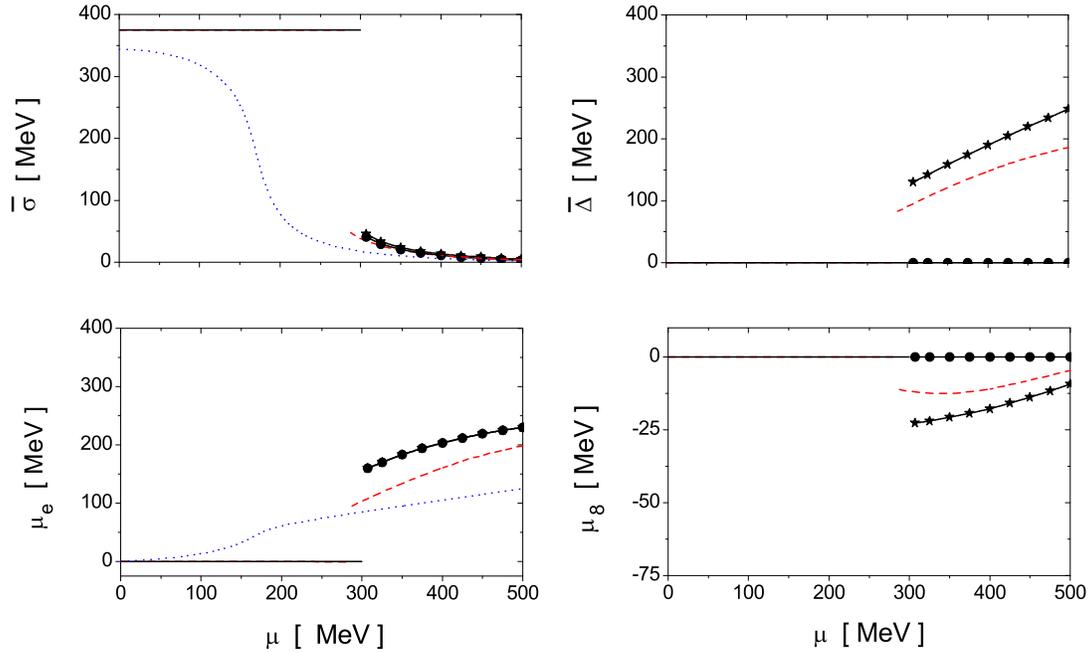}
\caption{(Color online) Behavior of $\bar{\sigma}$, $\bar{\Delta}$, $\mu_e$
and $\mu_8$ as functions of the chemical potential, for three different
values of the temperature. Solid lines correspond to $T=0$, dashed lines to
$T = 40$~MeV and dotted lines to $T = 100$~MeV. In the case of $T=0$, lines
marked with stars and dots correspond to 2SC and NQM phases respectively.
The nlNJL model parameters used here are $m_c = 5.12$~MeV, $\Lambda =
827$~MeV and $G_S \Lambda_0^2= 18.78$~\cite{GomezDumm:2005hy}.}
    \label{fig5.1.1}
\end{center}
\end{figure}

In Fig.~\ref{fig5.1.1} we quote the numerical results for the
behavior of the mean field values $\bar\sigma$ and $\bar\Delta$,
as well as the effective chemical potentials $\mu_e$ and $\mu_8$,
as functions of $\mu$. The plots correspond to a ratio $G_D/G_S =
0.75$. Solid, dashed and dotted lines correspond to $T=0$, 40 and
100~MeV respectively. For $T=0$, at low chemical potentials the
system lies in a chiral symmetry broken phase (CSB), where the
quarks acquire large dynamical masses. By increasing the chemical
potential one reaches a first order phase transition, in which the
chiral symmetry is approximately restored, and a certain fraction
of the quark matter undergoes a transition to the 2SC phase,
coexisting with the remaining normal quark matter (NQM) phase. The
jump of $\mu_8$ at the transition is {related to} that of
$\bar\Delta$, which governs the amount of breakdown of the color
symmetry {arising from} quark pairing. Moreover, it is seen that
the chemical potential $\mu_e$ (which for $T=0$ vanishes in the
CSB region) also shows a discontinuity across the transition.

The new 2SC-NQM mixed phase is a way in which the constraint of electric
neutrality is globally fulfilled: the coexisting phases have opposite
electric charges which neutralize eac other, at a common equilibrium
pressure. In its simplest realization, this mixed phase is treated within an
approximation in which Coulomb and surface energies are
neglected~\cite{Glendenning:1992vb}. For color superconducting quark matter
this realization of charge neutrality has been considered e.g.\ in
Refs.~\cite{Neumann:2002jm} and~\cite{Aguilera:2004ag} within the NJL model
and an instantaneous nonlocal quark model, respectively. On the other hand,
notice that color neutrality has been imposed as a local
constraint~\cite{Shovkovy:2003ce,Reddy:2005}. This is based on the fact that
the color Debye screening length is expected to be short and comparable to
the inter-particle distance in the regime of interest. As a consequence, the
color chemical potential $\mu_8$ turns out to be different in the two
components of the mixed phase.

\begin{figure}[h!bt]
\begin{center}
\includegraphics[width=0.4\textwidth]{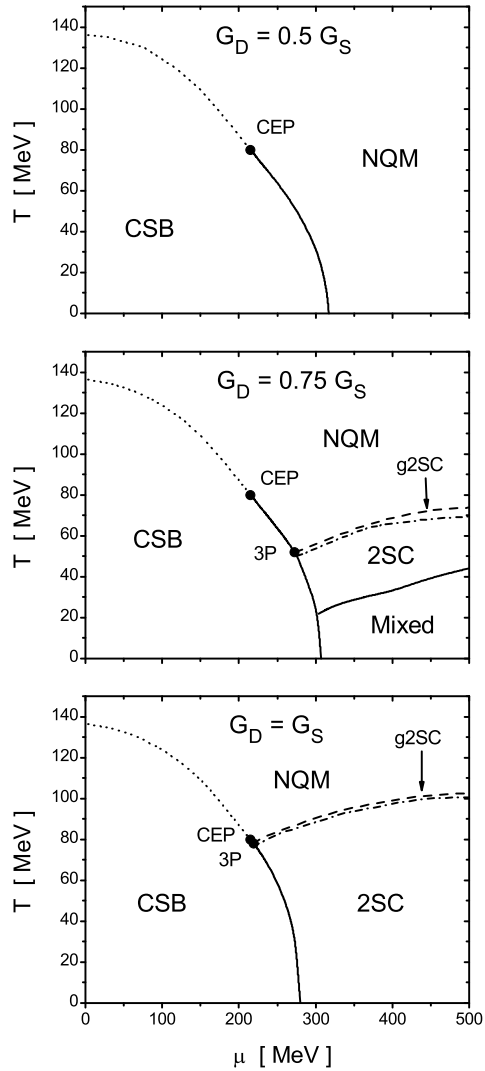}
\caption{Phase diagrams for $G_D/G_S= 0.5$, 0.75 and 1. Solid and dashed
lines indicate first and second order phase transition curves respectively,
dotted lines correspond to crossover-like transitions, and dash-dotted lines
delimit the gapless 2SC band. Different phases are denoted as NQM (normal
quark matter phase), CSB (chiral symmetry broken phase) and 2SC (two-flavor
superconducting phase), while the region marked as ``Mixed'' corresponds to
a NQM-2SC mixed phase. CEP and 3P denote the critical end points and
triple points, respectively.}
\label{fig5.1.2}
\end{center}
\end{figure}

When the temperature is increased up to $\sim 20$~MeV, the
transition to a mixed phase is no longer favored and the system
goes into a pure 2SC phase. As can be seen from the dashed curves
in Fig.~\ref{fig5.1.1}, for $T=40$ MeV this still shows up as a
first order phase transition. For even larger temperatures, it is
seen that the 2SC phase is no longer present, and the system
undergoes a transition from the CSB phase to a normal quark matter
(NQM) phase, in which the chiral symmetry is approximately
restored and there is no color superconductivity. For $T =
100$~MeV this transition occurs as a smooth crossover, as shown by
the dotted lines in Fig.~\ref{fig5.1.1}. The full phase diagram
for $G_D/G_S = 0.75$ is displayed in the central panel of
Fig.~\ref{fig5.1.2}. If one moves along the first order transition
line from $T=0$ towards higher temperatures, {at $T\simeq 50$~MeV}
a triple point (3P) is reached. At this point the CSB and 2SC
phases coexist with a third NQM phase. Finally, if $T$ is still
increased, one reaches a critical end point (CEP) where the first
order transition from the CSB to the NQM phase becomes a smooth
crossover. For comparison, in Fig.~\ref{fig5.1.2} the phase
diagrams for $G_D/G_S= 0.5$ and $G_D/G_S=1$ are also shown.

Another feature to be discussed is the presence of the so-called
gapless 2SC (g2SC) phases. As it is shown in Fig.~\ref{fig5.1.2},
g2SC phases are favored in a narrow band close to the 2SC$-$NQM
phase border. In this region, in addition to the two gapless modes
corresponding to the unpaired blue quarks, the presence of flavor
asymmetric chemical potentials $\mu_{dc}-\mu_{uc}\neq 0$ gives
rise to another two gapless fermionic quasiparticles~\cite{SH03}.
Although the corresponding dispersion relations cannot be derived
analytically owing to the nonlocality of the interactions, the
border of the g2SC region can be found numerically. This is done
by determining whether for some value of {the quark momentum}
$|\vec p|$ the imaginary part of some of the poles of the
Euclidean quark propagator vanish in the complex $p_4$ plane.
Although for the cases shown in the figure the g2SC region is
given by a narrow band, it appears to be considerable enlarged
depending on the model parametrization~\cite{GomezDumm:2005hy}.

The above phase diagrams can be compared with those obtained for isospin
symmetric quark matter. In general, it is seen that the 2SC region becomes
reduced when one imposes color and electric charge neutrality conditions.
This is what one would expect, since the condition of electric charge
neutrality leads in general to unequal $u$ and $d$ quark densities,
disfavoring the $u$-$d$ pairing. In any case, the effect is found to be
relatively small, and the positions of triple points and critical end
points, as well as the shapes of critical lines, remain approximately
unchanged~\cite{GomezDumm:2005hy}. Qualitatively similar results for isospin
symmetric quark matter have been obtained in the context of a chiral
two-flavor quark-meson-diquark model in Ref.~\cite{Braun:2018svj}.

\subsection{Astrophysical applications}
\label{sec5.2}

After having analyzed {the thermodynamic features of quark
matter}, it is natural to investigate if such a state can exist in
the cores of cold compact stars, where {particles} appear to be
compressed up to densities several times larger than that of
nuclear matter ($n_0\sim 0.16$~fm$^{-3}$). Unfortunately, a
consistent relativistic approach to the quark-hadron phase
transition, in which hadrons are treated as bound states of
quarks, has not been developed up to now. In this situation,
hadronic models, as e.g.\ the Walecka model or the relativistic
Dirac-Brueckner-Hartree-Fock (DBHF) approach, have been proposed
for the description of the hadronic phase, {and quark models have
been considered in order to account for the quark matter phase. In
particular, several works have employed nonlocal NJL-like schemes
to provide an effective theoretical description of quark matter
states inside compact stellar cores.}

From the observational side, early measurements of the pulsar J0751+1807 and
other neutron stars (NS) in X-ray binaries have indicated the existence of
large mass compact stars. This has been  confirmed by the
{further} observation of the binary pulsars PSR J1614-2230, PSR
J0348+0423, PSR J2215+5135 and PSR J0740+6620, with masses of the order of
$2\,M_\odot$~\cite{Demorest:2010bx,Antoniadis:2013pzd,Fonseca:2016tux,
Arzoumanian:2017puf,Linares:2018ppq,Cromartie:2019kug}. The evidence of
compact objects of such a large mass can be used as a test for the viability
of hadronic and/or quark matter equations of state (EoS) arising from
theoretical models~\cite{Klahn:2006ir}. Given the EoS of a neutron star
(i.e., the functional relation between pressure and energy density), the
corresponding mass vs.\ radius relation can be obtained through the
Tolman-Oppenheimer-Volkoff (TOV) equations of hydrodynamic stability for
self-gravitating matter. In general, it is found that a rather stiff EoS is
required in order to satisfy observational constraints for large mass
neutron stars.

In the context of covariant nonlocal NJL models (instantaneous nlNJL models
have been also analyzed, see Ref.~\cite{Grigorian:2003vi}), the study of a
possible quark matter phase in the core of compacts stars has been firstly
addressed in Ref.~\cite{Blaschke:2007ri}, considering a two-phase
description of the NS interior. In this ``hybrid star'' scheme, the nuclear
matter phase is treated within the relativistic DBHF
approach~\cite{vanDalen:2004pn} and the transition to quark matter is
obtained through a Maxwell construction. The two-flavor nlNJL model
considered in Ref.~\cite{Blaschke:2007ri} includes a diquark coupling of the
form proposed in Eq.~(\ref{qqint}), as well as a current-current vector
coupling in the {$I=0$} channel, as that in Eq.~(\ref{ch3.1.1-eq1}).
The presence of a nonvanishing diquark condensate reduces the critical
low-temperature transition density (allowing for a quark matter phase in the
NS core), whereas the nonzero mean field $\bar \omega$ {arising from
the vector coupling} has the effect of stiffening the EoS. This leads to an
increase of the maximum accessible NS masses up to a magnitude of about
$2M_\odot$. The model parameters are similar to those considered in
Sec.~\ref{sec5.1} for the study of phase diagrams under compact star
conditions, while the proposed additional interaction in the vector channel
is relatively weak, and does not affect the qualitative features of the
phase transitions~\cite{Blaschke:2007ri}. The numerical results for this
model are found to be in agreement with observational constraints for a
given range of the model parameters. Moreover, the corresponding isospin
symmetric equations of state are shown to be consistent with flow data
analyses from heavy ion collisions.

In Refs.~\cite{Orsaria:2012je,Orsaria:2013hna} the existence of quark matter
in the core of NSs has been studied using three-flavor nonlocal NJL models,
as those described in Sec.~\ref{sec4}. Once again, some repulsive
current-current interaction in the $I(J^P) = 0(1^-)$ channel is included, so
as to increase the critical $T=0$ chemical potential and stiffen the quark
matter EoS. In turn, the description of the confined, hadronic phase, is
carried out by considering the parametrizations
GM1~\cite{Glendenning:1991es} and NL3~\cite{Lalazissis:1996rd}, within the
relativistic mean field theory. The phase regions are determined by assuming
a smooth Gibbs transition, in which the condition of pressure equilibrium is
imposed together with the requirement of global electric charge neutrality
and baryon number conservation~\cite{Glendenning:1992vb} in the mixed
quark-hadronic phase. The volume fraction of quark matter is given by a
parameter $\chi$, which ranges continuously from 0 to 1 along the
transition. Form the corresponding numerical analysis, it is found that the
results are compatible with the existence of extended regions of mixed
quark-hadron matter in neutron stars with masses of about $2M_\odot$ and
radii in the canonical range of 12 to 13~km. This holds also for local
versions of NJL-like models. Furthermore, it is seen that pure quark matter
can exist in stellar cores for certain parametrizations of the effective
quark and hadronic interactions. These works have been complemented by the
study of the thermal evolution of NSs, obtained from balance and transport
equations under the assumption of spherical
symmetry~\cite{deCarvalho:2015lpa}. It is found that high-mass NSs may
contain a fraction of 35 to 40\% of deconfined quark-hybrid matter in their
cores, while for stars with canonical masses of around $1.4\,M_\odot$ (for
which cooling curves show a good agreement with experimental data) a pure
hadronic composition is predicted. In this same theoretical context, the
possible existence of a crystalline quark-hadron mixed phase and its effects
on neutrino emissivity have been studied in
Refs.~\cite{Spinella:2015ksa,Spinella:2018bdq}. Moreover, in
Ref.~\cite{Mellinger:2017dve} the existence of quark matter in the cores of
rotating NSs has been also considered. On the other hand, in
Ref.~\cite{Ranea-Sandoval:2015ldr} it is claimed that if the surface tension
at the boundary separating neutral hadronic and neutral quark matter phases
is larger than a critical value of about 10 to
40~MeV/fm$^2$~\cite{Alford:2001zr,Endo:2011em,Lugones:2013ema}, the mixed
phase turns out to be disfavored and there should be a sharp interface
between both regions (Maxwell construction). If this is the case, the
analysis of Ref.~\cite{Ranea-Sandoval:2015ldr} concludes that large mass
hybrid stars would not be allowed within the three-flavor nlNJL approach for
quark matter. Similar results are found for the three-flavor local NJL
model.

Turning back to astrophysical observations, in the past few years important
new constraints for the description of compact stars have come out from the
direct detection of gravitational waves emitted from the binary NS merger
GW170817~\cite{TheLIGOScientific:2017qsa}. In particular, the analysis of
GW170817 data leads to an upper limit for the tidal deformability ---which
measures the NS deformation due to the gravitational field of its companion
object--- and this translates into an upper limit for the NS radius. For a
NS mass of 1.4~$M_\odot$, it is found that the radius cannot exceed a limit
of approximately 13.6~km~\cite{Raithel:2018ncd}. Moreover, it is argued that
GW170817 data also imply a general upper limit of about $2.3 M_\odot$ for
the mass of a cold spherical NS~\cite{Shibata:2019ctb}. These constraints
have been taken into account for the analysis of EoS arising from hybrid
star models that include nlNJL approaches for quark
matter~\cite{Alvarez-Castillo:2018pve,Ranea-Sandoval:2019miz,
Orsaria:2019ftf,Malfatti:2019tpg,Shahrbaf:2019vtf,Shahrbaf:2020uau,Malfatti:2020onm}.

\begin{figure}[hbt]
\begin{center}
\includegraphics[width=0.59\textwidth]{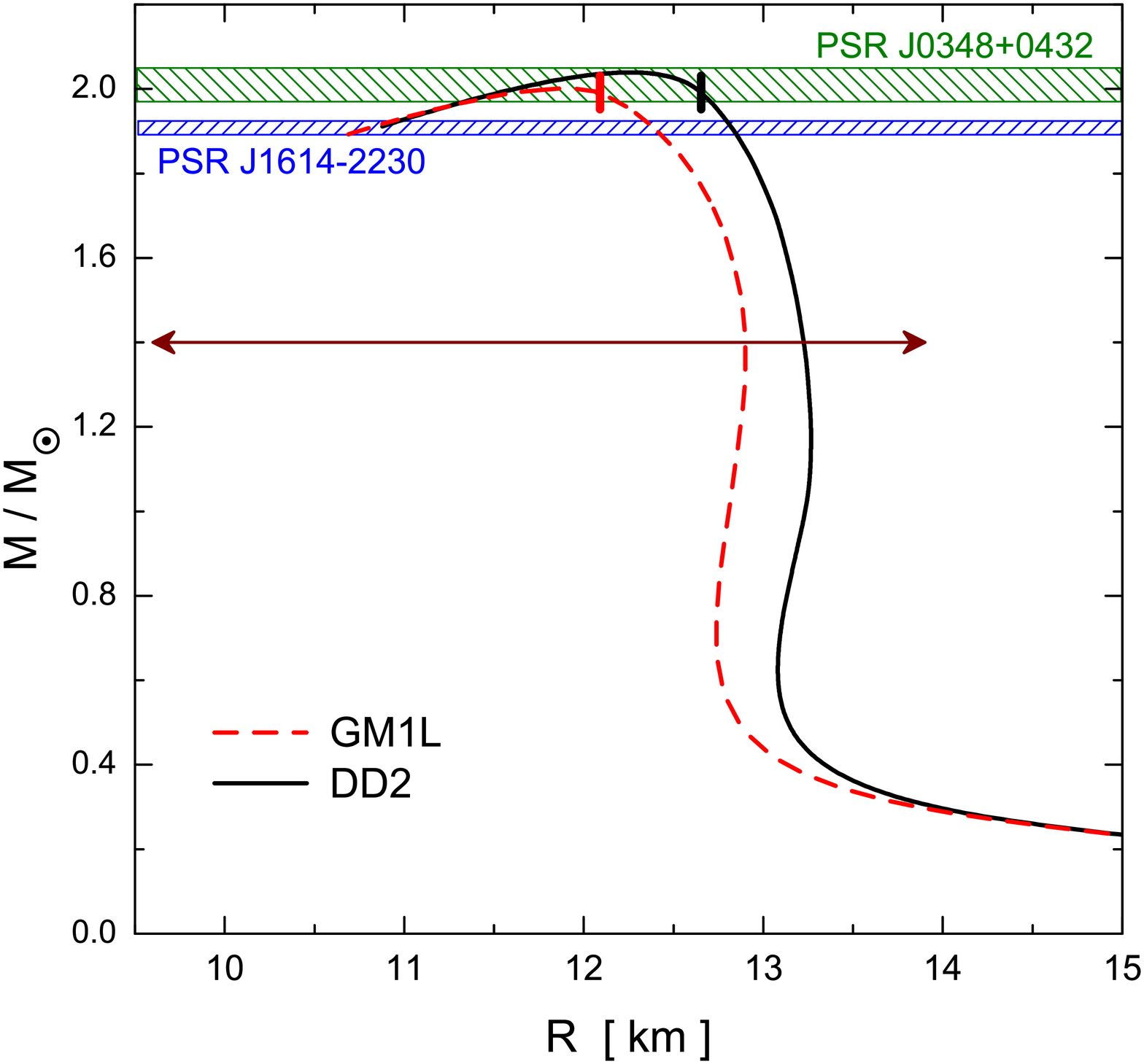}
\caption{(Color online) Gravitational mass as a function of the stellar
radius for two parametrizations of the hadronic phase. The shaded bands
indicate the constraints given by the observed PSR J1614-2230 and PSR
J0348+0423 masses, while the horizontal line shows the bounds for a
$1.4\,M_\odot$ NS radius arising from the analysis of GW170817 data. The
vertical bars on the curves denote the onset of the transition from hadronic
to quark matter.}
\label{fig5.2.1}
\end{center}
\end{figure}

In Refs.~\cite{Ranea-Sandoval:2019miz,Malfatti:2019tpg} these new
constraints are considered in the context of the above described
three-flavor nonlocal NJL models, at both zero and finite temperature. In
the finite temperature region, the treatment of quark matter includes the
coupling between the quarks and the Polyakov loop, as described in
Sec.~\ref{sec2.2}. In addition, new parametrizations called DD2 and GM1L are
used for the description of the purely hadronic phase. They are based on the
standard relativistic mean field approach, taking into account medium
effects through the inclusion of explicit density dependent meson-baryon
couplings~\cite{Typel:2009sy,Ranea-Sandoval:2019miz}. The hadron-quark
transition is treated according to a Maxwell construction in which there is
a sharp interface and no mixed phase regions. For the case of cold neutron
stars, it is found that the presence of a quark matter core is allowed for a
coupling constant ratio $G_0/G_S$ lying in a range from $\sim 0.33$ to 0.38
in order to satisfy observational constraints. This is illustrated in
Fig.~\ref{fig5.2.1}, where we show the curves for NS mass vs.~radius
obtained in Ref.~\cite{Malfatti:2019tpg} after solving the TOV equations,
for both GM1L and DD2 hadronic parametrizations. The shaded bands indicate
the mass region to be reached in order to {fulfill the constraints} from the
the measurements of PSR J1614-2230 and PSR J0348+0423 masses, while the
brown horizontal line indicates the bounds for the radius of a NS star of $M
= 1.4\,M_\odot$ arising from the analysis of GW170817 data. The vertical
bars on the curves denote the onset of the transition from hadronic to quark
matter. On the other hand, from the study of the evolution of proto-neutron
stars to neutron stars it is found that in these models the existence of
hybrid stars is allowed only for temperatures not higher than 15 to 30~MeV.
Even in the case in which the model parameters are compatible with a quark
matter core at $T=0$, this deconfined phase region would disappear for hot
neutron or proto-neutron stars. In a more recent
work~\cite{Malfatti:2020onm} the analysis has been extended (at $T=0$) to a
three-flavor nlNJL model that includes also a 2SC+s superconducting phase.
Depending on the stability criteria for the hadron-quark interface, once
again it is found that the existence of neutron stars with a superconducting
quark matter core can be compatible with observational data. The study of
hybrid star constraints has been also carried out in the context of the
quark-meson model for quark matter, considering the DD2 parametrization for
the hadronic phase~\cite{Otto:2019zjy}.

Finally, it is worth mentioning that recent
works~\cite{Shahrbaf:2019vtf,Shahrbaf:2020uau} have addressed, in the
context of nonlocal NJL-like models, the so-called ``hyperon puzzle''. In
principle, the core of a NS should include hyperonic degrees of freedom,
which soften the EoS in such a way that NS masses larger than $\sim
1.5\,M_\odot$ could become unreachable. A possible solution of this puzzle
can arise from hybrid NS models, provided that quark deconfinement occurs at
low enough densities so as to prevent the existence of hypernuclear matter
in the stellar core. In Refs.~\cite{Shahrbaf:2019vtf,Shahrbaf:2020uau} it is
seen that such a scenario can be obtained within two-flavor nlNJL models
that include color superconductivity and vector current-current couplings,
which lead to a stiff EoS~\cite{Alvarez-Castillo:2018pve}. The hadron-quark
phase transition is obtained by a Maxwell construction, considering either
fixed or density-dependent couplings. The numerical analysis shows that only
in the case of density-dependent couplings an intermediate hypernuclear
phase can be found in the cores of hybrid stars. In addition, it is seen
that in both situations the onset of deconfinement occurs for compact stars
of about 1 to $1.14\,M_\odot$. The observational radius and mass constraints
are found to be satisfied, circumventing the hyperon puzzle. There are also
alternative ways of dealing with this problem, e.g.~by considering the
effects of three-body forces that involve hyperons, such as
nucleon-nucleon-$\Lambda$ interactions. Analyses in this sense have been
carried out in Refs.~\cite{Lonardoni:2014bwa,Logoteta:2019utx}.

\subsection{Inhomogenous phases}
\label{sec5.3}

In the past few years, it has been claimed that the phase diagram of
strong-interaction matter at low temperatures and high densities could
include spatially nonuniform phases~(for a review see e.g.\
Ref.~\cite{Buballa:2014tba}). Within the framework of the local two-flavor
NJL model in the mean field approximation, it has been shown that the
critical end point of the first order chiral restoration transition exactly
coincides with the so-called Lifshitz point (LP), where two homogeneous
phases and one inhomogeneous phase meet~\cite{Nickel:2009ke,Nickel:2009wj}.
This result has been obtained in the chiral limit, where the critical end
point becomes a tricritical point (TCP). If vector-like interactions are
added, it is seen that the LP remains at the same temperature, while the TCP
moves to a lower $T$, in such a way that it gets covered by the
inhomogeneous phase and disappears from the phase
diagram~\cite{Carignano:2010ac}. In addition, this issue has been studied in
the context of a quark-meson model with vacuum
fluctuations~\cite{Carignano:2014jla}, where it is found that the LP might
coincide or not with the TCP depending on the model parametrization. In
fact, both in the framework of the NJL model and quark-meson models, it is
found that the TCP tends to becomes covered by the inhomogeneous phase and
does not show up in the phase diagram. The analysis of inhomogeneous phase
regions in the context of the Polyakov-NJL model~\cite{Braun:2015fva}
and the three-flavor NJL model~\cite{Carignano:2019ivp}, as well as the
effects of finite current quark masses~\cite{Buballa:2018hux}, have also
been addressed. Moreover, indications of the existence of an inhomogeneous
phase have been found in a recent Functional Renormalization Group study of
the QCD phase diagram~\cite{Fu:2019hdw}.

In this subsection we discuss the possible presence of inhomogeneous phases
in the context of nlNJL models. We consider in particular the analyses in
Refs.~\cite{Carlomagno:2014hoa,Carlomagno:2015nsa}, which correspond to a
model similar to the one introduced in Sec.~\ref{sec2.1}. Given the
complexity of the problem, the derivative currents $j_R(x)$ have not been
included and the chiral limit $m_c = 0$ has been taken.

As discussed in Sec.~\ref{sec2.1.1}, it is convenient to bosonize the
fermionic theory, introducing scalar and pseudoscalar fields that can be
expanded around the mean field values $\bar \sigma (\vec x)$ and $\bar \pi_i
(\vec x)$. These mean field values are now allowed to be inhomogeneous,
hence the explicit dependence on spatial coordinates. The resulting mean
field Euclidean action reads
\begin{equation}
S^{\rm MFA}_E \ = \ - \, \ln \det\ {\cal D}_0 \; + \;
\frac{1}{2G_S} \int d^3 x \; \phi^a(\vec x)\, \phi^a(\vec x)\ ,
\end{equation}
where we have defined a chiral four-vector $\phi^a = (\bar \sigma (\vec x) ,
\vec{\bar \pi}(\vec x))$. In this expression the operator ${\cal D}_0$ is
given by
\begin{eqnarray}
{\cal D}_0(x,y) \; = \;\delta^{(4)}(x-y) \; (-i \rlap/\partial_y) \; +
\; {\cal G}(x-y) \; \Gamma^a \, \phi^a \left((\vec x + \vec
y)/2\right) \ ,
\end{eqnarray}
where $\Gamma^a = (1 , \vec \tau)$. The extension to finite temperature $T$
and {quark} chemical potential $\mu$ can be performed by following the
usual Matsubara procedure (see e.g.\ Sec.~\ref{sec2.2}).

The relative locations of the TCP and LP in the {$\mu - T$} plane can
be studied in general through the so-called Ginzburg-Landau (GL) approach,
which does not require to specify the explicit form of the
inhomogeneity~\cite{Nickel:2009ke,Abuki:2011pf}. Following the analysis
proposed in Ref.~\cite{Nickel:2009ke}, the mean field thermodynamic
potential can be expanded around the symmetric ground state in powers of the
order parameters and their spatial gradients. Up to sixth order in this
double expansion, the GL functional has the general form~\cite{Iwata:2012bs}
\begin{eqnarray}
\Omega(T,\mu,\phi^a(\vec x)) & = & \frac{\alpha_2}{2} \ \phi^2
+ \frac{\alpha_4}{4} (\phi^2)^2 + \frac{\alpha_{4b}}{4} (\nabla \phi)^2 +
\nonumber \\
& &
\frac{\alpha_6}{6} (\phi^2)^3 + \frac{\alpha_{6b}}{6} (\phi, \nabla \phi)^2
+ \frac{\alpha_{6c}}{6} \left[ \phi^2 (\nabla \phi)^2 - (\phi, \nabla \phi)^2 \right]
+ \frac{\alpha_{6d}}{6} \ (\triangle \phi)^2 \ ,
\label{genome}
\end{eqnarray}
where $\phi^2 = (\phi,\phi) = \phi^a \phi^a = \bar \sigma^2 + \vec{\bar \pi}
\ \! ^2$, $(\phi, \nabla_{\! i} \;  \phi) = \phi^a \nabla_{\! i} \; \phi^a =
\bar \sigma \nabla_{\! i}\; \sigma + \vec{\bar \pi} \cdot \nabla_{\! i}\;
\vec{\bar \pi}$, etc. By looking at this functional, it is seen that for
$\alpha_{4b} > 0$ the system is in the usual homogeneous phase. Now if in
addition one has $\alpha_4 > 0$, the system undergoes a first order chiral
restoration transition when $\alpha_2=0$ ($\phi^2=0$ for $\alpha_2>0$,
$\phi^2\neq 0$ for $\alpha_2<0$), which defines a first order transition
line in the $\mu-T$ plane. This line ends at the tricritical point, where
also $\alpha_4=0$ is satisfied. Thus, the position of the TCP can be
determined by solving the set of equations
\begin{equation}
\alpha_2 = 0 \ , \qquad \alpha_4 = 0 \ .
\label{ch5.3-eq5}
\end{equation}
On the other hand, for $\alpha_{4b} < 0$ inhomogeneous solutions are
favored. Hence, the Lifshitz point (i.e., the point where the onset of the
inhomogeneous phase meets the chiral transition line) is obtained
from~\cite{Buballa:2014tba}
\begin{equation}
\alpha_2 = 0 \ , \qquad \alpha_{4b} = 0 \ .
\label{ch5.3-eq6}
\end{equation}

This general discussion can also be applied to the NJL model, where it can
be shown that the coefficients of the quartic terms are equal to each other,
i.e.\ $\alpha_4^{\rm NJL} = \alpha_{4b}^{\rm
NJL}$~\cite{Nickel:2009ke,Iwata:2012bs,Carlomagno:2014hoa}. Therefore,
Eqs.~(\ref{ch5.3-eq5}) and (\ref{ch5.3-eq6}) are simultaneously satisfied,
and, {as stated above,} the TCP and LP are predicted to coincide. This
is in general not true in the context of nlNJL models, where the {relation
becomes} modified owing to the presence of the nonlocal form
factor~\cite{Carlomagno:2014hoa}. In fact, a numerical analysis carried out
in Ref.~\cite{Carlomagno:2014hoa} shows that for various phenomenologically
acceptable parametrizations the TCP is located at a higher temperature and a
lower chemical potential in comparison with the LP. As a consequence, it is
seen that nlNJL models favor a scenario in which the TCP {\it is not}
covered by the inhomogeneous phase.

\hfill

Let us investigate, in the framework of nlNJL models, the possible shape of
the inhomogeneous phase regions. In principle, a full analysis would require
to consider general spatial-dependent condensates, looking for the
configurations that minimize the mean field thermodynamic potential at each
value of the temperature and chemical potential. Since for an arbitrary
3-dimensional configuration this turns out to be a very difficult task, even
in the case of local models it is customary to consider one-dimensional
modulations, expecting that the qualitative features of the inhomogeneous
phases will not be significantly affected by the specific form of the
spatial dependence carried by the condensates~\cite{Buballa:2014tba}. It is
worth noticing that, as stated in Ref.~\cite{Nickel:2009wj}, the theoretical
problem of finding inhomogeneous phases with a lower dimensional modulation
can be reduced to a problem in a lower dimensional theory. Thus, the results
from analytically solvable 1+1 dimensional chiral
models~\cite{Schon:2000qy,Schon:2000he,Thies:2006ti,Basar:2009fg,Bringoltz:2009ym}
can be used to study the crystalline phase structure in 3+1 dimensions.
Here, due to the additional difficulties introduced by the presence of
nonlocal quark currents, we consider a simple one-dimensional configuration,
namely the so-called dual chiral density wave (DCDW)~\cite{Nakano:2004cd},
in which the chiral condensate rotates along the chiral circle, carrying a
constant three-momentum $\vec Q$. The spatial dependence of the quark
condensates is given in this case by
\begin{equation}
\langle \bar q(\vec x) q(\vec x)\rangle \ \propto \ \cos(\vec Q\cdot\vec x)\ ,
\qquad \qquad
\langle \bar q(\vec x) i \gamma_5 q(\vec x)\rangle \ \propto \ \sin (\vec Q\cdot\vec x)\ ,
\label{ch5.3-eq7}
\end{equation}
for both $q = u$ and $d$ quark flavors. This behavior of the chiral
condensates can be obtained by considering an adequate ansatz for the mean
field configuration of the chiral four-vector in momentum
space~\cite{Muller:2013tya}.

{The values of} $\phi$ and $Q\equiv |\vec Q|$ can be obtained, as
usual, by minimization of the mean field thermodynamic potential
$\Omega^{\rm MFA}$. A region in which the absolute minimum is reached for a
nonzero $Q$ will correspond to an inhomogeneous phase. As expected, if
chiral symmetry is not dynamically broken ($\phi = 0$) the regularized
thermodynamic potential reduces to {the free quark piece $\Omega^{\rm
free}_q$}, which does not depend on $Q$.

\begin{figure}[hbt]
\centering
\includegraphics[width=0.9\textwidth]{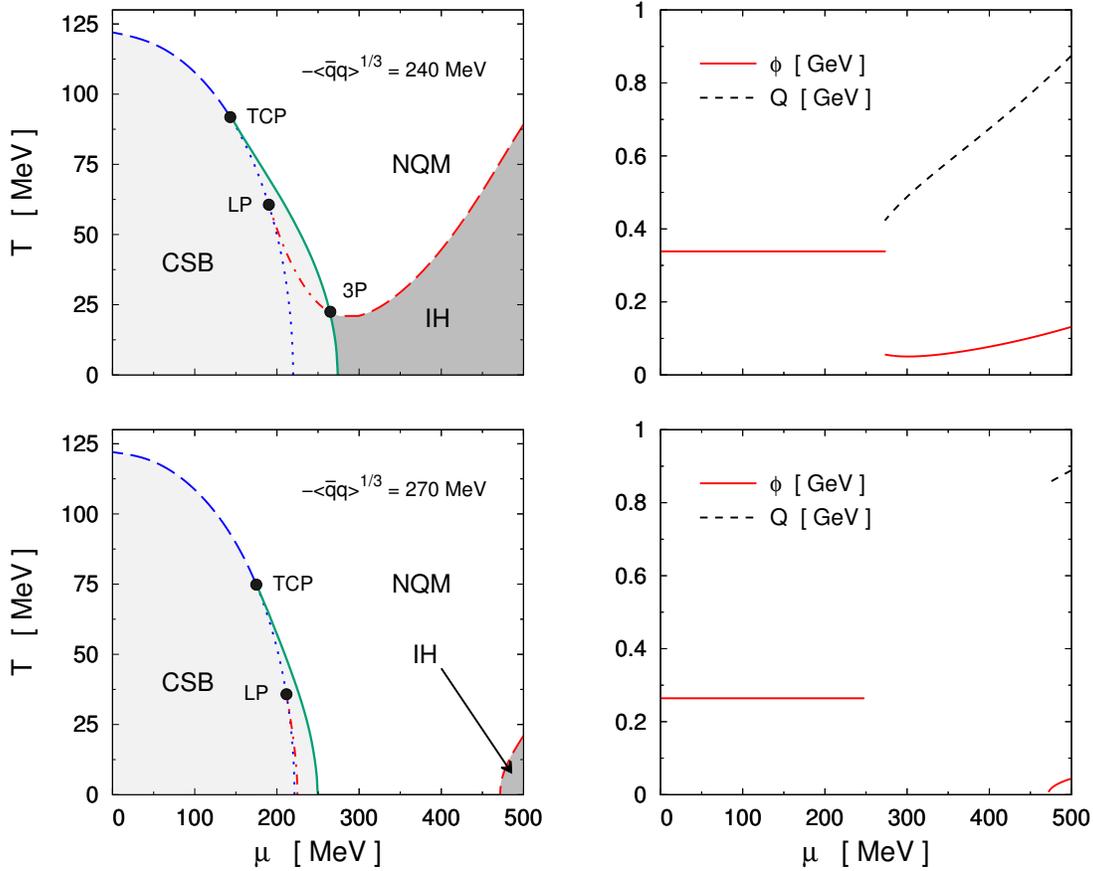}
\caption{(Color online) Left: $\mu-T$ phase diagrams. Solid (dashed) lines indicate first
(second) order phase transitions. The dotted line is the lower spinodal
corresponding to the homogeneous chiral restoration transition, while the
dashed-dotted line is a boundary of a region in which there exists a local
inhomogeneous minimum of the thermodynamic potential. TCP, LP and 3P stand
for tricritical, Lifshitz and triple points. Right: values of $\phi$ and $Q$
as functions of the chemical potential, for $T=0$.}
\label{fig:5.3.1}
\end{figure}

Numerical results for the corresponding phase diagrams within nlNJL models
have been quoted in Ref.~\cite{Carlomagno:2015nsa}, considering a Gaussian
form factor $g(p) = \exp(-p^2/\Lambda_0^2)$. Although this analysis does not
include the couplings between the quarks and the Polyakov loop, the qualitative
features of the phase diagram should not be significantly affected, since
one is mostly interested in the low temperature region. In
Fig.~\ref{fig:5.3.1} we quote the results for two parametrizations, which
correspond to a pion decay constant $f_\pi^{\rm ch} = 86$~MeV (we recall
that the chiral limit is taken) and quark condensate values $(-\qq^{\rm
ch})^{1/3} = 240$ and 270~MeV, respectively, at zero $T$ and $\mu$. The
various regions of the phase diagram, as well as the corresponding
transition curves and critical points, are shown in the left panels of
Fig.~\ref{fig:5.3.1}. Solid and dashed lines correspond to first and second
order transitions, respectively.

Let us analyze in detail the phase diagram in the upper left panel, which
corresponds to $\qq^{\rm ch} = -(240$~MeV$)^3$. For low chemical potentials,
at temperatures below $\sim 120$~MeV the system lies in an homogeneous
chiral symmetry broken (CSB) phase ({notice that the corresponding}
critical temperature should be higher if couplings with the Polyakov loop
are included). Taking a fixed temperature of e.g.\ 100~MeV, by increasing
the chemical potential one finds at some critical value $\mu_c$($T =
100$~MeV) a second order phase transition to an homogeneous normal quark
matter (NQM) phase in which chiral symmetry is restored. If the temperature
is lowered, the second order transition curve $\mu_c(T)$ ends at a
tricritical point, beyond which it becomes a first order transition line.
Now, by following this line, at a temperature $T_{3P} \simeq 20$~MeV one
arrives at a triple point. For $T < T_{3P}$, at a critical chemical
potential the system undergoes a first order transition from the CSB phase
into an inhomogeneous (IH) phase, in which {$Q\neq 0$ and} the chiral
symmetry is found to be only approximately restored. On the other hand, if
one starts with a system in the IH phase and increases the temperature at
constant chemical potential, at some critical value of $T$ one arrives at a
second order phase transition into the NQM phase. As shown in the figure,
the corresponding second order transition line continues beyond the triple
point with a dash-dotted line inside the CSB area. The latter is the
boundary of a region in which the thermodynamic potential has a local
minimum that corresponds to an (unstable) IH phase. Finally, the dotted line
in the phase diagram shows the lower spinodal corresponding to the
homogeneous chiral restoration transition.

\begin{figure}[htb]
\centering
\includegraphics[scale=0.3]{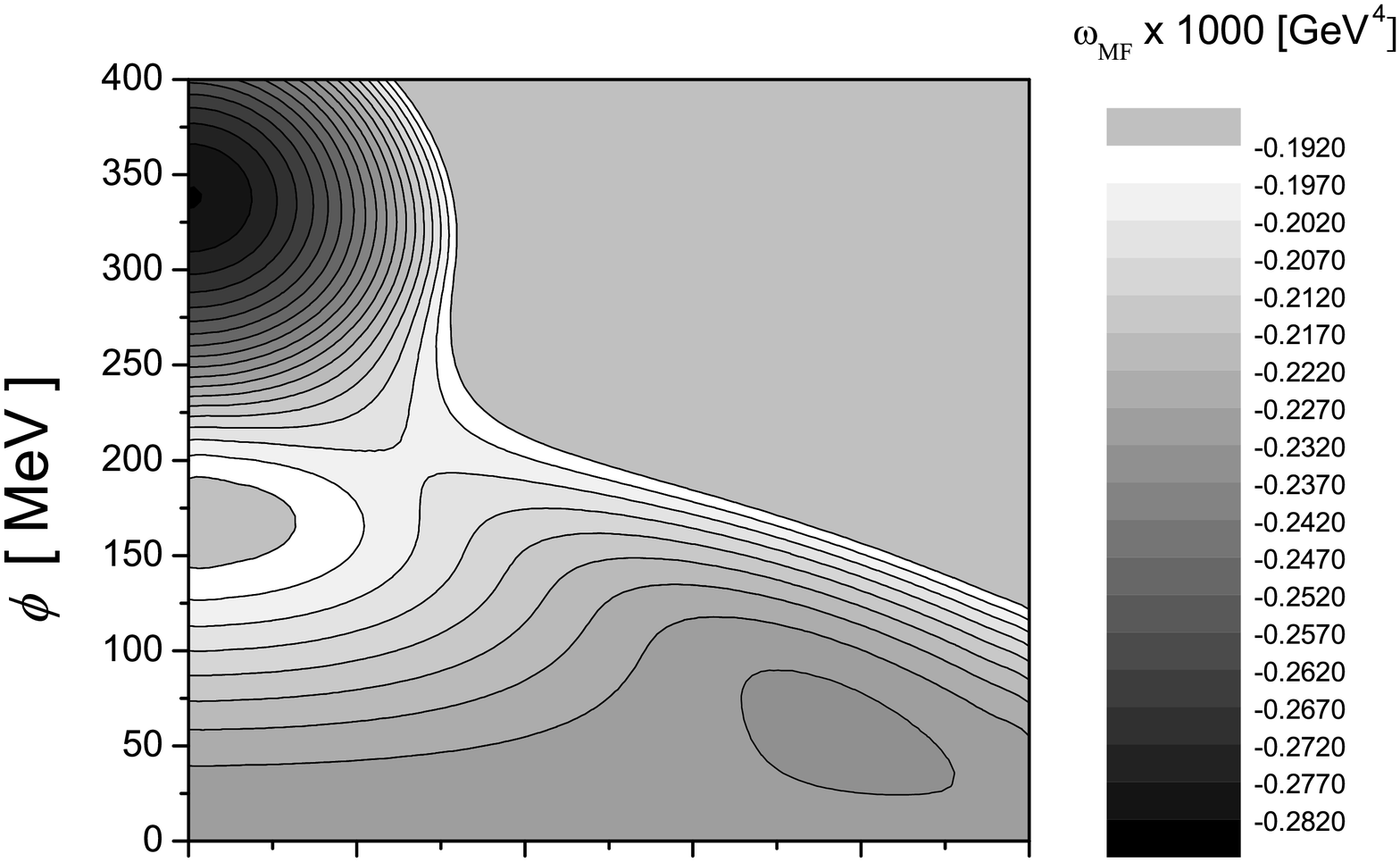}
\includegraphics[scale=0.3]{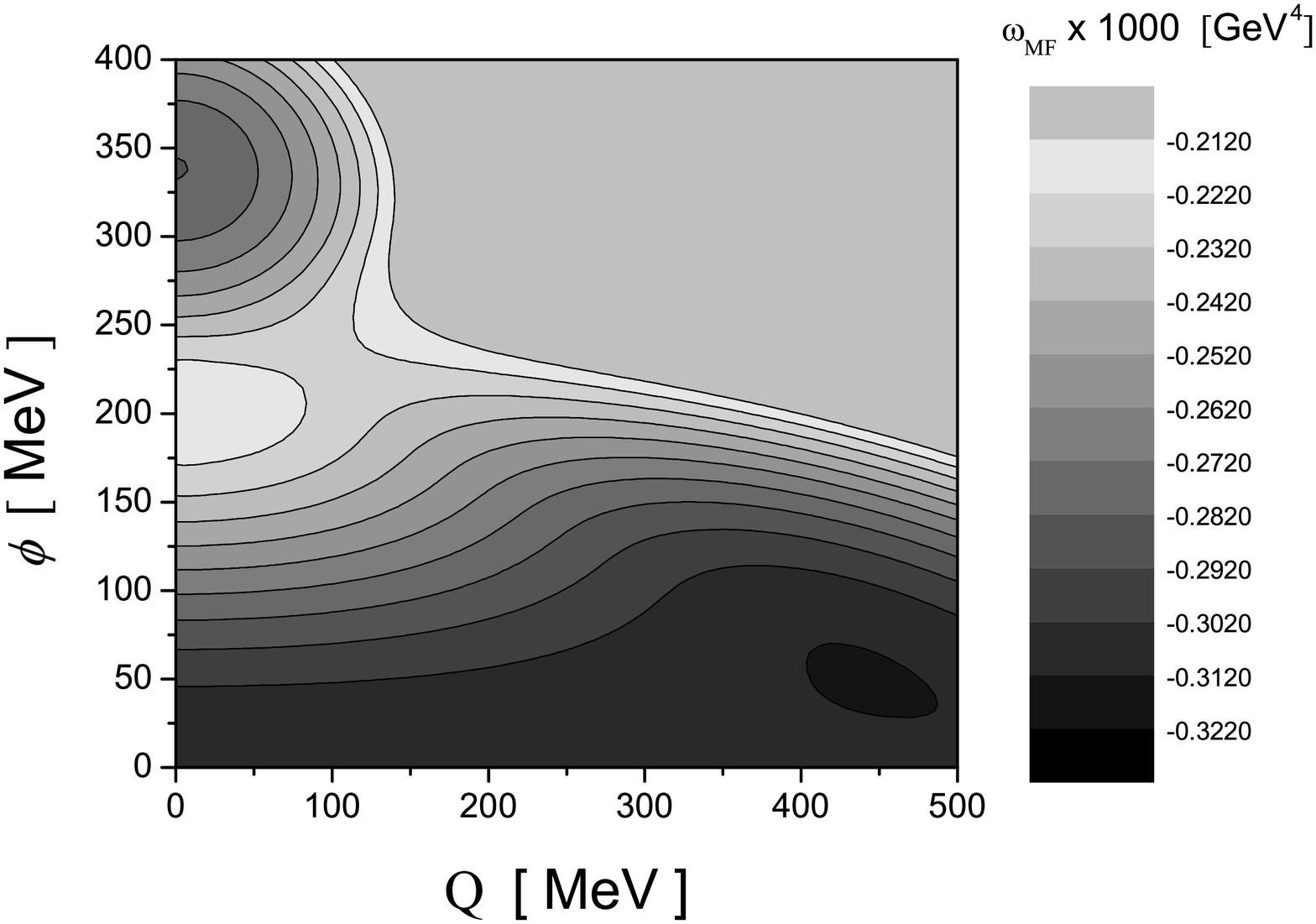}
\caption{Contour plots of the thermodynamic potential {$\Omega^{\rm MFA}$}
at zero temperature, close to the first order transition between CSB and IH
phases~\cite{Carlomagno:2015nsa}. The plots correspond to a condensate
$\qq^{\rm ch} = -(240$~MeV$)^3$, for chemical potentials $\mu = 260$~MeV
(up) and $\mu = 280$~MeV (down).} \label{fig:5.3.2}
\end{figure}

The previously described first order transition from the CSB to the IH phase
is illustrated in Fig.~\ref{fig:5.3.2}~\cite{Carlomagno:2015nsa}. Upper and
lower panels show contour plots of the mean field thermodynamic potential at
zero temperature for $\mu = 260$ and $\mu = 280$~MeV, respectively, which
correspond to both sides of the transition point $\mu_c(0) = 274$~MeV. The
plots show the transition from an absolute minimum at $\phi \simeq 340$~MeV,
$Q = 0$, to another one in which $\phi$ reduces to about 50~MeV, while the
chiral condensates get spatial dependences as those given by
Eq.~(\ref{ch5.3-eq7}), with $Q \simeq 450$~MeV. These features are also
shown in the upper right panel of Fig.~\ref{fig:5.3.1}, where the curves for
$\phi$ and $Q$ at $T=0$ as functions of the chemical potential are quoted.
Notice that on the CSB side (upper panel of Fig.~\ref{fig:5.3.2}) there also
exists a local minimum at $(\phi,Q)\sim (50$~MeV,$\,400$~MeV).

If the dimensionless product $G_S\Lambda_0^2$ is decreased, the absolute
value of the condensate $|\qq^{\rm ch}|$ gets increased, and the onset of
the inhomogeneous phase is pushed up to larger values of the chemical
potential, leaving a region of homogeneous NQM phase even at
$T=0$~\cite{Carlomagno:2015nsa}. This is shown in the lower left panel of
Fig.~\ref{fig:5.3.1}, where we quote the phase diagram for a
parameterization leading to $\qq^{\rm ch} = -(270$~MeV$)^3$ at zero $T$ and
$\mu$. In this case, it is seen that the onset of the IH phase (this region
is sometimes called a ``continent'') occurs at a chemical potential
$\mu_c(0)$ of the order of 500~MeV. The discontinuity of $Q$ at this
transition for $T=0$ becomes increased, as it is shown in the lower right
panel of Fig.~\ref{fig:5.3.1}.

It is worth pointing out that for the considered nlNJL models the would-be
Lifshitz point appears to be hidden inside the CSB phase
region~\cite{Carlomagno:2015nsa}. Instead, a triple point can be found if the CSB and
inhomogeneous phases meet. It is also worth mentioning that the second order
phase transition curves, as well as both the TCP and would-be LP, can be
calculated for these models through a quite precise semianalytical
approach~\cite{GomezDumm:2004sr,Carlomagno:2014hoa}. In addition, it is
interesting to notice that, according to the analyses in
Refs.~\cite{Carignano:2011gr,Carignano:2014jla,Buballa:2014tba}, for both
the NJL and quark-meson models some parameterizations lead to phase diagrams
that include IH ``continents'' which extend to arbitrarily high chemical
potentials. In fact, it is a matter of discussion whether the
{emergence of these continents is just a} regularization artifact. In
nonlocal models the ultraviolet convergence of loop integrals follows from
the behavior of form factors, which effectively embrace the underlying QCD
interactions (indeed, as discussed in previous sections, the form factors
can be fitted from lattice QCD calculations for the effective quark
propagators~\cite{Noguera:2008cm,Carlomagno:2013ona,Carlomagno:2019yvi}).
The fact that various quark models including different regularization
procedures lead to similar qualitative features of the phase diagram seems
to indicate that these features are rather robust. However, it is necessary
to mention that in the described works the effects of color
superconductivity have not been considered. As discussed in
Sec.~\ref{sec5.1}, the latter are expected to be important at intermediate
and large chemical potentials and could have a significant impact on the
phase diagram.

\subsection{Effects of external strong magnetic fields on
{phase transitions and meson properties}}
\label{sec5.4}

The study of the behavior of strongly interacting matter under intense
external magnetic fields is another subject that has gained significant
interest in the past years. Once again, the theoretical analysis requires in
general to deal with quantum chromodynamics in nonperturbative regimes,
therefore most studies are based either in the predictions of effective
models or in the results obtained from lattice QCD calculations. In fact, in
view of the theoretical difficulty, most works concentrate on the situations
in which one has a uniform and static external magnetic field. Recent
reviews on this subject can be found e.g.\ in
Refs.~\cite{Andersen:2014xxa,Miransky:2015ava}. In this subsection we
discuss, in the framework of nonlocal NJL-like models, the features of QCD
phase transitions under an intense homogeneous external magnetic field $\vec
B$~\cite{Pagura:2016pwr, GomezDumm:2017iex}. In addition, we show some
results on the behavior of $\pi^0$ and $\sigma$ meson properties, both at
zero and finite temperature~\cite{GomezDumm:2017jij,Dumm:2020muy}.

At zero temperature, the results of low-energy effective models of QCD as
well as LQCD calculations indicate that {the size of} light
quark-antiquark condensates should {get increased with} the magnetic
field. Thus, the external field appears to favor the breakdown of chiral
symmetry, which is usually known as ``magnetic catalysis''. On the contrary,
close to the chiral restoration temperature, LQCD calculations carried out
with realistic quark masses~\cite{Bali:2011qj,Bali:2012zg} show that the
condensates behave as nonmonotonic functions of $B$, and this leads to a
decrease in the transition temperature when the magnetic field is increased.
This effect is known as ``inverse magnetic catalysis'' (IMC). In addition,
LQCD calculations predict an entanglement between the chiral restoration and
deconfinement critical temperatures~\cite{Bali:2011qj}. The observation of
IMC has become a challenge for effective models. Indeed, most naive
effective approaches to low energy QCD predict that the chiral transition
temperature should grow with $B$, i.e., they do not find IMC. Interestingly,
the corresponding studies carried out in the context of nlNJL models show
that the latter are able to describe, at the mean field level, not only the
IMC effect but also the entanglement between chiral restoration and
deconfinement transition temperatures. Moreover, it is found that the
behavior of the mass and decay constant of the $\pi^0$ meson as functions of
the external magnetic field are also in agreement with LQCD
results~\cite{GomezDumm:2017jij}.

Here we concentrate on the two-flavor nlPNJL model introduced in
Sec.~\ref{sec2}, considering a parametrization of the type of PA, in which
quark-antiquark derivative currents are not included. In fact, the general
picture is expected to be similar for parametrizations in which such a
coupling is also taken into account. To account for the interaction with the
magnetic field one can proceed as described in previous sections for the
case of axial vector gauge fields. As usual, a coupling between the fermions
and the external electromagnetic gauge field $\mathcal{A}_{\mu}$ is obtained
by introducing a covariant derivative in the kinetic term in
Eq.~(\ref{ch2.1.1-eq1}), i.e.~by changing
\begin{equation}
\partial_{\mu}\ \rightarrow\ D_\mu\equiv\partial_{\mu}-i\,\hat Q
\mathcal{A}_{\mu}(x)\ ,
\label{covdev}
\end{equation}
where $\hat Q=\mbox{diag}(Q_u,Q_d)$, with $Q_u=2e/3$, $Q_d = -e/3$, is the
electromagnetic quark charge operator. In addition, as discussed in
Sec.~\ref{sec2.1.3}, gauge symmetry requires a further change in the
nonlocal currents $j_S(x)$ and $\vec \jmath_P(x)$ in
Eq.~(\ref{ch2.1.1-eq1}), namely
\begin{equation}
\psi(x-z/2) \ \rightarrow\ W_{\rm em}\left(  x,x-z/2\right)  \; \psi(x-z/2)\ ,
\label{transport}
\end{equation}
and the corresponding change for $\bar \psi(x+z/2)$~\cite{Noguera:2008cm}. The
function $W_{\rm em}({x,y})$ is given by
\begin{equation}
W_{\rm em}(x,y)\ =\ \mathrm{P}\;\exp\left[ -\, i \int_{x}^{y}ds_{\mu}\,
\hat Q\,\mathcal{A}_{\mu}(s) \right]  \ ,
\label{intpath}%
\end{equation}
where $s$ runs over a path connecting $x$ with $y$. As it is usually done,
we take it to be a straight line. As stated, we consider the case of a
constant and homogenous magnetic field, which, without loss of generality,
can be taken to be orientated along the 3-axis. For definiteness the
analysis can be carried out using the Landau gauge, in which one has
$\mathcal{A}_\mu = B\, x_1\, \delta_{\mu 2}$.

As {discussed in previous sections, it is convenient to carry out a
bosonization of the fermionic theory}, introducing scalar and pseudoscalar
meson fields and integrating out the fermions. Next, within the mean field
approximation, we assume that the scalar field $\sigma$ has a nontrivial
translational invariant mean field value $\bar \sigma$, while the mean field
values of pseudoscalar fields $\vec \pi$ are zero. It should be stressed at
this point that the assumption stating that $\bar \sigma$ is independent of
$x$ does not imply that the resulting quark propagator will be translational
invariant. In fact, as discussed below, one can show that {this}
invariance is broken by the appearance of the so-called Schwinger phase. Our
assumption just states that the deviations from translational invariance
driven by the magnetic field are not affected by the dynamics of the theory.
In this way, the {mean field} bosonized action can be written as
\begin{equation}
S^{\rm MFA}_E \ = \ -\ln\det\,\mathcal{D}_0 + V^{(4)} \frac{\bar
\sigma^2}{2G_S} \ ,
\label{sbosx}
\end{equation}
where
\begin{equation}
\mathcal{D}_0(x,x') \ = \ {\rm diag}
(\mathcal{D}_0^{u}(x,x')\ , \
 \mathcal{D}_0^{d}(x,x') )\ ,
\end{equation}
with
\begin{eqnarray}
 \mathcal{D}_0^{f}(x,x') & = &
\ \delta^{(4)}(x-x') \left( -i \rlap/\partial - Q_f B
\, x_1 \gamma_2 + m_c \right)
\nonumber \\
& &
+\; \bar\sigma \, {\cal G}(x-x') \,
\exp\left[i  \, \frac{Q_f B}{2}  (x_1+x'_1) \, (x_2-x'_2)\right]\, .
\label{df}
\end{eqnarray}
Here, the function ${\cal G}(z)$ is the nonlocal form factor in the
quark currents defined in Eq.~(\ref{ch2.1.1-eq1}). Notice that, contrary to
the $B=0$ case, in the presence of the magnetic field the charged particles
cannot be in states of definite momentum; hence, it is not adequate to
transform the action to momentum space. Instead, to deal with the operators
$ \mathcal{D}_0^{f}$ it is convenient to introduce Ritus
transforms $ \mathcal{D}_0^{f}(\bar p,\bar p\,')$, defined by
\begin{equation}
 \mathcal{D}_0^{f}(\bar p,\bar p\,') \ = \ \int d^4x \ d^4x' \
\bar{\mathbb{E}}_{\bar p} (x)  \  \mathcal{D}_0^{f}(x,x')  \
\mathbb{E}_{\bar p\,'} (x')\ ,
\label{dpp}
\end{equation}
where $\mathbb{E}_{\bar p} (x)$ and $\bar{\mathbb{E}}_{\bar p} (x)$ are
Ritus functions~\cite{Ritus:1978cj}. Here $\bar p=(k,p_2,p_3,p_4)$, $k$
being an integer quantum number that labels the so-called Landau energy
levels. Using the properties of Ritus functions, after some calculation it
can be shown that the operators $\mathcal{D}_0^{f}$ are
diagonal in this basis. One obtains~\cite{Pagura:2016pwr,GomezDumm:2017iex}
\begin{equation}
\mathcal{D}_0^{f}(\bar p,\bar p\,') \ = \
(2\pi)^4\,\delta_{kk'}\,\delta(p_2-
p_2^{\;\prime})\,\delta(p_3-p_3^{\;\prime})\, \delta(p_4-p_4^{\;\prime}) \ \mathcal{D}^f_{k, p_\parallel}\ ,
\label{diag}
\end{equation}
where
\begin{equation}
\mathcal{D}^f_{k, p_\parallel} \ = \ P_{k,{s_f}} \, \Big(\! -\! s_f\sqrt{2 k
|Q_f B|}\; \gamma_2 +   p_\parallel \cdot \gamma_\parallel \Big) +
\sum_{\lambda=\pm} M^{\lambda,f}_{k,p_\parallel}\, \Delta^\lambda\ ,
\label{twopoint}
\end{equation}
with
\begin{equation}
M^{\lambda,f}_{k,p_\parallel} \ = \
\frac{4\pi}{|Q_fB|}\,(-1)^{k_\lambda}
\int \frac{d^2p_\perp}{(2\pi)^2}\
\left[ m_c + \bar \sigma\ g(p) \right]
\, \exp(-p_\perp^2/|Q_fB|) \, L_{k_\lambda}(2p_\perp^2/|Q_fB|)\ .
\label{mpk}
\end{equation}
In these equations the definitions $s_{f} = {{\rm sgn}}(Q_f B)$, $p_\parallel
= (p_3,p_4)$, $p_\perp = (p_1,p_2)$, $\gamma_\parallel =
(\gamma_3,\gamma_4)$, $\Delta^+=\mbox{diag}(1,0,1,0)$,
$\Delta^-=\mbox{diag}(0,1,0,1)$, $P_{k,\pm
1}=(1-\delta_{k0})\, {I}+\delta_{k0}\,\Delta^\pm$ and $k_\pm = k -
1/2 \pm s_f/2\,$ have been used. The function $g(p)$ stands for the Fourier
transform of ${\cal G}(x)$, while $L_m(x)$ are Laguerre polynomials, with
the usual convention $L_{-1}(x) =0$.

Using the fact that $\mathcal{D}_0^{f}$ is diagonal in Ritus space, the
corresponding contribution to the mean field action can be readily
calculated. One obtains
\begin{equation}
\frac{S^{\rm MFA}_{E}}{V^{(4)}} \ = \ - N_c \sum_{f=u,d} \frac{  |Q_f B|}{2 \pi} \int \frac{d^2p_\parallel}{(2\pi)^2}
\bigg[ \ln\left(p_\parallel^2 + {M^{\,\lambda_{\!
f},f}_{0,p_\parallel}\,}^2\,\right)
+  \sum_{k=1}^\infty \ \ln \Delta^f_{k,p_\parallel}\bigg]  + \frac{ \bar \sigma^2}{2 G}\ ,
\label{smfa}
\end{equation}
where $\lambda_f = +\, (-)$ for $s_f = +1\,(-1)$, and
$\Delta^f_{k,p_\parallel}$ is defined by
\begin{equation}
\Delta^f_{k,p_\parallel} \ = \ \left( 2 k |Q_f B| + p_\parallel^2 +
M^{+,f}_{k,p_\parallel}\, M^{-,f}_{k,p_\parallel} \right)^2 + p_\parallel^2
\left( M^{+,f}_{k,p_\parallel} - M^{-,f}_{k,p_\parallel} \right)^2\ .
\end{equation}
By regarding at these equations [compare with Eq.~(\ref{ch2.1.1-eq9})] it is
seen that the functions $M^{\pm,f}_{k,p_\parallel}$ play the role of
constituent quark masses in the presence of the external magnetic field.

As done in Sec.~\ref{sec2.2}, the analysis can be extended to a system at
finite temperature using the Matsubara formalism, and a coupling of fermions
to the Polyakov loop can be included to account for confinement effects. In
what follows we present the results corresponding to the polynomial PL
potential quoted in Eq.~(\ref{ch2.2-eq2})~\cite{Ratti:2005jh}. The full
expression of the corresponding mean field thermodynamic potential can be
found in Ref.~\cite{GomezDumm:2017iex}. As in the $B=0$ case, this quantity
is divergent and can be regularized using the prescription in which one
subtracts a free contribution and adds it in a regularized form. In fact,
this ``free'' contribution corresponds to the {mean field} potential
obtained in absence of the {effective four-quark} coupling (i.e.\
setting $\bar \sigma = 0$), but keeping the interaction with the magnetic
field and the PL.

By minimizing the regularized mean field thermodynamic potential one can
obtain the values of $\bar\sigma$ and the traced Polyakov loop $\Phi$ as
functions of the temperature $T$ and the magnetic field. Then, the magnetic
field dependent quark condensates $\langle \bar q q\rangle$, $q=u,d$, can be
calculated, as usual, by taking the derivatives with respect to the corresponding
current quark masses. To make contact with LQCD results given in
Ref.~\cite{Bali:2012zg} it is convenient to define the quantities
\begin{equation}
\Sigma^q_{B,T} \ = \ -\frac{2\, m_c}{S^4} \, \Big[ \langle \bar q q \rangle_{B,T}
 - \langle \bar q q\rangle_{0,0} \Big] \, + \, 1 \ ,
\label{defi}
\end{equation}
where $S= (135 \times 86)^{1/2}$~MeV. We also introduce the definitions
$\Delta \Sigma^q_{B,T} = \Sigma^q_{B,T} - \Sigma^q_{0,T}$, $\bar
\Sigma_{B,T} = (\Sigma^u_{B,T}+\Sigma^d_{B,T})/2$ and $\Delta \bar
\Sigma_{B,T} = (\Delta \Sigma^u_{B,T}+\Delta \Sigma^d_{B,T})/2\,$, which
correspond to a subtracted normalized flavor condensate, a normalized
flavor average condensate and a subtracted normalized flavor average
condensate, respectively.

In what follows we quote the numerical results obtained for the case of a
Gaussian form factor, i.e.\ for some parametrizations similar to PA (see
Sec.~\ref{sec2.3}). For comparison we consider parameter sets leading to
quark-antiquark condensates $(-\langle \bar q q\rangle_{0,0})^{1/3} = 240$,
230 and 220~MeV. The corresponding model parameters can be found e.g.\ in
Ref.~\cite{GomezDumm:2017iex}.

\begin{figure}[hbt]
\begin{center}
\includegraphics[width=0.8\textwidth]{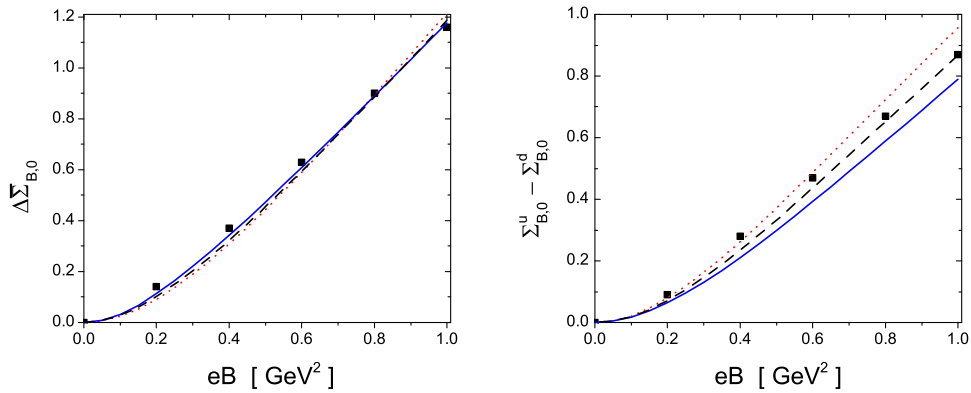}
\end{center}
\caption{(Color online) Normalized quark condensates as functions of the
magnetic field at $T = 0$. Solid (blue), dashed (black) and dotted (red)
curves correspond to parameterizations leading to $(- \langle\bar q q
\rangle_{0,0})^{1/3} = 240$, 230 and 220~MeV, respectively. Full square
symbols indicate LQCD results given in Ref.~\cite{Bali:2012zg}.}
\label{fig5.4.1}
\end{figure}

The behavior of quark condensates at zero temperature within this nlNJL
framework has been calculated in Ref.~\cite{Pagura:2016pwr}. The results are
shown in Fig.~\ref{fig5.4.1}, where we include the predictions for $\Delta
\bar \Sigma_{B,0}$ and $\Sigma^u_{B,0}-\Sigma^d_{B,0}$ as functions of $eB$
together with the corresponding LQCD data given in Ref.~\cite{Bali:2012zg}.
Solid, dashed and dotted curves correspond to $(-\langle \bar q
q\rangle_{0,0})^{1/3} = 240$, 230 and 220 MeV, respectively. The growth of
the condensates clearly show the effect known as ``magnetic catalysis''. It
can be seen that the predictions for $\Delta \bar \Sigma_{B,0}$ are very
similar for all parameter sets, while for the difference
$\Sigma^u_{B,0}-\Sigma^d_{B,0}$ there is some dependence on the
parametrization. In both cases the predictions show a good agreement with
LQCD results.

\begin{figure}[hbt]
\begin{center}
\subfloat{\includegraphics[width=0.39\textwidth]{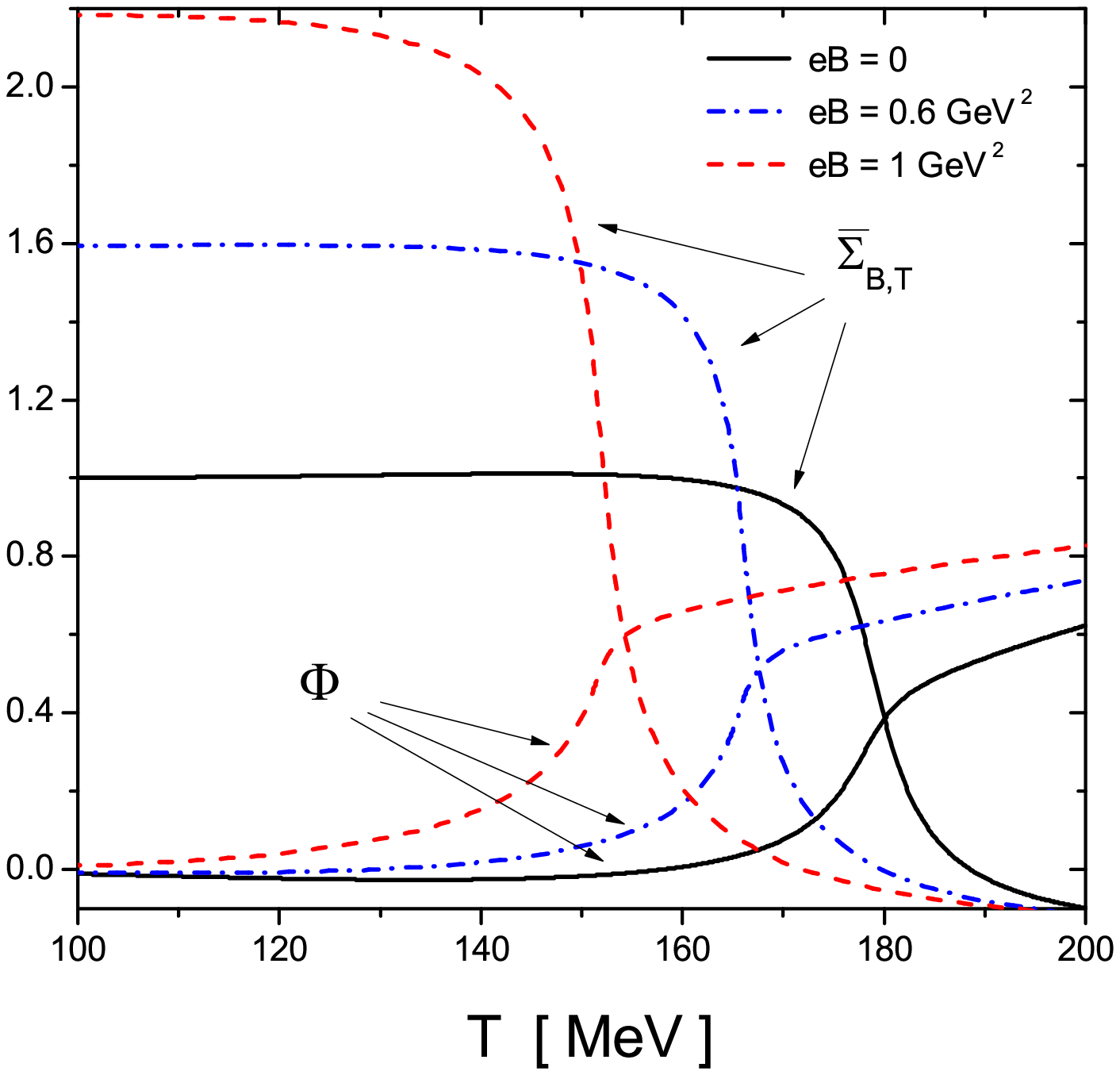}}
\hspace*{.4cm}
\subfloat{\includegraphics[width=0.42\textwidth]{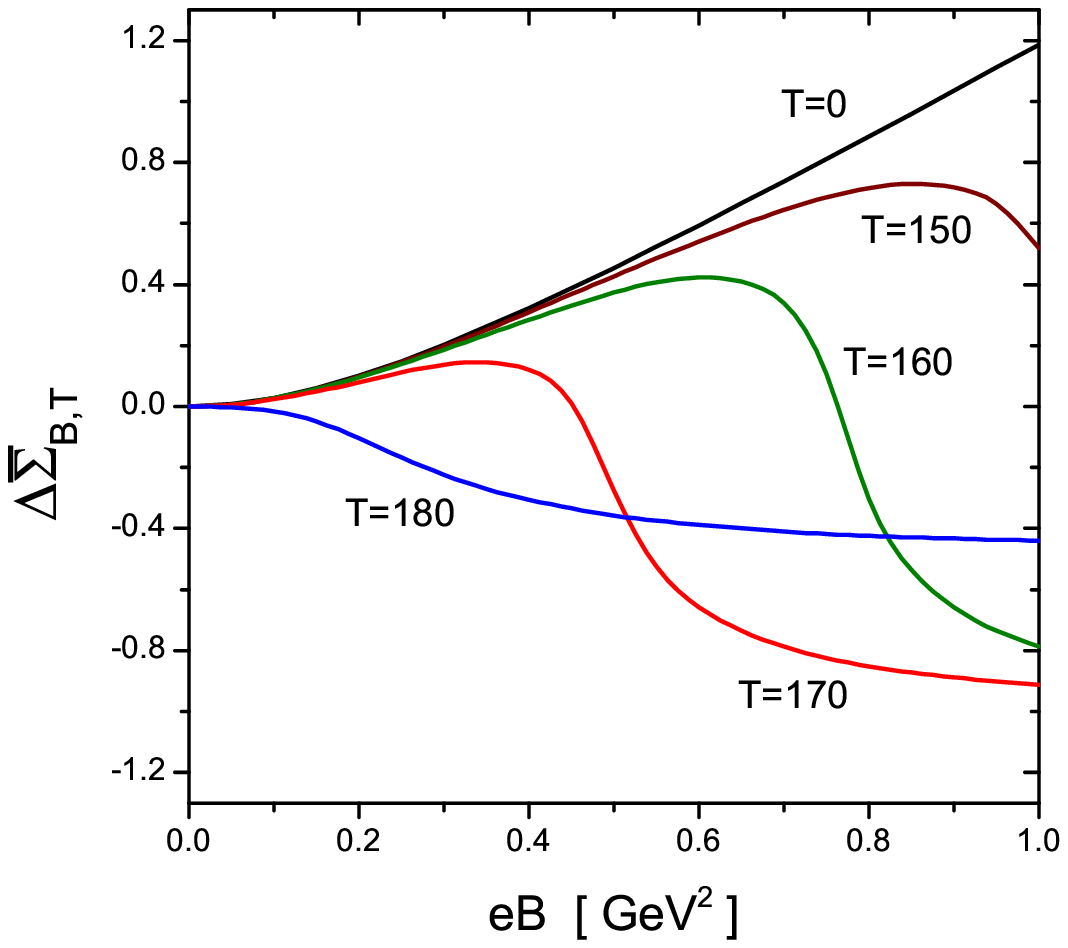}}
\end{center}
\caption{(Color online) Left: normalized flavor average condensate and
traced Polyakov loop as functions of the temperature, for three
representative values of $eB$. Right: subtracted normalized flavor average
condensate as a function of $eB$ for various representative temperatures.}
\label{fig5.4.2}
\end{figure}

Let us now look at the results for a system at finite temperature, which
have been obtained in Ref.~\cite{GomezDumm:2017iex}. In the left panel of
Fig.~\ref{fig5.4.2} we show the behavior of the averaged chiral condensate
$\bar\Sigma_{B,T}$ and the traced Polyakov loop $\Phi$ as functions of the
temperature, for three representative values of the external magnetic field,
namely $eB = 0$, 0.6 and 1 GeV$^2$. The curves correspond to a quark
condensate $(-\langle\bar q q \rangle_{0,0})^{1/3} = 230$~MeV. Given a value
of $B$, it is seen from the figure that chiral restoration and deconfinement
transitions proceed as smooth crossovers occurring at approximately the same
critical temperature $T_{c}(B)$ (chiral restoration and deconfinement
critical temperatures are defined here from the peaks in the derivatives
{$-d\bar\Sigma_{B,T}/dT$ and $d\Phi/dT$,} respectively). For $B=0$,
$T_{c}(0)$ is found to be approximately equal to 180~MeV, {with a
variation not larger than} a few percent within the above considered
parametrization range. This temperature compares well with the value
$T_{c}(0) = 173\pm 8$~MeV obtained from $N_f=2$ LQCD
calculations~\cite{Karsch:2003jg}. It is worth recalling that in absence of
the interaction with the Polyakov loop the value of $T_{c}(0)$ is found to
drop down to about 130~MeV~\cite{Pagura:2016pwr}.

It is interesting to discuss the effect of the magnetic field on the phase
transition features. On one hand, from the left panel of Fig.~\ref{fig5.4.2}
it is seen that the splitting between the chiral restoration and
deconfinement critical temperatures remains very small even for nonzero $B$.
On the other hand, the curves for the normalized flavor average condensate
clearly show the inverse magnetic catalysis effect. Indeed, contrary to what
happens e.g.~in the local NJL model~\cite{Andersen:2014xxa,Miransky:2015ava}
or in the quark-meson model~\cite{Fraga:2013ova}, within the nlNJL approach
the chiral restoration critical temperature becomes lower as the external
magnetic field is increased. This is related with the fact that the
condensates do not show in general a monotonic increase with $B$ for a fixed
value of the temperature. The situation is illustrated in the right panel of
Fig.~\ref{fig5.4.2}, where we show the behavior of the averaged difference
$\Delta \bar \Sigma_{B,T}$ as a function of $eB$, for $T=0$ and for values
of the temperature in the critical region. While the value of $\Delta \bar
\Sigma_{B,T=0}$ shows a monotonic growth with the external magnetic field,
it is seen that when the temperatures get closer to critical values
$T_{c}(B)$ the curves show a maximum and then start to decrease for
increasing $B$. This is a typical behavior associated to IMC and nicely
agrees with the results obtained from lattice QCD, see e.g.~Fig.~2 of
Ref.~\cite{Bali:2012zg}.

\begin{figure}[hbt]
\begin{center}
\includegraphics[width=0.5\textwidth]{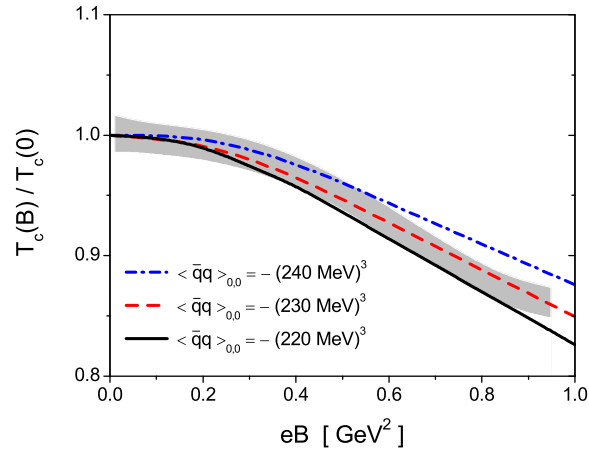}
\end{center}
\caption{(Color online) Normalized critical temperatures as functions of $eB$. For
comparison, LQCD results of Ref.~\cite{Bali:2012zg} are indicated by the
gray band.}
\label{fig5.4.4}
\end{figure}

In Fig.~\ref{fig5.4.4} we show the results given in
Ref.~\cite{GomezDumm:2017iex} for the normalized chiral restoration critical
temperature, $T_{c}(B)/T_{c}(0)$, as a function of $eB$. The curves
correspond to the above mentioned nlNJL model parametrizations. For
comparison, LQCD results quoted in Ref.~\cite{Bali:2012zg} are indicated by
the gray band. From the figure it is clearly seen that the IMC effect is
sizeable, and fully compatible with LQCD results for phenomenologically
adequate values of the chiral condensate. It is also worth mentioning that
some effective approaches to low energy QCD are able to obtain IMC by
assuming some explicit dependence of the effective coupling parameters on
$B$ and/or $T$~\cite{Ayala:2014iba,Farias:2014eca}, or by considering the
effect of anomalous magnetic moments in the quark energy dispersion
relations~\cite{Fayazbakhsh:2014mca}. In turn, it is seen that within nlNJL
models one gets IMC in a fully natural way. This can be understood by
noticing that for a given Landau level the associated nonlocal form factor
carries a dependence on the external magnetic field, which arises from the
convolution in Eq.~(\ref{mpk}).

\hfill

In what follows we quote some results concerning the properties of the
$\pi^0$ meson in the presence of the external magnetic field. At zero
temperature, the theoretical expression for the $\pi^0$ mass can be obtained
by expanding the corresponding bosonic action in powers of the fluctuations
{$\delta\pi_3 = \delta\pi^0$}. In momentum space the corresponding quadratic
piece of the bosonized action can be written as {
\begin{eqnarray}
S^{\rm quad}_E\Big|_{(\delta\pi^0)^2} & = &
\frac{1}{2}\int \frac{d^4q}{(2\pi)^4} \ G_{\pi^0}(q_\perp^2,q_\parallel^2) \
\delta\pi^0(q)\,\delta\pi^0(-q)\nonumber \\
& = & \frac{1}{2}\int \frac{d^4q}{(2\pi)^4} \ \left[ \frac{1}{G_S} \, +
F_{\pi^0}(q_\perp^2,q_\parallel^2) \right] \,
\delta\pi^0(q)\,\delta\pi^0(-q)\ ,
\label{sbospi}
\end{eqnarray}
where the polarization function $F_{\pi^0}(q_\perp^2,q_\parallel^2)$ is
given by a quark loop integral that involves the external field $B$.} Its
explicit form can be found in Refs.~\cite{GomezDumm:2017jij,Dumm:2020muy}.
Choosing the frame in which the pion is at rest, its mass can be obtained as
a solution of the equation
\begin{equation}
G_{\pi^0}(0,- m_{\pi^0}^2) = \ 0\ .
\label{pimass}
\end{equation}
As shown in Ref.~\cite{Fayazbakhsh:2013cha}, a relevant feature induced by
the presence of the external magnetic field is the fact that the $\pi^0$
dispersion relation turns out to be anisotropic, implying that the movement
along the direction perpendicular to the magnetic field is characterized by
a directional refraction index $u_{\pi^0}$ which is in general different
from one. To normalize the $\pi^0$ field one can expand the action in
Eq.~(\ref{sbospi}) around the pion pole ($q_\perp=0$,
$q_\parallel^2=-m_{\pi^0}^2$) up to first order in momentum squared.
Following Ref.~\cite{Dumm:2020muy} one can define
\begin{equation}
    Z_{\parallel}^{-1} \ = \ g_{\pi^0 q\bar q}^{-2} \ = \
\dfrac{dG_{\pi^0}(q_\perp^2,q_\parallel^2)}{dq_\parallel^2}
      \bigg\rvert_{q_\perp^2 = 0,\,q_\parallel^2=-m_{\pi^0}^2}
      , \qquad
    Z_{\perp}^{-1} \ = \ \dfrac{dG_{\pi^0}(q_\perp^2,q_\parallel^2)}{dq_\perp^2}
          \bigg\rvert_{q_\perp^2 = 0,\,q_\parallel^2=-m_{\pi^0}^2} \ , \label{defzperp}
\end{equation}
renormalizing the pion field according to $\pi^0(q)=Z_{\parallel}^{1/2} \,
\tilde{\pi}^0(q)$. Thus,
around the pion pole one has
\begin{equation}
    S_E^{\rm quad}\Big|_{(\delta\pi^0)^2} \ = \ \dfrac{1}{2} \int \frac{d^4q}{(2\pi)^4}
        \, \Big( u_{\pi^0}^2\, q_\perp^2 + q_\parallel^2+m_{\pi^0}^2 \Big)\,
        \delta\tilde{\pi}^0(q) \delta\tilde{\pi}^0(-q) \ ,
\label{actionquadpi0p_2}
\end{equation}
where
\begin{equation}
u_{\pi^0}^2 \ = \ \dfrac{Z_{\parallel}}{Z_{\perp}}\ .
\label{u_pi}
\end{equation}

\begin{figure}[h!bt]
\begin{center}
\includegraphics[width=0.5\textwidth]{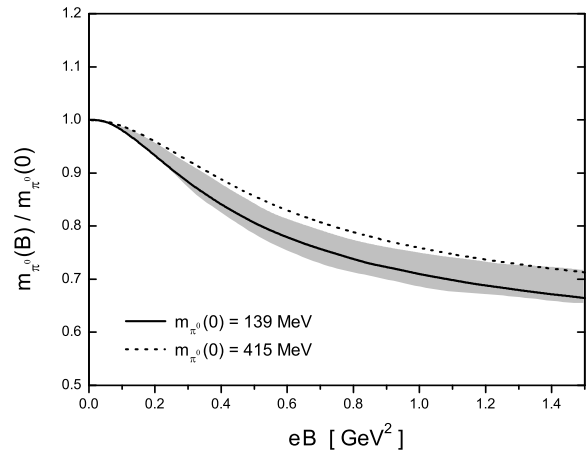}
\end{center}
\caption{Mass of the $\pi^0$ meson as a function of $eB$, normalized to its
value for $B=0$~\cite{GomezDumm:2017jij}. The dotted line is obtained for a
parameterization in which $m_\pi = 415$~MeV, while the gray band corresponds
to the results of lattice QCD calculations quoted in
Ref.~\cite{Bali:2017ian}.}
\label{fig5.4.8}
\end{figure}

The behavior of the pion mass $m_{\pi^0}(B)$ predicted by the nlNJL approach
is shown in Fig.~\ref{fig5.4.8}~\cite{GomezDumm:2017jij}. The results,
normalized to the empirical $\pi^0$ mass value at $B=0$, correspond to the
parameter set leading to $(-\langle\bar q q \rangle_{0,0})^{1/3} = 230$~MeV.
It is found that the $\pi^0$ mass decreases when the magnetic field gets
increased, reaching a value of about 65\% of $m_{\pi^0}(0)$ at $eB\simeq
1.5$ GeV$^2$, which corresponds to a magnetic field of about $2.5\times
10^{20}$~G. The figure also includes a gray band that corresponds to lattice
QCD results given in Ref.~\cite{Bali:2017ian}. The latter have been obtained
from a continuum extrapolation of lattice spacing, considering a relatively
large {current} quark mass for which $m_\pi = 415$~MeV. For comparison,
we also quote in the figure the results obtained within the nlNJL model by
shifting $m_c$ to 56.3~MeV, which leads to this enhanced pion mass. In
general, it can be seen that nlNJL model predictions ---for which no ad-hoc
adjustments or extra parameters have been required--- turn out to be in good
agreement with LQCD calculations. It is also worth mentioning that the
curves in Fig.~\ref{fig5.4.8} remain practically unchanged when the value of
the $B=0$ condensate used to fix the parametrization is varied within the
range from $-$(220 MeV)$^3$ to $-$(250 MeV)$^3$.

It is also interesting to study the behavior of the $\pi^0$ meson ``decay
constants''. As shown in Ref.~\cite{Coppola:2018ygv}, in the presence of an
external magnetic field $\vec B$ the pion-to-vacuum vector and axial vector
amplitudes can be in general parameterized in terms of three form factors.
{Two of them, $f_{\pi^0}^{(A||)}$ and $f_{\pi^0}^{(A\perp)}$, in the
$B=0$ limit reduce to the pion decay constant usually denoted by $f_\pi$.
The third one, $f_{\pi^0}^{(V)}$, is associated to the vector piece of the
quark weak current, and vanishes for $B=0$.} Following the notation in
Ref.~\cite{Coppola:2019uyr}, these form factors can be defined by the
relations
\begin{eqnarray}
H_4^{V}(x,\vec q)\pm H_3^{V}(x,\vec q) & = & \mp\, f^{(V)}_{\pi^0}(q_4\mp q_3)\, e^{iq\,\cdot x} \ , \nonumber\\
H_1^{V}(x,\vec q)\pm i H_2^{V}(x,\vec q) & = & 0 \ , \nonumber\\
H_4^{A}(x,\vec q)\pm H_3^{A}(x,\vec q) & = & -i f^{(A\parallel)}_{\pi^0}(q_4\pm q_3)\, e^{i q\,\cdot x} \ , \nonumber\\
H_1^{A}(x,\vec q)\pm i H_2^{A}(x,\vec q) & = & -i f^{(A\perp)}_{\pi^0}(q_1\pm i q_2)\, e^{i q\,\cdot x} \
,
\label{fdefs}
\end{eqnarray}
where $H_\mu^{V,A}$ are the $\pi^0$-to-vacuum amplitudes for vector and axial vector quark currents,
\begin{eqnarray}
H_\mu^{V} (x,\vec q) & = & \langle 0 |\bar\psi(x)\,\gamma_\mu
\frac{\tau^3}{2}\,\psi(x)|\tilde \pi^0(\vec q)\rangle \ , \nonumber\\
H_\mu^{A} (x,\vec q) & = & \langle 0 |\bar\psi(x)\,\gamma_\mu\gamma_5
\frac{\tau^3}{2}\,\psi(x)|\tilde \pi^0(\vec q)\rangle\ .
\label{fpidef}
\end{eqnarray}

The matrix elements in Eq.~(\ref{fpidef}) can be obtained by introducing
couplings between the quark currents and auxiliary vector and axial vector
gauge fields, and then taking the corresponding functional derivatives of
the effective action. Once again, gauge invariance requires the couplings to
these auxiliary gauge fields to be introduced through the covariant
derivative and the parallel transport of the fermion fields. The analytic
calculations of the form factors require to carry out combined
Laguerre-Fourier transformations of the form factors. Explicit expressions
can be found in Ref.~\cite{Dumm:2020muy}.

It is interesting to study the relations involving form factors and
renormalization constants in the chiral limit, $m_c\to 0$. As expected, in
this limit one gets ${F_{\pi^0}(0,0)} = - 1/G_S$~\cite{Dumm:2020muy},
which implies $m_{\pi^0}=0$ according to Eq.~(\ref{pimass}). In addition,
from the calculations in Refs.~\cite{GomezDumm:2017jij,Dumm:2020muy} it is
seen that the Goldberger-Treiman relation
\begin{equation}
f_{\pi^0,0}^{(A\parallel)}\; Z_{\parallel,0}^{1/2} \ = \ \bar\sigma_0
\label{fpicero}
\end{equation}
and the Gell-Mann-Oakes-Renner relation
\begin{equation}
{f_{\pi^0,0}^{(A\parallel)}}^2 \, m_{\pi^0}^2 \ = \ -\,m_c\,\langle \bar uu + \bar dd\rangle_0
\label{gor}
\end{equation}
remain valid in the presence of the external magnetic field (subindices 0
indicate that the involved quantities are evaluated in the chiral limit).
The above equations are complemented by the relation~\cite{Dumm:2020muy}
{\begin{equation}
f_{\pi^0,0}^{(A\perp)} \,
Z_{\parallel,0}^{-1/2}\,Z_{\perp,0} \ = \ \bar\sigma_0\ ,
\label{fpiperp}
\end{equation}}
which implies
\begin{equation}
\frac{f_{\pi^0,0}^{(A\perp)}}{f_{\pi^0,0}^{(A\parallel)}} \ = \
\frac{Z_{\parallel,0}}{Z_{\perp,0}}\ = \ u_{\pi^0,0}^2 \ .
\label{ratio}
\end{equation}
This result has been also found in the framework of the local NJL model in
Ref.~\cite{Coppola:2019uyr}, and (using a different notation) in
Ref.~\cite{Fayazbakhsh:2013cha}, where it is obtained from a modified
partially-conserved-axial-current (PCAC) relation.

\begin{figure}[h!bt]
\begin{center}
\includegraphics[width=0.75\textwidth]{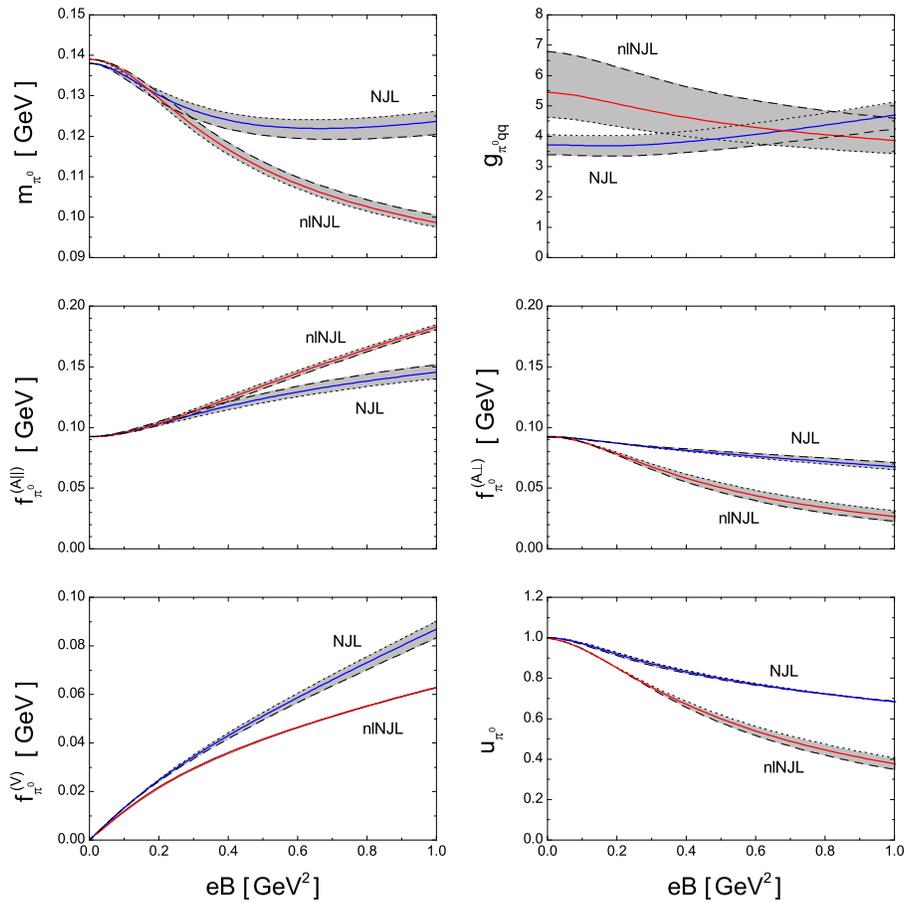}
\end{center}
\caption{(Color online) Neutral pion properties as functions of $eB$. Solid red and blue
lines correspond nlNJL and local NJL results, respectively. The gray bands
account for model parameter ranges indicated in the text.}
\label{fig5.4.9}
\end{figure}

In Fig.~\ref{fig5.4.9} we show the numerical results obtained within the
nlNJL model for various quantities associated with the neutral pion at zero
temperature, as functions of $eB$~\cite{Dumm:2020muy}. Dashed, solid and
dotted red lines correspond to quark condensate values $(-\langle\bar q q
\rangle_{0,0})^{1/3} = 220$, 230 and 240~MeV, respectively. For comparison
we also include in the figure the numerical results obtained within the
local NJL model, quoted in Ref.~\cite{Coppola:2019uyr}. Solid blue lines
correspond to a parametrization leading to a constituent quark mass
$M=350$~MeV (for $B=0$), while the limits of the gray bands correspond to
$M=320$~MeV (dashed lines) and $M=380$~MeV (dotted lines). The values of the
$B=0$ quark-antiquark condensates for these parametrizations of the NJL
model are $\langle \bar qq\rangle \simeq (- 243~\rm{MeV})^3$,
$(-236~\rm{MeV})^3$ and $(-250~\rm{MeV})^3$, respectively. As shown in the
figure, in general nlNJL results do not show a large dependence with the
model parametrization. On the other hand, in most cases the dependence with
the external field is significantly stronger for the nlNJL model than for
the local NJL approach. In the case of the effective coupling constant
$g_{\pi^0 q \bar q}$, the behavior is found to be opposite for both models.
Concerning the axial form factors, for $B=0$ one has spacial rotation
symmetry and both $f_{\pi^0}^{(A\parallel)}$ and $f_{\pi^0}^{(A\perp)}$
reduce to the usual pion decay constant $f_\pi$ [see Eqs.~(\ref{fpidef})].
As the magnetic field increases, $f_{\pi^0}^{(A\parallel)}$ gets enhanced
and $f_{\pi^0}^{(A\perp)}$ gets reduced, {both in the case of the nlNJL
and the local} NJL model. The vector form factor $f_{\pi^0}^{(V)}$, shown in
the lower left panel, is zero at vanishing external field and shows a
monotonic growth with $eB$, with little dependence on the parametrization
within the nlNJL model.

\hfill

\begin{figure}[h!bt]
\begin{center}
\includegraphics[width=0.5\textwidth]{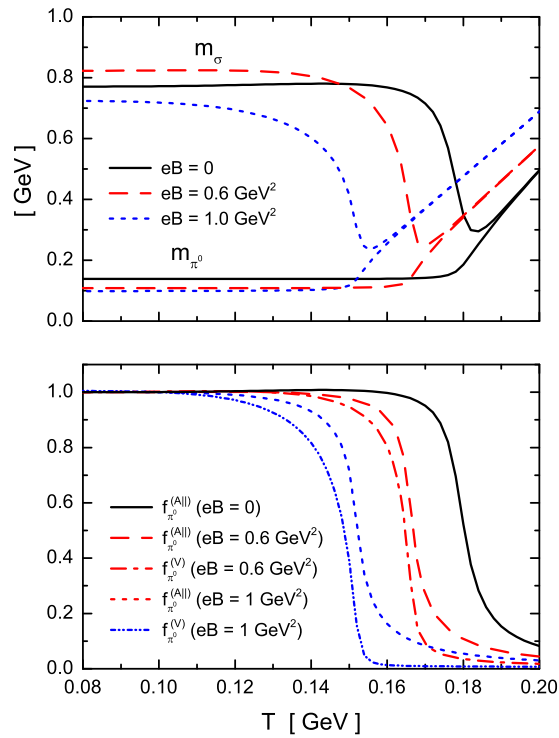}
\end{center}
\caption{(Color online) $\pi^0$ and $\sigma$ meson masses (upper panel) and normalized
$\pi^0$ decay form factors (lower panel) as functions of the temperature,
for three representative values of $eB$.}
\label{fig5.4.10}
\end{figure}

Finally, let us discuss the behavior of $\pi^0$ and $\sigma$ meson
properties in the context of nlNJL models for a system at finite temperature
$T$. As in the case of the quark condensates, we quote the results for the
model parametrization corresponding to $(-\langle\bar q q
\rangle_{0,0})^{1/3} = 230$~MeV and a polynomial PL potential. In
Fig.~\ref{fig5.4.10} we show the behavior of the $\pi^0$ and $\sigma$ meson
masses (upper panel), and the normalized $\pi^0$ axial and vector decay form
factors (lower panel) as functions of the temperature, for three
representative values of the external magnetic field, namely $eB=0$,
$eB=0.6$~GeV$^2$ and $eB=1$~GeV$^2$. These results have been obtained in
Ref.~\cite{Dumm:2020muy}. It can be seen that for nonzero $B$ the masses
show a similar qualitative behavior with $T$ as in the $B=0$ case. The
$\pi^0$ mass remains approximately constant up to the critical temperature,
and $\pi^0$ and $\sigma$ masses match above $T_{c}(B)$, as expected from
chiral symmetry. For large temperatures it is seen that the masses get
steadily increased, the growth being dominated by pure thermal effects. As
stated, the IMC effect is observed, i.e., $T_{c}(B)$ gets lower for
increasing $B$. In the case of the form factors, the curves for
$f_{\pi^0}^{(A\parallel)}$ and $f_{\pi^0}^{(V)}$ show sudden drops at the
critical temperatures, exhibiting once again a qualitatively similar
behavior for zero and nonzero external magnetic field. The curves for
$f_{\pi^0}^{(A\perp)}$ overlap with those corresponding to
$f_{\pi^0}^{(A\parallel)}$ and are not displayed in the figure. We recall
that, at any temperature, $f_{\pi^0}^{(V)}$ is zero for vanishing external
field.

\begin{figure}[h!bt]
\begin{center}
\subfloat{\includegraphics[width=0.48\textwidth]{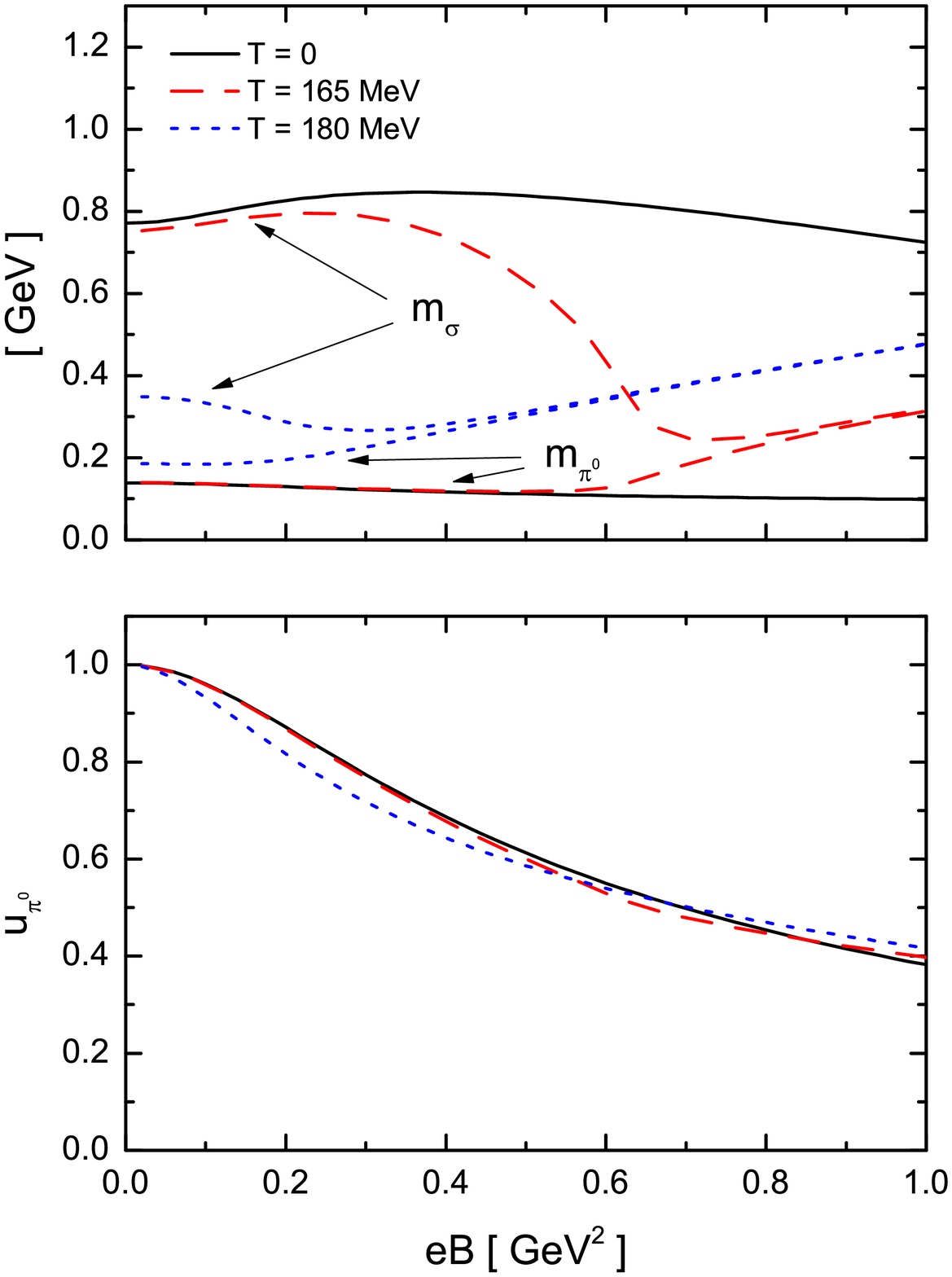}}
\hspace*{0cm}
\subfloat{\includegraphics[width=0.49\textwidth]{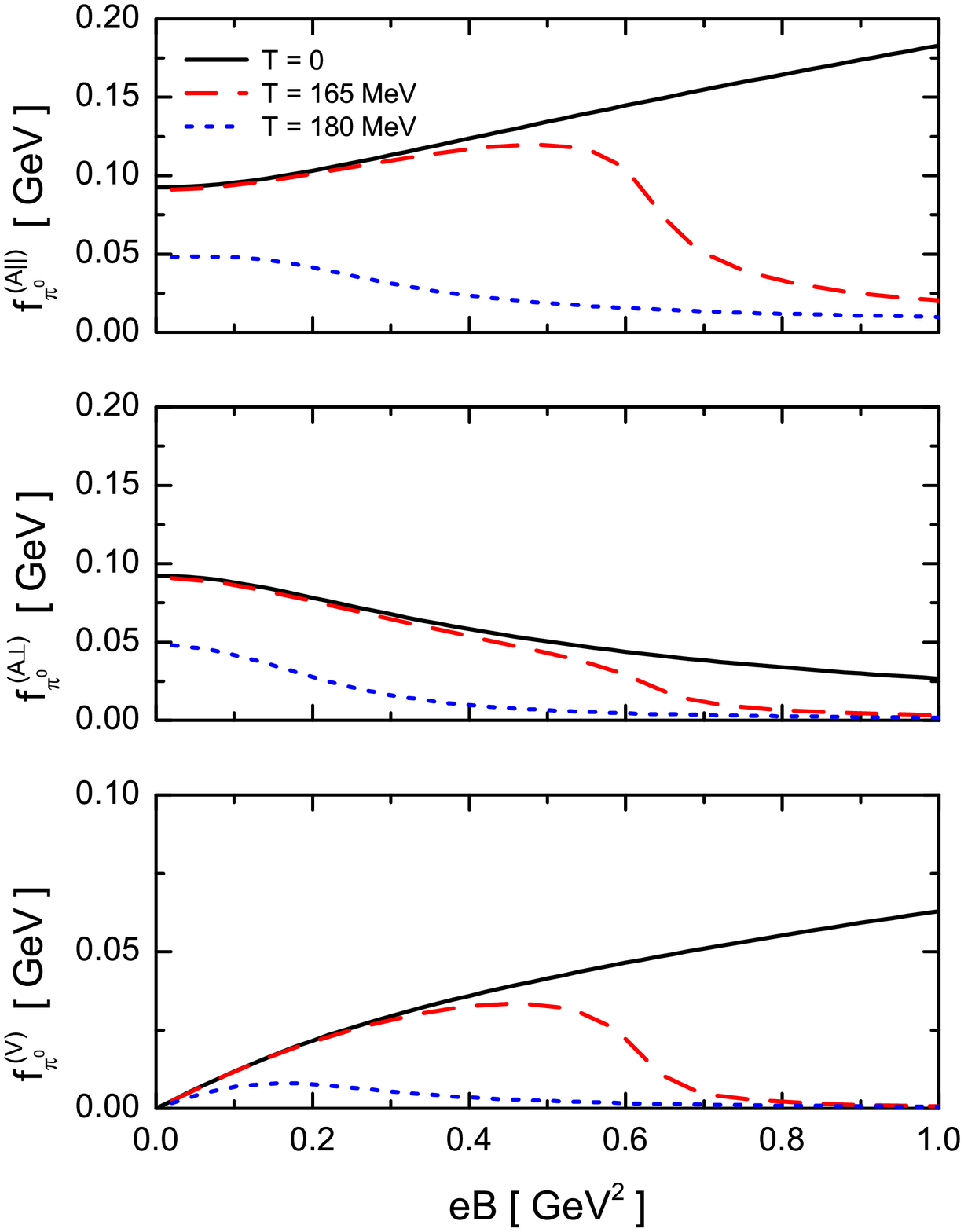}}
\end{center}
\caption{(Color online) Left: $\pi^0$ and $\sigma$ meson masses and
directional refraction index as functions of $eB$, for three representative
values of the temperature. Right: from top to bottom, decay form factors
$f_{\pi^0}^{(A\parallel)}$, $f_{\pi^0}^{(A\perp)}$ and $f_{\pi^0}^{(V)}$ as
functions of $eB$, for three representative values of the temperature.}
\label{fig5.4.11}
\end{figure}

For completeness, in Fig.~\ref{fig5.4.11} we show the behavior of meson
properties as functions of $eB$ for three representative values of the
temperature, namely $T=0$, 165~MeV and 180~MeV~\cite{Dumm:2020muy}. The
results for $T=0$, same as those previously shown in Fig.~\ref{fig5.4.10},
are included just for comparison. The curves for $T=165$~MeV can be
understood by looking at the results {in Fig.~\ref{fig5.4.2},} which
show that this is the critical temperature that corresponds to $eB\simeq
0.6$~GeV$^2$. Thus, {for this temperature,} the pion mass and form
factors in Fig.~\ref{fig5.4.11} {are expected to} show approximately
the same behavior as for $T=0$ up to $eB\sim 0.5-0.6$~GeV$^2$. Beyond these
values, as expected from the results in Fig.~\ref{fig5.4.10}, one finds an
enhancement of the pion mass and a decrease of the axial and vector form
factors. On the other hand, it is seen that for $T=180$~MeV the values of
the pion mass and axial form factors are well separated from the $T=0$
values already at $B=0$. This reflects the fact that at $T\sim 180$~MeV the
system is undergoing the chiral restoration transition for vanishing
magnetic field.

%% file: Sec6-rev.tex

\section{Summary \& concluding remarks}
\label{sec6}

In this work we have presented an overview of the current status of the
research on effective nonlocal NJL-like chiral quark models. Within this
framework, we have analyzed the description of hadron properties in vacuum,
as well as the features of deconfinement and chiral restoration transitions
for systems at finite temperature and/or density. In addition, we have
addressed other related subjects, such as the study of phase transitions for
imaginary chemical potentials, the possible existence of inhomogeneous phase
regions, the presence of color superconductivity, the effects produced by
strong external magnetic fields and the applications to the description of
compact stellar objects.

The reliability of nlNJL models and their advantage over the
standard, local NJL model can be evaluated by comparing the
obtained predictions against vacuum hadron phenomenology and
existing results from lattice calculations, in particular, in the
region of low or vanishing chemical potentials. Here, we have
started by considering a two-flavor model (see Sec.~\ref{sec2}),
which has been extended by including interaction channels that
account for vector and axial vector meson physics
(Sec.~\ref{sec3}). Then, the extension to three flavors has been
also reviewed (Sec.~\ref{sec4}). In all these approaches the
underlying QCD interactions are effectively taken into account
through a few coupling constants and nonlocal form factors. In the
simplest versions of the {nonlocal NJL-like} models (see
Sec.~\ref{sec2}), these form factors can be obtained from a fit to
lattice QCD results for the momentum dependence of effective quark
propagators. In fact, a general analyses of the results presented
throughout this article shows that most qualitative predictions of
nlNJL models do not depend significantly on the ultraviolet
behavior of the form factors. Hence, in many cases it is
convenient to use Gaussian functions, which guarantee a rapid
convergence of quark loop integrals and simplify analytical
calculations. It is also shown that ---as expected--- a more
accurate fit to LQCD results for quark propagators leads to a
better agreement between nlNJL model and LQCD predictions for
thermodynamical quantities (see Sec.~\ref{sec4}). After
determining the values of the coupling constants from a few input
empirical quantities, various particle properties can be obtained
as predictions of the models. As shown by Tables~\ref{tab2},
\ref{tab:propb} and \ref{tab:2-su3}, these results show in general
a reasonable agreement with the observed phenomenology. This
includes both the $J=1$ and the strange meson sectors. On the
other hand, there is a good agreement with most lattice QCD
results for finite temperature and low chemical potentials, and
also for the theoretical case of imaginary chemical potentials. {A
point that deserves further investigation is the thermal behavior
of the traced Polyakov loop, for which the comparison between
quark effective models and LQCD data has to be taken with some
care (see discussion in Sec.~\ref{sec4.4}).}

Concerning the comparison with standard local NJL-like approaches,
there are several reasons for which nonlocal models are expected
to be more adequate for the description of low energy QCD
phenomenology than local effective theories. In fact, as stated in
Sec.~\ref{sec2}, nonlocality arises naturally in the context of
various successful approaches to strong interactions in the
nonperturbative regime, covering from the instanton liquid model
to Dyson-Schwinger resummation methods. Moreover, the presence of
well behaved form factors in nonlocal models ensures the
convergence of quark loop integrals, which make nlNJL predictions
more stable against changes in model parameters in comparison with
models that include sharp cutoff regularizations. At the same
time, the assumption of a separable interaction makes these
nlNJL-like schemes capable of retaining much of the simplicity of
the local NJL model, in comparison with more rigorous analytical
approaches to nonperturbative QCD. While for many physical
quantities the results are similar for both local and nonlocal
approaches, there are some cases in which significant qualitative
differences show up. One of them is the entanglement between
chiral restoration and deconfinement critical temperatures. This
{feature is observed from} LQCD results and arises naturally in
the framework of nlPNJL models (see Secs.~\ref{sec2.4} and
\ref{sec4.4}), while local models predict in general a splitting
between critical temperatures. Another example is the dependence
of the pion decay constant, $f_\pi$, on the value of the explicit
chiral symmetry breaking parameter $m_c$ (or, equivalently, on the
pion mass). As shown in Fig.~\ref{rev-ch2.4-fig3} (right panel),
the results from nlNJL models show a much better agreement with
LQCD results than those obtained within local NJL schemes. In
addition, special attention should be deserved to the effects of
an external strong magnetic field on phase transitions and meson
properties. As shown in Sec.~\ref{sec5.4}, nlNJL models predict
the existence of inverse magnetic catalysis, according to which
the chiral restoration transition temperature behaves as a
decreasing function of the external magnetic field. It is worth
pointing out that this effect, which has been found by LQCD
calculations, arises naturally in the context of nlNJL approaches
due to the magnetic field dependence of nonlocal form factors. In
the local NJL the inverse magnetic catalysis can be obtained only
after the inclusion of some ad-hoc mechanism, such as e.g.~the
assumption of an explicit dependence of the coupling constants on
the magnetic field. With no ad-hoc adjustments, a good agreement
between nlNJL models and lattice QCD results for the behavior of
critical temperatures is found, as shown in Fig.~\ref{fig5.4.4}.
Moreover, the predictions of nlNJL models for the behavior of the
$\pi^0$ mass as a function of the magnetic field also show
excellent agreement with LQCD data, as shown in
Fig.~\ref{fig5.4.8}. It is seen that, in general, there are
significant differences between local and nonlocal model
predictions for the magnetic field dependence of various $\pi^0$
properties (see Fig.~\ref{fig5.4.9}).

{We also find it important to state some remarks on $\mu-T$ phase
diagrams.} For almost all studied parametrizations, nonlocal
models predict the existence of a critical end point separating
crossover-like and first order transition critical temperatures.
This is in accordance with the general belief from most approaches
to effective models for low energy strong interactions. From the
phase diagrams in Figs.~\ref{rev-ch2.5-fig2}, \ref{fig:QCDpd} and
\ref{fig:ch4.5-fig2} it can be seen that, in general, the CEP
location is found to lie within a range of temperatures between
120 and 170~MeV and (quark) chemical potentials from 150 to
250~MeV. The precise position depends on the model parametrization
and the form of the Polyakov loop potential, which {accounts for}
confinement features and is a major source of uncertainty for both
local and nonlocal NJL-like approaches. It is worth remarking
that, as discussed in Secs.~\ref{sec5.1} and \ref{sec5.3}, nlNJL
models allow for the presence of two-flavor superconducting states
and mixed phases, as well as inhomogeneous phase regions. In
particular, as discussed in Sec.~\ref{sec5.2}, several works show
that the existence of mixed and/or quark matter phases in the deep
core of compact stars is compatible with present observational
constraints. Regarding inhomogeneous (or crystalline) phases, it
is worth mentioning that local two-flavor NJL models predict the
existence of a so-called Lifshitz point ---which in the chiral
limit coincides with the tricritical point--- whereas in the nlNJL
models studied here the would-be Lifshitz point appears to be
hidden into the hadronic phase in which chiral symmetry is
spontaneously broken (see Fig.~\ref{fig:5.3.1}). Depending on the
parametrization, one can find instead a triple point, in which
hadronic, quark matter and inhomogeneous phases meet.

{As a general conclusion, it could be stated that nonlocal NJL-like
models provide a scenario in which several of the traditional problems of
the NJL model are overcome, whilst much of the simplicity and predictive
power of the standard local approach is retained. Moreover, the general good
agreement with lattice QCD results for both zero and finite temperature
indicates that the nlNJL model approach serves as a successful tool to deal
with the description of nonperturbative QCD features, and can be used to get
predictions for the behavior of hadronic and quark matter systems at finite
densities and/or in the presence of external fields.}

\section*{Acknowledgments}

The authors have benefited from scientific exchange with many researchers
over the years, including former students and collaborators. Though we skip
quoting a full list, which could certainly be incomplete, we want in
particular to acknowledge fruitful discussions with D.B.\ Blaschke,  M.\
Buballa, A.G.\ Grunfeld, M.\ Loewe, D.P.\ Menezes, S.\ Noguera, M.G.\
Orsaria and W.\ Weise. This work has been supported in part by Consejo
Nacional de Investigaciones Cient\'ificas y T\'ecnicas and Agencia Nacional
de Promoci\'on Cient\'ifica y Tecnol\'ogica (Argentina), under Grants
No.~PIP17-700 and No.~PICT17-03-0571, respectively, and by the National
University of La Plata (Argentina), Project No.~X824.

%% file: App1-rev.tex
\section*{Appendix: Basic notation and conventions}
\label{secapp1}

Throughout this article we use Euclidean space-time coordinates $x_\mu =
(\vec x , x_4)$, where $x_4 = i x_0$. Then, the associated gradient
four-vector is $\partial_\mu  = (\vec \nabla , \partial_4)$. The scalar
product of two Euclidean four-vectors $V$ and $W$ is denoted by
\begin{eqnarray}
V \cdot W \ = \ V_\mu W_\mu \ = \ \vec V \cdot \vec W + V_4 W_4 \ .
\end{eqnarray}
The Euclidean Dirac matrices are defined as $\gamma_\mu = (\vec \gamma ,
\gamma_4)$, where $\gamma_4 = i \gamma_0$. They satisfy the commutation
relations $\{ \gamma_\mu , \gamma_\nu\} = -2 \delta_{\mu\nu}$. The Dirac
matrix $\gamma_5$ is given by $\gamma_5 = -
\gamma_1\gamma_2\gamma_3\gamma_4$. Given a four-vector $V_\mu$, the
``slash'' notation is defined as
\begin{eqnarray}
\rlap/V \ = \ \vec \gamma \cdot \vec V + \gamma_4 V_4 \ .
\end{eqnarray}

For the Fourier transform we use
\begin{equation}
f(x) = \int \frac{d^4p}{(2\pi^4)} \exp(i p \cdot x) \ f(p)\ ,   \qquad
\qquad f(p) = \int d^4x \exp(-i p \cdot x) \ f(x)\ ,
\end{equation}
which is consistent with $p_\mu = - i  \partial_\mu$. For finite-temperature
calculations within the Matsubara formalism we use the transformations
\begin{eqnarray}
f(\tau) \ = \ T \sum_{n=-\infty}^\infty \ \exp(i \omega_n \tau) f(\omega_n) \ ,
\qquad \qquad
f(\omega_n) \ = \ \int_0^{1/T} d\tau \ \exp(-i \omega_n \tau) f(\tau)\ ,
\end{eqnarray}
where $\tau = i t = x_4$. Here, $\omega_n$ are the Matsubara frequencies,
i.e., $\omega_n = (2 n + 1) \pi T$ for fermions and $\omega_n = 2 n \pi T$
for bosons.

%% file: review_main_rev.bbl
\begin{thebibliography}{999}


\bibitem{Schwarz:2003du}
D.~J.~Schwarz,
Annalen Phys. \textbf{12}, 220-270 (2003)
[arXiv:astro-ph/0303574 [astro-ph]]

\bibitem{Page:2006ud}
D.~Page and S.~Reddy,
Ann. Rev. Nucl. Part. Sci. \textbf{56}, 327-374 (2006)
[arXiv:astro-ph/0608360 [astro-ph]]

\bibitem{Lattimer:2015nhk}
J.~M.~Lattimer and M.~Prakash,
Phys. Rept. \textbf{621}, 127-164 (2016)
[arXiv:1512.07820[astro-ph.SR]]

\bibitem{Braun-Munzinger:2015hba}
P.~Braun-Munzinger, V.~Koch, T.~Sch\"afer and J.~Stachel,
Phys. Rept. \textbf{621}, 76-126 (2016)
[arXiv:1510.00442 [nucl-th]]

\bibitem{Busza:2018rrf}
W.~Busza, K.~Rajagopal and W.~van der Schee,
Ann. Rev. Nucl. Part. Sci. \textbf{68}, 339-376 (2018)
[arXiv:1802.04801[hep-ph]]

\bibitem{Bzdak:2019pkr}
A.~Bzdak, S.~Esumi, V.~Koch, J.~Liao, M.~Stephanov and N.~Xu,
Phys. Rept. \textbf{853}, 1-87 (2020)
[arXiv:1906.00936 [nucl-th]]

\bibitem{Baym:2017whm}
G.~Baym, T.~Hatsuda, T.~Kojo, P.~D.~Powell, Y.~Song and
T.~Takatsuka,
Rept. Prog. Phys. \textbf{81}, 056902 (2018)
[arXiv:1707.04966 [astro-ph.HE]]

\bibitem{Karsch:2001vs}
F.~Karsch,
Nucl. Phys. A \textbf{698}, 199-208 (2002)
[arXiv:hep-ph/0103314 [hep-ph]]

\bibitem{Ding:2015ona}
H.~T.~Ding, F.~Karsch and S.~Mukherjee,
Int. J. Mod. Phys. E \textbf{24}, 1530007 (2015)
[arXiv:1504.05274 [hep-lat]]

\bibitem{Karsch:2001cy}
F.~Karsch,
Lect. Notes Phys. \textbf{583}, 209-249 (2002)
[arXiv:hep-lat/0106019 [hep-lat]]

\bibitem{Nambu:1961tp}
Y.~Nambu and G.~Jona-Lasinio,
Phys. Rev. \textbf{122}, 345-358 (1961)

\bibitem{Nambu:1961fr}
Y.~Nambu and G.~Jona-Lasinio,
Phys. Rev. \textbf{124}, 246-254 (1961)

\bibitem{Vogl:1991qt}
U.~Vogl and W.~Weise,
Prog. Part. Nucl. Phys. \textbf{27}, 195-272 (1991)

\bibitem{Klevansky:1992qe}
S.~P.~Klevansky,
Rev. Mod. Phys. \textbf{64}, 649-708 (1992)

\bibitem{Hatsuda:1994pi}
T.~Hatsuda and T.~Kunihiro,
Phys. Rept. \textbf{247}, 221-367 (1994)
[arXiv:hep-ph/9401310 [hep-ph]]

\bibitem{Polyakov:1978vu}
A.~M.~Polyakov,
Phys. Lett. B \textbf{72}, 477-480 (1978)

\bibitem{Fukushima:2017csk}
K.~Fukushima and V.~Skokov,
Prog. Part. Nucl. Phys. \textbf{96}, 154-199 (2017)
[arXiv:1705.00718 [hep-ph]]

\bibitem{Meisinger:1995ih}
P.~N.~Meisinger and M.~C.~Ogilvie,
Phys. Lett. B \textbf{379}, 163-168 (1996)
[arXiv:hep-lat/9512011[hep-lat]]

\bibitem{Fukushima:2003fw}
K.~Fukushima,
Phys. Lett. B \textbf{591}, 277-284 (2004)
[arXiv:hep-ph/0310121[hep-ph]]

\bibitem{Megias:2004hj}
E.~Megias, E.~Ruiz Arriola and L.~L.~Salcedo,
Phys. Rev. D \textbf{74}, 065005 (2006)
[arXiv:hep-ph/0412308 [hep-ph]]

\bibitem{Ratti:2005jh}
C.~Ratti, M.~A.~Thaler and W.~Weise,
Phys. Rev. D \textbf{73}, 014019 (2006)
[arXiv:hep-ph/0506234 [hep-ph]]

\bibitem{Roessner:2006xn}
S.~Roessner, C.~Ratti and W.~Weise,
Phys. Rev. D \textbf{75}, 034007 (2007)
[arXiv:hep-ph/0609281 [hep-ph]]

\bibitem{Mukherjee:2006hq}
S.~Mukherjee, M.~G.~Mustafa and R.~Ray,
Phys. Rev. D \textbf{75}, 094015 (2007)
[arXiv:hep-ph/0609249 [hep-ph]]

\bibitem{Sasaki:2006ww}
C.~Sasaki, B.~Friman and K.~Redlich,
Phys. Rev. D \textbf{75}, 074013 (2007)
[arXiv:hep-ph/0611147 [hep-ph]]

\bibitem{Schmidt:1994di}
S.~M.~Schmidt, D.~Blaschke and Y.~L.~Kalinovsky,
Phys. Rev. C \textbf{50}, 435-446 (1994)

\bibitem{Burden:1996nh}
C.~J.~Burden, L.~Qian, C.~D.~Roberts, P.~C.~Tandy and
M.~J.~Thomson,
Phys. Rev. C \textbf{55}, 2649-2664 (1997)
[arXiv:nucl-th/9605027 [nucl-th]]

\bibitem{Bowler:1994ir}
R.~D.~Bowler and M.~C.~Birse,
Nucl. Phys. A \textbf{582}, 655-664 (1995)
[arXiv:hep-ph/9407336 [hep-ph]]

\bibitem{Ripka:1997zb}
G.~Ripka, ``Quarks bound by chiral fields: The quark-structure of the vacuum and of light mesons and baryons,''
(Oxford University Press, Oxford, 1997)

\bibitem{Schafer:1996wv}
T.~Sch\"afer and E.~V.~Shuryak,
Rev. Mod. Phys. \textbf{70}, 323-426 (1998)
[arXiv:hep-ph/9610451 [hep-ph]]

\bibitem{Roberts:1994dr}
C.~D.~Roberts and A.~G.~Williams,
Prog. Part. Nucl. Phys. \textbf{33}, 477-575 (1994)
[arXiv:hep-ph/9403224 [hep-ph]]

\bibitem{Noguera:2005ej}
S.~Noguera,
Int. J. Mod. Phys. E \textbf{16}, 97-132 (2007)
[arXiv:hep-ph/0502171 [hep-ph]]

\bibitem{Noguera:2008cm}
S.~Noguera and N.~N.~Scoccola,
Phys. Rev. D \textbf{78}, 114002 (2008)
[arXiv:0806.0818 [hep-ph]]

\bibitem {Parappilly:2005ei}
M.~B.~Parappilly, P.~O.~Bowman, U.~M.~Heller,
D.~B.~Leinweber, A.~G.~Williams and J.~B.~Zhang,
Phys. Rev. D \textbf{73}, 054504 (2006)
[arXiv:hep-lat/0511007 [hep-lat]]

\bibitem{Furui:2006ks}
S.~Furui and H.~Nakajima,
Phys. Rev. D \textbf{73}, 074503 (2006)

\bibitem{RuizArriola:1998zi}
E.~Ruiz Arriola and L.~L.~Salcedo,
Phys. Lett. B \textbf{450}, 225-233 (1999)
[arXiv:hep-th/9811073 [hep-th]]

\bibitem{Blaschke:1995gr}
D.~Blaschke, Y.~L.~Kalinovsky, G.~Roepke, S.~M.~Schmidt and M.~K.~Volkov,
Phys. Rev. C \textbf{53}, 2394-2400 (1996)
[arXiv:nucl-th/9511003 [nucl-th]]

\bibitem{Plant:2000ty}
R.~S.~Plant and M.~C.~Birse,
Nucl. Phys. A \textbf{703}, 717-744 (2002)
[arXiv:hep-ph/0007340 [hep-ph]]

\bibitem{Plant:1997jr}
R.~S.~Plant and M.~C.~Birse,
Nucl. Phys. A \textbf{628}, 607-644 (1998)
[arXiv:hep-ph/9705372 [hep-ph]]

\bibitem{Broniowski:1999dm}
W.~Broniowski,
AIP Conf. Proc. \textbf{508}, 380 (2000)
[arXiv:hep-ph/9911204 [hep-ph]]

\bibitem {Praszalowicz:2001wy}
M.~Praszalowicz and A.~Rostworowski,
Phys. Rev. D \textbf{64}, 074003 (2001)
[arXiv:hep-ph/0105188 [hep-ph]]

\bibitem{Praszalowicz:2001pi}
M.~Praszalowicz and A.~Rostworowski,
Phys. Rev. D \textbf{66}, 054002 (2002)
[arXiv:hep-ph/0111196 [hep-ph]]

\bibitem{Praszalowicz:2003pr}
M.~Praszalowicz and A.~Rostworowski,
Acta Phys. Polon. B \textbf{34}, 2699-2730 (2003)
[arXiv:hep-ph/0302269 [hep-ph]]

\bibitem{Dorokhov:2003kf}
A.~E.~Dorokhov and W.~Broniowski,
Eur. Phys. J. C \textbf{32}, 79-96 (2003)
[arXiv:hep-ph/0305037 [hep-ph]]

\bibitem{Kotko:2008gy}
P.~Kotko and M.~Praszalowicz,
Acta Phys. Polon. B \textbf{40}, 123-152 (2009)
[arXiv:0803.2847 [hep-ph]]

\bibitem{Kotko:2009ij}
P.~Kotko and M.~Praszalowicz,
Phys. Rev. D \textbf{80}, 074002 (2009)
[arXiv:0907.4044 [hep-ph]]

\bibitem{Kotko:2009mb}
P.~Kotko and M.~Praszalowicz,
Phys. Rev. D \textbf{81}, 034019 (2010)
[arXiv:0912.0029 [hep-ph]]

\bibitem{Nam:2012vm}
S.~i.~Nam,
Phys. Rev. D \textbf{86}, 074005 (2012)
[arXiv:1205.4156 [hep-ph]]

\bibitem{Nam:2017gzm}
S.~i.~Nam,
Mod. Phys. Lett. A \textbf{32}, 1750218 (2017)
[arXiv:1704.03824 [hep-ph]]

\bibitem{Dumm:2013zoa}
D.~Gomez~Dumm, S.~Noguera, N.~N.~Scoccola and S.~Scopetta,
Phys. Rev. D \textbf{89}, 054031 (2014)
[arXiv:1311.3595 [hep-ph]]

\bibitem{GomezDumm:2012qh}
D.~Gomez Dumm, S.~Noguera and N.~N.~Scoccola,
Phys. Rev. D \textbf{86}, 074020 (2012)
[arXiv:1205.2730 [hep-ph]]

\bibitem{Golli:1998rf}
B.~Golli, W.~Broniowski and G.~Ripka,
Phys. Lett. B \textbf{437}, 24-28 (1998)
[arXiv:hep-ph/9807261 [hep-ph]]

\bibitem{Broniowski:2001cx}
W.~Broniowski, B.~Golli and G.~Ripka,
Nucl. Phys. A \textbf{703}, 667-701 (2002)
[arXiv:hep-ph/0107139 [hep-ph]]

\bibitem{Rezaeian:2004nf}
A.~H.~Rezaeian, N.~R.~Walet and M.~C.~Birse,
Phys. Rev. C \textbf{70}, 065203 (2004)
[arXiv:hep-ph/0408233 [hep-ph]]

\bibitem{Szczerbinska:1999iz}
B.~Szczerbinska and W.~Broniowski,
Acta Phys. Polon. B \textbf{31} (2000), 835-845
[arXiv:hep-ph/9911514 [hep-ph]]

\bibitem{Blaschke:2000gd}
D.~Blaschke, G.~Burau, Y.~L.~Kalinovsky, P.~Maris and P.~C.~Tandy,
Int. J. Mod. Phys. A \textbf{16}, 2267-2291 (2001)
[arXiv:nucl-th/0002024 [nucl-th]]

\bibitem{General:2000zx}
I.~General, D.~Gomez Dumm and N.~N.~Scoccola,
Phys. Lett. B \textbf{506}, 267-274 (2001)
[arXiv:hep-ph/0010034 [hep-ph]]

\bibitem{GomezDumm:2001fz}
D.~Gomez Dumm and N.~N.~Scoccola,
Phys. Rev. D \textbf{65}, 074021 (2002)
[arXiv:hep-ph/0107251 [hep-ph]]

\bibitem{GomezDumm:2004sr}
D.~Gomez Dumm and N.~N.~Scoccola,
Phys. Rev. C \textbf{72}, 014909 (2005)
[arXiv:hep-ph/0410262 [hep-ph]]

\bibitem{Blaschke:2007np}
D.~Blaschke, M.~Buballa, A.~E.~Radzhabov and M.~K.~Volkov,
Yad. Fiz. \textbf{71}, 2012-2018 (2008)
[arXiv:0705.0384 [hep-ph]]

\bibitem{Contrera:2007wu}
G.~A.~Contrera, D.~Gomez Dumm and N.~N.~Scoccola,
Phys. Lett. B \textbf{661}, 113-117 (2008)
[arXiv:0711.0139 [hep-ph]]

\bibitem{Hell:2008cc}
T.~Hell, S.~Roessner, M.~Cristoforetti and W.~Weise,
Phys. Rev. D \textbf{79}, 014022 (2009)
[arXiv:0810.1099 [hep-ph]]

\bibitem{Kondo:2010ts}
K.~I.~Kondo,
Phys. Rev. D \textbf{82}, 065024 (2010)
[arXiv:1005.0314 [hep-th]]



\bibitem{GomezDumm:2006vz}
D.~Gomez Dumm, A.~G.~Grunfeld and N.~N.~Scoccola,
Phys. Rev. D \textbf{74}, 054026 (2006)
[arXiv:hep-ph/0607023 [hep-ph]]

\bibitem{Broniowski:1999bz}
W.~Broniowski,
Mini-Workshop Bled 1999, 17-26 [arXiv:hep-ph/9909438 [hep-ph]]


\bibitem{Kapusta:2006pm}
J.~I.~Kapusta and C.~Gale,
``Finite-temperature field theory: Principles and applications''
(Cambridge University Press, New York, 2006)

\bibitem{Bellac:2011kqa}
M.~L.~Bellac, ``Thermal Field Theory''
(Cambridge University Press, New York, 1996)

\bibitem{Loewe:2011qc}
M.~Loewe, P.~Morales and C.~Villavicencio,
Phys. Rev. D \textbf{83}, 096005 (2011)
[arXiv:1102.1778 [hep-ph]]

\bibitem{Loewe:2013zaa}
M.~Loewe, F.~Marquez and C.~Villavicencio,
Phys. Rev. D \textbf{88}, 056004 (2013)
[arXiv:1307.6764 [hep-ph]]

\bibitem{Carter:1998ji}
G.~W.~Carter and D.~Diakonov,
Phys. Rev. D \textbf{60}, 016004 (1999)
[arXiv:hep-ph/9812445 [hep-ph]]

\bibitem{Pisarski:2000eq}
R.~D.~Pisarski,
Phys. Rev. D \textbf{62}, 111501 (2000)
[arXiv:hep-ph/0006205 [hep-ph]]

\bibitem{Schaefer:2007pw}
B.~J.~Schaefer, J.~M.~Pawlowski and J.~Wambach,
Phys. Rev. D \textbf{76}, 074023 (2007)
[arXiv:0704.3234 [hep-ph]]

\bibitem{Schaefer:2009ui}
B.~J.~Schaefer, M.~Wagner and J.~Wambach,
Phys. Rev. D \textbf{81}, 074013 (2010)
[arXiv:0910.5628 [hep-ph]]

\bibitem{Dumitru:2005ng}
A.~Dumitru, R.~D.~Pisarski and D.~Zschiesche,
Phys. Rev. D \textbf{72}, 065008 (2005)
[arXiv:hep-ph/0505256 [hep-ph]]

\bibitem{Ratti:2006wg}
C.~Ratti, S.~Roessner, M.~A.~Thaler and W.~Weise,
Eur. Phys. J. C \textbf{49}, 213-217 (2007)
[arXiv:hep-ph/0609218 [hep-ph]]

\bibitem{Ratti:2007jf}
C.~Ratti, S.~Roessner and W.~Weise,
Phys. Lett. B \textbf{649}, 57-60 (2007)
[arXiv:hep-ph/0701091 [hep-ph]]

\bibitem{Fukushima:2006uv}
K.~Fukushima and Y.~Hidaka,
Phys. Rev. D \textbf{75}, 036002 (2007)
[arXiv:hep-ph/0610323 [hep-ph]]

\bibitem{Nishimura:2014kla}
H.~Nishimura, M.~C.~Ogilvie and K.~Pangeni,
Phys. Rev. D \textbf{91}, 054004 (2015)
[arXiv:1411.4959 [hep-ph]]

\bibitem{Tanizaki:2015pua}
Y.~Tanizaki, H.~Nishimura and K.~Kashiwa,
Phys. Rev. D \textbf{91}, 101701 (2015)
[arXiv:1504.02979 [hep-th]]

\bibitem{Pagura:2011rt}
V.~Pagura, D.~Gomez Dumm and N.~N.~Scoccola,
Phys. Lett. B \textbf{707}, 76-82 (2012)
[arXiv:1105.1739 [hep-ph]]

\bibitem{Pagura:2013rza}
V.~P.~Pagura, Ph.D. Thesis (2013) [https://inspirehep.net/literature/1309403]

\bibitem{Contrera:2010kz}
G.~A.~Contrera, M.~Orsaria and N.~N.~Scoccola,
Phys. Rev. D \textbf{82}, 054026 (2010)
[arXiv:1006.4639 [hep-ph]]



\bibitem {Bowman2003}
P.~O.~Bowman, U.~M.~Heller, D.~B.~Leinweber and A.~G.~Williams,
Nucl. Phys. B Proc. Suppl. \textbf{119}, 323-325 (2003)
[arXiv:hep-lat/0209129 [hep-lat]];\\
P.~O.~Bowman, U.~M.~Heller and A.~G.~Williams,
Phys. Rev. D \textbf{66}, 014505 (2002)
[arXiv:hep-lat/0203001 [hep-lat]]

\bibitem {RuizArriola:2003bs}
E.~Ruiz Arriola and W.~Broniowski,
Phys. Rev. D \textbf{67}, 074021 (2003)
[arXiv:hep-ph/0301202 [hep-ph]]

\bibitem{Dosch:1997wb}
H.~G.~Dosch and S.~Narison,
Phys. Lett. B \textbf{417}, 173-176 (1998)
[arXiv:hep-ph/9709215 [hep-ph]]

\bibitem{Giusti:1998wy}
L.~Giusti, F.~Rapuano, M.~Talevi and A.~Vladikas,
Nucl. Phys. B \textbf{538}, 249-277 (1999)
[arXiv:hep-lat/9807014 [hep-lat]]

\bibitem{Nakayama:1991ue}
K.~Nakayama and S.~Krewald,
Phys. Lett. B \textbf{273}, 199-204 (1991)

\bibitem{Zyla:2020zbs}
P.~A.~Zyla \textit{et al.} [Particle Data Group],
PTEP \textbf{2020}, 083C01 (2020)

\bibitem{Hell:2011ic}
T.~Hell, K.~Kashiwa and W.~Weise,
Phys. Rev. D \textbf{83}, 114008 (2011)
[arXiv:1104.0572 [hep-ph]]

\bibitem{Hell:2009by}
T.~Hell, S.~Rossner, M.~Cristoforetti and W.~Weise,
Phys. Rev. D \textbf{81}, 074034 (2010)
[arXiv:0911.3510 [hep-ph]]


\bibitem{Fu:2007xc}
W.~j.~Fu, Z.~Zhang and Y.~x.~Liu,
Phys. Rev. D \textbf{77} , 014006 (2008)
[arXiv:0711.0154 [hep-ph]]

\bibitem{Costa:2008dp}
P.~Costa, M.~C.~Ruivo, C.~A.~de Sousa, H.~Hansen and W.~M.~Alberico,
Phys. Rev. D \textbf{79} , 116003 (2009)
[arXiv:0807.2134 [hep-ph]]

\bibitem{Sakai:2009dv}
Y.~Sakai, K.~Kashiwa, H.~Kouno, M.~Matsuzaki and M.~Yahiro,
Phys. Rev. D \textbf{79} , 096001 (2009)
[arXiv:0902.0487 [hep-ph]]

\bibitem{Sakai:2010rp}
Y.~Sakai, T.~Sasaki, H.~Kouno and M.~Yahiro,
Phys. Rev. D \textbf{82} , 076003 (2010)
[arXiv:1006.3648 [hep-ph]]

\bibitem{Sasaki:2011wu}
T.~Sasaki, Y.~Sakai, H.~Kouno and M.~Yahiro,
Phys. Rev. D \textbf{84} , 091901 (2011)
[arXiv:1105.3959 [hep-ph]]

\bibitem{Karsch:2003jg}
F.~Karsch and E.~Laermann,
[arXiv:hep-lat/0305025 [hep-lat]]

\bibitem{Karsch:1994hm}
F.~Karsch and E.~Laermann,
Phys. Rev. D \textbf{50}, 6954-6962 (1994)
[arXiv:hep-lat/9406008 [hep-lat]]

\bibitem{Bernard:1996iz}
C.~W.~Bernard, T.~Blum, C.~E.~Detar, S.~A.~Gottlieb, U.~M.~Heller, J.~E.~Hetrick, K.~Rummukainen, R.~Sugar, D.~Toussaint and M.~Wingate,
Phys. Rev. Lett. \textbf{78}, 598-601 (1997)
[arXiv:hep-lat/9611031 [hep-lat]]

\bibitem{Iwasaki:1996ya}
Y.~Iwasaki, K.~Kanaya, S.~Kaya and T.~Yoshie,
Phys. Rev. Lett. \textbf{78}, 179-182 (1997)
[arXiv:hep-lat/9609022 [hep-lat]]

\bibitem{Aoki:1998wg}
S.~Aoki \textit{et al.} [JLQCD],
Phys. Rev. D \textbf{57}, 3910-3922 (1998)
[arXiv:hep-lat/9710048 [hep-lat]]

\bibitem{AliKhan:2000wou}
A.~Ali Khan \textit{et al.} [CP-PACS],
Phys. Rev. D \textbf{63}, 034502 (2000)
[arXiv:hep-lat/0008011 [hep-lat]]

\bibitem{DElia:2005nmv}
M.~D'Elia, A.~Di Giacomo and C.~Pica,
Phys. Rev. D \textbf{72}, 114510 (2005)
[arXiv:hep-lat/0503030 [hep-lat]]

\bibitem{Bonati:2009yg}
C.~Bonati, G.~Cossu, M.~D'Elia, A.~Di Giacomo and C.~Pica,
PoS \textbf{LATTICE2008}, 204 (2008)
[arXiv:0901.3231 [hep-lat]]

\bibitem{Berges:1997eu}
J.~Berges, D.~U.~Jungnickel and C.~Wetterich,
Phys. Rev. D \textbf{59}, 034010 (1999)
[arXiv:hep-ph/9705474 [hep-ph]]

\bibitem{Dumitru:2003cf}
A.~Dumitru, D.~Roder and J.~Ruppert,
Phys. Rev. D \textbf{70}, 074001 (2004)
[arXiv:hep-ph/0311119 [hep-ph]]

\bibitem{Braun:2005fj}
J.~Braun, B.~Klein, H.~J.~Pirner and A.~H.~Rezaeian,
Phys. Rev. D \textbf{73}, 074010 (2006)
[arXiv:hep-ph/0512274 [hep-ph]]

\bibitem{Pagura:2012ku}
V.~Pagura, D.~Gomez Dumm and N.~N.~Scoccola,
Phys. Rev. D \textbf{87}, 014027 (2013)
[arXiv:1210.1553 [hep-ph]]

\bibitem{Noaki:2008iy}
J.~Noaki {\it et al.} [JLQCD and TWQCD Collab.],
Phys.\ Rev.\ Lett.\  {\bf 101}, 202004 (2008)
[arXiv:0806.0894 [hep-lat]]

\bibitem{Kahara:2009sq}
T.~Kahara and K.~Tuominen,
Phys.\ Rev.\ D {\bf 80}, 114022 (2009)
[arXiv:0906.0890 [hep-ph]];\\
Phys.\ Rev.\ D {\bf 82}, 114026 (2010)
[arXiv:1006.3931 [hep-ph]]

\bibitem{Karsch:2000kv}
F.~Karsch, E.~Laermann and A.~Peikert,
Nucl. Phys. B \textbf{605}, 579-599 (2001)
[arXiv:hep-lat/0012023 [hep-lat]]

\bibitem{Bornyakov:2009qh}
V.~G.~Bornyakov, R.~Horsley, S.~M.~Morozov, Y.~Nakamura, M.~I.~Polikarpov, P.~E.~L.~Rakow, G.~Schierholz and T.~Suzuki,
Phys.\ Rev.\ D {\bf 82}, 014504 (2010)
[arXiv:0910.2392 [hep-lat]]

\bibitem{Karsch:2007dt}
F.~Karsch,
PoS LAT {\bf 2007}, 015 (2007)
[arXiv:0711.0661 [hep-lat]]

\bibitem{Bornyakov:2005dt}
V.~G.~Bornyakov, M.~N.~Chernodub, Y.~Mori, S.~M.~Morozov, Y.~Nakamura, M.~I.~Polikarpov, G.~Schierholz and A.~A.~Slavnov {\it et al.},
PoS LAT {\bf 2005}, 157 (2006)
[hep-lat/0509122]

\bibitem{Cheng:2006qk}
M.~Cheng, N.~H.~Christ, S.~Datta, J.~van der Heide, C.~Jung, F.~Karsch, O.~Kaczmarek and E.~Laermann {\it et al.},
Phys. Rev. D \textbf{74}, 054507 (2006)
[arXiv:hep-lat/0608013 [hep-lat]]

\bibitem{Ejiri:2009ac}
S.~Ejiri, F.~Karsch, E.~Laermann, C.~Miao, S.~Mukherjee, P.~Petreczky, C.~Schmidt and W.~Soeldner {\it et al.},
Phys.\ Rev.\ D {\bf 80}, 094505 (2009)
[arXiv:0909.5122 [hep-lat]]

\bibitem{Saito:2011fs}
H.~Saito \textit{et al.} [WHOT-QCD],
Phys. Rev. D \textbf{84}, 054502 (2011)
[erratum: Phys. Rev. D \textbf{85}, 079902 (2012)]
[arXiv:1106.0974 [hep-lat]]

\bibitem{Benic:2012ec}
S.~Benic, D.~Blaschke and M.~Buballa,
Phys. Rev. D \textbf{86}, 074002 (2012)
[arXiv:1206.6582 [hep-ph]]

\bibitem{Marquez:2014kla}
F.~Marquez,
Phys. Rev. D \textbf{89}, 076010 (2014)
[arXiv:1401.3722 [hep-ph]]

\bibitem{Benic:2013eqa}
S.~Benic, D.~Blaschke, G.~A.~Contrera and D.~Horvatic,
Phys. Rev. D \textbf{89}, 016007 (2014)
[arXiv:1306.0588 [hep-ph]]

\bibitem{Radzhabov:2010dd}
A.~E.~Radzhabov, D.~Blaschke, M.~Buballa and M.~K.~Volkov,
Phys. Rev. D \textbf{83}, 116004 (2011)
[arXiv:1012.0664 [hep-ph]]



\bibitem{McLerran:2007qj}
L.~McLerran and R.~D.~Pisarski,
Nucl. Phys. A \textbf{796}, 83-100 (2007)
[arXiv:0706.2191 [hep-ph]]

\bibitem{McLerran:2008ua}
L.~McLerran, K.~Redlich and C.~Sasaki,
Nucl. Phys. A \textbf{824}, 86-100 (2009)
[arXiv:0812.3585 [hep-ph]]

\bibitem{Abuki:2008nm}
H.~Abuki, R.~Anglani, R.~Gatto, G.~Nardulli and M.~Ruggieri,
Phys. Rev. D \textbf{78}, 034034 (2008)
[arXiv:0805.1509 [hep-ph]]

\bibitem{Herbst:2010rf}
T.~K.~Herbst, J.~M.~Pawlowski and B.~J.~Schaefer,
Phys. Lett. B \textbf{696}, 58-67 (2011)
[arXiv:1008.0081 [hep-ph]]

\bibitem{Herbst:2013ail}
T.~K.~Herbst, J.~M.~Pawlowski and B.~J.~Schaefer,
Phys. Rev. D \textbf{88}, no.1, 014007 (2013)
[arXiv:1302.1426 [hep-ph]]

\bibitem{Pawlowski:2014zaa}
J.~M.~Pawlowski and F.~Rennecke,
Phys. Rev. D \textbf{90}, no.7, 076002 (2014)
[arXiv:1403.1179 [hep-ph]]


\bibitem{Alford:1998sd}
M.~G.~Alford, A.~Kapustin and F.~Wilczek,
Phys. Rev. D \textbf{59}, 054502 (1999)
[arXiv:hep-lat/9807039 [hep-lat]]

\bibitem{D'Elia:2002gd}
M.~D'Elia, M.~-P.~Lombardo,
Phys.\ Rev.\  {\bf D67} , 014505 (2003)
[hep-lat/0209146]

\bibitem{deForcrand:2002hgr}
P.~de Forcrand and O.~Philipsen,
Nucl. Phys. B \textbf{642}, 290-306 (2002)
[arXiv:hep-lat/0205016 [hep-lat]]

\bibitem{Wu:2006su}
L.~-K.~Wu, X.~-Q.~Luo, H.~-S.~Chen,
Phys. Rev. D \textbf{76}, 034505 (2007)
[arXiv:hep-lat/0611035 [hep-lat]]

\bibitem{Braun:2009gm}
J.~Braun, L.~M.~Haas, F.~Marhauser, J.~M.~Pawlowski,
Phys.\ Rev.\ Lett.\  {\bf 106 },  022002 (2011)
[arXiv:0908.0008 [hep-ph]]

\bibitem{Karbstein:2006er}
F.~Karbstein and M.~Thies,
Phys. Rev. D \textbf{75}, 025003 (2007)
[arXiv:hep-th/0610243 [hep-th]]

\bibitem{Bonati:2015bha}
C.~Bonati, M.~D'Elia, M.~Mariti, M.~Mesiti, F.~Negro and F.~Sanfilippo,
Phys. Rev. D \textbf{92} , 054503 (2015)
[arXiv:1507.03571 [hep-lat]]

\bibitem{Bellwied:2015rza}
R.~Bellwied, S.~Borsanyi, Z.~Fodor, J.~G\"unther, S.~D.~Katz, C.~Ratti and K.~K.~Szabo,
Phys. Lett. B \textbf{751}, 559-564 (2015)
[arXiv:1507.07510 [hep-lat]]

\bibitem{Borsanyi:2020fev}
S.~Borsanyi, Z.~Fodor, J.~N.~Guenther, R.~Kara, S.~D.~Katz, P.~Parotto, A.~Pasztor, C.~Ratti and K.~K.~Szabo,
Phys. Rev. Lett. \textbf{125}, 052001 (2020)
[arXiv:2002.02821 [hep-lat]]

\bibitem{Roberge:1986mm}
A.~Roberge and N.~Weiss,
Nucl. Phys. B \textbf{275}, 734-745 (1986)

\bibitem{D'Elia:2007ke}
M.~D'Elia, F.~Di Renzo, M.~P.~Lombardo,
Phys.\ Rev.\ D {\bf 76},  114509 (2007)
[arXiv:0705.3814 [hep-lat]]

\bibitem{D'Elia:2009qz}
M.~D'Elia, F.~Sanfilippo,
Phys.\ Rev.\ D {\bf 80}, 111501 (2009)
[arXiv:0909.0254 [hep-lat]]

\bibitem{deForcrand:2010he}
P.~de Forcrand, O.~Philipsen,
Phys.\ Rev.\ Lett.\  {\bf 105}, 152001 (2010)
[arXiv:1004.3144 [hep-lat]]

\bibitem{Bonati:2010gi}
C.~Bonati, G.~Cossu, M.~D'Elia and F.~Sanfilippo,
Phys. Rev. D \textbf{83}, 054505 (2011)
[arXiv:1011.4515 [hep-lat]]

\bibitem{Bonati:2018fvg}
C.~Bonati, E.~Calore, M.~D'Elia, M.~Mesiti, F.~Negro, F.~Sanfilippo, S.~F.~Schifano, G.~Silvi and R.~Tripiccione,
Phys. Rev. D \textbf{99}, 014502 (2019)
[arXiv:1807.02106 [hep-lat]]

\bibitem{Goswami:2018qhc}
J.~Goswami, F.~Karsch, A.~Lahiri and C.~Schmidt,
PoS \textbf{LATTICE2018}, 162 (2018)
[arXiv:1811.02494 [hep-lat]]

\bibitem{Kashiwa:2011td}
K.~Kashiwa, T.~Hell and W.~Weise,
Phys. Rev. D \textbf{84}, 056010 (2011)
[arXiv:1106.5025 [hep-ph]]

\bibitem{Sakai:2008um}
Y.~Sakai, K.~Kashiwa, H.~Kouno, M.~Yahiro,
Phys.\ Rev.\ D {\bf 77},  051901 (2008)
[arXiv:0801.0034 [hep-ph]];\\
Phys.\ Rev.\ D {\bf 78}, 036001 (2008)
[arXiv:0803.1902 [hep-ph]]

\bibitem{Morita:2011jva}
K.~Morita, V.~Skokov, B.~Friman and K.~Redlich,
Phys. Rev. D \textbf{84}, 074020 (2011)
[arXiv:1108.0735 [hep-ph]]

\bibitem{Bonati:2016pwz}
C.~Bonati, M.~D'Elia, M.~Mariti, M.~Mesiti, F.~Negro and F.~Sanfilippo,
Phys. Rev. D \textbf{93}, 074504 (2016)
[arXiv:1602.01426 [hep-lat]]


\bibitem{Florkowski:1993br}
W.~Florkowski and B.~L.~Friman,
Acta Phys. Polon. B \textbf{25}, 49-71 (1994)

\bibitem{Eletsky:1988an}
V.~L.~Eletsky and B.~L.~Ioffe,
Sov. J. Nucl. Phys. \textbf{48}, 384 (1988)



\bibitem{Villafane:2016ukb}
M.~F.~Izzo Villafa\~ne, D.~Gomez Dumm and N.~N.~Scoccola,
Phys. Rev. D \textbf{94}, 054003 (2016)
[arXiv:1602.06984 [hep-ph]]



\bibitem{Contrera:2012wj}
G.~A.~Contrera, A.~G.~Grunfeld and D.~B.~Blaschke,
Phys. Part. Nucl. Lett. \textbf{11}, 342-351 (2014)
[arXiv:1207.4890 [hep-ph]]

\bibitem{Contrera:2016rqj}
G.~A.~Contrera, A.~G.~Grunfeld and D.~Blaschke,
Eur. Phys. J. A \textbf{52},  231 (2016)
[arXiv:1605.08430 [hep-ph]]

\bibitem{Carlomagno:2019yvi}
J.~P.~Carlomagno and M.~F.~Izzo Villafa\~ne,
Phys. Rev. D \textbf{100}, 076011 (2019)
[arXiv:1906.04257 [hep-ph]]

\bibitem{Ebert:1985kz}
D.~Ebert and H.~Reinhardt,
Nucl. Phys. B \textbf{271}, 188-226 (1986)

\bibitem{Bernard:1993rz}
V.~Bernard, U.~G.~Meissner and A.~A.~Osipov,
Phys. Lett. B \textbf{324}, 201-208 (1994)
[arXiv:hep-ph/9312203 [hep-ph]]






\bibitem{Scarpettini:2003fj}
A.~Scarpettini, D.~Gomez Dumm and N.~N.~Scoccola,
Phys. Rev. D \textbf{69} , 114018 (2004)
[arXiv:hep-ph/0311030 [hep-ph]]

\bibitem{Contrera:2009hk}
G.~A.~Contrera, D.~Gomez~Dumm and N.~N.~Scoccola,
Phys. Rev. D \textbf{81}, 054005 (2010)
[arXiv:0911.3848 [hep-ph]]

\bibitem{Carlomagno:2013ona}
J.~P.~Carlomagno, D.~Gomez Dumm and N.~N.~Scoccola,
Phys. Rev. D \textbf{88}, 074034 (2013)
[arXiv:1305.2969 [hep-ph]]


\bibitem{Leutwyler:1997yr}
H.~Leutwyler,
Nucl. Phys. B Proc. Suppl. \textbf{64}, 223-231 (1998)
[arXiv:hep-ph/9709408 [hep-ph]]

\bibitem{Feldmann:1999uf}
T.~Feldmann,
Int. J. Mod. Phys. A \textbf{15}, 159-207 (2000)
[arXiv:hep-ph/9907491 [hep-ph]]





\bibitem{Carlomagno:2018tyk}
J.~P.~Carlomagno,
Phys. Rev. D \textbf{97}, 094012 (2018)
[arXiv:1803.03235 [hep-ph]]




\bibitem{Bazavov:2010sb}
A.~Bazavov \textit{et al.} [HotQCD],
J. Phys. Conf. Ser. \textbf{230}, 012014 (2010)
[arXiv:1005.1131 [hep-lat]]

\bibitem{Borsanyi:2010bp}
S.~Borsanyi {\it et al.} [Wuppertal-Budapest Collaboration],
JHEP {\bf 1009}, 073 (2010)
[arXiv:1005.3508 [hep-lat]]

\bibitem{Bazavov:2018mes}
A.~Bazavov \textit{et al.} [HotQCD],
Phys. Lett. B \textbf{795}, 15-21 (2019)
[arXiv:1812.08235 [hep-lat]]

\bibitem{Bazavov:2009zn}
A.~Bazavov, T.~Bhattacharya, M.~Cheng, N.~H.~Christ, C.~DeTar, S.~Ejiri,
S.~Gottlieb, R.~Gupta, U.~M.~Heller and K.~Huebner, \textit{et al.}
Phys. Rev. D \textbf{80}, 014504 (2009)
[arXiv:0903.4379 [hep-lat]]

\bibitem{Borsanyi:2010cj}
S.~Borsanyi, G.~Endrodi, Z.~Fodor, A.~Jakovac, S.~D.~Katz, S.~Krieg,
C.~Ratti and K.~K.~Szabo,
JHEP \textbf{11}, 077 (2010) 
[arXiv:1007.2580 [hep-lat]]

\bibitem{Braun:2007bx}
J.~Braun, H.~Gies and J.~M.~Pawlowski,
Phys. Lett. B \textbf{684}, 262-267 (2010)
[arXiv:0708.2413 [hep-th]]

\bibitem{Marhauser:2008fz}
F.~Marhauser and J.~M.~Pawlowski,
[arXiv:0812.1144 [hep-ph]]

\bibitem{Herbst:2013ufa}
T.~K.~Herbst, M.~Mitter, J.~M.~Pawlowski, B.~J.~Schaefer and R.~Stiele,
Phys. Lett. B \textbf{731}, 248-256 (2014)
[arXiv:1308.3621 [hep-ph]]

\bibitem{Haas:2013qwp}
L.~M.~Haas, R.~Stiele, J.~Braun, J.~M.~Pawlowski and J.~Schaffner-Bielich,
Phys. Rev. D \textbf{87}, no.7, 076004 (2013)
[arXiv:1302.1993 [hep-ph]]

\bibitem{Pagura:2016rit}
V.~P.~Pagura, D.~Gomez Dumm, S.~Noguera and N.~N.~Scoccola,
Phys. Rev. D \textbf{94}, 054038 (2016)
[arXiv:1605.04675 [hep-ph]]

\bibitem{Bali:2012jv}
G.~S.~Bali, F.~Bruckmann, M.~Constantinou, M.~Costa, G.~Endrodi,
S.~D.~Katz, H.~Panagopoulos and A.~Schafer,
Phys. Rev. D \textbf{86}, 094512 (2012)
[arXiv:1209.6015 [hep-lat]]

\bibitem{Nam:2013wja}
S.~i.~Nam,
Phys. Rev. D \textbf{87},  116003 (2013)
[arXiv:1304.1265 [hep-ph]]

\bibitem{Rennecke:2016tkm}
F.~Rennecke and B.~J.~Schaefer,
Phys. Rev. D \textbf{96}, no.1, 016009 (2017)
[arXiv:1610.08748 [hep-ph]]

\bibitem{Schaefer:2011ex}
B.~J.~Schaefer and M.~Wagner,
Phys. Rev. D \textbf{85}, 034027 (2012)
[arXiv:1111.6871 [hep-ph]]





\bibitem{Iida:2000ha}
K.~Iida and G.~Baym,
Phys. Rev. D \textbf{63}, 074018 (2001)
[erratum: Phys. Rev. D \textbf{66}, 059903 (2002)]
[arXiv:hep-ph/0011229 [hep-ph]].

\bibitem{AR02}
M.~Alford and K.~Rajagopal,
JHEP \textbf{06}, 031 (2002)
[arXiv:hep-ph/0204001 [hep-ph]]

\bibitem{ARW98}
M.~G.~Alford, K.~Rajagopal and F.~Wilczek,
Phys. Lett. B \textbf{422}, 247-256 (1998)
[arXiv:hep-ph/9711395 [hep-ph]]

\bibitem{Alford:2007xm}
M.~G.~Alford, A.~Schmitt, K.~Rajagopal and T.~Sch\"afer,
Rev. Mod. Phys. \textbf{80}, 1455-1515 (2008)
[arXiv:0709.4635 [hep-ph]]

\bibitem{Casalbuoni:2003wh}
R.~Casalbuoni and G.~Nardulli,
Rev. Mod. Phys. \textbf{76}, 263-320 (2004)
[arXiv:hep-ph/0305069 [hep-ph]]

\bibitem{Buballa:2003qv}
M.~Buballa,
Phys. Rept. \textbf{407}, 205-376 (2005)
[arXiv:hep-ph/0402234 [hep-ph]]

\bibitem{Ru05}
S.~B.~Ruester, V.~Werth, M.~Buballa, I.~A.~Shovkovy and D.~H.~Rischke,
Phys. Rev. D \textbf{72}, 034004 (2005)
[arXiv:hep-ph/0503184 [hep-ph]]

\bibitem{Bla05}
D.~Blaschke, S.~Fredriksson, H.~Grigorian, A.~M.~\"Oztas and F.~Sandin,
Phys. Rev. D \textbf{72}, 065020 (2005)
[arXiv:hep-ph/0503194 [hep-ph]]

\bibitem{Abuki:2005ms}
H.~Abuki and T.~Kunihiro,
Nucl. Phys. A \textbf{768}, 118-159 (2006)
[arXiv:hep-ph/0509172 [hep-ph]]

\bibitem{Baldo:2002ju}
M.~Baldo, M.~Buballa, F.~Burgio, F.~Neumann, M.~Oertel and H.~J.~Schulze,
Phys. Lett. B \textbf{562}, 153-160 (2003)
[arXiv:nucl-th/0212096 [nucl-th]]

\bibitem{SH03}
I.~Shovkovy and M.~Huang,
Phys. Lett. B \textbf{564}, 205 (2003)
[arXiv:hep-ph/0302142 [hep-ph]]

\bibitem{Aguilera:2004ag}
D.~N.~Aguilera, D.~Blaschke and H.~Grigorian,
Nucl. Phys. A \textbf{757}, 527-542 (2005)
[arXiv:hep-ph/0412266 [hep-ph]]

\bibitem{BuSh05}
M.~Buballa and I.~A.~Shovkovy,
Phys. Rev. D \textbf{72}, 097501 (2005)
[arXiv:hep-ph/0508197 [hep-ph]]

\bibitem{GomezDumm:2005hy}
D.~Gomez Dumm, D.~B.~Blaschke, A.~G.~Grunfeld and N.~N.~Scoccola,
Phys. Rev. D \textbf{73}, 114019 (2006)
[arXiv:hep-ph/0512218 [hep-ph]]

\bibitem{Duhau:2004pq}
R.~S.~Duhau, A.~G.~Grunfeld and N.~N.~Scoccola,
Phys.\ Rev.\ D {\bf 70}, 074026 (2004)
[arXiv:hep-ph/0409072]

\bibitem{Glendenning:1992vb}
N.~K.~Glendenning,
Phys. Rev. D \textbf{46}, 1274-1287 (1992)

\bibitem{Neumann:2002jm}
F.~Neumann, M.~Buballa and M.~Oertel,
Nucl. Phys. A \textbf{714}, 481-501 (2003)
[arXiv:hep-ph/0210078 [hep-ph]]

\bibitem{Shovkovy:2003ce}
I.~Shovkovy, M.~Hanauske and M.~Huang,
Phys. Rev. D \textbf{67}, 103004 (2003)
[arXiv:hep-ph/0303027 [hep-ph]]

\bibitem{Reddy:2005}
S.~Reddy and G.~Rupak,
Phys. Rev. C \textbf{71}, 025201 (2005)
[arXiv:nucl-th/0405054 [nucl-th]]

\bibitem{Braun:2018svj}
J.~Braun, M.~Leonhardt and J.~M.~Pawlowski,
SciPost Phys. \textbf{6} (2019) no.5, 056
[arXiv:1806.04432 [hep-ph]]


\bibitem{Demorest:2010bx}
P.~Demorest, T.~Pennucci, S.~Ransom, M.~Roberts and J.~Hessels,
Nature \textbf{467}, 1081 (2010)
[arXiv:1010.5788 [astro-ph.HE]]

\bibitem{Antoniadis:2013pzd}
J.~Antoniadis, P.~C.~C.~Freire, N.~Wex, T.~M.~Tauris, R.~S.~Lynch, M.~H.~van Kerkwijk, M.~Kramer, C.~Bassa, V.~S.~Dhillon and T.~Driebe, \textit{et al.}
Science \textbf{340}, 6131 (2013)
[arXiv:1304.6875 [astro-ph.HE]]

\bibitem{Fonseca:2016tux}
E.~Fonseca, T.~T.~Pennucci, J.~A.~Ellis, I.~H.~Stairs, D.~J.~Nice, S.~M.~Ransom, P.~B.~Demorest, Z.~Arzoumanian, K.~Crowter and T.~Dolch, \textit{et al.}
Astrophys. J. \textbf{832},  167 (2016)
[arXiv:1603.00545 [astro-ph.HE]]

\bibitem{Arzoumanian:2017puf}
Z.~Arzoumanian \textit{et al.} [NANOGrav],
Astrophys. J. Suppl. \textbf{235},  37 (2018)
[arXiv:1801.01837 [astro-ph.HE]]

\bibitem{Linares:2018ppq}
M.~Linares, T.~Shahbaz and J.~Casares,
Astrophys. J. \textbf{859},  54 (2018)
[arXiv:1805.08799 [astro-ph.HE]]

\bibitem{Cromartie:2019kug}
H.~T.~Cromartie \textit{et al.} [NANOGrav],
Nature Astron. \textbf{4}, 72-76 (2019)
[arXiv:1904.06759 [astro-ph.HE]]

\bibitem{Klahn:2006ir}
T.~Klahn, D.~Blaschke, S.~Typel, E.~N.~E.~van Dalen, A.~Faessler,
C.~Fuchs, T.~Gaitanos, H.~Grigorian, A.~Ho and E.~E.~Kolomeitsev,
\textit{et al.}
Phys. Rev. C \textbf{74}, 035802 (2006)
[arXiv:nucl-th/0602038 [nucl-th]]

\bibitem{Grigorian:2003vi}
H.~Grigorian, D.~Blaschke and D.~N.~Aguilera,
Phys. Rev. C \textbf{69}, 065802 (2004)
[arXiv:astro-ph/0303518 [astro-ph]]

\bibitem{Blaschke:2007ri}
D.~B.~Blaschke, D.~Gomez Dumm, A.~G.~Grunfeld, T.~Klahn and
N.~N.~Scoccola,
Phys. Rev. C \textbf{75}, 065804 (2007)
[arXiv:nucl-th/0703088 [nucl-th]]

\bibitem{vanDalen:2004pn}
E.~N.~E.~van Dalen, C.~Fuchs and A.~Faessler,
Nucl. Phys. A \textbf{744}, 227-248 (2004)
[arXiv:nucl-th/0407070 [nucl-th]]

\bibitem{Orsaria:2012je}
M.~Orsaria, H.~Rodrigues, F.~Weber and G.~A.~Contrera,
Phys. Rev. D \textbf{87},  023001 (2013)
[arXiv:1212.4213 [astro-ph.SR]]

\bibitem{Orsaria:2013hna}
M.~Orsaria, H.~Rodrigues, F.~Weber and G.~A.~Contrera,
Phys. Rev. C \textbf{89}, 015806 (2014)
[arXiv:1308.1657 [nucl-th]]

\bibitem{Glendenning:1991es}
N.~K.~Glendenning and S.~A.~Moszkowski,
Phys. Rev. Lett. \textbf{67}, 2414-2417 (1991)

\bibitem{Lalazissis:1996rd}
G.~A.~Lalazissis, J.~Konig and P.~Ring,
Phys. Rev. C \textbf{55}, 540-543 (1997)
[arXiv:nucl-th/9607039 [nucl-th]]

\bibitem{deCarvalho:2015lpa}
S.~M.~de Carvalho, R.~Negreiros, M.~Orsaria, G.~A.~Contrera,
F.~Weber and W.~Spinella,
Phys. Rev. C \textbf{92},  035810 (2015)
[arXiv:1601.02938 [nucl-th]]

\bibitem{Spinella:2015ksa}
W.~M.~Spinella, F.~Weber, G.~A.~Contrera and M.~G.~Orsaria,
Eur. Phys. J. A \textbf{52},  61 (2016)
[arXiv:1507.06067 [nucl-th]]

\bibitem{Spinella:2018bdq}
W.~M.~Spinella, F.~Weber, M.~G.~Orsaria and G.~A.~Contrera,
Universe \textbf{4},  64 (2018)
[arXiv:1805.05772 [nucl-th]]

\bibitem{Mellinger:2017dve}
R.~D.~Mellinger, F.~Weber, W.~Spinella, G.~A.~Contrera and M.~G.~Orsaria,
Universe \textbf{3}, 5 (2017)
[arXiv:1701.07343 [astro-ph.HE]]

\bibitem{Ranea-Sandoval:2015ldr}
I.~F.~Ranea-Sandoval, S.~Han, M.~G.~Orsaria, G.~A.~Contrera,
F.~Weber and M.~G.~Alford,
Phys. Rev. C \textbf{93}, 045812 (2016)
[arXiv:1512.09183 [nucl-th]]

\bibitem{Alford:2001zr}
M.~G.~Alford, K.~Rajagopal, S.~Reddy and F.~Wilczek,
Phys. Rev. D \textbf{64}, 074017 (2001)
[arXiv:hep-ph/0105009 [hep-ph]]

\bibitem{Endo:2011em}
T.~Endo,
Phys. Rev. C \textbf{83}, 068801 (2011)
[arXiv:1105.2445 [astro-ph.SR]]

\bibitem{Lugones:2013ema}
G.~Lugones, A.~G.~Grunfeld and M.~Al Ajmi,
Phys. Rev. C \textbf{88}, no.4, 045803 (2013)
[arXiv:1308.1452 [hep-ph]]

\bibitem{TheLIGOScientific:2017qsa}
B.~P.~Abbott \textit{et al.} [LIGO Scientific and Virgo],
Phys. Rev. Lett. \textbf{119},  161101 (2017)
[arXiv:1710.05832 [gr-qc]]

\bibitem{Raithel:2018ncd}
C.~Raithel, F.~\"Ozel and D.~Psaltis,
Astrophys. J. Lett. \textbf{857}, L23 (2018)
[arXiv:1803.07687 [astro-ph.HE]]

\bibitem{Shibata:2019ctb}
M.~Shibata, E.~Zhou, K.~Kiuchi and S.~Fujibayashi,
Phys. Rev. D \textbf{100}, 023015 (2019)
[arXiv:1905.03656 [astro-ph.HE]]

\bibitem{Alvarez-Castillo:2018pve}
D.~E.~Alvarez-Castillo, D.~B.~Blaschke, A.~G.~Grunfeld and V.~P.~Pagura,
Phys. Rev. D \textbf{99},  063010 (2019)
[arXiv:1805.04105 [hep-ph]]

\bibitem{Ranea-Sandoval:2019miz}
I.~F.~Ranea-Sandoval, M.~G.~Orsaria, G.~Malfatti, D.~Curin, M.~Mariani, G.~A.~Contrera and O.~M.~Guilera,
Symmetry \textbf{11},  425 (2019)
[arXiv:1903.11974 [nucl-th]]

\bibitem{Orsaria:2019ftf}
M.~G.~Orsaria, G.~Malfatti, M.~Mariani, I.~F.~Ranea-Sandoval, F.~Garc\'\i{}a, W.~M.~Spinella, G.~A.~Contrera, G.~Lugones and F.~Weber,
J. Phys. G \textbf{46},  073002 (2019)
[arXiv:1907.04654 [astro-ph.HE]]

\bibitem{Malfatti:2019tpg}
G.~Malfatti, M.~G.~Orsaria, G.~A.~Contrera, F.~Weber and I.~F.~Ranea-Sandoval,
Phys. Rev. C \textbf{100},  015803 (2019)
[arXiv:1907.06597 [nucl-th]]

\bibitem{Shahrbaf:2019vtf}
M.~Shahrbaf, D.~Blaschke, A.~G.~Grunfeld and H.~R.~Moshfegh,
Phys. Rev. C \textbf{101}, 025807 (2020)
[arXiv:1908.04740 [nucl-th]]

\bibitem{Shahrbaf:2020uau}
M.~Shahrbaf, D.~Blaschke and S.~Khanmohamadi,
J. Phys. G \textbf{47}, no.11, 115201 (2020)
[arXiv:2004.14377 [nucl-th]]

\bibitem{Malfatti:2020onm}
G.~Malfatti, M.~G.~Orsaria, I.~F.~Ranea-Sandoval, G.~A.~Contrera and F.~Weber,
Phys. Rev. D \textbf{102}, 063008 (2020)
[arXiv:2008.06459 [astro-ph.HE]]

\bibitem{Typel:2009sy}
S.~Typel, G.~Ropke, T.~Klahn, D.~Blaschke and H.~H.~Wolter,
Phys. Rev. C \textbf{81}, 015803 (2010)
[arXiv:0908.2344 [nucl-th]]

\bibitem{Otto:2019zjy}
K.~Otto, M.~Oertel and B.~J.~Schaefer,
Phys. Rev. D \textbf{101}, no.10, 103021 (2020)
[arXiv:1910.11929 [hep-ph]]

\bibitem{Lonardoni:2014bwa}
D.~Lonardoni, A.~Lovato, S.~Gandolfi and F.~Pederiva,
Phys. Rev. Lett. \textbf{114}, no.9, 092301 (2015)
[arXiv:1407.4448 [nucl-th]]

\bibitem{Logoteta:2019utx}
D.~Logoteta, I.~Vidana and I.~Bombaci,
Eur. Phys. J. A \textbf{55}, no.11, 207 (2019)
[arXiv:1906.11722 [nucl-th]]


\bibitem{Buballa:2014tba}
M.~Buballa and S.~Carignano,
Prog. Part. Nucl. Phys. \textbf{81}, 39-96 (2015)
[arXiv:1406.1367 [hep-ph]]

\bibitem{Nickel:2009ke}
D.~Nickel,
Phys.\ Rev.\ Lett.\  {\bf 103}, 072301 (2009)
[arXiv:0902.1778 [hep-ph]]

\bibitem{Nickel:2009wj}
D.~Nickel,
Phys.\ Rev.\ D {\bf 80}, 074025 (2009)
[arXiv:0906.5295 [hep-ph]]

\bibitem{Carignano:2010ac}
S.~Carignano, D.~Nickel and M.~Buballa,
Phys. Rev. D \textbf{82}, 054009 (2010)
[arXiv:1007.1397 [hep-ph]]

\bibitem{Carignano:2014jla}
S.~Carignano, M.~Buballa and B.~J.~Schaefer,
Phys.\ Rev.\ D {\bf 90}, 014033 (2014)
[arXiv:1404.0057 [hep-ph]]

\bibitem{Braun:2015fva}
J.~Braun, F.~Karbstein, S.~Rechenberger and D.~Roscher,
Phys. Rev. D \textbf{93}, 014032 (2016)
[arXiv:1510.04012 [hep-ph]]

\bibitem{Carignano:2019ivp}
S.~Carignano and M.~Buballa,
Phys. Rev. D \textbf{101}, 014026 (2020)
[arXiv:1910.03604 [hep-ph]]

\bibitem{Buballa:2018hux}
M.~Buballa and S.~Carignano,
Phys. Lett. B \textbf{791}, 361-366 (2019)
[arXiv:1809.10066 [hep-ph]]

\bibitem{Fu:2019hdw}
W.~j.~Fu, J.~M.~Pawlowski and F.~Rennecke,
Phys. Rev. D \textbf{101}, 054032 (2020)
[arXiv:1909.02991 [hep-ph]]

\bibitem{Carlomagno:2014hoa}
J.~P.~Carlomagno, D.~Gomez Dumm and N.~N.~Scoccola,
Phys. Lett. B \textbf{745}, 1-4 (2015)
[arXiv:1411.0909 [hep-ph]]

\bibitem{Schon:2000qy}
V.~Schon and M.~Thies,
[arXiv:hep-th/0008175 [hep-th]]

\bibitem{Schon:2000he}
V.~Schon and M.~Thies,
Phys. Rev. D \textbf{62}, 096002 (2000)
[arXiv:hep-th/0003195 [hep-th]]

\bibitem{Thies:2006ti}
M.~Thies,
J. Phys. A \textbf{39}, 12707-12734 (2006)
[arXiv:hep-th/0601049 [hep-th]]

\bibitem{Basar:2009fg}
G.~Basar, G.~V.~Dunne and M.~Thies,
Phys. Rev. D \textbf{79}, 105012 (2009)
[arXiv:0903.1868 [hep-th]]

\bibitem{Bringoltz:2009ym}
B.~Bringoltz,
Phys. Rev. D \textbf{79}, 125006 (2009)
[arXiv:0901.4035 [hep-lat]].

\bibitem{Carlomagno:2015nsa}
J.~P.~Carlomagno, D.~Gomez Dumm and N.~N.~Scoccola,
Phys. Rev. D \textbf{92}, 056007 (2015)
[arXiv:1507.01560 [hep-ph]]

\bibitem{Abuki:2011pf}
H.~Abuki, D.~Ishibashi and K.~Suzuki,
Phys.\ Rev.\ D {\bf 85}, 074002 (2012)
[arXiv:1109.1615 [hep-ph]]

\bibitem{Iwata:2012bs}
Y.~Iwata, H.~Abuki and K.~Suzuki,
[arXiv:1206.2870 [hep-ph]]

\bibitem{Nakano:2004cd}
E.~Nakano and T.~Tatsumi,
Phys. Rev. D \textbf{71}, 114006 (2005)
[arXiv:hep-ph/0411350 [hep-ph]]

\bibitem{Muller:2013tya}
D.~M\"uller, M.~Buballa and J.~Wambach,
Phys. Lett. B \textbf{727}, 240-243 (2013)
[arXiv:1308.4303 [hep-ph]]

\bibitem{Carignano:2011gr}
S.~Carignano and M.~Buballa,
Acta Phys.\ Polon.\ Supp.\  {\bf 5}, 641 (2012)
[arXiv:1111.4400 [hep-ph]]

\bibitem{Andersen:2014xxa}
J.~O.~Andersen, W.~R.~Naylor and A.~Tranberg,
Rev.\ Mod.\ Phys.\  {\bf 88}, 025001 (2016)
[arXiv:1411.7176 [hep-ph]]

\bibitem{Miransky:2015ava}
V.~A.~Miransky and I.~A.~Shovkovy,
Phys. Rept. \textbf{576}, 1-209 (2015)
[arXiv:1503.00732 [hep-ph]]

\bibitem{Pagura:2016pwr}
V.~P.~Pagura, D.~Gomez Dumm, S.~Noguera and N.~N.~Scoccola,
Phys. Rev. D \textbf{95}, 034013 (2017)
[arXiv:1609.02025 [hep-ph]]

\bibitem{GomezDumm:2017iex}
D.~Gomez Dumm, M.~F.~Izzo Villafa\~ne, S.~Noguera, V.~P.~Pagura and
N.~N.~Scoccola,
Phys. Rev. D \textbf{96}, 114012 (2017)
[arXiv:1709.04742 [hep-ph]]

\bibitem{GomezDumm:2017jij}
D.~Gomez Dumm, M.~F.~Izzo Villafa\~ne and N.~N.~Scoccola,
Phys. Rev. D \textbf{97},  034025 (2018)
[arXiv:1710.08950 [hep-ph]]

\bibitem{Dumm:2020muy}
D.~Gomez Dumm, M.~F.~Izzo Villafa\~ne and N.~N.~Scoccola,
Phys. Rev. D \textbf{101},  116018 (2020)
[arXiv:2004.10052 [hep-ph]]

\bibitem{Bali:2011qj}
G.~S.~Bali, F.~Bruckmann, G.~Endr\H odi, Z.~Fodor, S.~D.~Katz, S.~Krieg, A.~Sch\"afer and K.~K.~Szabo,
JHEP {\bf 1202}, 044 (2012)
[arXiv:1111.4956 [hep-lat]]

\bibitem{Bali:2012zg}
G.~S.~Bali, F.~Bruckmann, G.~Endr\H odi, Z.~Fodor, S.~D.~Katz and A.~Sch\"afer,
Phys.\ Rev.\ D {\bf 86}, 071502 (2012)
[arXiv:1206.4205 [hep-lat]]

\bibitem{Ritus:1978cj}
V.~I.~Ritus,
Sov.\ Phys.\ JETP {\bf 48}, 788 (1978)

\bibitem{Fraga:2013ova}
E.~S.~Fraga, B.~W.~Mintz and J.~Schaffner-Bielich,
Phys. Lett. B \textbf{731}, 154-158 (2014)
[arXiv:1311.3964 [hep-ph]]

\bibitem{Ayala:2014iba}
A.~Ayala, M.~Loewe, A.~J.~Mizher and R.~Zamora,
Phys.\ Rev.\ D {\bf 90}, 036001 (2014)
[arXiv:1406.3885 [hep-ph]]

\bibitem{Farias:2014eca}
R.~L.~S.~Farias, K.~P.~Gomes, G.~I.~Krein and M.~B.~Pinto,
Phys.\ Rev.\ C {\bf 90},  025203 (2014)
[arXiv:1404.3931 [hep-ph]]

\bibitem{Fayazbakhsh:2014mca}
S.~Fayazbakhsh and N.~Sadooghi,
Phys. Rev. D \textbf{90}, no.10, 105030 (2014)
[arXiv:1408.5457 [hep-ph]]

\bibitem{Fayazbakhsh:2013cha}
S.~Fayazbakhsh and N.~Sadooghi,
Phys.\ Rev.\ D {\bf 88}, 065030 (2013)
[arXiv:1306.2098 [hep-ph]]

\bibitem{Bali:2017ian}
G.~S.~Bali, B.~B.~Brandt, G.~Endr\H odi and B.~Gl\"assle,
Phys.\ Rev.\ D {\bf 97}, 034505 (2018)
[arXiv:1707.05600 [hep-lat]]

\bibitem{Coppola:2018ygv}
M.~Coppola, D.~Gomez Dumm, S.~Noguera and N.~N.~Scoccola,
Phys.\ Rev.\ D {\bf 99}, 054031 (2019)
[arXiv:1810.08110 [hep-ph]]

\bibitem{Coppola:2019uyr}
M.~Coppola, D.~Gomez Dumm, S.~Noguera and N.~N.~Scoccola,
Phys.\ Rev.\ D {\bf 100}, 054014 (2019)
[arXiv:1907.05840 [hep-ph]]

\end{thebibliography}
